\documentclass[12pt,oneside]{report}

\usepackage[a4paper, top=3cm, bottom=2cm, inner=3cm, outer=2cm]{geometry}

\usepackage{setspace}
\usepackage{amsmath,amssymb,amsfonts,mathtools}
\usepackage{bm}
\usepackage{physics}
\usepackage{slashed}

\usepackage{graphicx}
\usepackage{caption}
\usepackage{subcaption}
\usepackage{float}
\usepackage{lastpage}
\usepackage{pdfpages}

\usepackage[numbers,sort&compress]{natbib}

\usepackage{hyperref}
\hypersetup{
  colorlinks=true,
  linkcolor=blue,
  citecolor=blue,
  urlcolor=blue
}

\usepackage{fancyhdr}
\makeatletter
\def\cleardoublepage{%
  \clearpage
  \if@twoside
    \ifodd\c@page
    \else
      \hbox{}%
      \thispagestyle{empty}%
      \newpage
    \fi
  \fi
}
\makeatother

\renewcommand{\chaptermark}[1]{%
  \markboth{Chapter \thechapter.\quad #1}{}%
}

\AtBeginDocument{%
  \fancypagestyle{fancy}{%
    \fancyhf{}%
    \fancyhead[L]{\itshape\nouppercase{\leftmark}}%
    \fancyhead[R]{\thepage}%
}%
  \fancypagestyle{plain}{%
    \fancyhf{}%
    \fancyhead[R]{\thepage}%
  }%
  \fancypagestyle{frontmatter}{%
    \fancyhf{}%
    \fancyhead[R]{\thepage}%
  }%
}

\newenvironment{changemargin}[2]{%
\begin{list}{}{%
\setlength{\leftmargin}{#1}%
\setlength{\rightmargin}{#2}%
}%
\item[]}{\end{list}}

\newcommand{\indep}{\perp \!\!\! \perp}

\newcommand{\appchapter}[2]{%
  \refstepcounter{chapter}%
  \chapter*{Appendix \thechapter\\#1}%
  \addcontentsline{toc}{chapter}{Appendix \thechapter\quad #1}%
  \label{#2}%
  \markboth{Appendix \thechapter.\quad #1}{}%
  \setcounter{section}{0}%
  \setcounter{equation}{0}%
  \renewcommand{\thesection}{\thechapter.\arabic{section}}%
}

\begin{document}

\pagestyle{empty}

\begin{changemargin}{0cm}{0cm}
\thispagestyle{empty}
\baselineskip25pt
\begin{center}
{\Large {\bf RELATIVISTIC MAGNETOHYDRODYNAMICS FROM KINETIC THEORY}}\\
\end{center}

\vfill
\baselineskip15pt
\begin{center}
{\em A thesis Submitted} \\
in Partial Fulfilment of the Requirements \\
for the Degree of \
\vskip .30\baselineskip
{\large{\bf PhD}}
\end{center}
\baselineskip25pt

\vfill
\begin{center} {\bf {\em by}} \\
{\large{\bf KHWAHISH KUSHWAH}} \\
\end{center}

\vfill
\begin{center}
\begin{figure}[h!]
\centering
\includegraphics[scale=0.1]{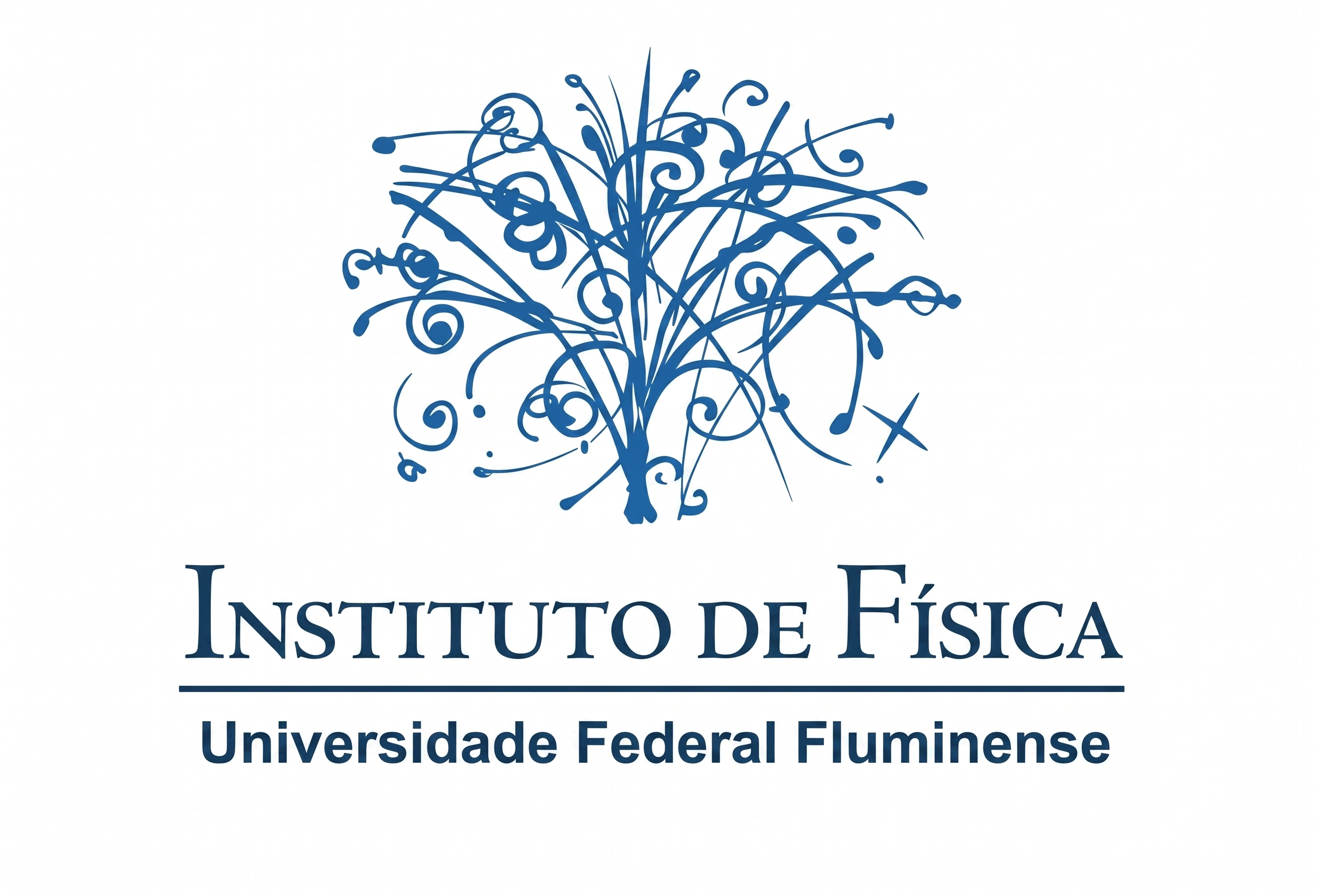}
\end{figure}
 {\bf {\em to the }} \\
{\bf {\large Institute of Physics}} \\
{\bf {\large Universidade Federal Fluminense}} \\
{\bf Niter\'oi} \\
{\bf 28/May/2026} 
\end{center}
\end{changemargin}

\cleardoublepage

\includepdf[pages=1]{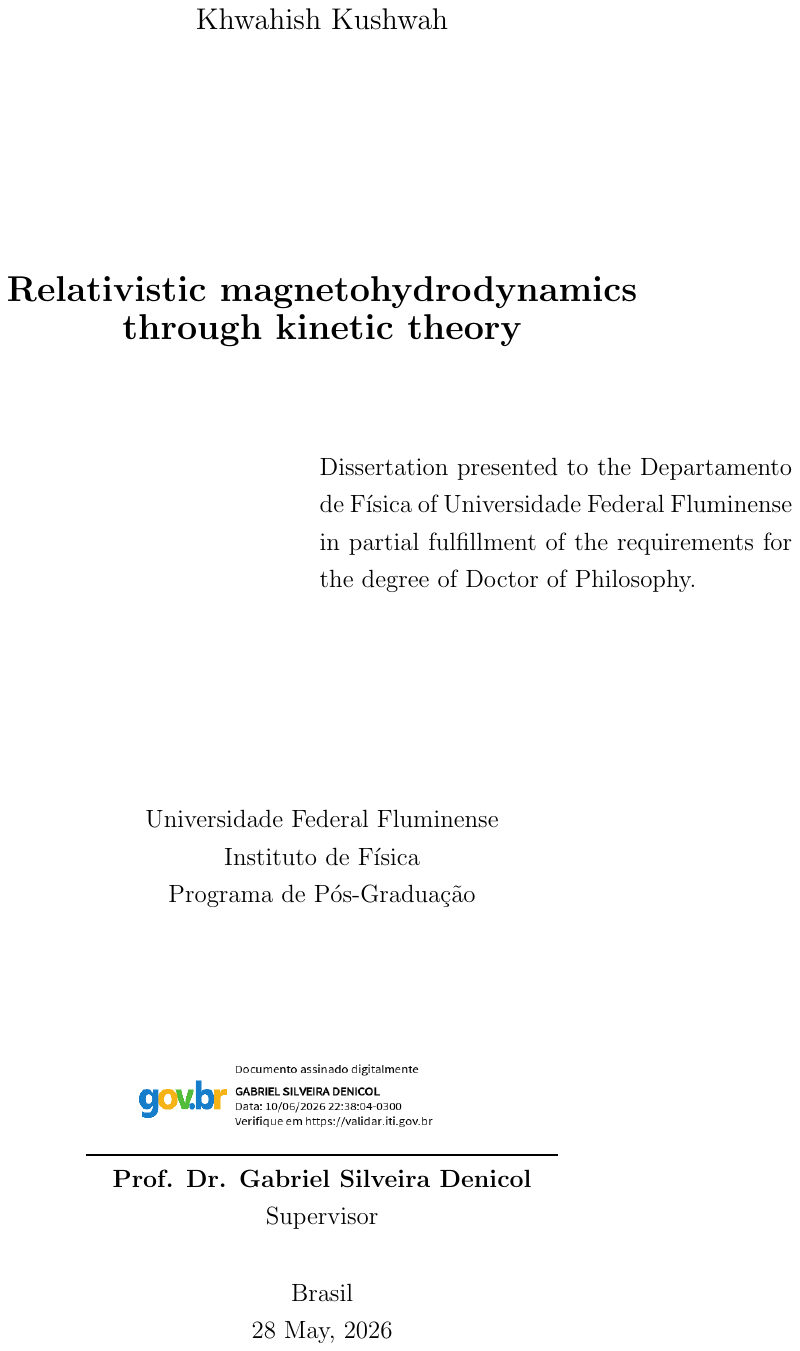}

\includepdf[pages=1]{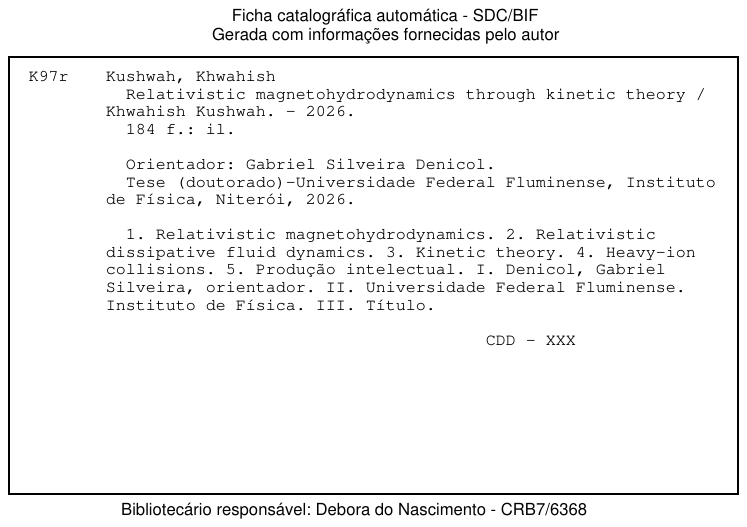}
\cleardoublepage

\includepdf[pages=1]{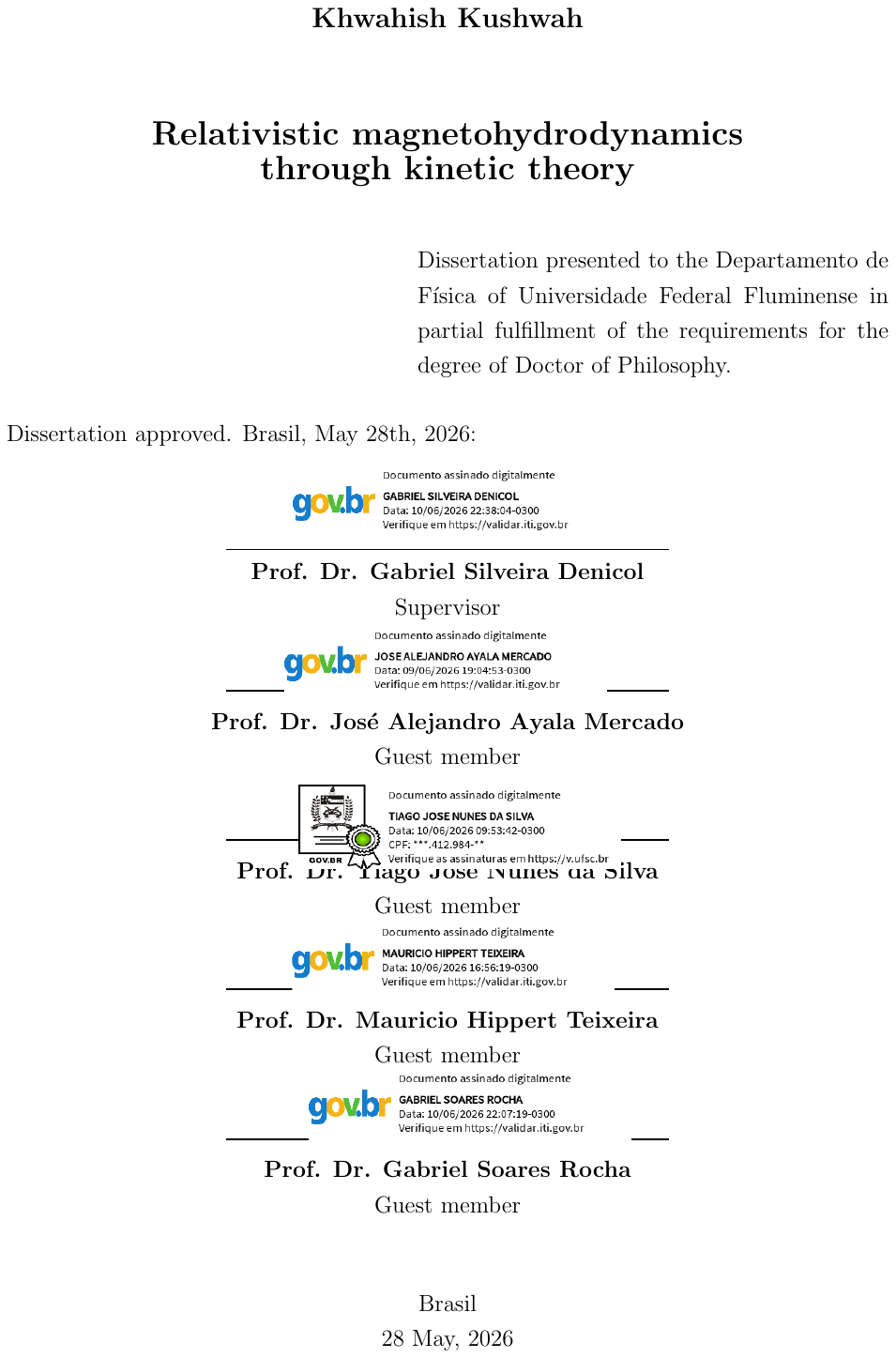}
\cleardoublepage

\pagenumbering{roman}
\pagestyle{frontmatter}

\vspace*{0.65\textheight}

\begin{flushright}
\begin{minipage}{0.65\textwidth}
\itshape
``If ever there is a tomorrow when we are not together, 
always remember that you are braver than you believe, 
stronger than you seem, and smarter than you think.''

\vspace{0.5cm}

\begin{flushright}
--- Christopher Robin, in A. A. Milne's \textit{Winnie-the-Pooh}
\end{flushright}
\end{minipage}
\end{flushright}
\cleardoublepage


\begin{center}
{\bf ACKNOWLEDGEMENTS}
\end{center}

I began this PhD expecting that by the end of it, I would have a deep and confident command of the field -- that expertise would mean having the answers. What I found instead was humbling and, I think, more beautiful: the further I went, the larger the horizon became. I know more now than when I started, and yet I have never felt more aware of how much remains unexplored. That sense of standing at the edge of something vast and being \emph{excited} rather than afraid is perhaps the most important thing this PhD gave me.

None of it would have been the same without my supervisor, Prof. Gabriel S. Denicol. He gave me the freedom to explore while ensuring I never got lost. Under his mentorship, I learned not only physics but also something far more difficult to teach: how to think critically, how to be patient with a problem, and how to persevere. His mentorship extended beyond physics and research. The life lessons I took from our time working together have shaped my development as a researcher, but even more importantly, as a person. For his patience, trust and support, I will always be sincerely grateful to him.

I have also been fortunate to be surrounded by exceptional colleagues during my time at Universidade Federal Fluminense. They showed me support that I can never forget -- from taking me to the hospital when I first arrived in Brazil and fell seriously ill, to simply always checking in to make sure I was doing okay. These were not just small gestures, but the kind of care that stays with you. Others gave me something equally irreplaceable: a warmth and closeness that quickly grew into genuine friendship. They looked after me in ways that would not have even occurred to me to ask for, shared their own lives with me, and trusted me to be there for them too. The friends I made beyond the university walls, through our shared world of academia, added their own colour to those years. All the good memories I have of Brazil can, without exception, be traced back to one or several of these people. It is no exaggeration to say that my experience of Brazil was shaped entirely by the friendships I found here. Their support and camaraderie meant more than they perhaps know.

Moving to Brazil to pursue my PhD was one of the most challenging experiences of my life. It tested me in ways I had not anticipated and demanded a version of myself I did not know I was capable of becoming. It also shaped me quietly, stubbornly, and irreversibly. I am a different person for having lived through it, and I believe a better one. Looking back, I am proud of the person I chose to become through this journey. Throughout it all, my friends scattered across the globe kept me sane. I am especially grateful to my friends from my undergraduate days, who remained a constant presence despite the distance. There was not a single happy moment I did not want to share with them. They were always there to listen, to share in the frustrations and triumphs of this journey. That kind of steadiness is something you don't forget. Sometimes, simply having them on the other end of a call was enough to make ordinary days more enjoyable and difficult days easier to endure.

My family has been my foundation throughout all of this. Their love and care gave me a sense of security that never depended on success or failure. They supported me even when they did not fully understand the challenges I was facing, trusted me to make my own decisions, respected the choices I made and gave me the space to find my own way. Knowing that I \emph{always} have them and a home to return to gave me the courage to take risks, make difficult decisions, and follow paths whose outcomes were far from certain. I would not be where I am today without their unconditional love and support.

This journey began with physics, but it became much more than that. Research taught me that the most honest answer to any question is often that we do not understand it completely yet, and that understanding is never a destination but a process. Along the way, I learned that every answer uncovers new questions, and that the feeling of not knowing never truly disappears. If anything, research teaches you to become comfortable with it. This thesis is one contribution to that ongoing process of understanding. It reflects years of hard work, resilience and strength, and is shaped by the people who shared parts of this journey with me. To \emph{everyone} who helped me reach this point, thank you.


\cleardoublepage

\begin{center}
{\large {\bf RESUMO}}
\end{center}

Esta tese desenvolve uma teoria de magnetohidrodinâmica dissipativa relativística na presença de campos eletromagnéticos intensos a partir de teoria cinética, motivado principalmente pelo plasma de quarks e glúons produzido em colisões de íons pesados. Partindo da equação relativística de Boltzmann–Vlasov e empregando o método de momentos, são derivadas equações hidrodinâmicas causais de segunda ordem para plasmas relativísticos com níveis crescentes de generalidade física.

Inicialmente, o trabalho revisita os fundamentos da hidrodinâmica dissipativa relativística e sua formulação microscópica em teoria cinética, destacando a necessidade de teorias transientes do tipo Israel–Stewart para garantir causalidade e estabilidade. Em seguida, os campos eletromagnéticos são incorporados à descrição microscópica, mostrando que a força de Lorentz modifica a estrutura da hierarquia de momentos e introduz fenômenos de transporte anisotrópicos ausentes em fluidos sem campos externos. A tese desenvolve então a magnetohidrodinâmica dissipativa relativística para um plasma não resistivo de dois componentes composto por partículas com cargas opostas. Nesse contexto, o campo magnético induz novos acoplamentos entre os setores dissipativos das duas espécies, levando ao surgimento de correntes dissipativas relativas e de dinâmicas acopladas do tensor de cisalhamento. A aplicação do formalismo a sistemas em expansão de Bjorken revela um comportamento oscilatório amortecido no setor transversal do cisalhamento, associado ao movimento ciclotrônico das partículas carregadas.

Por fim, é formulado o caso mais geral considerado neste trabalho: um plasma relativístico resistivo de dois componentes, no qual o campo elétrico evolui dinamicamente e se acopla diretamente tanto à corrente de difusão de carga quanto ao tensor de cisalhamento. A teoria resultante prevê mecanismos não triviais de retroalimentação entre corrente e cisalhamento, geração transitória de anisotropia de momento induzida eletromagneticamente e oscilações dissipativas subamortecidas em sistemas homogêneos. Aplicações aos casos homogêneo e em expansão de Bjorken demonstram como efeitos resistivos e eletromagnéticos modificam a evolução de plasmas relativísticos para além do quadro hidrodinâmico convencional.

O arcabouço desenvolvido nesta tese fornece uma extensão sistemática da hidrodinâmica dissipativa relativística para plasmas magnetizados e resistivos, estabelecendo uma fundamentação microscópica para futuros estudos fenomenológicos do plasma de quarks e glúons fortemente magnetizado, bem como para aplicações em sistemas astrofísicos relacionados.

\vspace{1cm}

\noindent
\textbf{Palavras-chave}: Hidrodinâmica dissipativa relativística, magnetohidrodinâmica relativística, teoria cinética, equação de Boltzmann--Vlasov, aproximação de 14 momentos, plasma de dois componentes, magnetohidrodinâmica resistiva.

\cleardoublepage

\begin{center}
{\large {\bf ABSTRACT}}
\end{center}  

This thesis develops a kinetic-theory framework for relativistic dissipative magnetohydrodynamics in the presence of strong electromagnetic fields, motivated primarily by the quark–gluon plasma produced in relativistic heavy-ion collisions. Starting from the relativistic Boltzmann–Vlasov equation and employing the method of moments formalism within the 14-moment approximation, causal second-order hydrodynamic equations are derived for relativistic plasmas with increasing levels of physical generality. 

The work first revisits relativistic dissipative hydrodynamics and its kinetic-theory foundations, emphasizing the necessity of Israel–Stewart-type transient theories to ensure causality and stability. Electromagnetic fields are then incorporated into the microscopic description, showing that the Lorentz force modifies the structure of the moment hierarchy and generates anisotropic transport phenomena absent in field-free fluids. The thesis then develops relativistic dissipative magnetohydrodynamics for a non-resistive two-component plasma composed of oppositely charged particles. In this framework, the magnetic field induces new couplings between the dissipative sectors of the two species, leading to the emergence of relative dissipative currents and coupled shear dynamics. Application to Bjorken-expanding systems reveals damped oscillatory behavior in the transverse shear sector associated with the cyclotron motion of the charged components.

Finally, the most general case considered in this work is developed: a resistive two-component relativistic plasma in which the electric field evolves dynamically and couples directly to both the charge diffusion current and the shear-stress tensor. The resulting theory predicts nontrivial current–shear feedback, transient electromagnetic generation of momentum anisotropy, and underdamped dissipative oscillations in homogeneous systems. Applications to homogeneous and Bjorken-expanding configurations demonstrate how resistive and electromagnetic effects modify the evolution of relativistic plasmas beyond the standard hydrodynamic picture.

The framework derived in this thesis provides a systematic extension of relativistic dissipative hydrodynamics to magnetized and resistive plasmas and establishes a microscopic foundation for future phenomenological studies of strongly magnetized quark–gluon plasma and related astrophysical systems.

\vspace{1cm}

\noindent 
\textbf{Keywords}: Relativistic dissipative hydrodynamics, relativistic magnetohydrodynamics, kinetic theory, Boltzmann--Vlasov equation, 14-moment approximation, two-component plasma, resistive magnetohydrodynamics.
\cleardoublepage

\tableofcontents
\cleardoublepage

\listoffigures
\cleardoublepage

\pagenumbering{arabic}
\pagestyle{fancy}

\chapter{Introduction}
\label{ch:introduction}

One of the most profound questions in physics is: what is matter made of, and how does it behave under extreme conditions? For most of human history, the answer stopped at atoms. Then, in the early twentieth century, it became clear that atoms have nuclei, and those nuclei contain protons and neutrons. For a while, it seemed that the story ended there. But during the 1950s and 1960s, experiments at particle accelerators began producing hundreds of new particles~\cite{yagi2005quark, Griffiths2008, Perkins2000, BrownDresdenHoddeson1989}: pions, kaons, delta resonances, lambda baryons, and many more, all interacting through the strong nuclear force. The sheer number of them made it impossible to believe they were all elementary. Something smaller had to be inside.

That suspicion was confirmed in 1968, when deep inelastic scattering experiments at the Stanford Linear Accelerator Center (SLAC) revealed point-like constituents inside the proton~\cite{PhysRevLett.23.935}. These were eventually identified as the quarks that Gell-Mann~\cite{Gell-Mann:1964ewy} and Zweig~\cite{Zweig:1964jf} had proposed a few years earlier. Today we know that there are six flavours of quarks: up, down, strange, charm, bottom, and top, and that they are bound together by a force mediated by massless particles called gluons. The theory describing this interaction is Quantum Chromodynamics (QCD)~\cite{1973PhLB...47..365F, PhysRevLett.31.494}, a non-abelian gauge field theory based on the symmetry group $SU(3)$.

QCD has two remarkable properties that distinguish it sharply from the more familiar theory of electromagnetism. The first is \emph{confinement}~\cite{Rischke:2003mt}: at low temperatures and densities, quarks and gluons are permanently trapped inside colour-neutral bound states called hadrons and can never be observed in isolation. The second is \emph{asymptotic freedom}~\cite{Gross:1973id,Politzer:1973fx}: as the energy scale increases, the coupling between quarks becomes progressively weaker. This means that at sufficiently high energies or temperatures, quarks and gluons stop interacting with each other strongly and are freed from their hadronic prisons.

\begin{figure}
    \centering
    \includegraphics[width=0.8\linewidth]{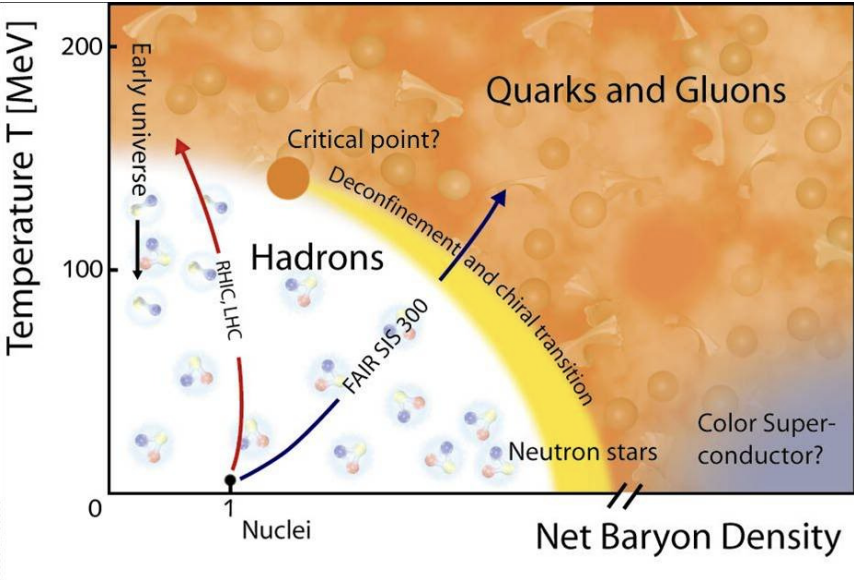}
    \caption[QCD phase diagram]{Schematic phase diagram of QCD matter in the plane of temperature and net baryon density. At low temperature and density, quarks and gluons are confined inside hadrons, while at sufficiently high temperature or density the system enters the deconfined quark-gluon plasma phase. The diagram also indicates the regions probed by heavy-ion collision experiments and the possible location of the critical point. Adapted from~\cite{yagi2005quark}.}
    \label{fig:qcd_phase_diagram}
\end{figure}

These two properties together imply that matter must undergo a phase transition. At low temperatures, nuclear matter exists as a gas of hadrons. But if the temperature is raised above a critical value of roughly $T_c \sim 155\,\mathrm{MeV}$~\cite{Borsanyi:2010bp}, the hadrons dissolve and their constituents (quarks and gluons) are liberated into a new state of matter known as the \emph{Quark-Gluon Plasma} (QGP)~\cite{Shuryak:1980tp,Collins:1974ky}.This transition from confined to deconfined matter is one of the central subjects of modern high-energy nuclear physics, and is illustrated schematically in Fig.~\ref{fig:qcd_phase_diagram}.

\section{The quark-gluon plasma and where to find it}

The QGP is not merely a theoretical curiosity. According to our best understanding of cosmology, the entire observable universe existed in the QGP phase for the first few microseconds after the Big Bang, before the temperature dropped below $T_c$ and hadrons condensed out~\cite{yagi2005quark}. It may also exist today in the interiors of the densest neutron stars, where the pressure is so extreme that nucleons may dissolve into their quark constituents~\cite{Annala:2019puf}. Understanding the QGP therefore has implications not only for laboratory nuclear physics but also for the history of the universe and the astrophysics of compact objects.

\begin{figure}
    \centering
    \includegraphics[width=0.7\linewidth]{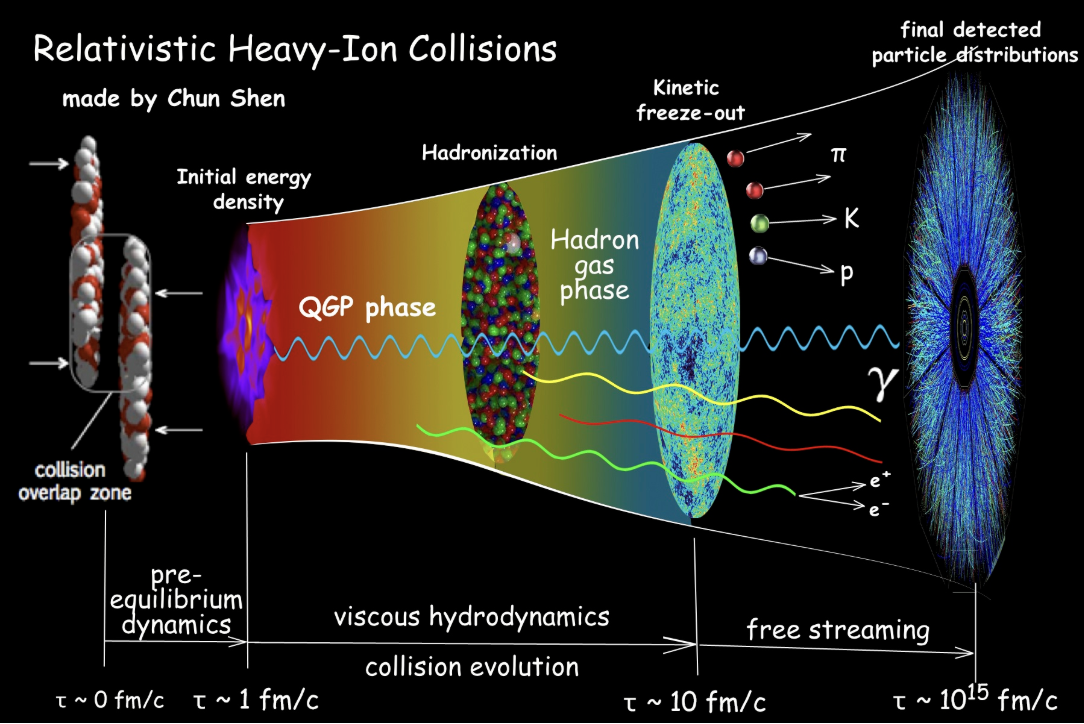}
    \caption[Space-time evolution of a heavy-ion collision]{Schematic space-time evolution of an ultrarelativistic heavy-ion collision. After the initial impact, the produced matter passes through a pre-equilibrium stage, forms a short-lived quark-gluon plasma, expands hydrodynamically, hadronizes, and finally freezes out into the particles measured by detectors.}
    \label{fig:heavy_ion_evolution}
\end{figure}

On Earth, the QGP can be recreated in the laboratory by colliding heavy atomic nuclei at ultrarelativistic energies. When two gold or lead nuclei travelling at nearly the speed of light collide head-on, the kinetic energy deposited in the collision zone heats a small volume of nuclear matter to temperatures several times $T_c$. For a very brief moment $\sim$ of order $10\,\mathrm{fm}/c$, roughly $10^{-23}$ seconds, that tiny fireball contains deconfined quarks and gluons. Then the fireball expands, cools, and the quarks recombine into the hadrons that detectors ultimately measure; the full space-time evolution is depicted schematically in Fig.~\ref{fig:heavy_ion_evolution}.

This experimental programme has been pursued at the Relativistic Heavy Ion Collider (RHIC) at Brookhaven National Laboratory, where Au+Au collisions at $\sqrt{s_{NN}} = 200\,\mathrm{GeV}$ began in 2000, and subsequently at the Large Hadron Collider (LHC) at CERN, where Pb+Pb collisions at $\sqrt{s_{NN}} = 2.76\,\mathrm{TeV}$ began in 2010~\cite{ALICE:2010spa}. The combined results of the four RHIC experiments such as BRAHMS, PHENIX, PHOBOS, and STAR~\cite{BRAHMS:2004adc,PHENIX:2004vcz,PHOBOS:2004zne,STAR:2005gfr}, established beyond reasonable doubt that a new state of matter had been created. What was entirely unexpected, however, was its character.

\section{The near-perfect fluid}

Before the RHIC experiments, the prevailing expectation was that the QGP, like any other gas at very high temperature, would be weakly coupled and would behave as a nearly ideal gas of quarks and gluons. The measurements, however, told a completely different story. The key observable was the \emph{elliptic flow coefficient} $v_2$~\cite{Voloshin:1994mz,Poskanzer:1998yz}, which measures the azimuthal asymmetry in the momentum distribution of the produced hadrons.

\begin{figure}
    \centering
    \includegraphics[width=0.6\linewidth]{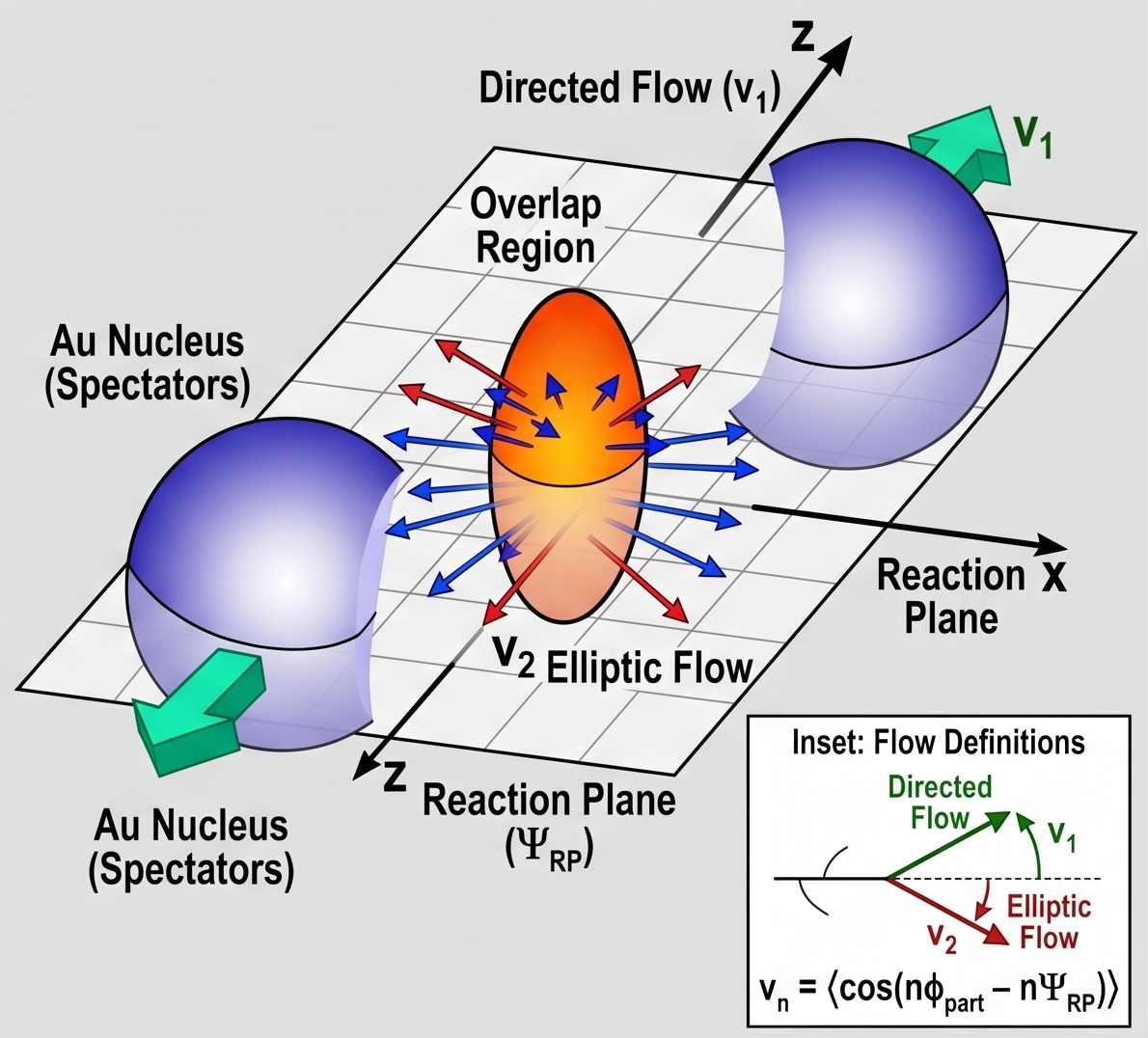}
    \caption[Geometric origin of elliptic flow]{Geometric origin of anisotropic flow in a non-central heavy-ion collision. The almond-shaped overlap region generates anisotropic pressure gradients, which convert the initial spatial asymmetry into momentum-space anisotropy of the final particles. The elliptic flow coefficient $v_2$ quantifies the dominant second-harmonic component of this azimuthal anisotropy. }
    \label{fig:elliptic_flow_geometry}
\end{figure}

\begin{figure}
    \centering
    \includegraphics[width=0.7\linewidth]{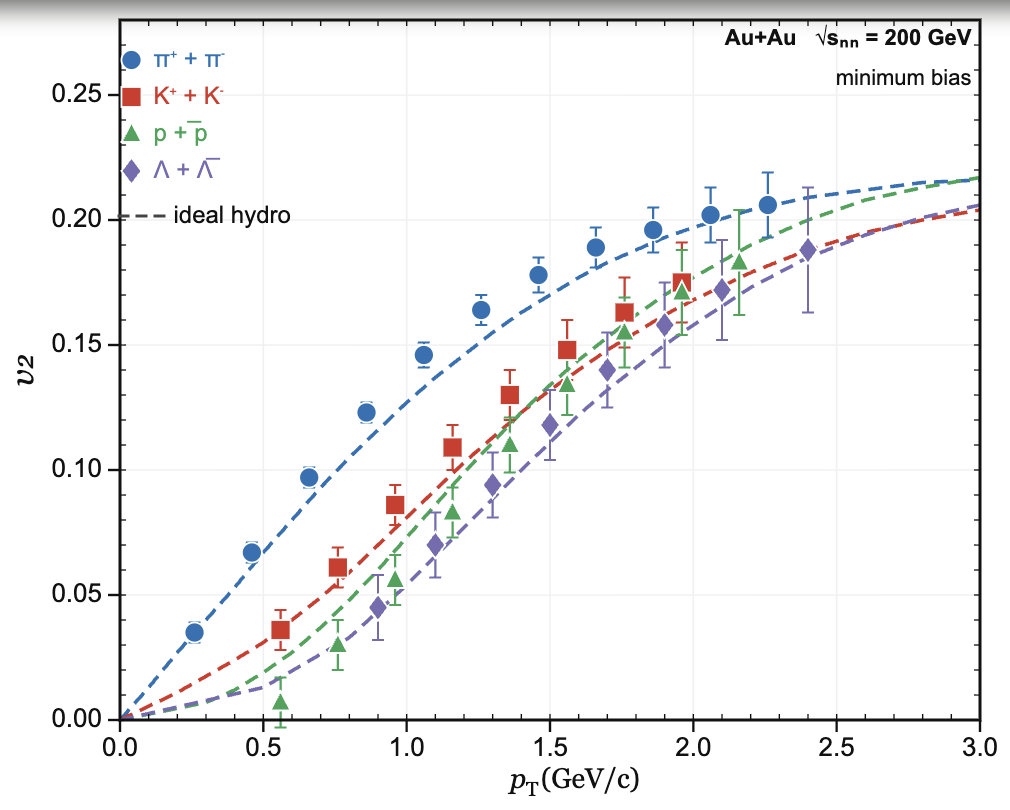}
    \caption[Elliptic flow coefficient at RHIC]{Elliptic flow coefficient $v_2$ as a function of transverse momentum $p_T$ for identified hadrons in Au+Au collisions at $\sqrt{s_{NN}}=200\,\mathrm{GeV}$. The sizeable measured values of $v_2$ and their comparison with ideal hydrodynamic calculations provide evidence for strong collective behaviour and very small shear viscosity in the quark-gluon plasma. Data from Ref.~\cite{STAR:2005gfr}.}
    \label{fig:v2_star}
\end{figure}

In a non-central collision, the two nuclei overlap in an almond-shaped region in the transverse plane, as illustrated in Fig.~\ref{fig:elliptic_flow_geometry}. If the created matter interacts strongly enough, this spatial asymmetry is converted by pressure gradients into a momentum asymmetry, i.e., more particles emerge along the short axis of the almond than along the long axis. The magnitude of $v_2$ is therefore a direct measure of how strongly the matter interacts and how efficiently it thermalises~\cite{Ollitrault:1992bk}. A weakly coupled gas would produce almost no $v_2$. What RHIC measured was a large value of $v_2$ (see Fig.~\ref{fig:v2_star}) and the agreement with relativistic ideal fluid dynamics was striking~\cite{Kolb:2003dz}: the QGP was not a gas at all! It was behaving as an almost perfect liquid, with negligible internal friction.

The theoretical backdrop for this conclusion was provided by a remarkable result from the AdS/CFT correspondence~\cite{Maldacena:1997re}. In 2005, Kovtun, Son, and Starinets~\cite{Kovtun:2004de} used string theory methods to compute the shear viscosity of a strongly coupled quantum field theory and conjectured that the ratio of shear viscosity to entropy density satisfies a universal lower bound,
\begin{equation}
\frac{\eta}{s} \geq \frac{1}{4\pi}.
\end{equation}
This KSS bound corresponds to the most perfect fluid theoretically conceivable. Remarkably, detailed viscous hydrodynamic analyses of RHIC data~\cite{Romatschke:2007mq,Luzum:2008cw} showed that the QGP has $\eta/s$ within a factor of two to three of this bound -- far smaller than any other known substance, including liquid helium. Which means that the QGP is the most perfect fluid ever observed. This discovery transformed the field: the primary mission of heavy-ion physics became extracting the precise value of $\eta/s$ and characterising the transport properties of the QGP.

\section{Relativistic hydrodynamics as the language of the QGP}

The fact that the QGP behaves as a fluid means that relativistic hydrodynamics is the correct effective framework for describing its space-time evolution between thermalization ($\tau_0 \sim 0.5\,\mathrm{fm}/c$ after the collision) and freeze-out (when the matter becomes too dilute to interact). The inputs to a hydrodynamic simulation are the initial energy density profile, the equation of state from lattice QCD, and the transport coefficients. And the outputs are the final spectra and anisotropic flow of produced hadrons which can be directly compared to experiment.

The simplest version of relativistic hydrodynamics is \emph{ideal} fluid dynamics, which assumes that the system is always in perfect local thermodynamic equilibrium and carries no viscosity, heat conduction, or diffusion. Ideal hydrodynamics was successfully applied to heavy-ion collisions by Kolb and Heinz~\cite{Kolb:2003dz} and others, giving a reasonable first description of the data. However, it was clear from the beginning that dissipative corrections matter: the system is small, the gradients are large, and the mean free path, while short, is not zero.

\emph{Dissipative} relativistic fluid dynamics incorporates these effects through the bulk viscous pressure $\Pi$, the heat flux or particle diffusion current $V^\mu$, and the shear stress tensor $\pi^{\mu\nu}$. The most natural approach is a relativistic generalisation of the familiar Navier--Stokes equations, in which these dissipative quantities are taken to be proportional to the local gradients of the flow. This was attempted independently by Eckart~\cite{Eckart:1940te} in 1940 and by Landau and Lifshitz~\cite{Landau:1959} in 1959, yielding what are now called the first-order relativistic dissipative theories.

\section{The acausality crisis and its resolution}

Unfortunately, the Eckart and Landau--Lifshitz theories suffer from a fundamental and fatal problem. Since the dissipative currents are determined algebraically by the instantaneous local gradients, any perturbation anywhere in the fluid has an \emph{immediate} effect everywhere. The governing equations are \emph{parabolic} (i.e., like heat equation, they have no finite maximum signal speed and disturbances spread instantaneously), and a simple Fourier analysis reveals that perturbations can propagate with arbitrarily large speeds. This violates special relativity. Worse still, as Hiscock and Lindblom demonstrated in a pair of landmark papers in 1983 and 1985~\cite{Hiscock:1983zz,Hiscock:1985zz}, these first-order theories are not merely acausal but also \emph{linearly unstable}: small perturbations around any moving equilibrium state grow exponentially without bound, making numerical simulation completely impossible.

The resolution came from an insight by M\"{u}ller~\cite{Muller:1967zza} in 1967 and was made rigorous by Israel~\cite{Israel:1976tn} in 1976 and Israel and Stewart~\cite{Israel:1979wp} in 1979. The key idea is conceptually simple: instead of requiring the dissipative currents to instantaneously equal their Navier--Stokes values, one allows them to \emph{relax} toward those values over a finite microscopic timescale $\tau$. This is the same idea that in non-relativistic physics leads from Fick's instantaneous law of diffusion to the Cattaneo equation~\cite{cattaneo1948}. Concretely, the Navier--Stokes relation $J = J_{\rm NS}$ is replaced by the relaxation equation
\begin{equation}
\tau\dot{J} + J = J_{\rm NS},
\end{equation}
where $\dot{J}$ is the comoving time derivative\footnote{see Sec.~\ref{sec:israel_stewart} for the corresponding relativistic Israel--Stewart construction.}. This seemingly small change has a dramatic effect: the governing equations become \emph{hyperbolic} (i.e., like the wave equation, they confine information to a causal cone and admit a finite maximum signal speed), perturbations propagate at the finite speed $\sqrt{D/\tau}$, causality is restored, and the instabilities disappear.

This is the Israel--Stewart (IS) theory~\cite{Israel:1979wp}, and it has been the cornerstone of relativistic viscous hydrodynamics ever since. Israel and Stewart derived it both phenomenologically, from the requirement that the entropy current have non-negative divergence at second order in deviations from equilibrium, and microscopically, from the relativistic Boltzmann equation using Grad's 14-moment approximation~\cite{Grad:1949}. The microscopic derivation connects all the transport coefficients and relaxation times directly to the underlying collision dynamics, giving the theory a firm kinetic-theory foundation.

The Israel--Stewart theory was successfully applied to heavy-ion collisions by Muronga%
~\cite{Muronga:2001zk}, Heinz and Song~\cite{Song:2007fn}, Romatschke and Romatschke~\cite{Romatschke:2007mq}, and many others, leading to the modern state-of-the-art viscous hydrodynamic codes (MUSIC~\cite{Schenke:2010nt,Schenke:2011pb}, VISH2+1~\cite{Song:2007fn}, etc.) that are the primary theoretical tools for extracting $\eta/s$ from RHIC and LHC data.

\section{From Israel--Stewart to the method of moments}

The original Israel--Stewart derivation, while physically correct in its essential ideas, contains a certain ambiguity. The theory requires choosing which moment of the Boltzmann equation to use as the equation of motion for the dissipative quantities, and this choice affects the values of the second-order transport coefficients. Israel and Stewart used the second moment, but this is not the only possibility.

This issue was clarified in a series of important papers by Denicol, Niemi, Moln\'{a}r, and Rischke~\cite{Denicol:2012cn,Denicol:2012es} (hereafter DNMR), who developed a systematic method-of-moments approach to deriving relativistic dissipative fluid dynamics from the Boltzmann equation. The key insight of the DNMR framework is that one need not truncate the distribution function at the 14-moment level from the outset. Instead, one derives an exact, and formally infinite, hierarchy of coupled evolution equations for all the irreducible moments of the non-equilibrium deviation $\delta f_\mathbf{k}$, and then reduces this hierarchy by a systematic power-counting in Knudsen number (ratio of mean free path to gradient scale) and inverse Reynolds number (ratio of dissipative to ideal contributions). The result is that at second order in this expansion, the equations of motion can always be closed in terms of exactly 14 dynamical variables, regardless of which moments are used for the closure. The transport coefficients, however, retain the memory of all higher moments through integrals over the equilibrium distribution. This framework reproduces the Israel--Stewart structure but provides a uniquely well-defined set of transport coefficients and a clear prescription for computing them from any given microscopic model.

The DNMR framework represents the current state of the art in deriving causal and stable second-order relativistic dissipative fluid dynamics from kinetic theory. It is the primary theoretical tool employed in this thesis.

\section{A new ingredient: large electromagnetic fields in heavy-ion collisions}

Alongside the developments in viscous hydrodynamics, a different physical ingredient has come to dominate discussions of heavy-ion phenomenology in the past fifteen years: the electromagnetic field.

\begin{figure}
    \centering
    \includegraphics[width=\linewidth]{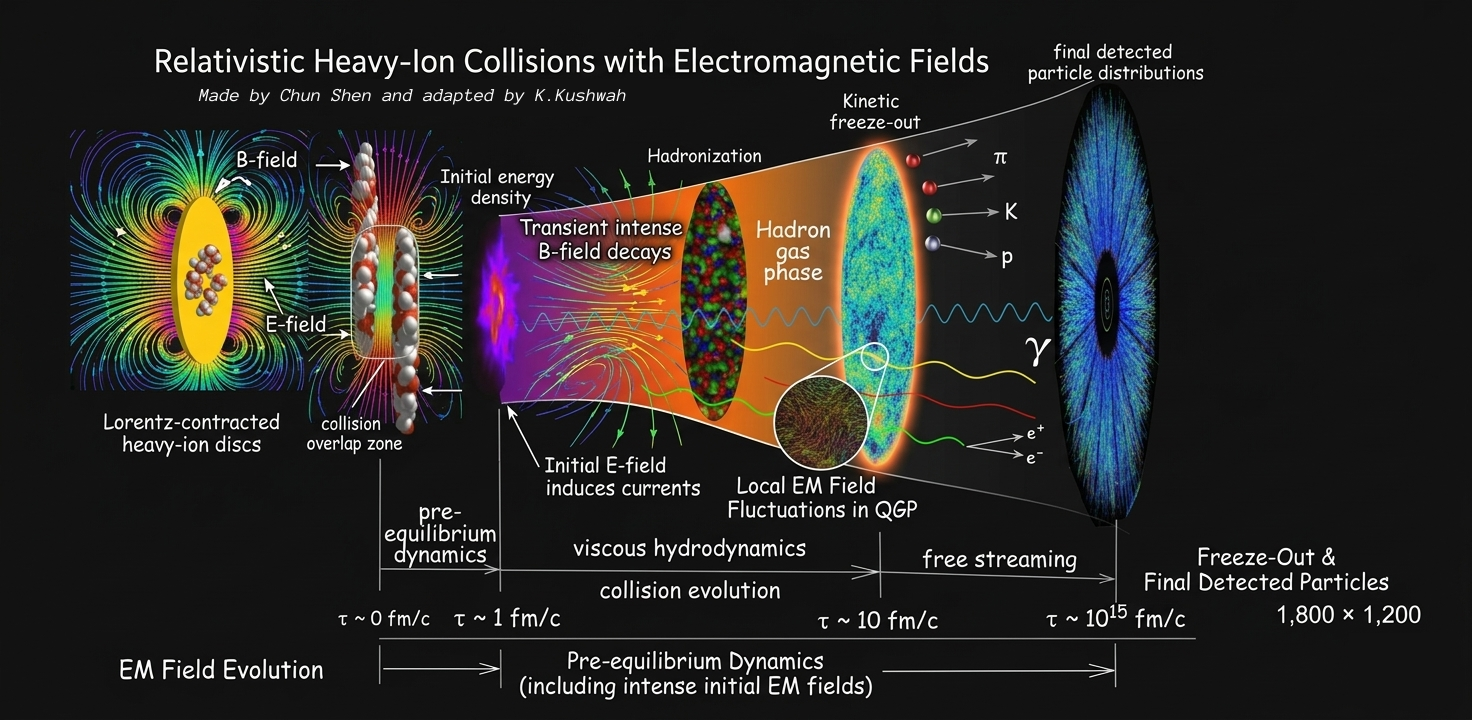}
    \caption[Magnetic field in a non-central heavy-ion collision]{Non-central heavy-ion collision with spectator protons moving past the interaction region. The rapidly moving spectator charges generate an intense, short-lived magnetic field approximately perpendicular to the reaction plane, providing an additional dynamical ingredient in the evolution and transport properties of the quark-gluon plasma.}
    \label{fig:heavy_ion_magnetic_field}
\end{figure}

In a non-central collision between two nuclei, the spectator protons, namely those that do not participate in the collision, move at nearly the speed of light in opposite directions on either side of the collision zone. These fast-moving charges generate an enormous transient magnetic field in the overlap region, as depicted in Fig.~\ref{fig:heavy_ion_magnetic_field}. Estimates by Skokov, Illarionov and Toneev~\cite{Skokov:2009qp} and subsequent calculations~\cite{Voronyuk:2011jd,Bzdak:2011yy,Deng:2012pc} showed that this field can reach strengths of order $eB \sim m_\pi^2 \sim 10^{18}\,\mathrm{G}$ at RHIC and up to $eB \sim 15\,m_\pi^2 \sim 10^{19}\,\mathrm{G}$ at the LHC. To put this in perspective, these are the strongest magnetic fields that have ever existed anywhere in the observable universe since the Big Bang. For comparison, the strongest sustained laboratory magnetic fields are of order $10^5\,\mathrm{G}$, and the surface fields of the most highly magnetized neutron stars (magnetars) reach about $10^{15}\,\mathrm{G}$~\cite{Duncan:1992hi}.

Such intense magnetic fields open entirely new physics. The most celebrated prediction is the \emph{chiral magnetic effect} (CME), proposed by Kharzeev, Levin, and Nardi~\cite{Kharzeev:2007jp} and elaborated by Fukushima, Kharzeev, and Warringa~\cite{Fukushima:2008xe}: in the presence of a strong magnetic field, local chirality imbalance generated by topological QCD fluctuations drives an electric current along the field direction, producing a charge separation perpendicular to the reaction plane. This is a macroscopic manifestation of quantum anomalies, and its experimental search has been one of the most active topics at RHIC and the LHC~\cite{Zhao:2019hta}. The magnetic field also modifies the QCD phase diagram through \emph{magnetic catalysis} and \emph{inverse magnetic catalysis}~\cite{Miransky:2015ava}, shifts the deconfinement temperature, generates the \emph{chiral separation effect} and \emph{chiral vortical effect}~\cite{Son:2009tf}, and alters the global polarization of strange quarks~\cite{STAR:2017ckg}.

But there is a more immediate and fundamental consequence: the magnetic field changes the transport properties of the QGP itself. Any charged particle moving in a magnetic field undergoes cyclotron motion, spiralling around the field lines. This microscopic physics has direct macroscopic consequences: diffusion across field lines is suppressed while diffusion along field lines is unaffected, generating anisotropic transport. And there are Hall-type off-diagonal responses in which the diffusion current flows in the direction transverse to both the gradient and the magnetic field. The shear viscosity similarly splits into multiple independent components. All of this was anticipated theoretically~\cite{Huang:2011dc,Lifshitz:1981} but was not incorporated into causal second-order theories until recently.


\section{The problem is that we need a theory!}

With magnetic fields this large, it is clear that a complete theoretical description of the QGP evolution in heavy-ion collisions must include both dissipation \emph{and} electromagnetic fields simultaneously. Yet for a long time, the two sides of this problem were treated separately. Ideal (non-dissipative) relativistic magnetohydrodynamics (MHD) was well-developed for astrophysical applications~\cite{Lichnerowicz:1967,Anile:1989} and was used in some heavy-ion modelling. Viscous hydrodynamics without magnetic fields was the workhorse of QGP phenomenology. But the combination of viscous MHD in a form that is both causal and stable and derived from kinetic theory was missing.

The stakes for having this theory are high. Without it, one cannot reliably compute how the magnetic field modifies the elliptic flow coefficient $v_2$, the directed flow $v_1$ (which is particularly sensitive to the early-time magnetic field~\cite{Gursoy:2014aka,STAR:2023jdd}), or the charge-dependent azimuthal correlations related to the CME. One cannot consistently model the simultaneous evolution of the QGP's momentum anisotropy and its electromagnetic properties. And one cannot correctly describe the resistive decay of the magnetic field, which depends on the plasma's electric conductivity $\sigma_E$ which in turn itself is a dissipative transport coefficient that must be derived on equal footing with shear viscosity $\eta$ and bulk viscosity $\zeta$ .

The theoretical challenge is substantial. The magnetic field breaks the isotropy of the three-dimensional space orthogonal to the fluid velocity. This proliferates the number of independent transport coefficients dramatically: the single shear viscosity $\eta$ of isotropic hydrodynamics becomes five independent viscosity components in a magnetized fluid, and the single diffusion coefficient $\kappa$ becomes three. More fundamentally, these anisotropic transport tensors first arise in the Navier--Stokes limit and are themselves acausal. A second-order Israel--Stewart-type extension is needed even in the magnetized case, and deriving it requires extending the DNMR method-of-moments framework to include the Lorentz force term in the Boltzmann equation. This is a non-trivial extension: the magnetic field does not merely add new transport coefficients to already known equations, but modifies the structure of the moment hierarchy itself, generating new couplings between tensor sectors that are absent in the field-free case. A first step in this direction was taken by Denicol, Huang, Moln\'{a}r, Monteiro, Niemi, Noronha, Rischke and Wang~\cite{Denicol:2018rbw}, who derived the full second-order relaxation equations for a single-component gas in the non-resistive limit. Their results showed that the magnetic field enters not only by splitting the transport coefficients into anisotropic components, but also by introducing qualitatively new terms in the relaxation equations themselves, namely terms that rotate the diffusion current and mix the components of the shear-stress tensor through the magnetic field direction, giving rise to two new transport coefficients with no analogue in the field-free theory. Yet this single-component non-resistive construction leaves two central aspects of a realistic plasma unaddressed: the multi-species nature of the medium, and the dynamical role of the electric field. These gaps define the scope of the present work.

\section{What this thesis does}

This thesis constructs, a complete kinetic-theory foundation for relativistic dissipative magnetohydrodynamics, progressing through two levels of generality.

The first level is a two-component plasma, basically a gas of positively and negatively charged particles, which is the minimal model of a realistic plasma. A two-component system has a richer structure than a single-species gas: both a total particle current and a net charge current can be formed, and these couple differently to the magnetic field and to the dissipative sector~\cite{Kushwah:2024zgd, Kushwah:2024csd}. New dynamical variables appear, including the relative shear-stress tensor between the two species, and the magnetic field induces new couplings between them. Application of the theory to the Bjorken expanding geometry reveals oscillatory dynamics in the transverse shear sector that is entirely absent in the single-component theory and is traceable to the cyclotron precession of the two species.

The second and most general level is the resistive theory, in which the electric field is no longer constrained to vanish but evolves dynamically. In this case, the charge diffusion current becomes a genuine dissipative degree of freedom with its own relaxation equation~\cite{Kushwah:2025jsb}. The electric field drives the current directly through an Ohmic term, but also couples to the shear-stress tensor, generating momentum anisotropy even in the absence of any flow gradients. The resulting theory is qualitatively richer than any of its predecessors: charge transport and viscous anisotropy are dynamically intertwined, and even simple situations like a homogeneous plasma in a uniform electric field exhibit non-trivial nonlinear dynamics involving underdamped oscillations and transient stress generation.

Taken together, these theories represent a systematic and complete generalisation of the Israel--Stewart framework to magnetized relativistic plasmas. They provide the theoretical foundation needed for quantitative modelling of the QGP in the presence of magnetic fields, and they are also directly relevant to astrophysical applications such as neutron star mergers and relativistic jets.

\section{Outline of the thesis}

This thesis is organized as follows.

\textbf{Chapter~\ref{ch:thermo_foundations}} establishes the thermodynamic foundations that underpin the entire hydrodynamic framework. Starting from the thermodynamic limit and the extensivity of entropy, the Euler relation, the Gibbs--Duhem relation, and the density form of the first law are derived. The notion of local thermodynamic equilibrium is introduced, and the ideal hydrodynamic equations are shown to follow from it.

\textbf{Chapter~\ref{chap:dissipative_hydro}} develops relativistic dissipative hydrodynamics at the macroscopic level. The tensor decomposition of the conserved currents, the Landau and Eckart frame choices, and the exact equations of motion in the Landau frame are presented. Relativistic Navier--Stokes theory is derived and shown to be acausal and linearly unstable. The Israel--Stewart transient theory is then derived from the entropy current, and the structure of the relaxation equations is identified. The chapter also analyses the Navier--Stokes constitutive relations in the presence of a magnetic field, identifying the anisotropic transport tensors that arise, and shows that the magnetic field does not resolve the acausality problem.

\textbf{Chapter~\ref{chap:kinetic_foundations}} provides the microscopic kinetic-theory
foundations. The relativistic Boltzmann equation, the collision term, and the linearisation
around local equilibrium are introduced. The macroscopic conservation laws are recovered as
moments of the Boltzmann equation, and the DNMR method-of-moments framework is presented in
detail, culminating in the derivation of the Israel--Stewart relaxation equations within the
14-moment approximation. The chapter then extends this framework to charged fluids by
introducing the Boltzmann--Vlasov equation in the presence of electromagnetic fields, states
the full hierarchy of moment equations, and applies the 14-moment truncation to obtain the
second-order relaxation equations for a single-component gas, including the new magnetic
transport coefficients and their resistive electric
counterparts.

\textbf{Chapter~\ref{ch:two_component_rmhd}} extends the kinetic-theory construction to a two-component relativistic plasma in the non-resistive limit~\cite{Kushwah:2024zgd}. The positively and negatively charged species are treated as independent microscopic components, which leads to coupled shear-stress equations and introduces a relative shear tensor with no analogue in the one-component theory. The chapter then develops the projection algebra adapted to the magnetic-field direction, decomposes the shear-stress tensor into longitudinal, semi-transverse, and transverse sectors, and analyzes the resulting dynamics in Bjorken flow. This application shows how the magnetic field can generate oscillatory transient behavior before the system approaches its late-time Navier--Stokes limit.

\textbf{Chapter~\ref{chap:resistive_rmhd_two_component}} derives the most general framework: resistive relativistic dissipative MHD for a two-component plasma~\cite{Kushwah:2025jsb}. The coupled relaxation equations for the charge diffusion current and the shear-stress tensor are derived from the Boltzmann--Vlasov equation, and the theory is applied to the homogeneous and Bjorken-expanding cases.

The original contributions of this thesis are contained entirely in Chapters~\ref{ch:two_component_rmhd} and~\ref{chap:resistive_rmhd_two_component} and 
Chapters~\ref{ch:thermo_foundations}--\ref{chap:kinetic_foundations} provide 
the necessary background.

\section{Conventions and notation}
\label{sec:notation}
In this thesis, unless stated otherwise, all physical quantities are expressed in natural units, $\hbar = c = k_B = 1$. The spacetime metric is taken to be Minkowskian with the signature $g^{\mu\nu} = \mathrm{diag}(+1,-1,-1,-1)$. We follow the Einstein summation convention and the Lorentz-invariant inner product of two four-vectors is written as $p\cdot q \equiv p_\mu q^\mu$.

The fluid four-velocity is denoted by $u^\mu$, normalised as $u^\mu u_\mu = 1$. The projector onto the three-space orthogonal to $u^\mu$ is $\Delta^{\mu\nu} \equiv g^{\mu\nu} - u^\mu u^\nu$, satisfying $\Delta^{\mu\nu}u_\nu = 0$ and $\Delta^{\mu}_{\ \mu} = 3$. The partial derivative decomposes as $\partial^\mu = \nabla^\mu + u^\mu D$, where $D \equiv u^\mu\partial_\mu$, and $\nabla^\mu \equiv \Delta^{\mu\nu}\partial_\nu$.
In the fluid rest frame, $D$ reduces to the time derivative and $\nabla^\mu$ to the spatial gradient. The shorthand $\dot{f} \equiv Df$ is used throughout. We also define the symmetric, antisymmetric, and angular-bracket index notations, respectively, as
\begin{equation}
A^{(\mu}B^{\nu)} \equiv \frac{1}{2}(A^\mu B^\nu + A^\nu B^\mu), \quad
A^{[\mu}B^{\nu]} \equiv \frac{1}{2}(A^\mu B^\nu - A^\nu B^\mu), \quad
A^{\langle\mu}B^{\nu\rangle} \equiv \Delta^{\mu\nu}_{\alpha\beta}\,A^\alpha B^\beta,
\end{equation}
where the traceless symmetric projection operator is
\begin{equation}
\Delta^{\mu\nu}_{\alpha\beta} \equiv \frac{1}{2}\!\left(\Delta^\mu_{\ \alpha}\Delta^\nu_{\ \beta} + \Delta^\mu_{\ \beta}\Delta^\nu_{\ \alpha}\right) - \frac{1}{3}\Delta^{\mu\nu}\Delta_{\alpha\beta}.
\end{equation}
The single-index version is $A^{\langle\mu\rangle} \equiv \Delta^\mu_{\ \nu}A^\nu$. The kinematic tensors constructed from gradients of $u^\mu$ are
$\theta \equiv \partial_\mu u^\mu$, $\sigma^{\mu\nu} \equiv \nabla^{\langle\mu}u^{\nu\rangle}$, $\omega^{\mu\nu} \equiv \nabla^{[\mu}u^{\nu]}$,
called the expansion scalar, shear tensor, and vorticity tensor respectively.

Further, the magnetic field, $B^\mu$, introduces a preferred spatial direction. We define the unit vector $b^\mu \equiv B^\mu/B$ along the field, with $B\equiv\sqrt{-B^\mu B_\mu}$. We can write 
\begin{align}
    b^{\mu\nu} &\equiv -\epsilon^{\mu\nu\alpha\beta}u_\alpha b_\beta, \qquad\Xi^{\mu\nu} \equiv \Delta^{\mu\nu} + b^\mu b^\nu, \\
    \Xi^{\mu\nu}_{\alpha\beta} & \equiv \frac{1}{2}\left(\Xi^{\mu}_\alpha \Xi^\nu_\beta + \Xi^\mu_\beta \Xi^\nu_\alpha\right) -\frac{1}{2}\Xi^{\mu\nu}\Xi_{\alpha\beta}
\end{align}
where $b^{\mu\nu}$ is the antisymmetric tensor and $\Xi^{\mu\nu}$ projects onto the two-dimensional plane transverse to both $u^\mu$ and $b^\mu$. $\Xi^{\mu\nu}_{\alpha\beta}$ is rank-four traceless symmetric projector analogous to $\Delta^{\mu\nu}_{\alpha\beta}$.

\chapter{Thermodynamical foundations of relativistic hydrodynamics}
\label{ch:thermo_foundations}

The previous chapter motivated relativistic hydrodynamics as the effective description of the long-wavelength evolution of the quark-gluon plasma. Before developing the microscopic kinetic-theory derivation used later in this thesis, it is useful to first establish the phenomenological hydrodynamic structures that the microscopic theory must reproduce.

Hydrodynamics is an effective macroscopic description valid when microscopic scales are well separated from the scales associated with macroscopic gradients~\cite{Landau:1959,deGroot:1980dk,Denicol_Rischke,Rezzolla:2013dea}. This implies that microscopic equilibration occurs over time and distance intervals much shorter than those characterizing the macroscopic evolution. As a result, the system can be treated as being close to thermodynamic equilibrium in the vicinity of each space-time point -- this is referred to as \textit{local} thermodynamic equilibrium. Thus, thermodynamics enters fluid dynamics at both equilibrium and non-equilibrium levels. At equilibrium, it supplies the equation of state that closes the ideal theory. Away from equilibrium, the second law of thermodynamics restricts the allowed macroscopic structures and provides a guiding principle for the formulation of dissipative theories~\cite{rocha2023theories}.

Therefore in this chapter, we do not develop thermodynamics as an independent subject, rather we aim to collect the thermodynamic identities that are used in relativistic hydrodynamics. The discussion begins with the thermodynamic limit and the density form of the first law, then moves to local thermodynamic equilibrium and the covariant ideal currents, and finally shows how these relations imply entropy conservation in ideal hydrodynamics. Dissipative constitutive relations are not introduced here and will be discussed in the following chapters.

\section{Thermodynamic relations for hydrodynamics}
\label{sec:thermo_limit}

Consider a macroscopic system described by the extensive variables
$(E,V,N)$, namely its total energy, volume, and conserved charge
or particle number~\cite{Callen:1985}. In the thermodynamic limit, the entropy is extensive,
\begin{equation}
S(\lambda E,\lambda V,\lambda N)=\lambda S(E,V,N),
\qquad \lambda>0.
\label{eq:extensivity_entropy}
\end{equation}
For quasi-static reversible changes between nearby equilibrium states, the first law is
\begin{equation}
dE = T\,dS - P\,dV + \mu\,dN,
\label{eq:first_law_E}
\end{equation}
which defines the intensive variables such as temperature $T$, isotropic pressure $P$, and chemical potential $\mu$.

It is often convenient to rewrite Eq.~\eqref{eq:first_law_E} in entropy form,
\begin{equation}
dS = \frac{1}{T}\,dE + \frac{P}{T}\,dV - \frac{\mu}{T}\,dN.
\label{eq:first_law_entropy_form}
\end{equation}

\subsection{Euler relation and Gibbs--Duhem relation}
\label{subsec:euler_gibbs_duhem}

Extensivity implies an Euler relation. Differentiating
Eq.~\eqref{eq:extensivity_entropy} with respect to $\lambda$ and then setting
$\lambda=1$, one obtains
\begin{equation}
S
=
\frac{\partial S}{\partial E}E
+
\frac{\partial S}{\partial V}V
+
\frac{\partial S}{\partial N}N.
\end{equation}
Using Eq.~\eqref{eq:first_law_entropy_form},
\begin{equation}
\frac{\partial S}{\partial E}=\frac{1}{T},
\qquad
\frac{\partial S}{\partial V}=\frac{P}{T},
\qquad
\frac{\partial S}{\partial N}=-\frac{\mu}{T},
\end{equation}
so that
\begin{equation}
TS = E + PV - \mu N.
\label{eq:euler_global}
\end{equation}
This is the Euler relation for a simple system~\cite{Callen:1985}.

Taking the differential of Eq.~\eqref{eq:euler_global},
\begin{equation}
d(TS)=dE+P\,dV+V\,dP-\mu\,dN-N\,d\mu,
\end{equation}
or equivalently,
\begin{equation}
T\,dS + S\,dT
=
dE+P\,dV+V\,dP-\mu\,dN-N\,d\mu.
\end{equation}
Substituting Eq.~\eqref{eq:first_law_E} and cancelling common terms gives
\begin{equation}
V\,dP = S\,dT + N\,d\mu.
\label{eq:gibbs_duhem_global}
\end{equation}
This is the Gibbs--Duhem relation~\cite{Callen:1985}. It shows that the intensive variables are not
independent, a fact that will later be important when reducing the hydrodynamic
equations.

\section{Density form and hydrodynamic thermodynamic variables}
\label{sec:thermo_densities}

Hydrodynamics is formulated in terms of local densities rather than global extensive
variables~\cite{deGroot:1980dk}. We therefore define the energy density, entropy density and particle density, respectively, as
\begin{equation}
\epsilon \equiv \frac{E}{V},
\qquad
s \equiv \frac{S}{V},
\qquad
n \equiv \frac{N}{V}.
\label{eq:density_defs}
\end{equation}
Dividing Eq.~\eqref{eq:euler_global} by $V$, one finds
\begin{equation}
\epsilon + P = Ts + \mu n.
\label{eq:euler_density}
\end{equation}
Likewise, dividing Eq.~\eqref{eq:gibbs_duhem_global} by $V$, one obtains
\begin{equation}
dP = s\,dT + n\,d\mu.
\label{eq:gibbs_duhem_density}
\end{equation}

To rewrite the first law in density form, use
\begin{equation}
E=\epsilon V,\qquad S=sV,\qquad N=nV,
\end{equation}
so that
\begin{equation}
dE=V\,d\epsilon+\epsilon\,dV,\qquad
dS=V\,ds+s\,dV,\qquad
dN=V\,dn+n\,dV.
\end{equation}
Substituting these into Eq.~\eqref{eq:first_law_E} gives
\begin{equation}
V\,d\epsilon+\epsilon\,dV
=
T\left(V\,ds+s\,dV\right)-P\,dV+\mu\left(V\,dn+n\,dV\right).
\end{equation}
Using Eq.~\eqref{eq:euler_density} to cancel the terms proportional to $dV$, one arrives at
\begin{equation}
d\epsilon = T\,ds + \mu\,dn.
\label{eq:first_law_density}
\end{equation}
Equivalently,
\begin{equation}
ds = \frac{1}{T}\,d\epsilon - \frac{\mu}{T}\,dn.
\end{equation}
Introducing the standard variables
\begin{equation}
\beta \equiv \frac{1}{T},
\qquad
\alpha \equiv \frac{\mu}{T},
\end{equation}
this becomes
\begin{equation}
ds = \beta\,d\epsilon - \alpha\,dn.
\label{eq:ds_beta_alpha}
\end{equation}
From Eq.~\eqref{eq:ds_beta_alpha}, one reads off
\begin{equation}
\beta = \left(\frac{\partial s}{\partial \epsilon}\right)_n,
\qquad
\alpha = -\left(\frac{\partial s}{\partial n}\right)_\epsilon.
\label{eq:beta_alpha_partials}
\end{equation}
Similarly, Eq.~\eqref{eq:gibbs_duhem_density} implies
\begin{equation}
s = \left(\frac{\partial P}{\partial T}\right)_\mu,
\qquad
n = \left(\frac{\partial P}{\partial \mu}\right)_T.
\label{eq:s_n_from_pressure}
\end{equation}

These are the basic thermodynamic relations that are used repeatedly in relativistic
hydrodynamics.

\subsection{Equation of state and response derivatives}
\label{subsec:eos_response}

The conservation laws alone do not close the hydrodynamic theory. Closure requires an
equation of state, for example in the form
\begin{equation}
P=P(\epsilon,n),
\label{eq:eos_local}
\end{equation}
or, equivalently, $P=P(T,\mu)$ or $s=s(\epsilon,n)$.
The equation of state contains the microscopic physics of the medium.

A particularly important derivative is the speed of sound,
\begin{equation}
c_s^2 \equiv \left(\frac{\partial P}{\partial \epsilon}\right)_{s/n}.
\label{eq:cs_def_entropy_per_particle}
\end{equation}
The choice of holding $s/n$ fixed is natural because, in ideal hydrodynamics,
this ratio is conserved along the flow. In situations where the net charge vanishes and
the equation of state reduces effectively to $P=P(\epsilon)$, one simply has
\begin{equation}
c_s^2 = \frac{dP}{d\epsilon}.
\end{equation}

\section{Local thermodynamic equilibrium and primary fluid variables}
\label{sec:LTE_primary_variables}

The hydrodynamic description is based on the assumption that, although the full system
is not in global equilibrium, it remains sufficiently close to equilibrium on
microscopic scales so that each fluid element can be assigned a local equilibrium state~\cite{Landau:1959,deGroot:1980dk,CercignaniKremer2002}. In this sense hydrodynamics treats the medium as a continuous set of infinitesimal fluid elements, each of which is characterized by local thermodynamic fields that vary smoothly in spacetime. The corresponding primary variables are the temperature $T(x)$, the chemical potential $\mu(x)$, and the fluid four-velocity $u^\mu(x)$. This is the basic content of local thermodynamic equilibrium. 

The fluid four-velocity is the timelike unit vector tangent to the worldline of a fluid
element. In Minkowski coordinates it is defined as
\begin{equation}
u^\mu \equiv \gamma \left(1,\mathbf{v}\right),
\qquad
\gamma \equiv \frac{1}{\sqrt{1-\mathbf{v}^2}},
\label{eq:u_def}
\end{equation}
where $\mathbf{v}$ is the ordinary three velocity of the fluid element and $\gamma$ is the
Lorentz factor. By construction, $u^\mu$ is normalized according to
\begin{equation}
u^\mu u_\mu = 1,
\label{eq:u_norm}
\end{equation}
for the metric convention $g^{\mu\nu}=\mathrm{diag}(1,-1,-1,-1)$. The projector onto the
three space orthogonal to the flow is then
\begin{equation}
\Delta^{\mu\nu} \equiv g^{\mu\nu} - u^\mu u^\nu,
\label{eq:Delta_def}
\end{equation}
which satisfies $\Delta^{\mu\nu}u_\nu=0$ and isolates the spatial part of tensors in the
frame comoving with the fluid. This is the standard definition used in relativistic
hydrodynamics~\cite{Landau:1959,Rezzolla:2013dea}. 

The \emph{local rest frame} of a fluid element is the frame moving instantaneously with
that element. In this frame, the fluid element is at rest, so that its spatial velocity
vanishes and the four-velocity reduces to
\begin{equation}
u^\mu_{\mathrm{LRF}} = (1,0,0,0).
\label{eq:u_lrf}
\end{equation}
Equivalently, the local rest frame is the frame in which the timelike direction is aligned
with $u^\mu$, and all tensors can be decomposed into temporal and purely spatial parts
relative to that direction. For an ideal fluid, this is also the frame in which there is
simultaneously no net energy flow and no net particle flow, so the local state can be
described entirely by the scalar quantities $\epsilon$, $n$, $P$, and $s$~\cite{Landau:1959,Rezzolla:2013dea}. In dissipative
fluids this identification is no longer unique, which is why the definition of the flow
velocity must later be fixed by an explicit frame choice such as Landau or Eckart (discussed in the following chapter)~\cite{Eckart:1940te,Landau:1959}.

Under the assumption of local thermodynamic equilibrium, the thermodynamic relations
derived in the previous sections can be applied point by point in spacetime. Thus one
may write
\begin{equation}
T=T(x),\qquad \mu=\mu(x),\qquad u^\mu=u^\mu(x),
\end{equation}
and regard the densities $\epsilon(x)$, $n(x)$, $s(x)$, and the pressure $P(x)$ as local
functions determined by these fields and by the equation of state. Hydrodynamics then
describes how these local equilibrium variables evolve under the conservation laws.

\section{Ideal hydrodynamics from local equilibrium}
\label{sec:ideal_hydro_from_lte}

With the fluid four-velocity and the local rest frame defined in the previous section, we can now introduce the central macroscopic quantities that characterize a relativistic fluid: the net-charge four-current $N^\mu$, the entropy four-current $S^\mu$, and the energy-momentum tensor $T^{\mu\nu}$. The current $N^\mu$ describes the flow of a conserved charge, such as baryon number, while $S^\mu$ represents the local entropy density and its transport. The energy-momentum tensor contains the local energy density, momentum density, energy flux, and momentum flux, and its conservation governs the exchange of energy and momentum in the fluid~\cite{Landau:1959,deGroot:1980dk,Denicol_Rischke,Rezzolla:2013dea}.

Since each fluid element is assumed to be in equilibrium in its own instantaneous rest frame, these currents must take their equilibrium form in that frame. In particular, there is no energy flow, no momentum density, and no particle or entropy flux in the local rest frame. The momentum flux is isotropic and is completely determined by the thermodynamic pressure. Therefore, the energy-momentum tensor in the local rest frame of the fluid has the following simple form
\begin{equation}
T^{\mu\nu}_{\mathrm{LRF}}=
\begin{pmatrix}
\epsilon & 0 & 0 & 0 \\
0 & P & 0 & 0 \\
0 & 0 & P & 0 \\
0 & 0 & 0 & P
\end{pmatrix},
\label{eq:T_LRF}
\end{equation}
together with the net-charge four-current and entropy four-current, which become
\begin{equation}
N^\mu_{\mathrm{LRF}}=(n,0,0,0),
\qquad
S^\mu_{\mathrm{LRF}}=(s,0,0,0).
\label{eq:NS_LRF}
\end{equation}
These expressions show that for an ideal fluid, in the local rest
frame, the state is fully specified by the scalar thermodynamic quantities
$\epsilon$, $P$, $n$, and $s$, whereas all dissipative fluxes vanish by construction~\cite{Landau:1959,Rezzolla:2013dea}.

The form of the ideal currents in an arbitrary frame is then obtained by boosting these
local rest frame expressions with the fluid four-velocity $u^\mu$. This gives~\cite{Landau:1959,deGroot:1980dk,Denicol_Rischke,Rezzolla:2013dea}
\begin{equation}
N^\mu_{(0)} = n\,u^\mu,
\qquad
S^\mu_{(0)} = s\,u^\mu,
\qquad
T^{\mu\nu}_{(0)} = \epsilon u^\mu u^\nu - P\Delta^{\mu\nu}.
\label{eq:ideal_currents_cov_thermo}
\end{equation}
Using this definition, the energy-momentum
tensor may also be written as
\begin{equation}
T^{\mu\nu}_{(0)}
=
(\epsilon+P)u^\mu u^\nu - P g^{\mu\nu}.
\label{eq:ideal_T_alt}
\end{equation}
Since the pressure is fixed by the equation of state and the normalization
$u^\mu u_\mu=1$ removes one component of $u^\mu$, the ideal fluid is described by
five independent hydrodynamic fields: $\epsilon$, $n$, and the three independent
components of $u^\mu$. These are matched exactly by the five independent
conservation equations.

The dynamics of the ideal fluid then follows from local conservation of energy-momentum and
charge,
\begin{equation}
\partial_\mu T^{\mu\nu}_{(0)}=0,
\qquad
\partial_\mu N^\mu_{(0)}=0.
\label{eq:conservation_laws_ideal}
\end{equation}
To extract their physical content, it is useful to re-introduce the comoving derivative
$D \equiv u^\mu \partial_\mu,$ and the spatial gradient $\nabla^\mu \equiv \Delta^{\mu\nu}\partial_\nu.$ With these definitions, the derivative $\partial^\mu$ can be decomposed as
\begin{equation}
\partial^\mu = u^\mu D + \nabla^\mu .
\end{equation}
We also define the expansion scalar as
\begin{equation}
\theta \equiv \partial_\mu u^\mu.
\end{equation}

Projecting Eq.~\eqref{eq:conservation_laws_ideal} parallel and orthogonal to the flow,
\begin{equation}
u_\nu \partial_\mu T^{\mu\nu}_{(0)}=0,
\qquad
\Delta^\lambda_{\ \nu}\partial_\mu T^{\mu\nu}_{(0)}=0,
\qquad
\partial_\mu N^\mu_{(0)}=0,
\end{equation}
one obtains the ideal hydrodynamic equations of motion in their usual form,
\begin{equation}
D\epsilon + (\epsilon+P)\theta = 0,
\label{eq:energy_ideal}
\end{equation}
\begin{equation}
(\epsilon+P)Du^\lambda - \nabla^\lambda P = 0,
\label{eq:euler_ideal}
\end{equation}
\begin{equation}
Dn + n\theta = 0.
\label{eq:charge_ideal}
\end{equation}
Equation \eqref{eq:energy_ideal} governs the evolution of the local energy density,
Eq.~\eqref{eq:euler_ideal} is the relativistic Euler equation, and
Eq.~\eqref{eq:charge_ideal} expresses conservation of the charge density.

At this stage, the thermodynamic identities derived earlier can be embedded into a
manifestly covariant form adapted to relativistic hydrodynamics. We define the thermal
vector~\cite{Israel:1976tn}
\begin{equation}
\beta^\mu \equiv \frac{u^\mu}{T},
\label{eq:beta_mu_def}
\end{equation}
and recall the thermal potential
\begin{equation}
\alpha \equiv \frac{\mu}{T}.
\end{equation}
Then the density relations
\eqref{eq:euler_density}, \eqref{eq:ds_beta_alpha}, and
\eqref{eq:gibbs_duhem_density} may be written covariantly as~\cite{Israel:1976tn,Israel:1979wp}
\begin{equation}
S^\mu_{(0)} = P\,\beta^\mu + T^{\mu\nu}_{(0)}\beta_\nu - \alpha\,N^\mu_{(0)},
\label{eq:covariant_euler}
\end{equation}
and
\begin{equation}
d(P\beta^\mu)=N^\mu_{(0)}\,d\alpha - T^{\mu\nu}_{(0)}\,d\beta_\nu.
\label{eq:covariant_gibbs_duhem}
\end{equation}
Contracting Eq.~\eqref{eq:covariant_euler} with $u_\mu$ reproduces
\begin{equation}
s+\alpha n-\beta(\epsilon+P)=0,
\end{equation}
which is equivalent to the Euler relation
$\epsilon+P=Ts+\mu n$. Likewise, contracting
Eq.~\eqref{eq:covariant_gibbs_duhem} with $u_\mu$ yields
\begin{equation}
d(\beta P)-n\,d\alpha+\epsilon\,d\beta=0,
\end{equation}
which is equivalent to the ordinary density relations. A direct consequence is the
covariant form of the first law,
\begin{equation}
dS^\mu_{(0)}=\beta_\nu\,dT^{\mu\nu}_{(0)}-\alpha\,dN^\mu_{(0)}.
\label{eq:covariant_first_law}
\end{equation}
This does not add any new thermodynamic information; it just packages the same equilibrium identities into a form suited to the covariant theory.

We further need an important consistency check: ideal hydrodynamics should not produce entropy. So we start from the density form of the first law,
\begin{equation}
d\epsilon = T\,ds + \mu\,dn,
\end{equation}
and take the comoving derivative to obtain
\begin{equation}
D\epsilon = T\,Ds + \mu\,Dn.
\end{equation}
Using Eqs.~\eqref{eq:energy_ideal} and \eqref{eq:charge_ideal}, one finds
\begin{equation}
-(\epsilon+P)\theta = T\,Ds - \mu n \theta.
\end{equation}
Substituting the Euler relation $\epsilon+P=Ts+\mu n$, this becomes
\begin{equation}
T\,Ds + Ts\,\theta = 0,
\end{equation}
or equivalently,
\begin{equation}
Ds + s\theta = 0.
\label{eq:entropy_evol_ideal}
\end{equation}
Since the ideal entropy current is $S^\mu_{(0)}=su^\mu$, this is equivalent to
\begin{equation}
\partial_\mu S^\mu_{(0)} = 0.
\label{eq:entropy_conserved_ideal}
\end{equation}
The same conclusion also follows directly from the covariant identity~\cite{Israel:1976tn,Israel:1979wp}
\begin{equation}
\partial_\mu S^\mu_{(0)}
=
\beta_\nu \partial_\mu T^{\mu\nu}_{(0)}
-
\alpha \partial_\mu N^\mu_{(0)},
\label{eq:entropy_div_identity}
\end{equation}
which vanishes by virtue of the ideal conservation laws. Comparing
Eq.~\eqref{eq:entropy_evol_ideal} with Eq.~\eqref{eq:charge_ideal}, one further obtains
\begin{equation}
D\left(\frac{s}{n}\right)=0,
\label{eq:s_over_n_conserved}
\end{equation}
showing that the entropy per particle is conserved along the ideal flow.

Together, these relations define the ideal hydrodynamic description associated with local equilibrium. Once the system departs from exact local equilibrium, the currents in Eq.~\eqref{eq:ideal_currents_cov_thermo} acquire
dissipative corrections and the entropy current is no longer conserved. Instead, the local form of the second law becomes~\cite{Israel:1976tn}
\begin{equation}
\partial_\mu S^\mu \ge 0.
\label{eq:second_law_local}
\end{equation}
This inequality will provide the basic thermodynamic constraint on the dissipative
constitutive relations introduced in the next chapter.

\section{Summary}
We established the thermodynamic and ideal hydrodynamic foundations required for the rest of the discussion. We began with the equilibrium relations most relevant for relativistic hydrodynamics, namely the first law of thermodynamics, the Euler relation, and the Gibbs--Duhem relation, and rewrote them in terms of local densities such as the energy density, charge density, entropy density, and pressure. Under the assumption of local thermodynamic equilibrium, the primary hydrodynamic fields $T(x)$, $\mu(x)$, and $u^\mu(x)$ were then introduced, and these were used to construct the ideal forms of the conserved currents and the energy-momentum tensor. From these structures, the ideal conservation equations were derived, and it was shown that, in the absence of dissipation, the entropy current is conserved. This chapter, therefore, established the equilibrium reference theory, clarified the meaning of the hydrodynamic variables, and fixed the notation and thermodynamic identities that will be used throughout what follows.

This equilibrium framework also makes clear why a more general theory is needed. Realistic relativistic fluids are not exactly in local equilibrium, so irreversible processes and entropy production cannot be neglected in general. Therefore, the ideal structures derived here provide a basis for the dissipative theory discussed in the next chapter, where bulk pressure, diffusion, and shear stress will be incorporated into the relativistic description.

\chapter{Relativistic dissipative hydrodynamics}
\label{chap:dissipative_hydro}

Chapter~\ref{ch:thermo_foundations} established the thermodynamic identities and ideal hydrodynamic structures that follow from local thermodynamic equilibrium. In realistic applications, however, ideal hydrodynamics is only an approximation. It neglects finite mean free path effects and assumes that each fluid element remains exactly in local equilibrium. Therefore, it cannot describe entropy production, charge diffusion, heat flow, bulk viscous pressure, or shear stresses. For this reason, ideal hydrodynamics is best understood as the zeroth-order limit of a more general dissipative theory~\cite{Denicol_Rischke,rocha2023theories,Rezzolla:2013dea}.

This limitation is especially important in relativistic systems, where microscopic and macroscopic scales are not always widely separated. In heavy-ion collisions, for example, the system expands rapidly and develops strong gradients~\cite{Heinz_2013,GALE_2013,shen2020recent,Rischke:2003mt}, while in astrophysical plasmas the transport of baryon number, heat, and momentum can play a central dynamical role~\cite{Rezzolla:2013dea}. Whenever the microscopic relaxation time is not negligible compared with the macroscopic evolution time, local equilibrium cannot be maintained exactly. The fluid then develops dissipative corrections associated with bulk expansion, momentum anisotropy, and charge or energy transport relative to the local flow. These corrections generate entropy and modify the stress tensor and conserved currents.

Thus, ideal hydrodynamics provides the reference state around which dissipative fluid dynamics is constructed, but it cannot serve as the final theory whenever gradients are non-zero. Many structures of the dissipative theory are still naturally defined relative to the ideal fluid: the equilibrium pressure $P(\epsilon,n)$, the equilibrium entropy density $s(\epsilon,n)$, and the four-velocity $u^\mu$ all enter the dissipative decomposition. The physical description of realistic relativistic fluids therefore requires a systematic treatment of deviations from local equilibrium.

This chapter develops the formulation of relativistic dissipative hydrodynamics used throughout the rest of this thesis. Far from equilibrium, the conserved currents must be decomposed into their dissipative components, together with matching conditions that define the hydrodynamic variables. This introduces the issue of frame choice, particularly the Landau and Eckart frames. Since the present work adopts the Landau frame, the conservation laws are projected accordingly to obtain the exact equations of motion for the dissipative fluid variables. These equations are not closed by themselves, so one must supplement them with constitutive relations. The first such closure is relativistic Navier–Stokes theory, obtained at first order in gradients. Although useful as a formal limit, it is acausal and unstable in the relativistic regime. A consistent causal description instead requires transient theories, most notably Israel--Stewart theory, where the dissipative currents are promoted to independent dynamical variables that relax toward their Navier--Stokes limits on finite timescales.

\section{Dissipative decomposition of the conserved currents}
\label{sec:dissipative_decomposition}
Away from exact local equilibrium, the energy-momentum tensor and net-charge current must be generalized beyond their ideal form. The most general decomposition with respect to a normalized timelike vector $u^\mu$ may be written as~\cite{Israel:1979wp,deGroot:1980dk,Denicol_Rischke,Rezzolla:2013dea}
\begin{equation}
N^\mu = n u^\mu + V^\mu,
\label{eq:Nmu_diss_general}
\end{equation}
\begin{equation}
T^{\mu\nu} = \epsilon u^\mu u^\nu -(P+\Pi)\Delta^{\mu\nu} + 2u^{(\mu}h^{\nu)} + \pi^{\mu\nu}.
\label{eq:Tmunu_diss_general}
\end{equation}
Here $n$ and $\epsilon$ are the local charge and energy densities, $P=P(\epsilon,n)$ is the equilibrium pressure obtained from the equation of state, and the new quantities $\Pi$, $h^\mu$, $V^\mu$, and $\pi^{\mu\nu}$ are the dissipative deviations from local equilibrium, where
\begin{equation}
\Pi \equiv -P - \frac{1}{3}\Delta_{\mu\nu}T^{\mu\nu},
\label{eq:Pi_def}
\end{equation}
is the bulk-viscous pressure, describing the scalar correction to the isotropic pressure.

\begin{equation}
h^\mu \equiv u_\alpha \Delta^\mu_{\ \beta} T^{\alpha\beta},
\label{eq:hmu_def}
\end{equation}
is the energy-diffusion current, also called heat-flow vector in some conventions.

\begin{equation}
V^\mu \equiv \Delta^\mu_{\ \alpha}N^\alpha,
\label{eq:nmu_def}
\end{equation}
is the net-charge diffusion current.

\begin{equation}
\pi^{\mu\nu} \equiv T^{\langle \mu\nu \rangle},
\label{eq:pimunu_def}
\end{equation}
is the shear-stress tensor, where angular brackets denote the symmetric, traceless, and
transverse projection,
\begin{equation}
A^{\langle \mu\nu \rangle}
\equiv
\Delta^{\mu\nu}_{\alpha\beta} A^{\alpha\beta},
\qquad
\Delta^{\mu\nu}_{\alpha\beta}
\equiv
\frac{1}{2}
\left(
\Delta^\mu_{\ \alpha}\Delta^\nu_{\ \beta}
+
\Delta^\mu_{\ \beta}\Delta^\nu_{\ \alpha}
\right)
-
\frac{1}{3}\Delta^{\mu\nu}\Delta_{\alpha\beta}.
\label{eq:double_projector}
\end{equation}

By construction,
\begin{equation}
u_\mu V^\mu = 0,
\qquad
u_\mu h^\mu = 0,
\qquad
u_\mu \pi^{\mu\nu} = 0,
\qquad
\pi^\mu_{\ \mu}=0,
\qquad
\pi^{\mu\nu}=\pi^{\nu\mu}.
\label{eq:diss_orthogonality}
\end{equation}

The tensor decomposition \eqref{eq:Nmu_diss_general} and \eqref{eq:Tmunu_diss_general} is exact. It reorganizes the 14 independent components of $N^\mu$ and $T^{\mu\nu}$ into scalars, vectors, and a rank-two traceless tensor relative to the chosen flow field $u^\mu$. The hydrodynamic problem is then to determine how these quantities evolve and how they are related to gradients of the hydrodynamic fields~\cite{Denicol_Rischke,Rezzolla:2013dea}.

\section{Matching conditions and the definition of the hydrodynamic variables}
\label{sec:matching_conditions}
Once dissipative corrections are included, the fluid can no longer be regarded as being in exact local thermodynamic equilibrium. As a result, the equilibrium variables that appeared in the ideal theory, such as $T$, $\mu$, $P$, and $s$, are no longer directly defined from the actual non-equilibrium state. The standard way to proceed is to associate each non-equilibrium fluid element with a \emph{fictitious local equilibrium state}~\cite{Israel:1976tn,Landau:1959,Denicol_Rischke}. The thermodynamic quantities $T$, $\mu$, $P$, and $s$ are then understood as properties of this reference state, not of the true non-equilibrium fluid.

The prescription that identifies this fictitious equilibrium state is known as the \emph{matching condition}~\cite{Israel:1976tn}. In practice, one requires that selected macroscopic densities of the non-equilibrium state coincide with those of the associated local equilibrium state. In relativistic dissipative hydrodynamics, the standard choice is to match the energy density and the net-charge density~\cite{Israel:1976tn,Landau:1959}. Thus one defines
\begin{equation}
\epsilon \equiv u_\mu u_\nu T^{\mu\nu},
\label{eq:matching_energy}
\end{equation}
\begin{equation}
n \equiv u_\mu N^\mu.
\label{eq:matching_charge}
\end{equation}
These conditions ensure that $\epsilon$ and $n$ retain their interpretation as the local energy density and local net-charge density in the fluid rest frame. Once these are fixed, all other equilibrium thermodynamic quantities are defined through the equation of state of the associated equilibrium state,
\begin{equation}
P = P(\epsilon,n).
\label{eq:matching_pressure}
\end{equation}

Therefore, the pressure $P$, temperature $T$, chemical potential $\mu$, and entropy density $s$ entering dissipative hydrodynamics should be understood as the thermodynamic quantities of the matched local equilibrium state, and not of the exact non-equilibrium state itself. This avoids double counting and provides a clean separation between equilibrium contributions and dissipative corrections. It also ensures that the thermodynamic identities established in Chapter~\ref{ch:thermo_foundations} remain applicable.

The matching conditions immediately imply orthogonality constraints on the dissipative corrections. Writing
\begin{equation}
T^{\mu\nu}=T^{\mu\nu}_{(0)}+\tau^{\mu\nu},
\qquad
N^\mu=N^\mu_{(0)}+V^\mu,
\end{equation}
with $T^{\mu\nu}_{(0)}$ and $N^\mu_{(0)}$ the ideal local-equilibrium contributions, and with $\tau^{\mu\nu}\equiv T^{\mu\nu}-T^{\mu\nu}_{(0)}$ denoting the non-equilibrium correction to the energy-momentum tensor, one obtains
\begin{equation}
u_\mu u_\nu \tau^{\mu\nu}=0,
\qquad
u_\mu V^\mu=0.
\label{eq:matching_orthogonality}
\end{equation}
These relations are the starting point for the irreducible decomposition of the dissipative currents into bulk, diffusion, and shear sectors.

\section{Choice of Frame}
\label{sec:landau_eckart}

For dissipative fluids, the local rest frame is no longer unique. In the ideal theory the rest frame was characterized simultaneously by vanishing energy flow and vanishing particle flow. Away from equilibrium this is impossible in general, because energy diffusion and charge diffusion need not vanish in the same frame. The choice of $u^\mu$ therefore requires an additional convention.

Two standard choices are used in relativistic hydrodynamics.

\subsection{Landau frame}

In the Landau frame, the flow velocity is defined as the timelike eigenvector of the energy-momentum tensor,
\begin{equation}
T^{\mu\nu}u_\nu = \epsilon u^\mu.
\label{eq:landau_definition}
\end{equation}
This implies
\begin{equation}
h^\mu = 0.
\label{eq:hmu_zero_landau}
\end{equation}
Thus the local rest frame is the frame in which there is no energy diffusion~\cite{Landau:1959}. In this frame, the dissipative correction to the current is carried entirely by $V^\mu$, $\Pi$, and $\pi^{\mu\nu}$, and the decompositions simplify to
\begin{equation}
N^\mu = n u^\mu + V^\mu,
\label{eq:Nmu_landau}
\end{equation}
\begin{equation}
T^{\mu\nu} = \epsilon u^\mu u^\nu - (P+\Pi)\Delta^{\mu\nu} + \pi^{\mu\nu}.
\label{eq:Tmunu_landau}
\end{equation}

\subsection{Eckart frame}
In the Eckart frame, the four-velocity is defined by the charge current,
\begin{equation}
N^\mu = n u^\mu.
\label{eq:eckart_definition}
\end{equation}
This implies
\begin{equation}
V^\mu = 0.
\label{eq:nmu_zero_eckart}
\end{equation}
In this frame, the local rest frame is the frame in which there is no particle diffusion, but energy diffusion $h^\mu$ is generally non-zero~\cite{Eckart:1940te}.

\subsection{Physical interpretation and frame preference}
The two frames only differ in convention, not in the underlying physics. Both describe the same underlying system, provided all quantities are transformed consistently. However, one frame may be more preferred than the other depending on the physical setting.

The Landau frame is the natural choice when the energy-momentum tensor is the primary macroscopic object and the charge current does not play a preferred role in defining the flow~\cite{Landau:1959}. This is precisely the situation in ultrarelativistic heavy-ion collisions, especially near mid-rapidity, where the net baryon density can be very small and where hydrodynamic initial conditions are often extracted directly from a pre-equilibrium energy-momentum tensor via the Landau eigenvalue condition \eqref{eq:landau_definition}~\cite{Muronga:2001zk,Song:2007fn,Romatschke:2007mq}. In such cases, the Landau frame is both physically natural and technically convenient.

The Eckart frame is more natural when a distinguished conserved particle flux exists and is itself central to the macroscopic description~\cite{Eckart:1940te}. This often happens in baryon-rich fluids or in astrophysical contexts where the transport of matter relative to the local flow is more directly tied to the physical interpretation of the fluid velocity. In that case, eliminating the charge diffusion current can be convenient.

Throughout this thesis, we work in the Landau frame~\cite{Landau:1959} since this choice is the most appropriate for the heavy-ion and kinetic-theory applications that motivate the present work. It is also the frame convention adopted in the later kinetic-theory derivation.

\section{Useful kinematic tensors}
\label{sec:kinematic_tensors}
 Before deriving the equations of motion, it is useful to redefine the irreducible gradients of the flow field. The covariant derivative of $u^\mu$ may be decomposed as
\begin{equation}
\partial_\mu u_\nu = u_\mu D u_\nu + \frac{1}{3}\theta \Delta_{\mu\nu} + \sigma_{\mu\nu} + \omega_{\mu\nu},
\label{eq:grad_u_decomp}
\end{equation}
where we recall, for sake convenience,  
$D \equiv u^\mu\partial_\mu,$ $ \nabla^\mu \equiv \Delta^{\mu\nu}\partial_\nu,
$ and $\theta \equiv \partial_\mu u^\mu$ are the comoving derivative, spatial gradient, and expansion scalar, respectively.

The shear tensor is
\begin{equation}
\sigma^{\mu\nu} \equiv \nabla^{\langle \mu}u^{\nu \rangle} =
\Delta^{\mu\nu}_{\alpha\beta}\partial^\alpha u^\beta,
\label{eq:shear_tensor_def}
\end{equation}
while the vorticity tensor is
\begin{equation}
\omega^{\mu\nu} \equiv \frac{1}{2} 
\left( \nabla^\mu u^\nu - \nabla^\nu u^\mu \right).
\label{eq:vorticity_tensor_def}
\end{equation}

These quantities organize the gradient structure of dissipative fluid dynamics~\cite{Denicol_Rischke,Rezzolla:2013dea}. The scalar $\theta$ describes isotropic expansion or compression, $\sigma^{\mu\nu}$ characterizes shear deformation of the flow, and $\omega^{\mu\nu}$ measures its local rotation.

\section{Exact equations of motion in the Landau frame}
\label{sec:exact_eom_diss}

The conservation laws remain exact even away from equilibrium,
\begin{equation}
\partial_\mu T^{\mu\nu}=0,
\qquad
\partial_\mu N^\mu=0.
\label{eq:exact_conservation_diss}
\end{equation}
Using the Landau-frame decomposition \eqref{eq:Nmu_landau} and
\eqref{eq:Tmunu_landau}, and projecting parallel and orthogonal to the flow, one obtains the exact fluid-dynamical equations of motion.

\paragraph{\textit{Energy equation: }}
Contracting $\partial_\mu T^{\mu\nu}=0$ with $u_\nu$ yields
\begin{equation}
D\epsilon + (\epsilon+P+\Pi)\theta - \pi^{\mu\nu}\sigma_{\mu\nu} = 0.
\label{eq:energy_eq_diss}
\end{equation}

\paragraph{\textit{Momentum equation :}}
Projecting orthogonally with $\Delta^\alpha_{\ \nu}$ gives
\begin{equation}
(\epsilon+P+\Pi)D u^\alpha - \nabla^\alpha(P+\Pi) + \Delta^\alpha_{\ \nu}\partial_\mu \pi^{\mu\nu} = 0.
\label{eq:momentum_eq_diss}
\end{equation}

\paragraph{\textit{Charge equation: }}
Using $N^\mu = nu^\mu+V^\mu$, one finds
\begin{equation}
Dn + n\theta + \nabla_\mu V^\mu - V^\mu D u_\mu = 0.
\label{eq:charge_eq_diss}
\end{equation}

These three equations are exact consequences of conservation laws and the chosen tensor decomposition~\cite{Denicol_Rischke,Rezzolla:2013dea}. However, they are not closed, because the dissipative quantities $\Pi$, $V^\mu$, and $\pi^{\mu\nu}$ appear as additional unknowns. One therefore needs constitutive relations.

\section{Entropy current and thermodynamic constraints}
\label{sec:entropy_current_diss}

In Chapter~\ref{ch:thermo_foundations}, the ideal entropy current was shown to obey $\partial_\mu S^\mu_{(0)}=0$. Away from equilibrium, the entropy current receives corrections,
\begin{equation}
S^\mu = s u^\mu + \Phi^\mu,
\label{eq:entropy_current_general}
\end{equation}
where $\Phi^\mu$ is at least first order in dissipative quantities.

At first order, and in the Landau frame, the natural extension is~\cite{Israel:1976tn,Israel:1979wp}
\begin{equation}
S^\mu = s u^\mu - \alpha V^\mu.
\label{eq:entropy_current_first_order}
\end{equation}
Using the thermodynamic identities derived previously, together with the exact equations of motion \eqref{eq:energy_eq_diss}--\eqref{eq:charge_eq_diss}, one obtains
\begin{equation}
\partial_\mu S^\mu
=
-\beta \Pi \theta
-\beta V^\mu \nabla_\mu \alpha
+\beta \pi^{\mu\nu}\sigma_{\mu\nu}.
\label{eq:entropy_production_first_order}
\end{equation}
The second law requires~\cite{Israel:1976tn}
\begin{equation}
\partial_\mu S^\mu \ge 0.
\label{eq:second_law_ch2}
\end{equation}

Equation \eqref{eq:entropy_production_first_order} is important as it identifies the thermodynamic forces conjugate to the dissipative fluxes~\cite{Israel:1976tn,Israel:1979wp}:
\begin{equation}
\Pi \leftrightarrow -\theta,
\qquad
V^\mu \leftrightarrow -\nabla^\mu \alpha,
\qquad
\pi^{\mu\nu} \leftrightarrow \sigma^{\mu\nu}.
\label{eq:flux_force_pairs}
\end{equation}
The simplest constitutive relations are then obtained by demanding that each contribution to $\partial_\mu S^\mu$ be positive semidefinite.

\section{Relativistic Navier--Stokes theory}
\label{sec:relativistic_NS}

The first-order relativistic closure is obtained by taking the dissipative currents to be instantaneously proportional to their corresponding gradients~\cite{Eckart:1940te,Landau:1959,pichon:65etude},
\begin{equation}
\Pi = -\zeta \theta,
\label{eq:NS_bulk}
\end{equation}
\begin{equation}
V^\mu = \kappa_n \nabla^\mu \alpha,
\label{eq:NS_diffusion}
\end{equation}
\begin{equation}
\pi^{\mu\nu} = 2\eta \sigma^{\mu\nu}.
\label{eq:NS_shear}
\end{equation}
Here $\zeta$, $\kappa_n$, and $\eta$ are the bulk viscosity, charge conductivity or diffusion coefficient, and shear viscosity, respectively. With these constitutive relations, Eq.~\eqref{eq:entropy_production_first_order} becomes
\begin{equation}
\partial_\mu S^\mu =
\beta \zeta \theta^2 -
\beta \kappa_n \nabla^\mu \alpha \nabla_\mu \alpha +
2\beta \eta \sigma^{\mu\nu}\sigma_{\mu\nu},
\label{eq:entropy_production_NS}
\end{equation}
which is non-negative provided
\begin{equation}
\zeta \ge 0,
\qquad
\kappa_n \ge 0,
\qquad
\eta \ge 0.
\label{eq:positive_transport}
\end{equation}

Formally, Relativistic Navier--Stokes theory is the leading term in a gradient expansion around local equilibrium~\cite{Eckart:1940te,Landau:1959}.  It is useful as a formal near-equilibrium limit and also as the asymptotic limit that any transient theory must reproduce at late times and long wavelengths.

However, in the relativistic regime this theory cannot be regarded as fundamental~\cite{Hiscock:1983zz,Hiscock:1985zz}. The constitutive relations \eqref{eq:NS_bulk}--\eqref{eq:NS_shear} make the dissipative currents respond instantaneously to gradients, rendering the full system parabolic rather than hyperbolic. This leads to propagation with infinite signal speed and, consequently, acausality. Moreover, the same instantaneous response is responsible for instabilities in relativistic settings.

\section{Relativistic Navier--Stokes theory in the presence of electromagnetic fields}
\label{sec:NS_EM}

The first-order constitutive relations discussed above are modified once the fluid is coupled to electromagnetic fields. At the level of the conservation laws, the energy-momentum tensor of the fluid is no longer conserved by itself, but satisfies the balance law~\cite{Landau:1959,Rezzolla:2013dea}
\begin{equation}
\partial_\mu T^{\mu\nu} = F^{\nu\lambda} J_\lambda,
\label{eq:matter_balance_NS_EM}
\end{equation}
where $J^\mu$ is the electric four-current and $F^{\mu\nu}$ is the Faraday tensor. Thus, electromagnetic fields exchange energy and momentum with the fluid through the Lorentz-force density. In addition, the charge-transport sector is modified because the electric field itself acts as a first-order driving force for dissipation. This is the relativistic generalization of the fact that an electric field induces conduction current in ordinary Ohmic media. The presence of a magnetic field further changes the tensorial structure of the constitutive relations by introducing a preferred spatial direction and breaking the isotropy that underlies the simple scalar transport coefficients of ordinary Navier--Stokes theory.

To describe these effects covariantly, it is convenient to introduce the Faraday tensor more explicitly. The tensor $F^{\mu\nu}$ provides the relativistic description of the electromagnetic field and is an antisymmetric rank-two tensor~\cite{jackson2012classical},
\begin{equation}
F^{\mu\nu} = -F^{\nu\mu},
\end{equation}
that unifies the electric and magnetic fields into a single geometric object. In a given Lorentz frame, its components contain the electric field $E^i$ and magnetic field $B^i$ as
\begin{equation}
F^{0i} = E^i,
\qquad
F^{ij} = -\epsilon^{ijk} B^k.
\end{equation}
Thus, rather than treating $\mathbf{E}$ and $\mathbf{B}$ as separate three-vectors from the outset, the relativistic formulation combines them into one tensorial object whose transformation properties under Lorentz transformations are manifest.

The Faraday tensor is defined in terms of the electromagnetic four-potential $A^\mu = (\phi,\mathbf{A})$ as~\cite{jackson2012classical}
\begin{equation}
F^{\mu\nu} = \partial^\mu A^\nu - \partial^\nu A^\mu,
\end{equation}
which ensures that it automatically satisfies the homogeneous Maxwell equations. This representation makes explicit that the electromagnetic field is derived from a potential and transforms covariantly under Lorentz transformations. It is also useful because the Lorentz-force term in the Boltzmann equation naturally appears in terms of $F^{\mu\nu}$, not in terms of the three-vector fields separately.

Given a fluid velocity $u^\mu$, the electric and magnetic fields measured in the local rest frame of the fluid are obtained from projections of the Faraday tensor~\cite{Landau:1959,Rezzolla:2013dea},
\begin{equation}
E^\mu \equiv F^{\mu\nu} u_\nu,
\qquad
B^\mu \equiv \frac{1}{2}\epsilon^{\mu\nu\alpha\beta} u_\nu F_{\alpha\beta}.
\end{equation}
so that
\begin{equation}
u_\mu E^\mu = 0,
\qquad
u_\mu B^\mu = 0.
\end{equation}
Thus, $E^\mu$ and $B^\mu$ are purely spatial in the local rest frame. In other words, $u^\mu$ selects the local observer, and the tensors $E^\mu$ and $B^\mu$ are simply the electromagnetic fields seen by that observer.

It is often useful to introduce the dual Faraday tensor as well,
\begin{equation}
\tilde{F}^{\mu\nu} \equiv \frac{1}{2}\epsilon^{\mu\nu\alpha\beta} F_{\alpha\beta}.
\end{equation}
Using the dual, the magnetic field can be written as $B^\mu = \tilde{F}^{\mu\nu} u_\nu$. The Faraday tensor itself can then be decomposed as
\begin{equation}
F^{\mu\nu}
=
E^\mu u^\nu - E^\nu u^\mu
+
\epsilon^{\mu\nu\alpha\beta} u_\alpha B_\beta.
\label{eq:Fmunu_decomp_NS_EM}
\end{equation}


Furthermore, the electric four-current $J^\mu$ is distinct from the conserved 
current $N^\mu$ introduced earlier. In general, $N^\mu$ represents the 
hydrodynamic conserved charge current used to define the fluid variables,
\begin{equation}
    N^\mu = n u^\mu + V^\mu,
\end{equation}
whereas the electric current weights the motion of the charged constituents 
by their electric charges. It is therefore written separately as
\begin{equation}
    J^\mu = n_q u^\mu + V_q^\mu,
    \label{eq:Jmu_decomp_NS_EM}
\end{equation}
where
\begin{equation}
    n_q \equiv u_\mu J^\mu, \qquad u_\mu V_q^\mu = 0.
\end{equation}
Thus, $n$ and $V^\mu$ characterize the conserved current $N^\mu$, while 
$n_q$ and $V_q^\mu$ characterize the electric current $J^\mu$. In a 
multi-component fluid, $n_q$ and $V_q^\mu$ appearing in 
Eq.~\eqref{eq:Jmu_decomp_NS_EM} are simply the charge-weighted sums of 
the individual species number densities and diffusion currents, 
$n_q = \sum_i q_i n_i$ and $V_q^\mu = \sum_i q_i V_i^\mu$, 
where $q_i$ is the electric charge of species $i$.

In the absence of electromagnetic fields, the first-order constitutive 
relation for the diffusion current is driven only by gradients of the 
thermal potential,
\begin{equation}
    V^\mu = \kappa_n \nabla^\mu \alpha.
\end{equation}
When an electric field is present, the entropy-production analysis is 
modified because the energy-momentum balance acquires a Lorentz-force 
source, Eq.~\eqref{eq:matter_balance_NS_EM}. The relevant current in 
the vector sector of the entropy production is the electric current 
$J^\mu$, and the thermodynamic force conjugate to it changes. 
Consequently, the constitutive relation for the electric diffusion 
current $V_q^\mu$ is modified. In a single-species fluid, 
$V_q^\mu = q V^\mu$, so this simultaneously determines the constitutive 
relation for $V^\mu$ itself. The thermodynamic force in the charge 
sector is replaced by the gauge-invariant electrochemical combination
\begin{equation}
    \mathcal{E}^\mu \equiv E^\mu - T \nabla^\mu \alpha.
\end{equation}
To see why, recall the entropy-production equation derived earlier, 
Eq.~\eqref{eq:entropy_production_first_order}. Repeating the 
entropy-current analysis with the modified balance law shows that the 
electric field contributes an additional term $\beta E_\mu J^\mu$, so 
the two contributions in the charge sector combine into 
$\beta J^\mu(E_\mu - T\nabla_\mu\alpha) = \beta J^\mu \mathcal{E}_\mu$. 
The full entropy production then reads
\begin{equation}
    \partial_\mu S^\mu = -\beta\Pi\,\theta + \beta J^\mu \mathcal{E}_\mu 
    + \beta\pi^{\mu\nu}\sigma_{\mu\nu},
    \label{eq:entropy_production_with_E}
\end{equation}
with the scalar, vector, and tensor sectors cleanly separated. The 
driving force in the charge sector is therefore $\mathcal{E}^\mu$, and 
the first-order constitutive law becomes~\cite{Hernandez_2017,
jackson2012classical}
\begin{equation}
    V_q^\mu = \sigma_E\,\mathcal{E}^\mu,
    \label{eq:ohm_rel_covariant}
\end{equation}
where $\sigma_E$ is the electrical conductivity. In the local rest 
frame, neglecting temperature and chemical-potential gradients, this 
reduces to the familiar Ohm's law, $\mathbf{J} = \sigma_E \mathbf{E}$.

Therefore, at first order, the only modification relative to the field-free theory is the replacement $\nabla^\mu \alpha \to \mathcal{E}^\mu$ in the vector sector. The scalar and tensor sectors 
retain the same sources $\theta$ and $\sigma^{\mu\nu}$, though their 
dynamics are indirectly affected by the fields through the modified 
conservation laws and, more generally, through anisotropic transport.

\subsection{Symmetry Breaking and the Origin of Anisotropic Transport}

The most important structural change occurs when a magnetic field is 
present. A non-zero $B^\mu$ picks up a preferred spatial direction, 
so the medium is no longer isotropic in the local rest frame~\cite{
Huang:2011dc,Hernandez_2017,Hattori_2022}. Defining
\begin{equation}
  B \equiv \sqrt{-B_\mu B^\mu},
  \qquad
  b^\mu \equiv \frac{B^\mu}{B},
  \qquad
  u_\mu b^\mu = 0,
  \qquad
  b_\mu b^\mu = -1,
  \label{eq:bmu_NS_EM}
\end{equation}
one can construct, in addition to $\Delta^{\mu\nu}$, the projector
\begin{equation}
  \Xi^{\mu\nu} \equiv \Delta^{\mu\nu} + b^\mu b^\nu,
  \label{eq:Xi_NS_EM}
\end{equation}
which projects onto the plane orthogonal to both $u^\mu$ and $b^\mu$.

Without a magnetic field, all spatial directions are equivalent and 
only $\Delta^{\mu\nu}$ is available. A non-zero $B^\mu$, however, reduces the local 
symmetry group from $SO(3)$ to $SO(2)$, making $b^\mu b^\nu$ and $\Xi^{\mu\nu}$ independent projectors, so that the fluid can 
respond differently along each spatial direction~\cite{Huang:2011dc,
Hernandez_2017,gedalin1991relativistic,gedalin1995generally}.

Because $b^\mu$ is now available, first-order constitutive relations 
can involve new tensor structures built from $\Delta^{\mu\nu}$, $b^\mu$, 
$\Xi^{\mu\nu}$, and $\epsilon^{\mu\nu\alpha\beta} u_\alpha b_\beta$. 
Any transverse vector $A^\mu$, satisfying $u_\mu A^\mu = 0$, admits the
decomposition
\begin{equation}
  A^\mu = A_\parallel^\mu + A_\perp^\mu,
  \qquad
  A_\parallel^\mu \equiv -(A \cdot b)\, b^\mu,
  \qquad
  A_\perp^\mu \equiv \Xi^\mu{}_\nu A^\nu,
  \label{eq:vector_split_b}
\end{equation}
with $A \cdot b = A_\mu b^\mu$. In an isotropic medium the response to 
$A_\parallel^\mu$ and $A_\perp^\mu$ must be the same, reducing to a 
single scalar coefficient. In a magnetized medium they are physically 
inequivalent and may carry different coefficients. In addition, the 
pseudotensor $\epsilon^{\mu\nu\alpha\beta}u_\alpha b_\beta$ allows a 
\emph{Hall-type} contribution. The most general linear relation between 
a transverse vector flux and a transverse vector force is therefore
\begin{equation}
  V_q^\mu = \sigma_E^{\mu\nu} \mathcal{E}_\nu,
  \label{eq:conductivity_tensor_relation}
\end{equation}
with
\begin{equation}
  \sigma_E^{\mu\nu}
  =
  \sigma_\perp\, \Xi^{\mu\nu}
  -
  \sigma_\parallel\, b^\mu b^\nu
  +
  \sigma_H\, \epsilon^{\mu\nu\alpha\beta}u_\alpha b_\beta.
  \label{eq:conductivity_tensor}
\end{equation}
The coefficients $\sigma_\parallel$, $\sigma_\perp$, and $\sigma_H$ are 
the longitudinal, transverse, and Hall conductivities, 
respectively~\cite{Huang:2011dc,Hernandez_2017,Hattori:2016cnt,
Hattori:2016lqx}. The magnetic field splits the response into these 
three independent channels, with one coefficient for each sector that 
the broken symmetry renders independent.

The same logic applies to the shear-stress tensor. In an isotropic 
fluid, the only allowed linear map between the symmetric, traceless, 
transverse tensors $\sigma^{\mu\nu}$ and $\pi^{\mu\nu}$ is the 
isotropic double projector, giving $  \pi^{\mu\nu}=2\eta\,\sigma^{\mu\nu}$.
In a magnetized fluid, the system becomes anisotropic and the most general linear relation between the shear-stress tensor and the shear tensor is,
\begin{equation}
  \pi^{\mu\nu}
  =
  2\,\eta^{\mu\nu\alpha\beta}\sigma_{\alpha\beta},
  \label{eq:shear_tensor_viscosity}
\end{equation}
and the shear viscosity becomes a fourth rank tensor. Expanding $\eta^{\mu\nu\alpha\beta}$ in the five rank-four projectors 
allowed by the residual symmetry around $b^\mu$,
\begin{align}
  \eta^{\mu\nu\alpha\beta}
  & =
   \eta_\| \left( \frac{\Xi^{\mu\nu}}{2} + b^\mu b^\nu \right) \left( \frac{\Xi^{\alpha\beta}}{2} + b^\alpha b^\beta \right) + \eta_{\indep} \Xi^{\mu\nu\alpha\beta} + \eta_\perp b^{(\mu}\Xi^{\nu)(\alpha} b^{\beta)}
   \nonumber \\
   & \qquad
   +\, \eta_{H_1}  b^{(\mu} b^{\nu)(\beta} b^{\alpha)}  + \eta_{H_2} \Big[b^{\mu(\alpha}\Xi^{\beta) \nu} + b^{\nu(\alpha}\Xi^{\beta)\nu}\Big], \label{eq:viscosity_tensor_projector_expansion}
\end{align}
Above we introduced 5 viscosities that quantify dissipative processes in the system. $\eta_\|$, $\eta_{\indep}$ and $\eta_{\perp}$ describe dissipative processes in combinations of directions parallel and orthogonal to the magnetic field. The coefficients
 $\eta_{H_1}$ and $\eta_{H_2}$ do not describe dissipative processes per se, and are analogous to the Hall conductivities discussed above.
In the limit $B\rightarrow 0$, this decomposition reduces to $\Delta^{\mu\nu\alpha\beta}$ and we recover the traditional expression for the shear-stress tensor.


\section{Why Relativistic Navier--Stokes fails}
\label{sec:NS_failure_and_transient}

The simplest way to understand why the relativistic Navier--Stokes theory is problematic is to study the diffusion of a scalar density, before any tensor structure or magnetic field enters the picture.
Suppose $\varphi$ satisfies a conservation law
\begin{equation}
    \partial_t \varphi + \nabla_i J^i = 0,
    \label{eq:scalar_conservation}
\end{equation}
together with the Fick-type constitutive relation
\begin{equation}
    J^i = -D\, \nabla^i \varphi.
    \label{eq:fick_law}
\end{equation}
Substituting \eqref{eq:fick_law} into \eqref{eq:scalar_conservation} yields
the diffusion equation
\begin{equation}
    \partial_t \varphi - D\, \nabla^2 \varphi = 0.
    \label{eq:diffusion_eq}
\end{equation}
This equation is \emph{parabolic}. Its fundamental solution for an initial
delta-function perturbation $\varphi(\mathbf{x}, 0) = \delta^{(3)}(\mathbf{x})$
is
\begin{equation}
    \varphi(\mathbf{x}, t)
    =
    \frac{1}{(4\pi D t)^{3/2}}\,
    \exp\!\left(-\frac{|\mathbf{x}|^2}{4Dt}\right),
    \qquad t > 0.
    \label{eq:diffusion_solution}
\end{equation}
For any $t > 0$, no matter how small, $\varphi(\mathbf{x}, t) \neq 0$ for all
$\mathbf{x}$. A localised perturbation becomes non-zero everywhere
instantaneously. There is no finite propagation speed which means information has
travelled an arbitrarily large distance in an arbitrarily short time.

In a non-relativistic context, this is an idealization that one typically accepts. In a relativistic theory, it is a fundamental violation though. No physical signal should propagate faster than the speed of light $c$. The parabolic structure of the diffusion equation is, therefore, incompatible with special relativity.

This is not a peculiarity of scalar diffusion but reflects a general property of any first-order gradient theory. In relativistic Navier--Stokes theory, the dissipative fluxes, namely, the bulk viscous pressure $\Pi$, the charge diffusion current $V^\mu$, and the shear-stress tensor $\pi^{\mu\nu}$, are given \emph{algebraically} as linear functions of first-order gradients of the hydrodynamic fields:
\begin{align}
    \Pi &= -\zeta\,\theta,
    \\
    V^\mu &= \kappa_n \nabla^\mu \alpha, 
    \\
    \pi^{\mu\nu} &= 2\eta\,\sigma^{\mu\nu}.
\end{align}
These are not evolution equations for the dissipative variables. They are constraints that hold at every instant. Substituting them back into the conservation equations $\partial_\mu T^{\mu\nu} = 0$ and $\partial_\mu N^\mu = 0$ produces a system of second-order partial differential equations for the thermodynamic fields $(T,\mu,u^\mu)$ that is, again, parabolic.

The consequences are two-fold and both incompatible with a relativistic theory:
\begin{enumerate}
    \item \textbf{Acausality.} Because the constitutive relations are
    algebraic, a gradient anywhere in the fluid instantaneously produces a
    dissipative flux everywhere. The propagation speed of perturbations in the
    linearised theory is formally infinite. One can verify this explicitly by
    performing a Fourier analysis: writing $\varphi \sim e^{-i\omega t + ik x}$
    and substituting into the linearised equations gives a dispersion relation
    \begin{equation}
        \omega = -i D k^2 + \mathcal{O}(k^4),
        \label{eq:NS_dispersion}
    \end{equation}
    for the diffusive mode. The phase velocity $v_\phi = \omega/k \sim -iDk$
    grows without bound as $k \to \infty$, confirming that arbitrarily
    short-wavelength perturbations travel at arbitrarily large speeds.

    \item \textbf{Linear instability.} Beyond the issue of propagation speed,
    the theory develops exponentially growing modes in the linearised
    analysis around a \emph{moving} background. Hiscock and Lindblom~\cite{Hiscock:1983zz,Hiscock:1985zz} showed
    in 1985 that the first-order Eckart and Landau frames are both linearly unstable for any nonzero background velocity, i.e., small perturbations
    grow without bound on short timescales, making any numerical simulation around a moving fluid cell ill-posed. Therefore, this instability is not an artifact of the choice of variables or frame, it remains present no matter which first-order frame is used.

\end{enumerate}

Both issues share the same origin: the dissipative variables have no dynamics of their own. They are dictated by the gradients of the thermodynamic fields at every instant, with no memory and no finite response time. The theory has no mechanism to prevent a sudden gradient from producing an instantaneous, unbounded dissipative response~\cite{Bemfica:2019cop,
Bemfica:2020xym,Gavassino:2021owo}. An alternative causal first-order framework, due to Bemfica, Disconzi, Noronha, and Kovtun (BDNK)~\cite{
Bemfica:2019cop,Bemfica:2020xym}, achieves causality through a careful out-of-equilibrium frame choice rather than relaxation; in this thesis we follow the Israel--Stewart route instead.

\section{Israel--Stewart theory}
\label{sec:israel_stewart}

A resolution of the causality problem is conceptually straightforward: one must
give the dissipative variables their own dynamics, allowing them to \emph{relax}
toward their Navier--Stokes values over a finite timescale $\tau$ rather than
being instantaneously constrained to match them. The simplest illustration is to
replace Fick's law by the Cattaneo equation~\cite{cattaneo1948},
\begin{equation}
    \tau\, \partial_t J^i + J^i = -D\, \nabla^i \varphi,
    \label{eq:cattaneo}
\end{equation}
which, upon substitution into the conservation law, yields the \emph{telegraph equation} -- a
damped wave equation that is hyperbolic with the finite propagation speed
\begin{equation}
    v = \sqrt{\frac{D}{\tau}}.
    \label{eq:propagation_speed}
\end{equation}
This is already visible in the dispersion relation of the telegraph equation,
\begin{equation}
    \omega = -\frac{i}{2\tau}
    \pm
    \sqrt{\frac{D}{\tau} k^2 - \frac{1}{4\tau^2}},
    \label{eq:telegraph_dispersion}
\end{equation}
which at large $k$ gives $|\omega/k| \to \sqrt{D/\tau}$, i.e., a finite, $k$-independent
propagation speed, in contrast to the Navier--Stokes result $|\omega/k|\sim Dk\to\infty$.
Thus the presence of relaxation time regulates the ultraviolet behaviour of the dispersion relation and restores causality.

Israel and Stewart~\cite{Israel:1976tn,Israel:1979wp} elevated this idea to the
full relativistic setting by deriving evolution equations for all dissipative
currents from the requirement that an extended entropy current have non-negative
divergence at second order in deviations from equilibrium.

\subsection{Derivation from the entropy current}

In the Navier--Stokes construction, the entropy current was taken in the form
\begin{equation}
S^\mu = s u^\mu - \alpha V^\mu,
\label{eq:entropy_current_first_order_repeat}
\end{equation}
which leads to constitutive relations that are algebraic in the gradients.
Israel and Stewart proposed that, away from equilibrium, the entropy current should
depend not only on the local equilibrium fields $(\alpha,\beta,u^\mu)$, but also on
the dissipative quantities themselves~\cite{Israel:1976tn,Israel:1979wp}. Keeping
terms up to second order, one writes~\cite{Israel:1979wp}
\begin{equation}
S^\mu = s u^\mu - \alpha V^\mu + Q^\mu,
\label{eq:IS_entropy_split}
\end{equation}
where $Q^\mu$ is quadratic in the dissipative variables. The simplest \textit{ansatz} that reduces the entropy density is,
\begin{equation}
Q^\mu = -\frac{u^\mu}{2}\left(\beta_0 \Pi^2 - \beta_1 V_\nu V^\nu
        + \beta_2 \pi_{\alpha\beta}\pi^{\alpha\beta}\right),
\label{eq:IS_entropy_current}
\end{equation}
with $\beta_0$, $\beta_1$, $\beta_2$ functions of the local equilibrium state. In principle, we could have also included second-order terms that combine dissipative currents, i.e., $\sim \Pi\, V^\mu$ and $\sim V_\nu \pi^{\mu\nu}$. These terms were omitted for the sake of simplicity and do not affect the following discussion.

Taking the divergence gives
\begin{equation}
\partial_\mu S^\mu = \partial_\mu (s u^\mu - \alpha V^\mu) + \partial_\mu Q^\mu.
\label{eq:divS_split}
\end{equation}
The first term is the first-order entropy production,
\begin{equation}
\partial_\mu (s u^\mu - \alpha V^\mu)
    = -\beta \Pi \theta - V^\mu \nabla_\mu \alpha + \beta \pi^{\mu\nu}\sigma_{\mu\nu}.
\label{eq:first_order_entropy_prod_repeat}
\end{equation}
The divergence of $Q^\mu$ introduces comoving derivatives of the dissipative
currents; using Eq.~\eqref{eq:IS_entropy_current} one finds schematically
\begin{equation}
\begin{split}
    \partial_\mu Q^\mu  = {} & -\beta_0 \Pi D\Pi
        + \beta_1 V_\mu D V^\mu - \beta_2 \pi_{\mu\nu} D\pi^{\mu\nu} \\
    & - \frac{1}{2}\left(\Pi^2 D\beta_0 - V_\mu V^\mu D\beta_1
        + \pi_{\mu\nu}\pi^{\mu\nu} D\beta_2\right) \\
    & - \frac{\theta}{2}\left(\beta_0 \Pi^2 - \beta_1 V_\mu V^\mu
        + \beta_2 \pi_{\mu\nu}\pi^{\mu\nu}\right).
    \label{eq:divQ_general}
\end{split}
\end{equation}
Unlike in first-order theory, the entropy production now contains the derivatives
$D\Pi$, $DV^\mu$, and $D\pi^{\mu\nu}$---this is the origin of relaxation dynamics.

Combining Eqs.~\eqref{eq:first_order_entropy_prod_repeat} and \eqref{eq:divQ_general}
and grouping by dissipative flux,
\begin{align}
\partial_\mu S^\mu ={}
    & -\Pi\left[\beta \theta + \beta_0 D\Pi
        + \tfrac{1}{2}\Pi \partial_\mu(\beta_0 u^\mu) \right] \nonumber\\
    &\quad -V^\mu\left[\nabla_\mu \alpha - \beta_1 \Delta_{\mu\nu}D V^\nu
        - \tfrac{1}{2}V_\mu \partial_\lambda(\beta_1 u^\lambda) \right] \nonumber\\
    &\quad +\pi^{\mu\nu}\left[\beta \sigma_{\mu\nu}
        - \beta_2 \Delta_{\mu\nu}^{\alpha\beta} D\pi_{\alpha\beta}
        - \tfrac{1}{2}\pi_{\mu\nu}\partial_\lambda(\beta_2 u^\lambda) \right].
\label{eq:entropy_prod_IS_grouped}
\end{align}
The second law $\partial_\mu S^\mu \ge 0$ is satisfied by requiring that the combined terms inside each bracket must
be proportional to the corresponding dissipative current,
\begin{align}
\beta \theta + \beta_0 D\Pi
    + \tfrac{1}{2}\Pi \partial_\mu(\beta_0 u^\mu)  &= -\frac{\Pi}{\zeta},
\label{eq:IS_bulk_condition}\\
\nabla_\mu \alpha - \beta_1 \Delta_{\mu\nu}D V^\nu
    - \tfrac{1}{2}V_\mu \partial_\lambda(\beta_1 u^\lambda) 
    &= \frac{V_\mu}{\kappa_n},
\label{eq:IS_diff_condition}\\
\beta \sigma_{\mu\nu} - \beta_2 \Delta_{\mu\nu}^{\alpha\beta} D\pi_{\alpha\beta}
    - \tfrac{1}{2}\pi_{\mu\nu}\partial_\lambda(\beta_2 u^\lambda) 
    &= \frac{\pi_{\mu\nu}}{2\eta},
\label{eq:IS_shear_condition}
\end{align}
which recovers the traditional positive semi-definite entropy production, 
\begin{equation}
\partial_\mu S^\mu = \frac{\Pi^2}{\zeta}
    - \frac{V^\mu V_\mu}{\kappa_n}
    + \frac{\pi^{\mu\nu}\pi_{\mu\nu}}{2\eta} \ge 0,
\label{eq:IS_entropy_positive}
\end{equation}
provided $\zeta,\kappa_n,\eta\ge 0$.
Introducing the relaxation times
\begin{equation}
\tau_\Pi \equiv \zeta \beta_0, \qquad
\tau_n \equiv \kappa_n \beta_1, \qquad
\tau_\pi \equiv 2\eta \beta_2,
\label{eq:relaxation_times_IS}
\end{equation}
one obtains equations for the dissipative
currents in their usual relaxation-type structure~\cite{Israel:1976tn,Israel:1979wp,Muller:1967zza},
\begin{align}
\tau_\Pi D\Pi + \Pi &= -\zeta \theta
    -\frac{\zeta}{2}\Pi\,\partial_\mu\!\left(\frac{\tau_\Pi}{\zeta}u^\mu\right),
\label{eq:IS_bulk_derived}\\
\tau_n \Delta^\mu_{\ \nu}D V^\nu + V^\mu &= \kappa_n \nabla^\mu \alpha
    -\frac{\kappa_n}{2}V^\mu\partial_\lambda\!\left(\frac{\tau_n}{\kappa_n}u^\lambda\right),
\label{eq:IS_diff_derived}\\
\tau_\pi \Delta^{\mu\nu}_{\alpha\beta}D\pi^{\alpha\beta} + \pi^{\mu\nu} &= 2\eta \sigma^{\mu\nu}
    -\eta\,\pi^{\mu\nu}\partial_\lambda\!\left(\frac{\tau_\pi}{2\eta}u^\lambda\right).
\label{eq:IS_shear_derived}
\end{align}
These reduce to the Navier--Stokes constitutive laws in the formal limit
$\tau_\Pi,\tau_n,\tau_\pi\to 0$, but for finite relaxation times describe a
delayed response of the dissipative currents to the thermodynamic forces. Taking this limit is equivalent to neglecting the higher-order contributions to the entropy four-current. These equations can be linearly causal and stable as long as the transport coefficients, in particular the relaxation times, satisfy some basic requirements \cite{DeGroot,Denicol_Rischke}. For instance, for a theory with only shear viscosity, one obtains that the shear viscosity and relaxation time must satisfy the inequality,
\begin{equation}
c_s^2 +\frac{4\eta}{3\tau_\pi\left(\epsilon + P\right)} \leq 1    .
\end{equation}

A more complete second-order formulation contains additional couplings required by
gradient power counting and kinetic-theory derivations~\cite{Denicol:2012cn,Denicol:2012es,Betz_2009,deBrito:2023tgb,Denicol_Rischke}.
In the Landau frame~\cite{Muronga:2001zk,Song:2007fn,Romatschke:2007mq,Niemi_2012,
Huovinen_2009,Heller:2014wfa,Tinti:2018qfb,Brito:2020nou},
\begin{align}
\tau_\Pi D\Pi + \Pi &= -\zeta\theta
    -\delta_{\Pi\Pi}\Pi\theta
    +\lambda_{\Pi n} n\cdot D u
    +\lambda_{\Pi\pi}\pi^{\mu\nu}\sigma_{\mu\nu} +\cdots,
\label{eq:IS_bulk_general}\\
\tau_n \Delta^\mu_{\ \nu}DV^\nu + V^\mu &= \kappa_n \nabla^\mu\alpha
    -\delta_{nn} V^\mu\theta
    -\tau_{n\pi}\pi^{\mu\nu}\nabla_\nu \alpha
    -\lambda_{n\omega} V_\nu \omega^{\nu\mu} +\cdots,
\label{eq:IS_diff_general}\\
\tau_\pi \Delta^{\mu\nu}_{\alpha\beta}D\pi^{\alpha\beta} + \pi^{\mu\nu} &=
    2\eta \sigma^{\mu\nu}
    -\delta_{\pi\pi}\pi^{\mu\nu}\theta
    -\tau_{\pi\pi}\pi^{\lambda\langle \mu}\sigma^{\nu\rangle}_{\ \ \lambda}
    +2\tau_\pi \pi^{\lambda\langle \mu}\omega^{\nu\rangle}_{\ \ \lambda}
    +\lambda_{\pi n} n^{\langle \mu}\nabla^{\nu\rangle}\alpha +\cdots.
\label{eq:IS_shear_general}
\end{align}
The precise values of the second-order transport coefficients depend on the
microscopic theory and approximation
scheme~\cite{Denicol:2012cn,Denicol:2012es,Betz_2009,deBrito:2023tgb,Denicol_Rischke}.
Israel--Stewart theory thus promotes the dissipative currents from algebraic constraints to dynamical variables, restoring causality. In the next Chapter, we will derive these equations from kinetic theory.

\section{The Problem in Magnetized Fluids \& Summary}
\label{sec:magnetized_NS_problem}

The failure of relativistic Navier--Stokes theory comes from the structure of
the first-order theory itself. The dissipative fluxes are fixed algebraically by
local gradients, with no independent dynamics and no finite response time. This
makes the equations parabolic and leads to acausality and instability. The same
issue remains when electromagnetic fields are included.

A magnetic field makes the constitutive relations anisotropic. The diffusion
current splits into longitudinal, transverse, and Hall sectors, while the shear
stress is controlled by a rank-four viscosity tensor. However, the first-order
relations still have the form
\begin{equation}
V_q^\mu = \sigma_E^{\mu\nu}\left(E_\nu - T\nabla_\nu \alpha\right),
\qquad
\pi^{\mu\nu} = 2\,\eta^{\mu\nu\alpha\beta}\sigma_{\alpha\beta}.
\label{eq:magnetized_NS_algebraic}
\end{equation}
Thus the dissipative fluxes are still determined instantaneously by the local
thermodynamic forces. The magnetic field changes the number and tensorial
structure of the transport coefficients, but the first-order equations remain
parabolic. For this reason, the same acausal signal propagation and instability
are expected to persist in relativistic magnetohydrodynamics
~\cite{Biswas:2020rps,Armas:2022wvb}.

In this chapter we showed how this problem can be addressed phenomenologically. We started
from the decomposition of the conserved currents and the entropy-current
analysis, derived the Navier--Stokes constitutive relations, and discussed their
failure in the relativistic regime. We then showed how Israel--Stewart theory
restores causality by promoting the dissipative currents to independent
dynamical variables with their own relaxation equations. The goal of this thesis
is to extend this causal framework to magnetized relativistic fluids, where the
magnetic field opens additional dissipative channels that also require transient
dynamics.

The next chapter develops the microscopic foundation for this construction. So far, the entire framework has been phenomenological: the constitutive relations and relaxation equations were derived by imposing the second law on a postulated entropy current, with no reference to the underlying microscopic dynamics. This changes in the next chapter where we start from the relativistic Boltzmann equation, applying the method of moments, and derive the Israel--Stewart theory from kinetic theory. This provides the microscopic basis on which the magnetized theory is then built in the chapters that follow.

\chapter{Microscopic foundations of relativistic hydrodynamics}
\label{chap:kinetic_foundations}

The macroscopic formulation of relativistic dissipative hydrodynamics introduced in the previous chapter is not yet tied to any particular microscopic dynamics. In order to give microscopic meaning to the conserved currents, dissipative quantities, transport coefficients, and relaxation times, one needs an underlying theory for the dynamics of many particles out of equilibrium. For dilute and moderately dilute systems, that role is played by relativistic kinetic theory~\cite{deGroot:1980dk, CercignaniKremer2002}, in which the state of the system is described in terms of the single-particle distribution function evolving in phase space.

Hydrodynamics emerges from kinetic theory after one identifies the macroscopic fields with low moments of the distribution function and systematically reduces the infinitely many microscopic degrees of freedom to a finite set of slow variables. The aim of this chapter is to construct that connection in a form suited for the later developments of this thesis. We first introduce the relativistic distribution function, the Boltzmann equation, and the macroscopic currents obtained from the moments. We then discuss the collision term, its relation to transition rates and differential cross sections, and the generalization to mixtures of gases. After that, we introduce local equilibrium, matching conditions, and the linearization of the collision operator around equilibrium. This provides the basis for the derivation of transient hydrodynamics through Israel--Stewart theory, the 14-moment approximation, and the method of moments~\cite{Israel:1976tn, Israel:1979wp,Denicol:2012es}. The chapter closes with a brief discussion of how electromagnetic fields modify the kinetic description and why these effects are treated in detail only in the next chapter.

\section{Single--particle distribution function and invariant phase space}
\label{sec:one_particle_distribution}

Relativistic kinetic theory describes a many-particle system in terms of a single-particle local momentum distribution function~\cite{deGroot:1980dk}. For each particle species $i$ in a fluid, one introduces a function
\begin{equation}
    f_i(x,k),
\end{equation}
defined at the spacetime point $x^\mu$ and four-momentum $k_i^\mu$. It specifies how the particles of species $i$ are distributed in momentum space at each spacetime point. More precisely, $f_i(x,k)$ is defined such that
\begin{equation}
    f_i(x,k)\, d^3x\, d^3k
\end{equation}
gives the number of particles of species $i$ contained in the phase-space cell around $(\mathbf{x},\mathbf{k})$.

In the relativistic theory, however, the momentum variables are not independent, since the particle four-momentum is constrained to the positive-energy mass shell,
\begin{equation}
    k_i^\mu k_{i\mu} = m_i^2,
    \qquad
    k_i^0 > 0.
\end{equation}
Thus, the distribution function is not defined on the full eight-dimensional phase space, but on the seven-dimensional on-shell phase space. In order for the moments of $f_i$ to transform covariantly, $f_i(x,k)$ itself must be a Lorentz scalar. It is therefore the relativistic generalization of the ordinary phase-space density used in non-relativistic kinetic theory. The Lorentz-invariant momentum-space measure for species $i$ is
\begin{equation}
    dK_i \equiv g_i\,\frac{d^3\mathbf{k}}{(2\pi)^3 k_i^0},
    \label{eq:dKi_measure}
\end{equation}
where $g_i$ is the degeneracy factor. Equivalently, one may write
\begin{equation}
    \int dK_i\,(\cdots)
    =
    g_i\int \frac{d^4k}{(2\pi)^3}\,
    \delta\!\left(k^\mu k_\mu - m_i^2\right)\Theta(k^0)\,(\cdots),
\end{equation}
where $\delta\!\left(k^\mu k_\mu - m_i^2\right)$ is the Dirac delta function, which enforces the on-shell condition $k^\mu k_\mu = m_i^2$, and $\Theta(k^0)$ is the Heaviside step function, which restricts the integration to positive-energy states. In this way, the Lorentz-invariant measure is seen explicitly to be confined to the positive-energy mass shell~\cite{deGroot:1980dk}.

Therefore, the role of the distribution function is that all macroscopic information accessible within kinetic theory is encoded in moments of $f_i$. However, for the purposes of hydrodynamics, one does not require the full microscopic content of the distribution function, but only a finite set of its lowest moments, which capture the slow, long-wavelength degrees of freedom relevant for a macroscopic description.

\section{Macroscopic currents as moments of the distribution function}
\label{sec:macroscopic_currents_from_moments}

In kinetic theory the hydrodynamic variables are constructed from moments of the distribution function \cite{deGroot:1980dk}. In order to understand this, we first describe how the basic conserved currents, i.e., the particle four-current, the net-charge four-current, and the energy-momentum tensor, are expressed in terms of $f_i$. Later in this chapter, these relations will be appropriately derived.

For each particle species one may define the particle four-current
\begin{equation}
    N_i^\mu \equiv \int dK_i\, k_i^\mu f_i,
\end{equation}
and the total particle current is
\begin{equation}
    N^\mu \equiv \sum_i N_i^\mu = \sum_i \int dK_i\, k_i^\mu f_i.
    \label{eq:net_conserved_current}
\end{equation}
Since in this thesis we only consider elastic collisions, the number of particles will be conserved -- this will be demonstrated later in this Chapter.  If the species carries electric charge $q_i$, the corresponding electric net-charge four-current is
\begin{equation}
    J^\mu \equiv \sum_i q_i N_i^\mu
    = \sum_i q_i\int dK_i\,k_i^\mu f_i .
    \label{eq:charge_current_kinetic}
\end{equation}
Similarly, the energy-momentum tensor of the fluid is given by
\begin{equation}
    T^{\mu\nu} \equiv \sum_i \int dK_i\,k_i^\mu k_i^\nu f_i .
    \label{eq:energymomentum_kinetic}
\end{equation}
In addition, one may define the entropy current as~\cite{deGroot:1980dk}
\begin{equation}
    S^\mu \equiv -\sum_i \int dK_i\,k_i^\mu
    \left[f_i\ln f_i - a_i^{-1}(1-a_i f_i)\ln(1-a_i f_i)\right],
\end{equation}
with $a_i = 1$ for fermions, $a_i=-1$ for bosons, and the classical limit understood as $a_i\to 0$. In the Maxwell Boltzmann case this reduces to the simpler expression
\begin{equation}
    S^\mu = -\sum_i \int dK_i\,k_i^\mu\left(f_i\ln f_i - f_i\right).
\end{equation}

These currents contain the microscopic information required by hydrodynamics. Their decomposition with respect to the local four-velocity $u^\mu$ yields the usual hydrodynamic variables. Since in the Landau frame,
\begin{equation*}
    n \equiv u_\mu N^\mu,
    \qquad
    \epsilon \equiv u_\mu u_\nu T^{\mu\nu},
\end{equation*}
while the dissipative parts are
\begin{equation*}
     V^\mu \equiv \Delta^\mu_{\ \nu}N^\nu,\qquad
     \Pi  \equiv -\frac{1}{3}\Delta_{\mu\nu}T^{\mu\nu} - P,\qquad
     \pi^{\mu\nu} \equiv T^{\langle\mu\nu\rangle}.
\end{equation*}
Thus, the microscopic meaning of the hydrodynamic decomposition is immediate. The particle density, energy density, diffusion current, bulk viscous pressure, and shear-stress tensor are all moments of the underlying distribution functions. Hydrodynamics amounts to evolving these moments without keeping track of the full phase space dynamics.

The definitions above identify the hydrodynamic variables as moments of the single particle distribution function, but they do not yet specify how those moments evolve in time. To obtain dynamics, one needs an equation of motion for the distribution itself. In relativistic kinetic theory this role is played by the Boltzmann equation, whose moments generate the conservation laws and, after suitable closure, the constitutive dynamics of dissipative hydrodynamics.

\section{Relativistic Boltzmann equation}
\label{sec:relativistic_boltzmann_equation}

\subsection{Collisionless evolution}
\label{subsec:collisionless_evolution}

In the absence of collisions and external forces, a particle distribution is transported along free streaming trajectories in phase space. In this case the distribution function satisfies
\begin{equation}
    k_i^\mu \partial_\mu f_i(x,k) = 0.
    \label{eq:collisionless_BE}
\end{equation}
This equation states that the distribution is conserved along the worldlines of free particles. It captures ballistic propagation, but it does not contain any mechanism that drives the system toward local equilibrium. Consequently, free streaming generally amplifies momentum space anisotropies rather than reducing them.

Although the collisionless equation is too simple to generate hydrodynamics by itself, it is still useful conceptually. It isolates the purely kinematic part of the phase space evolution and makes clear that equilibration requires interactions among particles.

\subsection{Boltzmann equation with collisions}
\label{subsec:boltzmann_with_collisions}
 
When binary interactions are included, the evolution equation becomes the relativistic Boltzmann equation~\cite{deGroot:1980dk, CercignaniKremer2002},
\begin{equation}
    k_i^\mu \partial_\mu f_i(x,k) = C_i[f],
    \label{eq:boltzmann_eq_general}
\end{equation}
where $C_i[f]$ is the collision term for species $i$. The left-hand side
describes free streaming in spacetime, while the right-hand side accounts
for the local redistribution of particles in momentum space caused by
microscopic scattering processes. The collision term is therefore the
object that encodes all interaction physics, and is derived from a small set of physical assumptions about
binary collisions, combined with a counting argument for how many
particles enter and leave a given phase-space cell.

\paragraph{\textit{a. Setting up the phase-space counting:}}
Consider a single-particle state $(x, k)$.
In the absence of collisions, particles stream freely and their phase-space density is conserved. When binary interactions are present, this
density changes. Denoting the net rate of change per unit spacetime volume per unit momentum-space volume by
$C(x,k)$,
the task is to determine the invariant function $C(x,k)$. The derivation rests on three conditions:
\begin{enumerate}
    \item \textit{Binary collisions only:} At any one spacetime point, at most two particles interact simultaneously. The process considered is
    \begin{equation}
        i(k) + j(k') \longleftrightarrow i(p) + j(p').
        \label{eq:binary_process}
    \end{equation}
    \item \textit{Molecular chaos (Stosszahlansatz):} The momenta of two
    particles about to collide are statistically uncorrelated~\cite{deGroot:1980dk}. The collision rate per unit spacetime volume with incoming momenta
    $(k, k')$ is therefore proportional to the product $f(x,k)\,f(x,k')$.
    \item \textit{Slow variation:} The distribution function varies slowly on the microscopic length and time scales of the interaction. All distribution functions may therefore be evaluated at the same
    spacetime point $x$, and the transition rate carries no $x$-dependence.
\end{enumerate}
\paragraph{\textit{b. Loss and gain:}}
A collision of type \eqref{eq:binary_process} with incoming momenta
$(k, k')$ and outgoing momenta $(p, p')$ removes a particle from the
state $k$. Applying the Stosszahlansatz and integrating over all possible
collision partners and final states, the number of particles
\emph{lost} per unit spacetime volume per unit momentum-space volume around $k$ is
\begin{equation}
    C_{\mathrm{loss}} = \frac{1}{2}
    \int
    \frac{\mathrm{d}^3 k'}{k'^0}\,
    \frac{\mathrm{d}^3 p}{p^0}\,
    \frac{\mathrm{d}^3 p'}{p'^0}\,
    f(x,k)\,f(x,k')\,W_{kk'\to pp'},
    \label{eq:loss_term}
\end{equation}
where the factor of $1/2$ corrects for the double counting of identical final-state pairs $(p,p')$ and $(p',p)$. The quantity $W_{kk'\to pp'}$ is the Lorentz-invariant transition rate for the scattering process $k,k' \to p,p'$. The inverse, or ``restituting'', process has initial momenta $(p, p')$ and final
momenta $(k, k')$, thereby adding a particle to the state $k$. Its
contribution, the \emph{gain} per unit spacetime volume per unit momentum-space volume, is
\begin{equation}
    C_{\mathrm{gain}} = \frac{1}{2}
    \int
    \frac{\mathrm{d}^3 k'}{k'^0}\,
    \frac{\mathrm{d}^3 p}{p^0}\,
    \frac{\mathrm{d}^3 p'}{p'^0}\,
    f(x,p)\,f(x,p')\,W_{pp'\to kk'}.
    \label{eq:gain_term}
\end{equation}
The net change will be given by the gain term minus the loss term, $C(x,k) = C_{\mathrm{gain}} - C_{\mathrm{loss}}$. Before calculating this, we first discuss the fundamental properties of the transition rate. 

\paragraph{\textit{c. The transition rate and the differential cross-section:}}
Further, in the collision term, the quantity $W_{kk'\to pp'}$ is the 
Lorentz-invariant transition rate~\cite{deGroot:1980dk}. It gives the 
probability per unit spacetime volume that an incoming pair with momenta 
$(k, k')$ scatters into an outgoing pair $(p, p')$. Being a Lorentz scalar 
and required to conserve total four-momentum in every collision, its general 
form is
\begin{equation}
    W_{kk'\to pp'}
    =
    \chi(s,\Theta_s)\,
    \delta^{(4)}(k+k'-p-p'),
    \label{eq:transition_rate_Xi}
\end{equation}
where $\chi(s,\Theta_s)$ is a Lorentz-scalar function of the Mandelstam 
variable $s = (k+k')^2$ and the CM scattering angle $\Theta_s$, and the 
four-dimensional delta function enforces energy-momentum conservation event 
by event. The physical content of $\chi$ will be identified below by 
connecting $W$ to the differential cross-section.

To determine this connection, one relates $W$ to the transition probability 
per unit incident flux. The invariant flux for the two incoming particles is
\begin{equation}
    F
    =
    \sqrt{(k_\mu k'^\mu)^2 - m^4}
    =
    \frac{1}{2}\sqrt{s(s-4m^2)},
    \label{eq:invariant_flux}
\end{equation}
and the differential cross-section element is formally defined as
\begin{equation}
    \mathrm{d}\sigma
    =
    \frac{1}{F}\,W_{kk'\to pp'}\,
    \frac{\mathrm{d}^3 p}{p^0}\,
    \frac{\mathrm{d}^3 p'}{p'^0}.
    \label{eq:dsigma_def}
\end{equation}
Integrating \eqref{eq:dsigma_def} over the final states and using the delta 
function in \eqref{eq:transition_rate_Xi}, one works in the CM frame 
where $\boldsymbol{k}+\boldsymbol{k}'=\boldsymbol{0}$, writes 
$\mathrm{d}^3 p = |\boldsymbol{p}|^2\,\mathrm{d}|\boldsymbol{p}|\,\mathrm{d}\Omega$, 
performs the $p'$-integration with the spatial delta function, and evaluates 
the remaining energy integral using
\begin{equation}
    \delta\!\left(\frac{1}{2}\sqrt{s}
    - \sqrt{|\boldsymbol{p}|^2 + m^2}\right)
    =
    \sqrt{\frac{s}{s-4m^2}}\;
    \delta\!\left(|\boldsymbol{p}| - \frac{1}{2}\sqrt{s-4m^2}\right).
    \label{eq:delta_identity}
\end{equation}
The result is
\begin{equation}
    \int\mathrm{d}\sigma = \int\mathrm{d}\Omega\;\frac{\chi(s,\Theta_s)}{(2\pi)^6\, s},
    \label{eq:total_dsigma}
\end{equation}
from which one identifies
\begin{equation}
    \frac{\mathrm{d}\sigma}{\mathrm{d}\Omega} = \frac{\chi(s,\Theta_s)}{(2\pi)^6\,s},
    \label{eq:Xi_identification}
\end{equation}
so that $\chi$ carries the full microscopic information of the interaction. 
Writing $\sigma(s,\Theta_s) \equiv \mathrm{d}\sigma/\mathrm{d}\Omega$, one has
\begin{equation}
    \chi(s,\Theta_s) = (2\pi)^6\, s\, \sigma(s,\Theta_s),
    \label{eq:Xi_explicit}
\end{equation}
and the transition rate \eqref{eq:transition_rate_Xi} takes its explicit form
\begin{equation}
    W_{kk'\to pp'}
    =
    (2\pi)^6\, s\, \sigma(s,\Theta_s)\,
    \delta^{(4)}(k+k'-p-p').
    \label{eq:transition_rate_sigma}
\end{equation}
The total cross-section follows by angular integration,
\begin{equation}
    \sigma_T(s) = \frac{2\pi}{\nu}\int\mathrm{d}\Theta_s\,\sin\Theta_s\,
    \sigma(s,\Theta_s).
    \label{eq:total_cross_section}
\end{equation}
Once $\sigma(s,\Theta_s)$ is specified, the entire collision term is 
determined.

From the representation \eqref{eq:transition_rate_Xi} an important 
symmetry follows at once. Interchanging the labels of initial and final 
states, $(k,k') \leftrightarrow (p,p')$, leaves both $\chi$ (which depends 
only on $s$ and $\Theta_s$) and the delta function unchanged. Therefore,
\begin{equation}
    W_{kk'\to pp'} = W_{pp'\to kk'}.
    \label{eq:detailed_balance_W}
\end{equation}
This is the \emph{detailed-balance property of the transition rate}. It is a 
consequence of the Lorentz scalar structure of $W$ alone and no separate 
assumption about time-reversal invariance of the interaction is required.

\paragraph{\textit{d. Collision term:}}
Using \eqref{eq:detailed_balance_W}, we can write down the collision term in its widely used form,
\begin{equation}
    C[f]
    =
    \frac{1}{\nu}\sum
    \int dK'\, dP\, dP'\; 
    W_{pp'\to kk'}\left(
        f_{p} f_{p'} 
        - 
        f_{k} f_{k'} 
    \right),
    \label{eq:binary_collision_term_classical}
\end{equation}
where we employed the notation $f_{k}=f(x,k)$. We can also extend this expression to describe a mixture of species, and, furthermore, include the effects of quantum statistics. We introduce the factors $\tilde{f}_i \equiv 1 - a_i f_i$ (with $a_i = +1$ for fermions, $a_i = -1$ for bosons, and $a_i = 0$ in the classical limit) that describe the fermi blocking or the boson enhancement effects for the outgoing states. In this case, the
collision term for species $i$ takes the form
\begin{equation}
    C_i[f]
    =
    \frac{1}{\nu_j}\sum_j
    \int dK'_j\, dP_i\, dP'_j\; 
    W_{pp'\to kk'}\left(
        f_{p_i} f_{p'_j} \tilde{f}_{k_i} \tilde{f}_{k'_j}
        - 
        f_{k_i} f_{k'_j} \tilde{f}_{p_i} \tilde{f}_{p'_j}
    \right),
    \label{eq:binary_collision_term}
\end{equation}
where $\nu_j$ is the symmetry factor for identical particles and
$dK_i \equiv g_i\,\mathrm{d}^3k/[(2\pi)^3 k^0]$ is the invariant
phase-space measure. The second term is the loss and the first is the
gain. Collisions do not create or destroy particles; they transfer
occupation between momentum cells. The competition between the two terms
is precisely what drives the system toward local equilibrium.

\paragraph{\textit{e. Collision invariants and conservation laws:}} For species $i$, the Boltzmann equation reads
\begin{equation}
    k_i^\mu \partial_\mu f_i(x,k) = C_i[f].
\end{equation} 
Taking the zeroth moment means integrating this equation over momentum. More generally, for a weight $\psi_i(k)$, we consider
\begin{equation}
    \sum_i \int dK_i\, \psi_i(k)\, k_i^\mu \partial_\mu f_i
    =  \sum_i \partial_\mu \left( \int dK_i\, \psi_i(k)\, k_i^\mu f_i \right) =
    \sum_i \int dK_i\, \psi_i(k)\, C_i[f].
    \label{eq:general_moment_BE}
\end{equation}
The second equality follows since the momentum variables are independent of the spacetime derivative. Thus, each choice of $\psi_i(k)$ generates an equation of motion for a corresponding moment of the distribution function. 

The structure of $C_i[f]$ encodes a fundamental property. If a quantity
$\psi_i(k)$ is conserved in every individual collision,
\begin{equation}
    \psi_i(k)+\psi_j(k')=\psi_i(p)+\psi_j(p'),
    \label{eq:collision_invariant_condition}
\end{equation}
then it is a \emph{collision invariant} and satisfies
\begin{equation}
    \sum_i \int dK_i\,\psi_i(k)\,C_i[f]=0.
    \label{eq:general_collision_invariant_identity}
\end{equation}
To see this, multiply \eqref{eq:binary_collision_term} by $\psi_i(k)$,
sum over species, and integrate over $k$. After relabelling integration
variables $(k,k')\leftrightarrow(p,p')$ in the gain term and
symmetrising between incoming and outgoing momenta, one finds
\begin{equation}
\begin{split}
    & \sum_i \int dK_i\,\psi_i(k)\,C_i[f]
     \\ & \quad=
    \frac{1}{4}
    \sum_{ij}
    \int dK_i\,dK'_j\,dP_i\,dP'_j\;
    W_{kk'\to pp'}
    \left[\psi_i(k)+\psi_j(k')-\psi_i(p)-\psi_j(p')\right]  \\
    & \qquad \times \left(f_{p_i}f_{p'_j}\tilde{f}_{k_i}\tilde{f}_{k'_j}
          -f_{k_i}f_{k'_j}\tilde{f}_{p_i}\tilde{f}_{p'_j}\right).
\end{split}
\end{equation}
The bracket in square parentheses vanishes identically when $\psi_i$ satisfies \eqref{eq:collision_invariant_condition}, and \eqref{eq:general_collision_invariant_identity} follows immediately. Note that the equation above clearly shows that $1$ is a collision invariant. This is not a general result; it happened because of our underlying assumption of elastic collisions. 

The most important collision invariants are the conserved charges and
the four-momentum. As already mentioned, $1$ is also a collision invariant in our case. If we then consider $\psi_i(k)=1$ in Eq.~\eqref{eq:general_moment_BE}, it then becomes
\begin{equation}
    \partial_\mu
    \left(
        \sum_i \int dK_i\, k_i^\mu f_i
    \right)
    =
    \sum_i \int dK_i\, C_i[f] = 0.
\end{equation}
Using the definition of the total particle current,
\begin{equation}
    N^\mu \equiv \sum_i \int dK_i\, k_i^\mu f_i,
\end{equation}
this may be written as the traditional continuity equation related to the conservation of the number of particles
\begin{equation}
    \partial_\mu N^\mu = 0.
\end{equation}
This equation also holds for individual particle species, that is, the number of particles per species is conserved. Naturally, if this is the case, the net-charge currents will also satisfy conservation laws of their own. For instance, the net-charge current will naturally satisfy,
\begin{equation}
    \partial_\mu J^\mu = 0.
\end{equation}

Likewise, to obtain energy-momentum conservation, we consider the collision invariant $\psi_i(k)=k_i^\nu$ and replace it in Eq.~\eqref{eq:general_moment_BE}. This leads to,
\begin{equation}
    \partial_\mu
    \left(
        \sum_i \int dK_i\, k_i^\mu k_i^\nu f_i
    \right)
    =
    \sum_i \int dK_i\, k_i^\nu C_i[f] = 0.
\end{equation}
Recognizing the energy-momentum tensor as
\begin{equation}
    T^{\mu\nu} \equiv \sum_i \int dK_i\, k_i^\mu k_i^\nu f_i,
\end{equation}
one obtains the continuity equation related to energy-momentum conservation already highlighted several times in this thesis,
\begin{equation}
    \partial_\mu T^{\mu\nu}
    = 0.
\end{equation}

These relations are exact consequences of the conservation laws on a microscopic level. These analyses demonstrate that the conserved currents are moments of single-particle distribution function, with weights given by collision invariants. Higher moments, involving additional powers of momentum, are not protected by collision invariance and therefore have non-vanishing collision integrals. These moments describe dissipative relaxation and provide the microscopic origin of viscous hydrodynamics.

\paragraph{\textit{f. The equilibrium distribution:}}
Having established what $C_i[f]$ is and what it conserves, it is
natural to ask: which distribution makes it vanish? A stationary
collisional state satisfies
\begin{equation}
    C_i[f_0] = 0.
\end{equation}
For the integrand of \eqref{eq:binary_collision_term} to vanish for
arbitrary momenta and an arbitrary transition rate, one needs
\begin{equation}
    f_{0,p_i}\,f_{0,p'_j}\,\tilde{f}_{0,k_i}\,\tilde{f}_{0,k'_j}
    =
    f_{0,k_i}\,f_{0,k'_j}\,\tilde{f}_{0,p_i}\,\tilde{f}_{0,p'_j}.
    \label{eq:detailed_balance_raw}
\end{equation}
Taking the logarithm of both sides,
\begin{equation}
    \ln\frac{f_{0,k_i}}{\tilde{f}_{0,k_i}}
    +
    \ln\frac{f_{0,k'_j}}{\tilde{f}_{0,k'_j}}
    =
    \ln\frac{f_{0,p_i}}{\tilde{f}_{0,p_i}}
    +
    \ln\frac{f_{0,p'_j}}{\tilde{f}_{0,p'_j}}.
\end{equation}
The quantity $\phi_i(k) \equiv \ln(f_{0,i}/\tilde{f}_{0,i})$ therefore
satisfies the same additive conservation law as a collision invariant.
Since the only additive invariants of binary collisions conserving charge
and four-momentum are linear combinations of $q_i$ and $k^\mu$, one must
have
\begin{equation}
    \phi_i(k) = \alpha_i(x) - \beta_\mu(x)\,k^\mu,
\end{equation}
where $\alpha_i(x)$ and $\beta^\mu(x)$ are spacetime-dependent
coefficients. Inverting the definition of $\phi_i$, one obtains the
local equilibrium distribution~\cite{Juttner:1911, deGroot:1980dk},
\begin{equation}
    f_{0,i}(x,k)
    =
    \frac{1}{\exp\!\left[\beta_\mu(x)\,k^\mu - \alpha_i(x)\right]+a_i}.
    \label{eq:local_equilibrium_distribution}
\end{equation}
Here $\beta^\mu = u^\mu/T$ is the inverse-temperature four-vector and
$\alpha_i = \mu_i/T$ is the thermal potential for species $i$. In the
classical Maxwell--Boltzmann limit $a_i\to 0$,
\eqref{eq:local_equilibrium_distribution} reduces to
\begin{equation}
    f_{0,i}(x,k)
    = \exp\!\left[\alpha_i(x) - \beta_\mu(x)\,k^\mu\right],
    \qquad
    \ln f_{0,i} = \alpha_i - \beta_\mu k^\mu.
\end{equation}

This is the central result. The equilibrium distribution is not an
ansatz; it is the \emph{unique} solution to $C_i[f_0]=0$ given that
binary collisions conserve charge and four-momentum.
Equation~\eqref{eq:detailed_balance_raw} is satisfied identically for
\eqref{eq:local_equilibrium_distribution}, which is therefore the
kinetic statement of detailed balance.

\subsection{Local equilibrium and matching conditions}
\label{subsec:local_equilibrium_and_matching}

The equilibrium distribution derived above is a solution to $C_i[f_0]=0$,
but a truly \emph{global} equilibrium state must simultaneously satisfy the
collisionless streaming equation, $k_i^\mu\partial_\mu f_i = 0$. Together,
these two requirements force the fields $T$, $\mu_i$, and $u^\mu$ to be
uniform constants throughout spacetime. Such a state is static,
homogeneous, and carries no currents: it is the thermodynamic equilibrium
of an isolated system. While it is an exact solution of the full Boltzmann
equation, it is clearly too restrictive to describe a real fluid undergoing
flow, diffusion, or heat transport.

The concept of \emph{local equilibrium} relaxes this constraint in a
physically motivated way. One retains the functional form of
\eqref{eq:local_equilibrium_distribution} but allows the thermodynamic
fields to vary slowly in spacetime,
\begin{equation}
    T = T(x), \quad \mu_i = \mu_i(x), \quad u^\mu = u^\mu(x).
\end{equation}
The resulting distribution $f_{0i}(x,k)$ is isotropic in the local rest
frame at each spacetime point and makes the collision term vanish,
$C_i[f_{0i}] = 0$, pointwise. However, it does \emph{not} satisfy the full
Boltzmann equation, because the streaming term $k_i^\mu\partial_\mu
f_{0i}$ no longer vanishes when the fields carry spacetime gradients. Local
equilibrium is therefore not an exact solution of the kinetic equation; it
is an idealized reference state around which the true non-equilibrium
distribution is expanded. It is this expansion that connects kinetic
theory to dissipative hydrodynamics.

Hydrodynamics assumes that the system remains close to this local equilibrium
state, characterized by a small number of slowly varying fields~\cite{deGroot:1980dk, Israel:1979wp}. In kinetic theory this is implemented by writing
\begin{equation}
    f_i = f_{0i} + \delta f_i,
    \label{eq:f_split_local_eq}
\end{equation}
where $f_{0i}$ is the local equilibrium distribution and $\delta f_i$ is the off-equilibrium correction.

The local equilibrium distribution is
\begin{equation}
    f_{0i}(x,k) = \left[\exp\left(\beta(x)u(x)\cdot k_i - \alpha_i(x)\right) + a_i\right]^{-1},
    \label{eq:local_eq_dist}
\end{equation}
with
\begin{equation}
    \beta \equiv \frac{1}{T},
    \qquad
    \alpha_i \equiv \frac{\mu_i}{T},
\end{equation}
with $\mu_i$ being the chemical potential of the species $i$ and $T$ the total temperature of the system. Because $f_{0i}$ depends on momentum only through the scalar $u\cdot k_i$, it is isotropic in the local rest frame. Substituting $f_{0i}$ into the kinetic definitions of the conserved currents then yields the ideal fluid forms,
\begin{equation*}
    N_{(0)}^\mu = nu^\mu,
    \qquad
    T_{(0)}^{\mu\nu} = \epsilon u^\mu u^\nu - P\Delta^{\mu\nu}.
\end{equation*}

The split in Eq.~\eqref{eq:f_split_local_eq} is not unique until one fixes the local equilibrium state associated with the full non-equilibrium distribution. This is done through matching conditions~\cite{Israel:1979wp, Denicol:2012cn}. In the Landau frame, the local energy density and conserved charge density of the exact state are required to equal those of the associated equilibrium state. Microscopically this means
\begin{equation}
    u_\mu u_\nu\,\delta T^{\mu\nu} = 0,
    \qquad
    u_\mu\,\delta N^\mu = 0,
    \label{eq:matching_conditions_micro}
\end{equation}
where
\begin{equation}
    \delta N^\mu \equiv \sum_i \int dK_i\,k_i^\mu\delta f_i,
    \qquad
    \delta T^{\mu\nu} \equiv \sum_i\int dK_i\,k_i^\mu k_i^\nu\delta f_i.
\end{equation}
These conditions determine the hydrodynamic fields $T$, $\mu_i$, and $u^\mu$ associated with a given non-equilibrium state. They are the microscopic implementation of the matching procedure already discussed at the macroscopic level in the previous chapter.

Once the matching is fixed, the dissipative quantities acquire direct kinetic expressions. In the Landau frame,
\begin{align}
    V^\mu & = \sum_i \int dK_i\,k_i^{\langle\mu\rangle}\delta f_i,\\
    \Pi & = -\frac{1}{3}\sum_i\int dK_i\,\Delta_{\alpha\beta}k_i^\alpha k_i^\beta\,\delta f_i, \\
    \pi^{\mu\nu} & = \sum_i\int dK_i\,k_i^{\langle\mu}k_i^{\nu\rangle}\delta f_i.
\end{align}
Thus, diffusion, bulk pressure, and shear correspond to specific momentum-space distortions of the local equilibrium distribution. These expressions make clear that computing the dissipative corrections reduces to determining $\delta f_i$, the off-equilibrium part of the distribution function. To obtain $\delta f_i$ systematically, one must go back to the Boltzmann equation itself and study how it constrains the evolution of deviations from local equilibrium. This is the subject of the following subsection.

\subsection{Linearization of the collision operator}
\label{sec:linearization_collision_operator}

Since $\delta f_i$ is the key quantity governing all dissipative phenomena, the natural next step is to derive an equation it must satisfy. The starting point is the full Boltzmann equation~\eqref{eq:boltzmann_eq_general}. Because local equilibrium sets the collision term to zero but not the streaming term, substituting the decomposition $f_i = f_{0i} + \delta f_i$ into the Boltzmann equation yields a non-trivial balance. The streaming of $f_{0i}$ through an inhomogeneous fluid, which acts as a source, is compensated by the collisional relaxation of $\delta f_i$ back toward equilibrium. To make this precise, one considers states close to local equilibrium and expands the collision term to first order in $\delta f_i$~\cite{deGroot:1980dk, CercignaniKremer2002}. Accordingly, one writes
\begin{equation}
    f_i = f_{0i} + \delta f_i,
\end{equation}
with $\delta f_i$ assumed to be small compared to $f_{0i}$. Now, we know that at equilibrium, the distribution can be written as
\begin{equation}
    f_{0i}
    =
    \frac{1}{e^{y_{0i}}+a_i},
    \qquad
    y_{0i}\equiv \beta_\mu k_i^\mu-\alpha_i,
\end{equation}
where $a_i=1, -1, 0$ for fermions, bosons, and in the classical limit, respectively. Away from equilibrium, one may still write the exact distribution in the same form,
\begin{equation}
    f_i
    =
    \frac{1}{e^{y_i}+a_i},
\end{equation}
but now considering that the variable $y_i$ may deviate from its equilibrium form,
\begin{equation}
    y_i = y_{0i}+\delta y_i,
\end{equation}
where $\delta y_i$ is small compared to $y_0i$. Expanding to first order in $\delta y_i$, one obtains
\begin{equation}
    \delta f_i
    =
    \left.\frac{\partial f_i}{\partial y_i}\right|_{y_i=y_{0i}} \delta y_i.
\end{equation}
Since
\begin{equation}
    \frac{\partial f_i}{\partial y_i}
    =
    -\frac{e^{y_i}}{(e^{y_i}+a_i)^2}
    =
    -f_i(1-a_i f_i),
\end{equation}
it follows that, to linear order,
\begin{equation}
    \delta f_i
    =
    -f_{0i}(1-a_i f_{0i})\,\delta y_i.
\end{equation}
Defining
\begin{equation}
    \phi_i \equiv -\delta y_i,
\end{equation}
one arrives at the convenient parametrization
\begin{equation}
    \delta f_i = f_{0i}\tilde f_{0i}\,\phi_i,
    \qquad
    \tilde f_{0i} \equiv 1-a_i f_{0i},
    \label{eq:deltaf_phi_def}
\end{equation}
so that
\begin{equation}
    f_i = f_{0i}\left(1+\tilde f_{0i}\phi_i\right).
    \label{eq:f_expanded_phi}
\end{equation}
Thus, this form is simply the linear expansion of the exact quantum-statistical distribution around local equilibrium.

The collision term for binary elastic collisions has the form
\begin{equation}
    C_i[f]
    =
    \frac{1}{\nu_j}\sum_j
    \int dK'_j\, dP_i\, dP'_j\,
    W_{kk'\to pp'}
    \left(
        f_{p_i} f_{p'_j}\tilde f_{k_i}\tilde f_{k'_j}
        -
        f_{k_i} f_{k'_j}\tilde f_{p_i}\tilde f_{p'_j}
    \right).
    \label{eq:collision_term_for_linearization}
\end{equation}
To linearize this expression, we substitute Eq.~\eqref{eq:f_expanded_phi} into the gain and loss terms and expand to first order in $\phi$. Using that $C_i[f_0]=0$ and, thus, that the equilibrium distribution satisfies the detailed balance relation,
\begin{equation}
    f_{0p_i}f_{0p'_j}\tilde f_{0k_i}\tilde f_{0k'_j}
    =
    f_{0k_i}f_{0k'_j}\tilde f_{0p_i}\tilde f_{0p'_j}.
\label{eq:detailed_balance_for_linearization}
\end{equation}
Carrying out the algebra, one obtains the following expression,
\begin{equation}
    C_i[f]
    =
    \hat L_i[\phi]
    + \mathcal{O}(\phi^2),
\end{equation}
with the linearized collision operator given by
\begin{equation}
    \hat L_i[\phi]
    =
    \frac{1}{\nu_j}\sum_j
    \int dK'_j\, dP_i\, dP'_j\,
    W_{kk'\to pp'}\,
    f_{0k_i}f_{0k'_j}\tilde f_{0p_i}\tilde f_{0p'_j}
    \left(
        \phi_{p_i}+\phi_{p'_j}-\phi_{k_i}-\phi_{k'_j}
    \right).
    \label{eq:linearized_collision_operator}
\end{equation}
This is the basic result of the linearization procedure. It shows that, near equilibrium, the collision term depends linearly on the combination
\begin{equation}
    \phi_{p_i}+\phi_{p'_j}-\phi_{k_i}-\phi_{k'_j},
\end{equation}
which measures the extent to which the non-equilibrium correction fails to be conserved across a microscopic collision.

The structure of Eq.~\eqref{eq:linearized_collision_operator} immediately reveals the special role of collision invariants. If $\phi_i$ is of the form
\begin{equation}
    \phi_i = \alpha_i + \beta_\mu k_i^\mu,
\end{equation}
then
\begin{equation}
    \phi_{p_i}+\phi_{p'_j}-\phi_{k_i}-\phi_{k'_j}=0,
\end{equation}
because charge and four-momentum are conserved in each binary collision. Therefore such distortions are zero modes of the linearized collision operator,
\begin{equation}
    \hat L_i[\phi_{\mathrm{inv}}]=0.
\end{equation}
These are precisely the modes associated with the conserved quantities, and so they cannot relax through collisions. All other distortions are not protected in this way and are damped by the action of $\hat L_i$. In other words, the conserved modes appear as zero modes of $\hat L_i$, while the non-conserved distortions relax with rates determined by the spectrum of the operator.

The linearized Boltzmann equation therefore takes the form
\begin{equation}
    k_i^\mu \partial_\mu f_{0i}
    +
    k_i^\mu \partial_\mu \delta f_i
    =
    \hat L_i[\phi]
    + \mathcal{O}(\phi^2).
    \label{eq:ch3_linearized_Boltzmann}
\end{equation}
It is the starting
point for deriving relativistic hydrodynamics from microscopic dynamics. The next step is to extract the macroscopic consequences of this kinetic equation. This is done by taking moments of the Boltzmann equation with weights corresponding to collision invariants, precisely the procedure established in paragraph~\ref{subsec:boltzmann_with_collisions}$d$. In this way, the conservation laws of hydrodynamics emerge directly from the microscopic dynamics, and the dissipative sector is organized through the Israel--Stewart construction~\cite{Israel:1976tn, Israel:1979wp}, in which the dissipative quantities themselves are promoted to independent dynamical variables and the hierarchy of moment equations is closed through the 14-moment approximation.

\section{Israel Stewart theory and the 14-moment approximation}
\label{sec:IS_and_14_moment}

One approach to dissipative hydrodynamics is to promote the dissipative quantities themselves to independent dynamical variables~\cite{Israel:1976tn, Israel:1979wp}. This is the basic idea behind Israel--Stewart theory. From the kinetic point of view, it is implemented by expanding the non-equilibrium correction $\delta f_i$ in a finite set of moments associated with the hydrodynamic degrees of freedom. Historically, this construction is the relativistic extension of Grad's moment method~\cite{Grad:1949}, in which the distribution function is approximated by a finite polynomial expansion in momentum around local equilibrium. In this section, we shall consider a single component gas for the sake of simplicity. For this reason, the particle species index $i$ shall be omitted from now on.  


\subsection{The 14-moment approximation}
\label{subsec:structure_14_moment}
The formulation of relativistic fluid dynamics from kinetic theory requires a controlled approximation of the single-particle distribution function away from equilibrium. In general, the non-equilibrium correction encodes an infinite number of microscopic degrees of freedom, making an exact treatment impractical. A systematic reduction becomes possible, however, when the system remains sufficiently close to local equilibrium, so that only a limited set of slow, long-wavelength modes governs the dynamics.

This idea can be traced back to the seminal work of Harold Grad \cite{Grad:1949}, who introduced a moment expansion method to approximate the distribution function in non-relativistic gases. Grad’s approach replaces the full momentum dependence by a finite set of moments, effectively projecting the kinetic description onto a small number of macroscopic variables. The 14-moment approximation \cite{Israel:1979wp} builds precisely on this philosophy. It assumes that near equilibrium the full momentum dependence of the deviation $\delta f$ need not be retained explicitly, but can instead be represented through a truncated expansion in momentum space. The truncation is chosen such that it captures all hydrodynamic degrees of freedom relevant for a system with a single conserved charge, while remaining minimal.

In this setting, the slow variables consist of the five equilibrium fields—energy density, conserved charge density, and the three independent components of the flow velocity—supplemented by the dissipative quantities: the bulk viscous pressure, the diffusion current, and the shear-stress tensor. Together, these account for fourteen independent moments, which gives the approximation its name and defines the minimal closure consistent with relativistic dissipative hydrodynamics.
The number 14 arises from a simple counting argument. The local
equilibrium state is specified by five fields: the energy density, one
conserved charge density, and the three independent components of
$u^\mu$. Out of equilibrium, the dissipative corrections introduce
nine additional degrees of freedom,
\begin{equation}
    \underbrace{1}_{\Pi}
    +
    \underbrace{3}_{V^\mu}
    +
    \underbrace{5}_{\pi^{\mu\nu}}
    =
    9,
\end{equation}
so that the total number of macroscopic degrees of freedom is
\begin{equation}
    5 + 9 = 14.
\end{equation}
The 14-moment approximation consists in retaining exactly these 14
degrees of freedom in $\delta f_k$ and setting all remaining moments to
zero.

\paragraph{\textit{The 14-moment ansatz:}}The starting point is to write the distribution function as
\begin{equation}
    f_k = \left[\exp(-y_k) + a\right]^{-1},
    \label{eq:IS_fk}
\end{equation}
where $a = 1, -1, 0$ for fermions, bosons, and a classical gas, respectively.
In local equilibrium, $y_k$ reduces to $y_{0k} = \alpha_0 - \beta_0 E_k$, so
the entire deviation from equilibrium is encoded in
$\delta y_k \equiv y_k - y_{0k}$. Linearizing \eqref{eq:IS_fk} in
$\delta y_k$ yields
\begin{equation}
    \delta f_k \equiv f_k - f_{0k}
    = f_{0k}\tilde{f}_{0k}\,\delta y_k + \mathcal{O}(\delta y_k^2).
    \label{eq:deltaf_IS}
\end{equation}
In the original Israel--Stewart approach, $\delta y_k$ is then expanded in a Taylor series in the particle 4-momentum~\cite{Israel:1976tn, Israel:1979wp},
\begin{equation}
    \delta y_k
    = \varepsilon
    + k^\mu\varepsilon_\mu
    + k^\mu k^\nu\varepsilon_{\mu\nu}
    + k^\mu k^\nu k^\lambda\varepsilon_{\mu\nu\lambda}
    + \cdots.
    \label{eq:deltay_full}
\end{equation}
The expansion is formally exact but
contains infinitely many unknown coefficients
$\varepsilon, \varepsilon_\mu, \varepsilon_{\mu\nu}, \dots$.

Inserting \eqref{eq:deltaf_IS} and \eqref{eq:deltay_full} into the kinetic
definitions of $N^\mu$~\eqref{eq:net_conserved_current} and $T^{\mu\nu}$~\eqref{eq:energymomentum_kinetic} gives
\begin{equation}
    N^\mu
    =
    I_0^\mu
    + \varepsilon\,\mathcal J_0^\mu
    + \mathcal J_0^{\mu\nu}\varepsilon_\nu
    + \mathcal J_0^{\mu\nu\lambda}\varepsilon_{\nu\lambda}
    + \cdots,
    \label{eq:Nmu_expanded}
\end{equation}
\begin{equation}
    T^{\mu\nu}
    =
    I_0^{\mu\nu}
    + \varepsilon\,\mathcal J_0^{\mu\nu}
    + \mathcal J_0^{\mu\nu\lambda}\varepsilon_\lambda
    + \mathcal J_0^{\mu\nu\lambda\rho}\varepsilon_{\lambda\rho}
    + \cdots,
    \label{eq:Tmunu_expanded}
\end{equation}
where the equilibrium tensors are defined as
\begin{equation}
    I_0^{\alpha_1\cdots\alpha_n}
    \equiv
    \int dK\,k^{\alpha_1}\cdots k^{\alpha_n}\,f_{0k},
    \qquad
    \mathcal{J}_0^{\alpha_1\cdots\alpha_n}
    \equiv
    \int dK\,k^{\alpha_1}\cdots k^{\alpha_n}\,f_{0k}\tilde{f}_{0k}.
    \label{eq:IJ_tensors}
\end{equation}
The first term in each of \eqref{eq:Nmu_expanded} and
\eqref{eq:Tmunu_expanded} is the ideal-fluid contribution,
\begin{equation}
    N^\mu_\mathrm{ideal} = I_0^\mu,
    \qquad
    T^{\mu\nu}_\mathrm{ideal} = I_0^{\mu\nu};
    \label{eq:equilibrium_currents}
\end{equation}
the rest are dissipative corrections. Since $I_0^{\alpha_1\cdots\alpha_n}$
and $\mathcal{J}_0^{\alpha_1\cdots\alpha_n}$ are moments of the isotropic
equilibrium distribution, they can be written entirely in terms of $u^\mu$
and $\Delta^{\mu\nu}$, with scalar coefficients given by the thermodynamic
integrals $\mathcal{I}_{nq}$ and $\mathcal{J}_{nq}$~\cite{deGroot:1980dk}.

The 14-moment approximation truncates the Taylor series \eqref{eq:deltay_full}
at second order in the 4-momentum, keeping only the tensors $1$, $k^\mu$, and
$k^\mu k^\nu$,
\begin{equation}
    \delta y_k
    \approx
    \varepsilon + k^\mu\varepsilon_\mu + k^\mu k^\nu\varepsilon_{\mu\nu}.
    \label{eq:phi_general_expansion}
\end{equation}
Without loss of generality, $\varepsilon_{\mu\nu}$ may be taken symmetric and
traceless: any antisymmetric part vanishes when contracted with
$k^\mu k^\nu$, and any trace can always be absorbed into the scalar
$\varepsilon$. The truncated expansion thus carries $1 + 4 + 9 = 14$
independent unknowns in $\{\varepsilon,\,\varepsilon_\mu,\,\varepsilon_{\mu\nu}\}$,
matching exactly the 14 components of $N^\mu$ and $T^{\mu\nu}$.

\paragraph{\textit{Matching procedure:}}

The 14 expansion coefficients can be uniquely related to the 14 components of $N^\mu$ and $T^{\mu\nu}$ through the so-called \emph{matching
procedure}~\cite{Israel:1979wp, deGroot:1980dk}. In the Landau frame, the
relevant constraints are
\begin{align}
    \Delta_{\mu\nu}\,N^\nu &= V_\mu,
    \label{eq:IS_match_n}
    \\
    \Delta^{\mu\nu}_{\alpha\beta}\,T^{\alpha\beta} &= \pi^{\mu\nu},
    \label{eq:IS_match_pi}
    \\
    -\frac{1}{3}\Delta_{\mu\nu}\!\left(T^{\mu\nu} - T^{\mu\nu}_\mathrm{ideal}\right) &= \Pi,
    \label{eq:IS_match_Pi}
    \\
    u_\mu\!\left(N^\mu - N^\mu_\mathrm{ideal}\right) &= 0,
    \label{eq:IS_Landau1}
    \\
    u_\nu\!\left(T^{\mu\nu} - T^{\mu\nu}_\mathrm{ideal}\right) &= 0.
    \label{eq:IS_Landau2}
\end{align}
The first three define the dissipative currents as projections of $N^\mu$
and $T^{\mu\nu}$, while the last two enforce the Landau matching conditions
\eqref{eq:matching_conditions_micro} on the energy and charge densities.
Together, they form a system of 14 equations for the 14 unknowns.

Lorentz covariance dictates the most general form the coefficients can
take. The only rank-0, rank-1, and symmetric traceless rank-2 tensors that
can be built from $u^\mu$, $\Delta^{\mu\nu}$, $\Pi$, $V^\mu$, and
$\pi^{\mu\nu}$ are
\begin{equation}
\begin{aligned}
    \varepsilon
    &=
    E_0\,\Pi,
    \\[3pt]
    \varepsilon_\lambda
    &=
    D_0\,u_\lambda\,\Pi
    + D_1\,V_\lambda,
    \\[3pt]
    \varepsilon_{\lambda\rho}
    &=
    B_0\!\left(\Delta_{\lambda\rho} - 3u_\lambda u_\rho\right)\Pi
    + B_1\,u_{(\lambda}V_{\rho)}
    + B_2\,\pi_{\lambda\rho},
\end{aligned}
\label{eq:phi_truncated}
\end{equation}
where $E_0$, $D_0$, $D_1$, $B_0$, $B_1$, $B_2$ are six scalar thermodynamic
functions to be determined.

Inserting the ansatz \eqref{eq:phi_truncated} into the matching equations
\eqref{eq:IS_match_n}--\eqref{eq:IS_Landau2} yields a closed linear system
for the six scalar coefficients~\cite{Israel:1979wp, deGroot:1980dk},
\begin{equation}
\begin{alignedat}{2}
    \mathcal{J}_{21}D_1 + \mathcal{J}_{31}B_1 &= -1,
    \qquad &
    2\mathcal{J}_{42}B_2 &= 1,
    \\[2pt]
    \mathcal{J}_{21}E_0 + \mathcal{J}_{31}D_0
      - (3\mathcal{J}_{41} + 5\mathcal{J}_{42})B_0 &= 1,
    \qquad &
    \mathcal{J}_{10}E_0 + \mathcal{J}_{20}D_0
      - 3(\mathcal{J}_{30} + \mathcal{J}_{31})B_0 &= 0,
    \\[2pt]
    \mathcal{J}_{31}D_1 + \mathcal{J}_{41}B_1 &= 0,
    \qquad &
    \mathcal{J}_{20}E_0 + \mathcal{J}_{30}D_0
      - 3(\mathcal{J}_{40} + \mathcal{J}_{41})B_0 &= 0.
\end{alignedat}
\label{eq:coeff_system}
\end{equation}
The first three equations come from the dissipative-current definitions
\eqref{eq:IS_match_n}--\eqref{eq:IS_match_Pi}, and the last three from the
Landau matching conditions \eqref{eq:IS_Landau1}--\eqref{eq:IS_Landau2}. The
solution is~\cite{Israel:1979wp, deGroot:1980dk}
\begin{align}
    \frac{E_0}{3B_0}
    & = m^2 + 4\,\frac{\mathcal{J}_{31}\mathcal{J}_{30}
     - \mathcal{J}_{41}\mathcal{J}_{20}}{\mathcal{D}_{20}}
    \equiv -\mathcal{C}_1,
    \qquad
    \frac{D_0}{3B_0}
    = -4\,\frac{\mathcal{J}_{31}\mathcal{J}_{20}
               - \mathcal{J}_{41}\mathcal{J}_{10}}{\mathcal{D}_{20}}
    \equiv -\mathcal{C}_2,
    \label{eq:C1C2_defs} \\
    B_0
    & = -\frac{1}{3\mathcal{C}_1\mathcal{J}_{21}
              + 3\mathcal{C}_2\mathcal{J}_{31}
              + 3\mathcal{J}_{41} + 5\mathcal{J}_{42}},
    \quad
    B_1 = \frac{\mathcal{J}_{31}}{\mathcal{D}_{31}},
    \quad
    D_1 = -\frac{\mathcal{J}_{41}}{\mathcal{D}_{31}},
    \quad
    B_2 = \frac{1}{2\mathcal{J}_{42}}
    \label{eq:ABC_explicit}
\end{align}
with $\mathcal{D}_{20} \equiv \mathcal{J}_{10}\mathcal{J}_{30} - \mathcal{J}_{20}^2$
and $\mathcal{D}_{31} \equiv \mathcal{J}_{21}\mathcal{J}_{41} - \mathcal{J}_{31}^2$.
The numerical values of all six coefficients depend on the equation of state
and on the particle statistics through the integrals $\mathcal{J}_{nq}$ defined as
\begin{equation}
\mathcal{J}_{ij}^\pm
=
\frac{(-1)^j}{(2j+1)!!}
\int dK \,
E_k^{\,i-2j}
\left(\Delta_{\alpha\beta}k^\alpha k^\beta\right)^j
f_{0k}^\pm\tilde{f}_{0k}^\pm.
\label{eq:Jij_general}
\end{equation}
This simplifies further in the classical limit as
\begin{equation}
\mathcal{I}_{ij}^\pm
=
\frac{(-1)^j}{(2j+1)!!}
\int dK \,
E_k^{\,i-2j}
\left(\Delta_{\alpha\beta}k^\alpha k^\beta\right)^j
f_{0k}^\pm.
\label{eq:Iij_general}
\end{equation}

\paragraph{\textit{The 14-moment distribution function:}} It is convenient to re-express the truncated moment expansion for $\phi_k$ explicitly in terms of the dissipative currents. In this case, it becomes,
\begin{equation}
    \phi_k
    =
    \mathcal{A}(E_k)\,\Pi
    +
    \mathcal{B}(E_k)\,k^{\langle\mu\rangle}V_\mu
    +
    \mathcal{C}(E_k)\,k^{\langle\mu}k^{\nu\rangle}\pi_{\mu\nu},
    \label{eq:phi_14moment_derived}
\end{equation}
where we defined the energy in the local rest frame, $E_k \equiv u_\mu k^\mu$. The energy dependence of $\mathcal{A}$, $\mathcal{B}$, $\mathcal{C}$ is
polynomial in $E_k$(at most quadratic in $E_k$), as expected from a truncation at second order in momentum.

The decisive feature of \eqref{eq:phi_14moment_derived} is that $\delta f_k$
depends on the macroscopic dynamics only through $\Pi$, $V^\mu$, and
$\pi^{\mu\nu}$. In order to see this, we define general irreducible moments of $f_k$ as~\cite{Denicol:2012es}
\begin{equation}
    \rho_r^{\mu_1\cdots\mu_\ell}
    \equiv
    \int dK\,E_k^r\,
    k^{\langle\mu_1}\cdots k^{\mu_\ell\rangle}\,
    \delta f_k,
    \label{eq:irred_moments_def}
\end{equation}
where $r$ is a non-negative integer labelling the power of $E_k$
included in the weight. The brackets in the indices of $k^{\langle\mu_1}\cdots k^{\mu_\ell\rangle}$ denote the traceless and symmetric projection orthogonal to the four-velocity. They become proportional to spherical harmonics in the local rest frame of the fluid and thus correspond to a complete and orthogonal basis on $S^\mathfrak{2}$. Thus each moment $\rho_r^{\mu_1\cdots\mu_\ell}$ is
symmetric and traceless in its free indices and orthogonal to $u^\mu$. We note that the
full collection $\{\rho_r^{\mu_1\cdots\mu_\ell}\}$ over all $r$ and
$\ell$ contains all the information of the full non-equilibrium correction $\delta f_k$, that is,
knowing all moments is equivalent to knowing the distribution function
itself.

The 14-moment approximation leads to a dramatic simplification of this collection of fields. Every irreducible moment of the
distribution function of rank 0, 1, and 2 becomes proportional to one of these dissipative
currents,
\begin{equation}
    \rho_r = a_r\,\Pi,
    \qquad
    \rho_r^{\mu} = b_r\,V^\mu,
    \qquad
    \rho_r^{\mu\nu} = c_r\,\pi^{\mu\nu},
    \label{eq:higher_moments_proportional}
\end{equation}
with coefficients $a_r$, $b_r$, $c_r$ given by simple equilibrium momentum
integrals, while moments of rank $\ell \geq 3$ vanish identically. Therefore, this infinite number of degrees of freedom is replaced by the dissipative quantities, $\Pi$, $V^\mu$, and
$\pi^{\mu\nu}$.

The 14-moment approximation specifies the
\emph{form of the distribution function} in terms of the dissipative currents, but \emph{not} their equations of
motion. Nevertheless, once \eqref{eq:phi_14moment_derived} is adopted, any moment of the
Boltzmann equation can be used to derive a closed set of relaxation-type
equations for $\Pi$, $V^\mu$, and $\pi^{\mu\nu}$ \cite{Denicol:2010xn,Denicol:2012es}. The general structure of
these equations will be the same in every case, but the numerical values of the
resulting transport coefficients depend on which moment is chosen for the
projection~\cite{Denicol:2012es}. With the truncation now established, the
remaining task is to extract these equations of motion explicitly. 

\section{Hydrodynamic equations from kinetic theory}
\label{sec:hydro_eqs_from_kinetic_theory}

Once the local equilibrium state has been specified and the
non-equilibrium correction has been truncated through the 14-moment
approximation, the evolution equations for the dissipative currents are
derived by projecting the linearized Boltzmann equation
\begin{equation}
    k^\mu \partial_\mu f_k =  \hat L[\phi]
    \label{eq:Boltzmann}
\end{equation}
onto each irreducible tensor sector. Multiplying both sides by
$E_k^r\,k^{\langle\mu_1}\cdots k^{\mu_\ell\rangle}$ and integrating
over the Lorentz-invariant momentum measure $dK$ generates an exact
equation of motion for each irreducible moment
$\rho_r^{\mu_1\cdots\mu_\ell}$. The projected comoving derivative and collision integral are defined as~\cite{Denicol:2012es}
\begin{equation}
 \dot{\rho}_r^{\langle\mu_1\cdots\mu_\ell\rangle}
    \equiv
    \Delta^{\mu_1\cdots\mu_\ell}_{\nu_1\cdots\nu_\ell}\,
    u^\alpha\partial_\alpha\rho_r^{\nu_1\cdots\nu_\ell},
    \qquad
    \mathcal{C}_r^{\langle\mu_1\cdots\mu_\ell\rangle}
    \equiv
    \Delta^{\mu_1\cdots\mu_\ell}_{\nu_1\cdots\nu_\ell}
    \int dK\,E_k^r\,k^{\nu_1}\cdots k^{\nu_\ell}\,\hat L[\phi].
    \label{eq:projected_dot_collision}
\end{equation}
Using the decomposition $k^\mu = E_k u^\mu + k^{\langle\mu\rangle}$
and substituting the Boltzmann equation in the form
$\delta\dot{f}_k = -\dot{f}_{0k} - E_k^{-1}k^\nu\nabla_\nu f_{0k}
- E_k^{-1}k^\nu\nabla_\nu\delta f_k + E_k^{-1}\hat{L}[\phi]$,
one obtains, after a lengthy but straightforward calculation, the
following exact equations of motion for the moments up to tensor rank
two~\cite{Denicol:2012es}:

\paragraph{\textit{Scalar sector ($\ell = 0$):}}
\begin{equation}
\begin{aligned}
    \dot{\rho}_r - \mathcal{C}_{r-1}
    & = \alpha_r^{(0)}\theta
    - \frac{G_{2r}}{D_{20}}\,\Pi\,\theta
    + \frac{G_{2r}}{D_{20}}\,\pi^{\mu\nu}\sigma_{\mu\nu}
    + \frac{G_{3r}}{D_{20}}\,\nabla_\mu V^\mu
    + r\,\rho_{r-1}^\mu\dot{u}_\mu
    \\[4pt]
    &\qquad 
    - \nabla_\mu\rho_{r-1}^\mu
    + (r-1)\,\rho_{r-2}^{\mu\nu}\sigma_{\mu\nu}
    - \frac{1}{3}
    \bigl[ (r+2)\,\rho_r - (r-1)\,m_0^2\,\rho_{r-2}
    \bigr] \theta,
\end{aligned}
\label{eq:rho_scalar_eom}
\end{equation}

\paragraph{\textit{Vector sector ($\ell = 1$):}}
\begin{equation}
\begin{aligned}
    \dot{\rho}_r^{\langle\mu\rangle} - \mathcal{C}_{r-1}^{\mu}
    &=
    \alpha_r^{(1)}\nabla^\mu\alpha_0
    + \rho_r^\nu\,\omega^\mu_{\ \nu}
    + \frac{1}{3}
    \bigl[ (r-1)\,m_0^2\,\rho_{r-2}^\mu - (r+3)\,\rho_r^\mu
    \bigr]\theta
    \\[4pt]
    &\quad
    + \frac{1}{5}
    \bigl[ (2r-2)\,m_0^2\,\rho_{r-2}^\nu - (2r+3)\,\rho_r^\nu
    \bigr]\sigma^\mu_{\ \nu}
    + r\,\rho_{r-1}^{\mu\nu}\dot{u}_\nu
    + (r-1)\,\rho_{r-2}^{\mu\nu\lambda}\sigma_{\lambda\nu} \\[4pt]
    &\quad
    + \frac{1}{3}
    \bigl[ m_0^2\,r\,\rho_{r-1} - (r+3)\,\rho_{r+1}
    \bigr]\dot{u}^\mu
    - \frac{1}{3}\nabla^\mu
    \bigl[ m_0^2\,\rho_{r-1} - \rho_{r+1} \bigr]
    \\[4pt]
    &\quad
    - \Delta^\mu_{\ \lambda}\nabla_\nu\rho_{r-1}^{\lambda\nu}
    + \frac{\beta_0 J_{r+2,1}}{\varepsilon_0+P_0}
    \bigl[
        \Pi\,\dot{u}^\mu
        -\nabla^\mu\Pi
        +\Delta^\mu_{\ \nu}\partial_\lambda\pi^{\lambda\nu}
    \bigr],
\end{aligned}
    \label{eq:rho_vector_eom}
\end{equation}

\paragraph{\textit{Tensor sector ($\ell = 2$):}}
\begin{equation}
\begin{aligned}
    \dot{\rho}_r^{\langle\mu\nu\rangle} - \mathcal{C}_{r-1}^{\mu\nu}
    & =
    2\alpha_r^{(2)}\sigma^{\mu\nu}
    + 2\rho_r^{\lambda\langle\mu}\omega^{\nu\rangle}_{\ \lambda}
    - \frac{2}{7}
    \bigl[ (2r+5)\,\rho_r^{\lambda\langle\mu}
        -
        2m_0^2(r-1)\,\rho_{r-2}^{\lambda\langle\mu}
    \bigr]\sigma^{\nu\rangle}_{\ \lambda}
    \\[4pt]
    &\quad
    + \frac{2}{15}
    \bigl[
        (r+4)\,\rho_{r+2}
        -
        (2r+3)\,m_0^2\,\rho_r
        +
        (r-1)\,m_0^4\,\rho_{r-2}
    \bigr]\sigma^{\mu\nu}
    \\[4pt]
    &\quad
    + \frac{2}{5}\nabla^{\langle\mu}
    \bigl[
        \rho_{r+1}^{\nu\rangle}
        -
        m_0^2\,\rho_{r-1}^{\nu\rangle}
    \bigr]
    - \frac{2}{5}
    \bigl[
        (r+5)\,\rho_{r+1}^{\langle\mu}
        -
        m_0^2\,r\,\rho_{r-1}^{\langle\mu}
    \bigr]\dot{u}^{\nu\rangle}
    \\[4pt]
    &\quad
    - \frac{1}{3}
    \bigl[
        (r+4)\,\rho_r^{\mu\nu}
        -
        m_0^2(r-1)\,\rho_{r-2}^{\mu\nu}
    \bigr]\theta
    + r\,\rho_{r-1}^{\mu\nu\lambda}\dot{u}_\lambda
    \\[4pt]
    &\quad
    + (r-1)\,\rho_{r-2}^{\mu\nu\lambda\rho}\sigma_{\lambda\rho}
    - \Delta^{\mu\nu}_{\alpha\beta}\nabla_\lambda\rho_{r-1}^{\alpha\beta\lambda}.
\end{aligned}
    \label{eq:rho_tensor_eom}
\end{equation}
Here $\omega^{\mu\nu} \equiv (\nabla^\mu u^\nu - \nabla^\nu u^\mu)/2$
is the vorticity tensor, $I^\mu \equiv \nabla^\mu\alpha_0$, and
$\alpha_r^{(0)}$, $\alpha_r^{(1)}$, $\alpha_r^{(2)}$ are
thermodynamic functions given by equilibrium momentum integrals,
\begin{align}
    \alpha_r^{(0)}
   &  =
    (1-r)\,\mathcal{I}_{r,1} - \mathcal{I}_{r,0}
    - \frac{G_{2r}(\varepsilon_0+P_0) - G_{3r}n_0}{D_{20}},
   \nonumber \\ 
    \alpha_r^{(1)}
   &  =
    \mathcal{J}_{r+1,1} - h_0^{-1}\mathcal{J}_{r+2,1},
    \qquad
    \alpha_r^{(2)}
    =
    \mathcal{I}_{r+2,1} + (r-1)\,\mathcal{I}_{r+2,2},
    \label{eq:alpha_coefficients}
\end{align}
with $G_{nm} \equiv \mathcal{J}_{n0}\mathcal{J}_{m0}
- \mathcal{J}_{n-1,0}\mathcal{J}_{m+1,0}$,
$D_{20} \equiv \mathcal{J}_{10}\mathcal{J}_{30} - \mathcal{J}_{20}^2$,
and $h_0 = (\varepsilon_0+P_0)/n_0$ the enthalpy per particle.
All comoving derivatives of $\alpha_0$ and $\beta_0$ that appear have
been eliminated using the conservation laws for particle number,
energy, and momentum. Deriving these moment equations is also a fundamental part of this thesis. In the next Chapter, we shall revisit this problem in the context of the Boltzmann--Vlasov equation and derive them more carefully.

Equations~\eqref{eq:rho_scalar_eom}--\eqref{eq:rho_tensor_eom}
constitute an infinite, exact hierarchy: the equation for
$\rho_r^{\mu_1\cdots\mu_\ell}$ involves $\rho_r^{\mu_1\cdots\mu_{\ell+1}}$
(through the gradient $\nabla_\lambda\rho_{r-1}^{\alpha\beta\lambda}$),
$\rho_{r\pm1}^{\mu_1\cdots\mu_{\ell\pm1}}$ (through velocity-gradient
couplings), and rank-$(\ell+2)$ moments (through the $\sigma_{\lambda\rho}$
and $\dot{u}_\lambda$ terms). This hierarchy is still completely
equivalent to the Boltzmann equation.

\subsection{Closure via the 14-moment approximation}

The hierarchy is closed by imposing the 14-moment ansatz~\eqref{eq:higher_moments_proportional},
\begin{equation}
    \rho_r = a_r\,\Pi,
    \qquad
    \rho_r^\mu = b_r\,V^\mu,
    \qquad
    \rho_r^{\mu\nu} = c_r\,\pi^{\mu\nu},
    \qquad
    \rho_r^{\mu_1\cdots\mu_\ell} = 0 \quad (\ell \geq 3),
    \label{eq:14moment_closure}
\end{equation}
with coefficients $a_r$, $b_r$, $c_r$ given by equilibrium momentum integrals
through~\eqref{eq:ABC_explicit}. The collision integrals are evaluated by inserting
the 14-moment distribution function~\eqref{eq:phi_14moment_derived} into the
definitions~\eqref{eq:projected_dot_collision}. For each sector this yields a term
proportional to the corresponding dissipative current,
\begin{equation}
    \mathcal{C}_{r-1} = -\mathcal{A}_{r0}^{(0)}\,\Pi,
    \qquad
    \mathcal{C}_{r-1}^{\mu} = -\mathcal{A}_{r0}^{(1)}\,V^\mu,
    \qquad
    \mathcal{C}_{r-1}^{\mu\nu} = -\mathcal{A}_{r0}^{(2)}\,\pi^{\mu\nu},
    \label{eq:collision_closed}
\end{equation}
where $\mathcal{A}_{r0}^{(\ell)}$ are the matrix elements of the linearised collision
operator projected onto the truncated basis~\cite{Denicol:2012es}.
Substituting~\eqref{eq:14moment_closure} and~\eqref{eq:collision_closed}
into~\eqref{eq:rho_scalar_eom}--\eqref{eq:rho_tensor_eom} and retaining only the
$r=0$ moment in each sector (which correspond to $\rho_0 = -(m^2/3) \Pi$, $\rho_0^\mu = V^\mu$ and $\rho_0^{\mu\nu}=\pi^{\mu\nu}$), one obtains a closed set of
relaxation equations for the dissipative currents. Within the 14-moment approximation
the relaxation times and first-order transport coefficients are expressed directly in
terms of the collision matrix elements $\mathcal{A}_{00}^{(\ell)}$ and the thermodynamic
integrals $\alpha_0^{(\ell)}$~\eqref{eq:alpha_coefficients},
\begin{equation}
    \tau_\Pi = \frac{m^2/3}{\mathcal{A}_{00}^{(0)}},
    \quad
    \zeta = -\frac{\alpha_0^{(0)}}{\mathcal{A}_{00}^{(0)}}\,\tau_\Pi,
    \qquad
    \tau_n = \frac{1}{\mathcal{A}_{00}^{(1)}},
    \quad
    \kappa = \frac{\alpha_0^{(1)}}{\mathcal{A}_{00}^{(1)}}\,\tau_n,
    \qquad
    \tau_\pi = \frac{1}{\mathcal{A}_{00}^{(2)}},
    \quad
    \eta = \frac{\alpha_0^{(2)}}{\mathcal{A}_{00}^{(2)}}\,\tau_\pi.
    \label{eq:transport_coeff_kinetic}
\end{equation}
The resulting second-order relaxation equations
read~\cite{Israel:1979wp, Denicol:2012es}
\begin{equation}
    \tau_\Pi\,\dot{\Pi} + \Pi
    =
    -\zeta\,\theta
    - \ell_{\Pi V}\,\nabla_\mu V^\mu
    - \tau_{\Pi V}\,V^\mu\dot{u}_\mu
    - \delta_{\Pi\Pi}\,\Pi\,\theta
    - \lambda_{\Pi V}\,V^\mu\nabla_\mu\alpha_0
    + \lambda_{\Pi\pi}\,\pi^{\mu\nu}\sigma_{\mu\nu},
    \label{eq:IS_bulk_relaxation}
\end{equation}
\begin{equation}
\begin{aligned}
    \tau_n\,\dot{V}^{\langle\mu\rangle} + V^\mu
    &=
    \kappa\,\nabla^\mu\alpha_0
    - V_\nu\,\omega^{\nu\mu}
    - \delta_{nn}\,V^\mu\,\theta
    - \ell_{n\Pi}\,\nabla^\mu\Pi
    + \ell_{n\pi}\,\Delta^\mu_{\ \lambda}\partial_\nu\pi^{\lambda\nu}
    \\[4pt]
    &\quad
    + \tau_{n\Pi}\,\Pi\,\dot{u}^\mu
    - \tau_{n\pi}\,\pi^{\mu\nu}\dot{u}_\nu
    - \lambda_{nn}\,V_\nu\sigma^{\mu\nu}
    + \lambda_{n\Pi}\,\Pi\,\nabla^\mu\alpha_0
    - \lambda_{n\pi}\,\pi^{\mu\nu}\nabla_\nu\alpha_0,
\end{aligned}
    \label{eq:IS_diffusion_relaxation}
\end{equation}
\begin{equation}
\begin{aligned}
    \tau_\pi\,\dot{\pi}^{\langle\mu\nu\rangle} + \pi^{\mu\nu}
    &=
    2\eta\,\sigma^{\mu\nu}
    + 2\pi^{\lambda\langle\mu}\omega^{\nu\rangle}_{\ \lambda}
    - \delta_{\pi\pi}\,\pi^{\mu\nu}\theta
    - \tau_{\pi\pi}\,\pi^{\lambda\langle\mu}\sigma^{\nu\rangle}_{\ \lambda}
    + \lambda_{\pi\Pi}\,\Pi\,\sigma^{\mu\nu}
    \\[4pt]
    &\quad
    - \tau_{\pi n}\,V^{\langle\mu}\dot{u}^{\nu\rangle}
    + \ell_{\pi n}\,\nabla^{\langle\mu}V^{\nu\rangle}
    + \lambda_{\pi n}\,V^{\langle\mu}\nabla^{\nu\rangle}\alpha_0.
\end{aligned}
    \label{eq:IS_shear_relaxation}
\end{equation}
As before, $\dot{A} \equiv u^\mu\partial_\mu A$,
$\theta \equiv \partial_\mu u^\mu$,
$\nabla^\mu \equiv \Delta^{\mu\nu}\partial_\nu$,
$\alpha_0 \equiv \mu/T$, and
$\sigma^{\mu\nu} \equiv \nabla^{\langle\mu}u^{\nu\rangle}$.
The coefficients $\zeta$, $\kappa$, $\eta$ are the bulk viscosity, particle diffusion, and shear viscosity, given by~\eqref{eq:transport_coeff_kinetic}.
The second-order coefficients $\delta_{\Pi\Pi}$, $\ell_{\Pi V}$, $\tau_{\Pi V}$,
$\lambda_{\Pi V}$, $\lambda_{\Pi\pi}$, $\delta_{nn}$, $\ell_{n\Pi}$, $\ell_{n\pi}$,
$\tau_{n\Pi}$, $\tau_{n\pi}$, $\lambda_{nn}$, $\lambda_{n\Pi}$, $\lambda_{n\pi}$,
$\delta_{\pi\pi}$, $\tau_{\pi\pi}$, $\tau_{\pi n}$, $\ell_{\pi n}$,
$\lambda_{\pi\Pi}$, and $\lambda_{\pi n}$ encode the couplings among the three
dissipative sectors and are likewise determined by equilibrium momentum integrals and
collision matrix elements~\cite{Denicol:2012es}. In contrast to relativistic
Navier--Stokes theory~\cite{Eckart:1940te, Landau:1959}, the dissipative currents
relax toward their Navier--Stokes values over finite microscopic timescales
$\tau_\Pi$, $\tau_n$, $\tau_\pi$ set by the collision kernel.

The complete hydrodynamic description is obtained by supplementing the conservation
laws $\partial_\mu N^\mu = 0$ and $\partial_\mu T^{\mu\nu} = 0$
with~\eqref{eq:IS_bulk_relaxation}--\eqref{eq:IS_shear_relaxation}, giving a closed,
causal, transient theory of relativistic dissipative
hydrodynamics~\cite{Israel:1979wp, Denicol:2012es}.

\section{Electromagnetic fields}
\label{sec:brief_note_em_rmhd}

When electromagnetic fields are present, the relativistic fluid framework must be extended
to magnetohydrodynamics: the fluid interacts with the field through the Lorentz force and
Maxwell's equations, while dissipative effects require a second-order Israel--Stewart-type
formulation~\cite{Israel:1976tn,Israel:1979wp} to avoid the acausality and linear instabilities
that afflict the relativistic Navier--Stokes description~\cite{Hiscock:1983zz,Hiscock:1985zz}.
In the non-resistive limit, where the electric conductivity is taken to infinity so that the
electric field vanishes in the local rest frame, the magnetic field still leaves a non-trivial
imprint on the dissipative structure, modifying the tensorial structure of the transport
coefficients and appearing explicitly in the second-order relaxation equations~\cite{Denicol:2018rbw}.

The most straightforward way to derive magnetohydrodynamic equations of motion is to simply repeat the procedure outlined in the previous section for the Boltzmann-Vlasov equation. This is the approach developed in Refs.~\cite{Denicol:2018rbw,Denicol:2019iyh}. This approach has a fundamental flaw: it describes a single-component gas of charged particles and thus can never achieve local charge neutrality -- as a consequence, this system can never achieve global equilibrium. Nevertheless, it serves as an interesting test case to derive the equations and understand their possible mathematical structure.

At the microscopic level, charged particles experience the Lorentz force, and the kinetic
equation becomes the Boltzmann--Vlasov equation~\cite{deGroot:1980dk,CercignaniKremer2002},
\begin{equation}
k^\mu\partial_\mu f_\mathbf{k}
+ q\,F^{\mu\nu}k_\nu\frac{\partial f_\mathbf{k}}{\partial k^\mu}
= C[f],
\label{eq:Boltzmann_EM}
\end{equation}
where the second term on the left-hand side describes the acceleration of particles in
momentum space due to the Lorentz force. The Lorentz is given in terms of the Faraday tensor, $F^{\mu\nu}$, which we express in terms of $u^\mu$, 
\begin{equation}
F^{\mu\nu}
= E^\mu u^\nu - E^\nu u^\mu + \epsilon^{\mu\nu\alpha\beta}\,u_\alpha B_\beta,
\label{eq:Fmunu_decomp}
\end{equation}
with $E^\mu$ and $B^\mu$ being the electric and magnetic fields in the local rest frame of the fluid,
\begin{equation}
E^\mu \equiv F^{\mu\nu}u_\nu, \qquad
B^\mu \equiv \tfrac{1}{2}\,\epsilon^{\mu\nu\alpha\beta}F_{\alpha\beta}\,u_\nu, \qquad
E^\mu u_\mu = 0, \qquad B^\mu u_\mu = 0.
\end{equation}
The collision term is the same one used in the previous section and does not contain any effects due to the electromagnetic fields.

Repeating the steps outlined in the previous section and taking moments of Eq.~\eqref{eq:Boltzmann_EM} with weight
$E_\mathbf{k}^r k^{\langle\mu_1}\cdots k^{\mu_\ell\rangle}$ generates once more an infinite hierarchy
of evolution equations~\cite{Denicol:2012cn,Denicol:2018rbw}. The structure of this hierarchy
mirrors the field-free case, but each sector acquires additional source terms proportional to
$F^{\mu\nu}$. The key results are as follows.

\paragraph{\textit{Scalar moments ($\ell=0$):}}
\begin{align}
\dot{\rho}_r - \mathcal{C}_{r-1}
&= \alpha_r^{(0)}\theta
+ \frac{G_{3r}}{D_{20}}\partial_\mu V_f^\mu
+ \frac{\theta}{3}\!\left[m_0^2(r-1)\rho_{r-2}-(r+2)\rho_r-3\frac{G_{2r}}{D_{20}}\Pi\right]
+ r\rho_{r-1}^\mu\dot{u}_\mu \nonumber\\
&\quad
-\nabla_\mu\rho_{r-1}^\mu
+\!\left[(r-1)\rho_{r-2}^{\mu\nu}+\frac{G_{2r}}{D_{20}}\pi^{\mu\nu}\right]\!\sigma_{\mu\nu}
-\frac{G_{2r}}{D_{20}}\,q E_\nu V_f^\nu
-(r-1)\,q E_\nu\rho_{r-2}^\nu.
\label{eq:scalar_moment_eq}
\end{align}
The last two terms are new relative to the field-free result ~\cite{Denicol:2012cn} and
encode the coupling of the electric field to the diffusion current and to lower-rank moments.
In the non-resistive limit $E^\mu=0$ these terms vanish and the scalar sector is
unaffected by the electromagnetic field.

\paragraph{\textit{Vector moments ($\ell=1$):}}
\begin{align}
\dot{\rho}_r^{\langle\mu\rangle} - \mathcal{C}_{r-1}^{\langle\mu\rangle}
&= \alpha_r^{(1)}\nabla^\mu\alpha_0
+ r\rho_{r-1}^{\mu\nu}\dot{u}_\nu
- \frac{1}{3}\nabla^\mu\!\left(m_0^2\rho_{r-1}-\rho_{r+1}\right)
- \Delta^\mu_{\ \alpha}\!\left(\nabla_\nu\rho_{r-1}^{\alpha\nu}+\alpha_r^h\partial_\kappa\pi^{\kappa\alpha}\right)
\nonumber\\
&\quad
+\frac{1}{3}\!\left[m_0^2(r-1)\rho_{r-2}^\mu-(r+3)\rho_r^\mu\right]\!\theta
+(r-1)\rho_{r-2}^{\mu\nu\lambda}\sigma_{\nu\lambda}
+\rho_{r,\nu}\omega^{\mu\nu} \nonumber\\
&\quad
+\frac{1}{5}\sigma^{\mu\nu}\!\left[m_0^2(2r-2)\rho_{r-2,\nu}-(2r+3)\rho_{r,\nu}\right]
+\alpha_r^h\nabla^\mu\Pi 
\nonumber\\
&\quad
+\frac{1}{3}\!\left[m_0^2 r\rho_{r-1}-(r+3)\rho_{r+1}-3\alpha_r^h\Pi\right]\!\dot{u}^\mu
-\alpha_r^h\,qB\,b^{\mu}_{\ \nu}V_{f}^{\nu}
-qB\,b^{\mu}_{\ \nu}\rho_{r-1}^{\nu}
\nonumber\\
&\quad
+\!\left(\alpha_r^h n_{f0}+\beta_0 J_{r+1,1}\right)q E^\mu
+\frac{1}{3}\!\left[(r+2)\rho_r - m_0^2(r-1)\rho_{r-2}\right]q E^\mu
-(r-1)\,\rho_{r-2}^{\mu\nu}\,q E_\nu.
\label{eq:vector_moment_eq}
\end{align}
The terms proportional to $qB$
are the magnetic Lorentz-force couplings and the terms proportional to $qE^\mu$ are new to the resistive case and vanish when $E^\mu=0$.

\paragraph{\textit{Rank-two tensor moments ($\ell=2$):}}
\begin{align}
\dot{\rho}_r^{\langle\mu\nu\rangle} - \mathcal{C}_{r-1}^{\langle\mu\nu\rangle}
&= 2\alpha_r^{(2)}\sigma^{\mu\nu}
+\frac{2}{15}\!\left[m_0^4(r-1)\rho_{r-2}-(2r+3)m_0^2\rho_r+(r+4)\rho_{r+2}\right]\!\sigma^{\mu\nu}
\nonumber\\
&\quad
+\frac{2}{5}\dot{u}^{\langle\mu}\!\left[m_0^2 r\rho_{r-1}^{\nu\rangle}-(r+5)\rho_{r+1}^{\nu\rangle}\right]
-\frac{2}{5}\nabla^{\langle\mu}\!\left[m_0^2\rho_{r-1}^{\nu\rangle}-\rho_{r+1}^{\nu\rangle}\right]
\nonumber\\
&\quad
+r\rho_{r-1}^{\mu\nu\gamma}\dot{u}_\gamma
-\Delta^{\mu\nu}_{\alpha\beta}\nabla_\lambda\rho_{r-1}^{\alpha\beta\lambda}
+(r-1)\rho_{r-2}^{\mu\nu\lambda\kappa}\sigma_{\lambda\kappa}
+2\rho_r^{\lambda\langle\mu}\omega^{\nu\rangle}_{\ \lambda}
\nonumber\\
&\quad
+\frac{1}{3}\!\left[m_0^2(r-1)\rho_{r-2}^{\mu\nu}-(r+4)\rho_r^{\mu\nu}\right]\!\theta
+\frac{2}{7}\!\left[m_0^2(2r-2)\rho_{r-2}^{\kappa\langle\mu}-(2r+5)\rho_r^{\kappa\langle\mu}\right]\!\sigma^{\nu\rangle}_{\ \kappa}
\nonumber\\
&\quad
-2\,qB\,b^{\alpha\beta}\Delta^{\mu\nu}_{\ \ \alpha\kappa}g_{\lambda\beta}\rho_{r-1}^{\kappa\lambda}
\nonumber\\
&\quad
+2\,q E^{\langle\mu}\rho_r^{\nu\rangle}
-(r-1)\,\Delta^{\mu\nu}_{\alpha\beta}\!\left[q E_\lambda\rho_{r-2}^{\alpha\beta\lambda}
+\frac{2}{5}\,q E^{\langle\alpha}\!\left(m_0^2\rho_{r-2}^{\beta\rangle}-\rho_r^{\beta\rangle}\right)\right].
\label{eq:tensor_moment_eq}
\end{align}
The term proportional to $qB$ is due to the magnetic Lorentz-force and the remaining terms proportional to $qE^\mu$ are due to the electric Lorentz force and vanish in the non-resistive limit.

As before, this infinite moment hierarchy is closed by the 14-moment
approximation ~\eqref{eq:IS_bulk_relaxation}--~\eqref{eq:IS_shear_relaxation}, which
is essentially unmodified by the inclusion of electromagnetic fields. Replacing these relations into the moment equations for $r=0$, one simply obtains:

\paragraph{\textit{Bulk viscous pressure}}
\begin{align}
\tau_\Pi\dot{\Pi}+\Pi
&= -\zeta\,\theta
-\ell_{\Pi V}\nabla_\mu V_f^\mu
-\tau_{\Pi V}V_f^\mu\dot{u}_\mu
-\delta_{\Pi\Pi}\Pi\,\theta
-\lambda_{\Pi V}V_f^\mu\nabla_\mu\alpha_0
+\lambda_{\Pi\pi}\pi^{\mu\nu}\sigma_{\mu\nu}
-\delta_{\Pi VE}\,q\,V_f^\mu E_\mu,
\label{eq:bulk_relaxation}
\end{align}
with relaxation time $\tau_\Pi=m^2/(3\,\mathcal{A}_{00}^{(0)})$ and bulk viscosity
$\zeta=-(\alpha_0^{(0)}/\mathcal{A}_{00}^{(0)})\tau_\Pi$. The last term is the sole new
contribution from the electric field in the scalar sector and it vanishes in the non-resistive
limit $E^\mu=0$, restoring the result of Ref.~\cite{Denicol:2018rbw}.

\paragraph{\textit{Particle diffusion current}}
\begin{align}
\tau_V\dot{V}_f^{\langle\mu\rangle}+V_f^\mu
&= \kappa\nabla^\mu\alpha_0
-\tau_V V_{f,\nu}\omega^{\nu\mu}
-\delta_{VV}V_f^\mu\theta
-\ell_{V\Pi}\nabla^\mu\Pi
+\ell_{V\pi}\Delta^{\mu\nu}\nabla^\lambda\pi_{\nu\lambda}
\nonumber\\
&\quad
+\tau_{V\Pi}\Pi\dot{u}^\mu
-\tau_{V\pi}\pi^{\mu\nu}\dot{u}_\nu
-\lambda_{VV}V_{f,\nu}\sigma^{\mu\nu}
+\lambda_{V\Pi}\Pi\nabla^\mu\alpha_0
-\lambda_{V\pi}\pi^{\mu\nu}\nabla_\nu\alpha_0
\nonumber\\
&\quad
-\delta_{VB}\,qB\,b^{\mu}_{\ \nu}V_f^\nu
+\delta_{VE}\,q\,E^\mu
+\delta_{V\Pi E}\,q\,\Pi\, E^\mu
+\delta_{V\pi E}\,q\,\pi^{\mu\nu}E_\nu,
\label{eq:diffusion_relaxation}
\end{align}
with $\tau_V=1/\mathcal{A}_{00}^{(1)}$ and $\kappa=(\alpha_0^{(1)}/\mathcal{A}_{00}^{(1)})\tau_V$.
The term $\delta_{VB}$ is the magnetic rotation of the diffusion current already present in
the non-resistive case~\cite{Denicol:2018rbw}. The three new terms proportional to $q E^\mu$
are specific to the resistive case and vanish when $E^\mu=0$.

\paragraph{\textit{Shear-stress tensor}}
\begin{align}
\tau_\pi\dot{\pi}^{\langle\mu\nu\rangle}+\pi^{\mu\nu}
&= 2\eta\,\sigma^{\mu\nu}
+2\tau_\pi\pi_\lambda^{\ \langle\mu}\omega^{\nu\rangle\lambda}
-\delta_{\pi\pi}\pi^{\mu\nu}\theta
-\tau_{\pi\pi}\pi^{\lambda\langle\mu}\sigma_\lambda^{\ \nu\rangle}
+\lambda_{\pi\Pi}\Pi\sigma^{\mu\nu}
\nonumber\\
&\quad
-\tau_{\pi V}V_f^{\langle\mu}\dot{u}^{\nu\rangle}
+\ell_{\pi V}\nabla^{\langle\mu}V_f^{\nu\rangle}
+\lambda_{\pi V}V_f^{\langle\mu}\nabla^{\nu\rangle}\alpha_0
\nonumber\\
&\quad
-\delta_{\pi B}\,qB\,\Delta^{\mu\nu}_{\ \ \alpha\kappa}b^\alpha_{\ \beta}\pi^{\kappa\beta}
+\delta_{\pi VE}\,q\,E^{\langle\mu}V_f^{\nu\rangle},
\label{eq:shear_relaxation}
\end{align}
with $\tau_\pi=1/\mathcal{A}_{00}^{(2)}$ and $\eta=(\alpha_0^{(2)}/\mathcal{A}_{00}^{(2)})\tau_\pi$.
The term $\delta_{\pi B}$ is the magnetic shear-mixing coefficient of the non-resistive
case~\cite{Denicol:2018rbw}. The last term couples the electric field to the diffusion current
in the tensor sector and is absent when $E^\mu=0$.

Equations~\eqref{eq:bulk_relaxation}, \eqref{eq:diffusion_relaxation},
and~\eqref{eq:shear_relaxation}, together with the conservation laws $\partial_\mu J^\mu=0$
and $\partial_\mu T^{\mu\nu}=F^{\nu\lambda}J_{\lambda}$ and the Maxwell´s equations, provide
a closed causal description of the resistive magnetized fluid. Setting $E^\mu=0$ recovers the
non-resistive theory of Ref.~\cite{Denicol:2018rbw}, confirming that the new coefficients
$\delta_{\Pi VE}$, $\delta_{VE}$, $\delta_{V\Pi E}$, $\delta_{V\pi E}$, and $\delta_{\pi VE}$
are purely electric in origin. In the present single-component setting, however, these coefficients are kept
symbolic. The reason being that a plasma with only one charged species cannot
equilibrate to a neutral state, since there is no oppositely charged component
through which the net charge density can relax. Therefore, the equations above should be understood as a motivating single-component construction rather than the complete resistive theory. The full kinetic derivation will be carried out for a two-component plasma in Chapter~\ref{chap:resistive_rmhd_two_component}, where the corresponding transport coefficients are obtained explicitly.

\section{Summary and Limitations of single-component framework}
\label{sec:summary_limitations}

This chapter built the kinetic-theory foundations of relativistic dissipative
magnetohydrodynamics. Starting from the relativistic Boltzmann equation,
all macroscopic currents were constructed as momentum-space averages of the single-particle
distribution function, and the conservation laws followed directly from the collision
invariants. Closing the resulting moment hierarchy at the 14-moment level yielded
Israel--Stewart relaxation equations for the bulk viscous pressure, the diffusion current,
and the shear-stress tensor, with transport coefficients and relaxation times tied explicitly
to the underlying microscopic interactions. Extending the framework to include electromagnetic
fields via the Boltzmann--Vlasov equation showed that the Lorentz force does more than add a
body force: it restructures the dissipative sector, introducing the magnetic coupling
coefficients $\delta_{VB}$ and $\delta_{\pi B}$ in the non-resistive limit, and the
additional electric coefficients $\delta_{\Pi VE}$, $\delta_{VE}$, $\delta_{V\Pi E}$,
$\delta_{V\pi E}$, and $\delta_{\pi VE}$ once the electric field is retained as a dynamical
variable. These are predictions of the kinetic approach that are invisible to a purely
macroscopic construction.

However, the single-component  framework presented here for the electromagnetic field case rests on several simplifying assumptions. The
particles are taken to be spinless and non-polarized, so magnetization and polarization
effects are absent. Landau quantization is neglected throughout, restricting the validity
of the theory to the regime $T^2 \gg qB$ where the thermal energy far exceeds the cyclotron
scale. The 14-moment truncation captures the essential causal dissipative structure, but
yields only approximate transport coefficients. Higher accuracy would require a higher-order
moment expansion or an alternative closure scheme.

\emph{The most significant physical restriction that remains is the single-component assumption.} The theory was constructed for a single charged species, which means that the relative dynamics between oppositely charged constituents, namely charge separation, inter-species
friction, and the independent response of each species to the Lorentz force, are entirely absent. In a realistic plasma these effects are far from negligible: the two species drift differently, a relative current develops, and the dissipative sector acquires new dynamical variables that a one-component treatment cannot accommodate. More fundamentally, a single-component charged fluid cannot reach thermal and chemical equilibrium, since there is no mechanism for the annihilation or creation of charge carriers and the net charge density is permanently non-zero. The theory is therefore qualitatively inadequate for any realistic plasma, and its use is limited to formal benchmarking against the two-component case. 

Lifting this restriction is the central task of this thesis and will be the topic of the next chapters, which extends the kinetic framework to a two-component plasma and derives the corresponding resistive dissipative magnetohydrodynamics for a system with both positive and negative charge carriers. We shall demonstrate that the equations are significantly modified when a more realistic system is considered and that the mathematical structure uncovered so far will not be sufficient to describe a general theory of magnetohydrodynamics.
\chapter{Relativistic non-resistive dissipative magnetohydrodynamics for a two-component gas}
\label{ch:two_component_rmhd}

The purpose of this chapter is to construct relativistic dissipative magnetohydrodynamics for the simplest plasma that can meaningfully approach equilibrium, namely a dilute gas composed of two massless species carrying opposite electric charges in the classical limit. The central question is how the presence of a magnetic field modifies the structure of dissipative relativistic fluid dynamics already at the microscopic level, and how a causal second-order theory emerges once the Boltzmann equation is projected onto a finite set of moments and truncated within the 14-moment approximation. Throughout this chapter the electric field is set to zero, so that the only magnetic field effects are retained. This restriction isolates the purely magnetic modifications to the dissipative structure and keeps the algebra simple. We will further include the effects of the electric field in the next chapter.

The logic of this chapter follows the microscopic construction of the two-component theory as discussed in Ref.~\cite{Kushwah:2024zgd}. We first describe the electromagnetic sector and fix the geometric tensors that will later be needed to project the equations. We then formulate kinetic theory for two oppositely charged species, discuss the structure of the collision term, and introduce the hydrodynamic moments for each component. After that, the 14-moment approximation~\cite{Grad:1949,Denicol:2012cn,Denicol:2012es} is implemented in the shear sector, which provides a closed description in terms of the species dependent shear tensors. This makes it possible to define the total and relative shear tensors and derive their coupled equations of motion. Finally, we show that the magnetic field breaks the degeneracy of the five independent components of the shear tensor and naturally leads to a decomposition into longitudinal, semi transverse, and fully transverse sectors. This decomposition is the key step that reveals why the standard Israel--Stewart-like picture becomes insufficient in the strong field regime.

\section{Kinetic formulation}

The microscopic starting point here is not a single Boltzmann equation, but rather one equation for each species~\cite{CercignaniKremer2002,DeGroot}. Denoting the distribution functions of the positively and negatively charged particles by $f_k^{+}$ and $f_k^{-}$, their dynamics is governed by
\begin{equation}
k^\mu \partial_\mu f_k^i + q_i F^{\mu\nu} k_\nu \frac{\partial f_k^i}{\partial k^\mu}
=
\sum_j C[f^i,f^j],
\qquad i=\pm.
\label{eq:ch5_BE_species}
\end{equation}
where $C[f^i,f^j]$ now denotes the collision kernel. 
The first term describes free streaming, while the second term is the Vlasov force term generated by the electromagnetic field. The right hand side contains collisions, and the sum over $j$ indicates that particles of one species can collide both with particles of the same species and with particles of the opposite species. This is precisely where the two-component nature of the system enters dynamically.

Equation \eqref{eq:ch5_BE_species} already shows the two sources of coupling between the species. First, because the charges are opposite, the Lorentz force acts differently on the two species. Second, because collisions occur both within each species and between distinct species, the relaxation of one component cannot be described independently of the other. The resulting dissipative dynamics is therefore matrix valued in species space even before any hydrodynamic truncation is introduced.

At the macroscopic level, the total energy-momentum tensor and the electric current are obtained by summing the corresponding moments of the distribution functions over the two species,
\begin{equation}
T^{\mu\nu}=\sum_{i=\pm} \langle k^\mu k^\nu\rangle_i,
\qquad
N^\mu=\sum_{i=\pm} q_i \langle k^\mu\rangle_i.
\end{equation}
Here $\langle \cdots \rangle_i$ $\equiv$ $\int dK_i (\cdots) f_i $. 

The collision term appearing in the Boltzmann equation for each species contains contributions from binary scattering processes involving both identical and opposite species. Denoting the particle species by indices $i,j \in \{+,-\}$, the collision operator acting on species $i$ can be written compactly as
\begin{equation}
C[f^i,f^j]
=
\gamma_{ij}\int dK' dP dP'\,
W_{kk'\rightarrow pp'}^{ij}
\Big[
f_i(p)f_j(p') - f_i(k)f_j(k')
\Big],
\end{equation}
where we defined $\gamma_{ij} \equiv 1-\delta_{ij}/2$ and we neglect the effect of quantum statistics. The first term is a gain term, describing scattering into the momentum state $k$, while the second is a loss term, describing scattering out of the same state. Note that the index $i$ labels the species whose distribution is being evolved, while the index $j$ labels the species with which it scatters.

Here, the transition rate is defined in terms of the total cross section, $\sigma_T$, as \cite{DeGroot}
\begin{equation}
   W_{kk^{'}\rightarrow pp^{'}}^{ij} = s\sigma_T^{ij} (2\pi)^5 \delta^{(4)}( k^{\mu} + k^{'\mu} - p^{\mu} - p^{'\mu}).
\end{equation}
The Dirac delta distribution imposes four-momentum conservation, while the factor $s$ indicates that the invariant transition rate depends on the center of mass energy of the binary collision. This expression hold for all the channels $++$, $--$, and $-+$ \footnote{Since transition rates are invariant under the permutation of the incoming or the outgoing particles, the cross-sections satisfy the relation, $\sigma_T^{+-}=\sigma_T^{-+}$}. 

Therefore, the full collision term for species $i$ is then obtained by summing over all scattering channels. Although all contributions share the same gain minus loss structure, the terms with $j=i$ describe self interactions within a given species, while the terms with $j\neq i$ describe interspecies scattering. The latter are responsible for transferring momentum anisotropy between the two components. Thus, each species evolves under the combined effect of its own self interactions and its interactions with the other species, which leads to a coupled dissipative dynamics.

Similar to earlier sections, we introduce the energy-momentum tensor, $T^{\mu\nu}_{\pm}$, and the net-charge current, $N^{\mu}_{\pm}$, of each particle species and decompose these tensors in terms of the fluid 4-velocity,
\begin{equation}
\begin{split}
T^{\mu \nu }_{\pm} &= \epsilon^{\pm} u^{\mu }u^{\nu }-\Delta ^{\mu \nu } P^{\pm} +h^{\mu}_{\pm}u^{\nu}+h^{\nu}_{\pm}u^{\mu}
+\pi ^{\mu \nu }_{\pm}, \\
N^{\mu}_{\pm} &= n^{\pm}u^{\mu }+V^{\mu }_{\pm}.
\end{split}
\end{equation}
This notation follows that of the single particle case, but now all quantities carry a species label. Therefore, $\epsilon^{\pm}$, $P^{\pm}$, $h^{\mu}_{\pm}$, $\pi^{\mu\nu}_{\pm}$, $n^{\pm}$, and $V^{\mu}_{\pm}$ are the energy density, isotropic pressure, energy diffusion current, shear-stress tensor, net-charge density, and diffusion current associated with the corresponding $\pm$ species.

In the two-component case, each variable is defined by a suitable projection of the corresponding current with $u^\mu$ and $\Delta^{\mu\nu}$,
\begin{eqnarray*}
\epsilon^{\pm} &\equiv&u_{\mu }u_{\nu }T^{\mu \nu }_{\pm}\text{,}\qquad P^{\pm}+\Pi^{\pm} \equiv-\frac{1}{3}%
\Delta _{\mu \nu }T^{\mu \nu}_{\pm}\text{,} \qquad \pi ^{\mu \nu }_{\pm} \equiv
T^{\langle\mu\nu\rangle }_{\pm}\text{,}\\
h^{\mu}_{\pm} &\equiv& u_{\alpha}T^{\langle\mu\rangle\alpha}_{\pm}  \text{, } \hspace{4.2 em} n^{\pm} \equiv u_{\mu }N^{\mu }_{\pm}\text{, } \hspace{4.2 em} V^{\mu }_{\pm}\equiv N^{\left\langle \mu
\right\rangle }_{\pm}.
\end{eqnarray*}
These definitions are completely analogous to those of standard relativistic hydrodynamics, but one must keep in mind that they now apply separately to each species before the total macroscopic currents are formed.

\subsection{Matching conditions}

The matching procedure follows the same logic as in the usual hydrodynamics as discussed in Sec.~\ref{sec:matching_conditions}. A reference local equilibrium state is characterized by a temperature $T(x)$, a chemical potential $\mu(x)$, and a fluid four-velocity $u^\mu(x)$, and suitable moments of the full distribution are required to coincide with those of the equilibrium state~\cite{Denicol_Rischke,Israel:1976tn}. In the two-component case, however, the two species carry independent chemical potentials and the matching conditions must be imposed consistently across both components simultaneously.

Therefore, for the two-component system considered here, we introduce a reference local equilibrium state for the positively and negatively charged species and decompose the scalar moments of each species as
\begin{equation}
\epsilon^\pm \equiv \epsilon_0^\pm + \delta\epsilon^\pm,
\qquad
P^\pm \equiv P_0^\pm + \Pi^\pm,
\qquad
n^\pm \equiv n_0^\pm + \delta n^\pm.
\label{eq:matching_decomp_species}
\end{equation}
Here, $\epsilon_0^\pm$, $P_0^\pm$, and $n_0^\pm$ denote the equilibrium energy density, equilibrium pressure, and equilibrium number density of the corresponding particle species, while $\delta\epsilon^\pm$, $\Pi^\pm$, and $\delta n^\pm$ describe the deviations from local equilibrium. Thus, the non-equilibrium state is written as an equilibrium background plus corrections that carry dissipative information.

Since we are considering a system of massless particles, the equation of state of each species is conformal,
\begin{equation}
\epsilon^\pm = 3P^\pm,
\qquad
\epsilon_0^\pm = 3P_0^\pm.
\label{eq:species_conformal_eos}
\end{equation}
Subtracting the equilibrium relation from the full one gives
\begin{equation}
\delta\epsilon^\pm = 3\Pi^\pm.
\label{eq:deltaeps_bulk_relation}
\end{equation}
This relation is important because it shows that, for massless particles, the non-equilibrium correction to the energy density is not an independent scalar quantity. It is directly tied to the bulk correction to the isotropic pressure. Even though the later derivation will focus on the shear sector, this relation already constrains the scalar sector strongly.

To define the hydrodynamic fields $u^\mu$, $T$, and $\mu$, we impose Landau matching conditions~\cite{Israel:1976tn}. In the present two-component system, the Landau frame must be defined with respect to the \emph{total} energy-momentum tensor, not the tensor of each species separately.%
\footnote{It is also possible to introduce one velocity for each particle-species by applying the Landau matching condition per particle species. This would lead to a two-fluid description, which is not what we aim to describe here.} Therefore, the common fluid velocity is chosen such that
\begin{equation}
u_\mu \left( T_+^{\mu\nu} + T_-^{\mu\nu} \right)
=
\epsilon_0(\mu,T)\,u^\nu .
\label{eq:landau_match_T}
\end{equation}
Likewise, since the electric current is obtained from the difference between the currents of the positively and negatively charged species, the corresponding matching condition for the net electric charge density is
\begin{equation}
u_\mu \left( N_+^\mu - N_-^\mu \right)
=
n_0(\mu,T).
\label{eq:landau_match_N}
\end{equation}
These two conditions determine the local temperature $T$ and electric charge chemical potential $\mu$ from the scalar moments of the microscopic theory.

Using the definitions introduced earlier for the species currents,
\begin{equation}
T_{\pm}^{\mu\nu}
=
\epsilon^\pm u^\mu u^\nu
-
\Delta^{\mu\nu} P^\pm
+
h_\pm^\mu u^\nu
+
h_\pm^\nu u^\mu
+
\pi_\pm^{\mu\nu},
\qquad
N_\pm^\mu
=
n^\pm u^\mu + V_\pm^\mu,
\label{eq:species_decomp_matching}
\end{equation}
we can evaluate the contractions appearing in the matching conditions explicitly. Contracting Eq.~\eqref{eq:species_decomp_matching} with $u_\mu$ gives
\begin{equation}
u_\mu T_\pm^{\mu\nu}
=
\epsilon^\pm u^\nu + h_\pm^\nu,
\label{eq:uT_species}
\end{equation}
where we used
\begin{equation}
u_\mu \Delta^{\mu\nu}=0,
\qquad
u_\mu \pi_\pm^{\mu\nu}=0,
\qquad
u_\mu h_\pm^\mu = 0.
\end{equation}
Therefore,
\begin{equation}
u_\mu \left(T_+^{\mu\nu}+T_-^{\mu\nu}\right)
=
(\epsilon^+ + \epsilon^-)\,u^\nu + h_+^\nu + h_-^\nu.
\label{eq:uT_total_expand}
\end{equation}
Comparing this with the Landau matching condition \eqref{eq:landau_match_T}, one obtains two independent consequences. First, projecting along $u_\nu$ yields
\begin{equation}
\epsilon^+ + \epsilon^- = \epsilon_0(\mu,T),
\label{eq:energy_matching_total}
\end{equation}
which implies
\begin{equation}
\delta\epsilon^+ + \delta\epsilon^- = 0.
\label{eq:deltaeps_sum_zero}
\end{equation}
Second, projecting orthogonally to $u^\nu$ with $\Delta^\alpha_{\ \nu}$ gives
\begin{equation}
h_+^\alpha + h_-^\alpha = 0.
\label{eq:total_energy_diffusion_zero}
\end{equation}
Thus, the total energy diffusion current vanishes in the Landau frame, even though the individual energy diffusion currents $h_\pm^\mu$ do not need to vanish separately.

A similar analysis follows from the charge matching condition. Contracting $N_\pm^\mu$ with $u_\mu$ yields
\begin{equation}
u_\mu N_\pm^\mu = n^\pm,
\end{equation}
since $u_\mu V_\pm^\mu = 0$. Therefore Eq.~\eqref{eq:landau_match_N} becomes
\begin{equation}
n^+ - n^- = n_0(\mu,T),
\label{eq:charge_matching_total}
\end{equation}
and hence
\begin{equation}
\delta n^+ - \delta n^- = 0.
\label{eq:deltan_difference_zero}
\end{equation}
This means that the non-equilibrium corrections to the number densities are also constrained: their difference must vanish once the local chemical potential and temperature have been fixed through matching.

Equations \eqref{eq:deltaeps_sum_zero} and \eqref{eq:deltan_difference_zero} play a central conceptual role. They show that the scalar non-equilibrium corrections of the two species are not all independent. The sum of the energy-density corrections must vanish, while the difference of the number-density corrections must vanish. In this sense, the matching conditions remove redundant scalar degrees of freedom and ensure that the local equilibrium state is uniquely tied to the conserved currents.

It is useful to make the role of $\mu$ and $T$ more explicit. The equilibrium state is parameterized by the local equilibrium distribution function,
\begin{equation}
f_{0k}^\pm = \exp\!\left(\alpha_0^\pm - \beta_0\, u_\mu k^\mu \right),
\label{eq:eq_dist_matching}
\end{equation}
where
\begin{equation}
\beta_0 \equiv \frac{1}{T},
\qquad
\alpha_0^\pm \equiv \frac{\mu_\pm}{T}.
\end{equation}
For a two-component gas with opposite charges, the equilibrium chemical potentials of the two species are related through the electric charge chemical potential,
\begin{equation}
\mu_+ = \mu,
\qquad
\mu_- = -\mu,
\end{equation}
so that
\begin{equation}
\alpha_0^+ = \frac{\mu}{T},
\qquad
\alpha_0^- = -\frac{\mu}{T}.
\label{eq:alpha_pm_matching}
\end{equation}
Hence, the matching conditions determine precisely the thermodynamic fields that enter the equilibrium distributions around which the kinetic theory is expanded.

We further emphasize that Eqs.~\eqref{eq:deltaeps_sum_zero} and \eqref{eq:deltan_difference_zero} define the local temperature $T$ and the electric charge chemical potential $\mu$. Once this is done, the bulk viscous pressure of the total system vanishes identically. To see this explicitly, note that the total bulk correction is
\begin{equation}
\Pi \equiv \Pi^+ + \Pi^-.
\end{equation}
Using Eq.~\eqref{eq:deltaeps_bulk_relation} for each species,
\begin{equation}
\delta\epsilon^+ + \delta\epsilon^- = 3(\Pi^+ + \Pi^-)=3\Pi.
\end{equation}
Combining this with Eq.~\eqref{eq:deltaeps_sum_zero}, one finds
\begin{equation}
\Pi = 0.
\label{eq:total_bulk_zero}
\end{equation}
Therefore, although the species dependent scalar corrections $\delta\epsilon^\pm$, $\Pi^\pm$, and $\delta n^\pm$ are not necessarily zero individually, the total bulk viscous pressure vanishes once Landau matching is imposed.

At this point, the simplifications specific to the massless two-component system become evident. The individual scalar dissipative quantities $\delta\epsilon^\pm$, $\Pi^\pm$, and $\delta n^\pm$ need not vanish exactly, but for a massless system they are at least of second order in gradients (Navier-Stokes terms vanish in the massless limit). Consequently, they do not contribute at the order considered in the derivation and may be consistently neglected. In other words, the scalar dissipative sector does not enter the first nontrivial equations derived in this chapter.

Finally, we impose the additional assumption
\begin{equation}
\mu = 0
\qquad \Longrightarrow \qquad
n_0(T)=0.
\label{eq:mu_zero_matching}
\end{equation}
This choice corresponds to a locally neutral plasma with vanishing equilibrium net electric charge density. Under this assumption, the net-charge diffusion current can also be neglected. Moreover, although the energy diffusion currents of the individual species do not vanish separately, Eq.~\eqref{eq:total_energy_diffusion_zero} ensures that their sum vanishes, and each of them becomes at least second order in gradients (Navier-Stokes term vanishes when the chemical potential is zero). Therefore, at the order relevant for the present derivation, they may also be neglected.

The conclusion of the matching analysis is therefore quite restrictive and very useful. Once Landau matching is imposed and the system is specialized to massless particles with $\mu=0$, the scalar and vector dissipative sectors become either exactly zero or higher order in gradients. As a result, the only dissipative quantity that remains relevant at leading nontrivial order is the shear-stress tensor. This is why the remainder of the chapter can focus entirely on the evolution of $\pi_+^{\mu\nu}$ and $\pi_-^{\mu\nu}$, and later on the corresponding total and relative shear tensors.

Thus, so far we have fixed the local equilibrium state through the matching conditions and clarified the meaning of the thermodynamic variables $T$ and $\mu$. We have also shown that all dissipative structures other than the shear sector can be consistently ignored at the order of interest. This is precisely what makes it possible to build a closed theory for the coupled shear dynamics of the two oppositely charged species.

\subsection{Exact equations of motion}

After the matching conditions have fixed the local equilibrium state and after the scalar and vector dissipative sectors have been shown to be negligible at the order of interest, the next step is to derive the exact equations of motion for the shear-stress tensor of each particle species. At this stage, no 14-moment truncation has yet been imposed. Therefore, the equations derived below are still exact consequences of the Boltzmann equation, although they are not yet closed in terms of the hydrodynamic fields.

We begin with the definition of the species-dependent shear-stress tensor,
\begin{equation}
\pi_\pm^{\mu\nu}
\equiv
\left\langle k^{\langle \mu} k^{\nu\rangle} \right\rangle_\pm
=
\int dK\, k^{\langle\mu}k^{\nu\rangle} f_k^\pm,
\label{eq:pi_species_definition_exact}
\end{equation}
where angled brackets denote the symmetric, traceless projection orthogonal to the fluid velocity. The exact equation of motion is obtained by taking the comoving derivative of this quantity,
\begin{equation}
\dot{\pi}_\pm^{\mu\nu}
\equiv
\frac{d}{d\tau}\pi_\pm^{\mu\nu}
=
\frac{d}{d\tau}
\int dK\, k^{\langle\mu}k^{\nu\rangle} f_k^\pm,
\qquad
\frac{d}{d\tau}\equiv u^\alpha \partial_\alpha .
\label{eq:pi_dot_start}
\end{equation}
Because the tensorial projector itself depends on the local fluid velocity, the comoving derivative acts not only on $f_k^\pm$ but also on the projected momentum structure $k^{\langle\mu}k^{\nu\rangle}$. Carrying out this differentiation carefully yields
\begin{equation}
\Delta^{\mu\nu}_{\alpha\beta}\dot{\pi}_\pm^{\alpha\beta}
=
\dot{\pi}_\pm^{\langle\mu\nu\rangle}
=
\int dK\, k^{\langle\mu}k^{\nu\rangle}\frac{d f_k^\pm}{d\tau}
-
2 \dot{u}^{\langle\mu} h_\pm^{\nu\rangle},
\label{eq:pi_dot_projected}
\end{equation}
where
\begin{equation}
h_\pm^\mu \equiv u_\alpha T_\pm^{\langle\mu\rangle\alpha}
\end{equation}
is the energy-diffusion current of each species. The second term in Eq.~\eqref{eq:pi_dot_projected} appears because the projection operator depends explicitly on $u^\mu$. This is an important detail-- even if one is ultimately interested only in the shear sector, the exact time evolution of $\pi_\pm^{\mu\nu}$ still knows about the acceleration of the fluid and about the energy-diffusion currents of the individual species.

The next step is to determine the comoving derivative of the single-particle distribution function. Starting from the Boltzmann equation for each species,
\begin{equation}
k^\mu \partial_\mu f_k^\pm
\pm |q| F^{\mu\nu}k_\nu \frac{\partial f_k^\pm}{\partial k^\mu}
=
 C[f^\pm,f^\pm]+C[f^\pm,f^\mp],
\label{eq:boltzmann_species_again}
\end{equation}
we rewrite the space-time derivative in terms of its decomposition along and orthogonal to the flow,
so that
$ k^\mu \partial_\mu =
E_k (d/d\tau) + k^\mu \nabla_\mu,
$ and as defined earlier, $
E_k \equiv k^\mu u_\mu
$ in local rest frame.
In the magnetically dominated regime considered here, the electromagnetic force term is written with the antisymmetric magnetic tensor,
\begin{equation}
B^{\mu\nu} \equiv  B\, b^{\mu\nu},
\qquad
b^{\mu\nu}\equiv -\epsilon^{\mu\nu\alpha\beta}u_\alpha b_\beta,
\end{equation}
which allows the Boltzmann equation to be recasted as
\begin{equation}
E_k \frac{d}{d\tau} f_k^\pm
=
-
k^\mu \nabla_\mu f_k^\pm
\mp |q|\, k_\nu B^{\mu\nu}\frac{\partial f_k^\pm}{\partial k^\mu}
+ C[f^\pm,f^\pm] +
C[f^\pm,f^\mp].
\label{eq:dfdtau_from_BE}
\end{equation}
Equation~\eqref{eq:dfdtau_from_BE} is the fundamental input in the derivation, because it transforms the kinetic equation into an expression directly usable in the time derivative of the shear tensor.

Substituting Eq.~\eqref{eq:dfdtau_from_BE} into Eq.~\eqref{eq:pi_dot_projected}, performing the irreducible projections, and organizing the result in terms of the standard kinematic tensors, $\sigma^{\mu\nu}$, $\theta$, and $\omega^{\mu\nu}$,
one obtains the exact equations of motion for the shear-stress tensor of each particle species:
\begin{align}
\dot{\pi}_+^{\langle\mu\nu\rangle}
&=
-
\Delta^{\mu\nu}_{\alpha\beta}
\nabla_\lambda
\left\langle
E_k^{-1}k^{\langle\alpha}k^\beta k^{\lambda\rangle}
\right\rangle_+
+
\frac{2}{5}
\Delta^{\mu\nu}_{\alpha\beta}
\nabla^\alpha
\left\langle
E_k^{-1}k^{\langle\beta\rangle}
\right\rangle_+
\nonumber\\
&\quad
-
\Delta^{\mu\nu}_{\alpha\beta}
\nabla_\lambda u_\kappa
\left\langle
E_k^{-2}
k^{\langle\alpha}k^\beta k^\lambda k^{\kappa\rangle}
\right\rangle_+
-
\frac{4}{3}\pi_+^{\mu\nu}\theta
+
\frac{8}{15}\epsilon_+ \sigma^{\mu\nu}
-
\frac{10}{7}
\Delta^{\mu\nu}_{\alpha\beta}
\sigma^\alpha_{\ \lambda}\pi_+^{\beta\lambda}
\nonumber\\
&\quad
-
2\Delta^{\mu\nu}_{\alpha\beta}\omega^\beta_{\ \lambda}\pi_+^{\alpha\lambda}
-
2|q|
\Delta^{\mu\nu}_{\alpha\beta}
B\, b^\alpha_{\ \lambda}
\left\langle
E_k^{-1}k^{\langle\beta}k^{\lambda\rangle}
\right\rangle_+
+
2|q| E^{\langle\mu} V_+^{\nu\rangle}
-
2\dot{u}^{\langle\mu}h_+^{\nu\rangle}
+
C_+^{\langle\mu\nu\rangle},
\label{eq:exact_pi_plus}
\\[0.3cm]
\dot{\pi}_-^{\langle\mu\nu\rangle}
&=
-
\Delta^{\mu\nu}_{\alpha\beta}
\nabla_\lambda
\left\langle
E_k^{-1}k^{\langle\alpha}k^\beta k^{\lambda\rangle}
\right\rangle_-
+
\frac{2}{5}
\Delta^{\mu\nu}_{\alpha\beta}
\nabla^\alpha
\left\langle
E_k^{-1}k^{\langle\beta\rangle}
\right\rangle_-
\nonumber\\
&\quad
-
\Delta^{\mu\nu}_{\alpha\beta}
\nabla_\lambda u_\kappa
\left\langle
E_k^{-2}
k^{\langle\alpha}k^\beta k^\lambda k^{\kappa\rangle}
\right\rangle_-
-
\frac{4}{3}\pi_-^{\mu\nu}\theta
+
\frac{8}{15}\epsilon_- \sigma^{\mu\nu}
-
\frac{10}{7}
\Delta^{\mu\nu}_{\alpha\beta}
\sigma^\alpha_{\ \lambda}\pi_-^{\beta\lambda}
\nonumber\\
&\quad
-
2\Delta^{\mu\nu}_{\alpha\beta}\omega^\beta_{\ \lambda}\pi_-^{\alpha\lambda}
+
2|q|
\Delta^{\mu\nu}_{\alpha\beta}
B\, b^\alpha_{\ \lambda}
\left\langle
E_k^{-1}k^{\langle\beta}k^{\lambda\rangle}
\right\rangle_-
-
2|q| E^{\langle\mu} V_-^{\nu\rangle}
-
2\dot{u}^{\langle\mu}h_-^{\nu\rangle}
+
C_-^{\langle\mu\nu\rangle}.
\label{eq:exact_pi_minus}
\end{align}

As discussed in the previous Chapter, these equations are exact but not closed: they involve higher-rank irreducible moments such as
\begin{equation}
\left\langle
E_k^{-1}k^{\langle\alpha}k^\beta k^{\lambda\rangle}
\right\rangle_\pm,
\qquad
\left\langle
E_k^{-2}k^{\langle\alpha}k^\beta k^\lambda k^{\kappa\rangle}
\right\rangle_\pm,
\end{equation}
as well as the projected second moments of the collision term, defined by
\begin{align}
C_+^{\langle\mu\nu\rangle}
& =
\frac{1}{2}\int dK\, E_k^{-1}k^{\langle\mu}k^{\nu\rangle} C[f^+,f^+] + \int dK\, E_k^{-1}k^{\langle\mu}k^{\nu\rangle} C[f^+,f^-],
\\
C_-^{\langle\mu\nu\rangle}
&=
\frac{1}{2}\int dK\, E_k^{-1}k^{\langle\mu}k^{\nu\rangle} C[f^-,f^-] + \int dK\, E_k^{-1}k^{\langle\mu}k^{\nu\rangle} C[f^-,f^+].
\label{eq:collision_moments_exact_again}
\end{align}
Thus, Eqs.~\eqref{eq:exact_pi_plus} and \eqref{eq:exact_pi_minus} are exact equations of motion, but they still involve moments beyond the dissipative fields we wish to evolve. It is also noteworthy that the terms proportional to $\sigma^{\mu\nu}$ are the usual shear-source terms that generate anisotropic stress from velocity gradients. The terms proportional to $\theta$, $\sigma^\alpha_{\ \lambda}\pi_\pm^{\beta\lambda}$, and $\omega^\beta_{\ \lambda}\pi_\pm^{\alpha\lambda}$ describe the nonlinear coupling of the shear tensor to the local expansion, shear, and vorticity of the flow. The terms involving $B\,b^\alpha_{\ \lambda}$ encode the direct action of the magnetic field on the tensor structure of the anisotropic stress. Finally, the terms involving $h_\pm^\mu$, $V_\pm^\mu$, and the higher moments show explicitly that, at the exact level, the shear sector is still coupled to additional kinetic information.

At this point, the need for a closure prescription becomes evident. This is precisely where the 14-moment approximation enters. In the following section, we generalize it for a multicomponent system.

\subsection{14-moment approximation for a two-component gas}

The role of the 14-moment approximation~\cite{Grad:1949,Denicol:2012cn,Denicol:2012es} is to truncate this hierarchy in a controlled way. Since the general idea of the method has already been discussed in earlier chapters, the point here is to explain how it is implemented for a gas composed of two oppositely charged species and how this differs from the single-component case. In the present problem, the crucial novelty is that the truncation must be applied separately to $f_k^+$ and $f_k^-$, after which the resulting dissipative tensors remain coupled through both the Lorentz force and the collision operator. 

For each particle species, we write the distribution function as a local-equilibrium contribution plus a deviation,
\begin{equation}
f_k^\pm = f_{0k}^\pm + \delta f_k^\pm .
\label{eq:ch5_fsplit_pm}
\end{equation}
The local-equilibrium distribution is parameterized by the same hydrodynamic fields $u^\mu$ and $T$ fixed through the matching conditions,
\begin{equation}
f_{0k}^\pm
=
\exp\!\left(- \beta_0\,u_\mu k^\mu\right),
\qquad
\beta_0 \equiv \frac{1}{T}.
\label{eq:ch5_f0_pm}
\end{equation}
Thus, the two components share the same collective flow and temperature. 

As usual in the method of moments, the deviation from equilibrium is expressed through a quantity $\delta y_k^\pm$ defined by
\begin{equation}
\delta f_k^\pm = f_{0k}^\pm \,\delta y_k^\pm ,
\label{eq:ch5_deltaf_deltay}
\end{equation}
and $\delta y_k^\pm$ is then expanded in irreducible tensors constructed from the particle four-momentum. For a given species, this expansion takes the form
\begin{equation}
\delta y_k^\pm
=
\varepsilon^\pm
+
\varepsilon_\alpha^\pm k^\alpha
+
\varepsilon_{\alpha\beta}^\pm k^\alpha k^\beta
+
\cdots .
\label{eq:ch5_irreducible_expansion_pm}
\end{equation}
The point of the 14-moment approximation is to truncate this expansion after the terms quadratic in the momentum. In the single-component case, this truncation leaves one scalar, one four-vector, and one rank-two traceless tensor sector. In the present two-component system, the same truncation is applied to each species separately. Therefore, before any further simplification is made, each distribution function carries its own scalar, vector, and tensor dissipative structures. 

At this stage, however, the matching conditions and the assumptions adopted in the chapter simplify the structure considerably. Because the particles are massless and the system is specialized to $\mu=0$, the scalar dissipative corrections are either constrained away or become higher order in gradients. Likewise, the vector sector can be neglected at the order relevant for the shear dynamics considered here. As a result, only the rank-two tensor part of the 14-moment ansatz is retained. Therefore, for each species, the deviation from equilibrium is approximated entirely in terms of the corresponding shear-stress tensor,
\begin{equation}
\delta f_k^\pm
=
f_{0k}^\pm\,\varepsilon_{\alpha\beta}^\pm k^\alpha k^\beta .
\label{eq:ch5_deltaf_tensor_only}
\end{equation}
This is the precise sense in which the 14-moment approximation is used in the present derivation: not as a fully general dissipative ansatz, but as a shear-sector truncation of the species-dependent moment expansion. 

The remaining task is to determine the coefficient $\varepsilon_{\alpha\beta}^\pm$ in terms of the hydrodynamic tensor $\pi_\pm^{\mu\nu}$. This is fixed by requiring that the irreducible second moment of the distribution reproduces the definition of the shear-stress tensor for each species,
\begin{equation}
\pi_\pm^{\mu\nu}
=
\left\langle k^{\langle\mu}k^{\nu\rangle}\right\rangle_\pm
=
\left\langle k^{\langle\mu}k^{\nu\rangle}\right\rangle_{0,\pm}
+
\left\langle k^{\langle\mu}k^{\nu\rangle}\right\rangle_{\delta,\pm}.
\label{eq:ch5_pi_match_species}
\end{equation}
Since the equilibrium contribution to a traceless, flow-orthogonal rank-two tensor vanishes by isotropy,
\begin{equation}
\left\langle k^{\langle\mu}k^{\nu\rangle}\right\rangle_{0,\pm}=0,
\end{equation}
only the non-equilibrium part contributes. Substituting Eq.~\eqref{eq:ch5_deltaf_tensor_only} into Eq.~\eqref{eq:ch5_pi_match_species} and using the isotropic tensor decomposition of the equilibrium momentum integrals fixes the coefficient uniquely. The result is the explicit 14-moment ansatz used in the chapter,
\begin{equation}
\delta f_k^\pm
=
f_{0k}^\pm
\frac{\pi_{\pm}^{\alpha\beta}k_\alpha k_\beta}
{2(\epsilon_0^\pm+P_0^\pm)T^2}.
\label{eq:ch5_14moment_ansatz_pm}
\end{equation}
This is the concrete form that must be inserted into both the higher irreducible moments and the collision integrals. 

It is worth stressing that Eq.~\eqref{eq:ch5_14moment_ansatz_pm} does not imply that the system has only one shear tensor. Rather, it says that after truncation, the deviation of each species from equilibrium is parameterized by its corresponding rank-two dissipative tensor $\pi_\pm^{\mu\nu}$. Thus, at the level of the two-component gas, the 14-moment approximation does not collapse the dissipative sector into a single tensorial variable. On the contrary, it naturally leads to two species-dependent shear-stress tensors, which will later be reorganized into total and relative combinations. This is the first point at which the larger dissipative state space of the two-component system becomes explicit. 

\subsection{Equations of motion for $\pi_+^{\mu\nu}$ and $\pi_-^{\mu\nu}$}

Once the 14-moment ansatz has been specified, the exact equations of motion obtained earlier can be simplified substantially. The reason is that all higher irreducible moments that appeared in the exact equations are now either expressed directly in terms of the shear-stress tensor or vanish identically within the truncation. In particular, after applying the 14-moment approximation, the following relations hold:
\begin{equation}
\left\langle E_k^{r} k^{\langle\mu\rangle} \right\rangle_\pm = 0,
\quad
\left\langle E_k^{-1} k^{\langle\beta}k^{\lambda\rangle} \right\rangle_\pm
=
\frac{\pi_\pm^{\beta\lambda}}{5T},
\quad
\left\langle E_k^{-1} k^{\langle\alpha}k^\beta k^{\lambda\rangle}\right\rangle_\pm = 0,
\quad
\left\langle E_k^{-2} k^{\langle\alpha}k^\beta k^\lambda k^{\kappa\rangle}\right\rangle_\pm = 0,
\label{eq:14m_moment_identities_again}
\end{equation}
where $r>-2$. 

The 14-moment approximation must then be applied consistently to the collision kernel. Substituting Eq.~\eqref{eq:ch5_14moment_ansatz_pm} into the projected collision moments yields expressions that are linear in $\pi_+^{\mu\nu}$ and $\pi_-^{\mu\nu}$. Retaining only the leading term in the expansion in $\delta y_k^\pm$, the projected collision moments become
\begin{align}
C_-^{\mu\nu}
&=
-\frac{6}{5}\sigma_{--} T n_0^-\, \pi_-^{\mu\nu}
+
\frac{2}{5}\sigma_{+-} T
\left(
n_0^- \pi_+^{\mu\nu}
-
4 n_0^+ \pi_-^{\mu\nu}
\right),
\nonumber\\
C_+^{\mu\nu}
&=
-\frac{6}{5}\sigma_{++} T n_0^+\, \pi_+^{\mu\nu}
+
\frac{2}{5}\sigma_{-+} T
\left(
n_0^+ \pi_-^{\mu\nu}
-
4 n_0^- \pi_+^{\mu\nu}
\right).
\label{eq:species_collision_after_14m}
\end{align}
This expression is derived in Appendix \ref{app:collision_integrals_shear} and considers only the linearized component of the collision kernel, as discussed in Chapter 4, Sec.~\ref{sec:linearization_collision_operator}.
Thus, after closure, the collision term of one species is still explicitly coupled to the shear tensor of the other species. This is already a major qualitative departure from the single-component case. 

We now substitute Eq.~\eqref{eq:14m_moment_identities_again} and Eq.~\eqref{eq:species_collision_after_14m} into the exact species equations. Since the vector moments vanish in the present truncation, the terms proportional to
$\langle E_k^{r}k^{\langle\mu\rangle}\rangle_\pm$,
$\langle E_k^{-1}k^{\langle\alpha}k^\beta k^{\lambda\rangle}\rangle_\pm$,
and
$\langle E_k^{-2}k^{\langle\alpha}k^\beta k^\lambda k^{\kappa\rangle}\rangle_\pm$
drop out. The only nonvanishing irreducible moment that survives is the projected second moment, which becomes proportional to $\pi_\pm^{\mu\nu}$. Therefore, the exact equations reduce to closed equations for the shear-stress tensors of each species. Finally, we also note that at vanishing electric-charge chemical potential, one has
\begin{equation}
n_0^+ = n_0^- \equiv \hat n_0.
\label{eq:n0_equal_hatn}
\end{equation}

The closed equations of motion for the two species can then be written schematically as
\begin{align}
\Delta^{\mu\nu}_{\alpha\beta}\dot{\pi}_+^{\alpha\beta}
+
\mathcal A_+\,\pi_+^{\mu\nu}
+
\mathcal B_+\,\pi_-^{\mu\nu}
+
\frac{2|q|B}{5T}\,b^{\lambda\langle\mu}\pi_{+\,\lambda}^{\nu\rangle}
& =
\frac{8}{15}\epsilon_+ \sigma^{\mu\nu}
-
\frac{4}{3}\pi_+^{\mu\nu}\theta
-
\frac{10}{7}\sigma^{\lambda\langle\mu}\pi_{+\,\lambda}^{\nu\rangle}
-
2\omega^{\lambda\langle\nu}\pi_{+\,\lambda}^{\mu\rangle},
\label{eq:species_plus_closed_schematic}
\\
\Delta^{\mu\nu}_{\alpha\beta}\dot{\pi}_-^{\alpha\beta}
+
\mathcal A_-\,\pi_-^{\mu\nu}
+
\mathcal B_-\,\pi_+^{\mu\nu}
-
\frac{2|q|B}{5T}\,b^{\lambda\langle\mu}\pi_{-\,\lambda}^{\nu\rangle}
& =
\frac{8}{15}\epsilon_- \sigma^{\mu\nu}
-
\frac{4}{3}\pi_-^{\mu\nu}\theta
-
\frac{10}{7}\sigma^{\lambda\langle\mu}\pi_{-\,\lambda}^{\nu\rangle}
-
2\omega^{\lambda\langle\nu}\pi_{-\,\lambda}^{\mu\rangle},
\label{eq:species_minus_closed_schematic}
\end{align}
where the coefficients $\mathcal A_\pm$ and $\mathcal B_\pm$ are 
\begin{align}
\mathcal{A}_+ &= \frac{2\,T\hat{n}_0}{5}
  \left(3\,\sigma_T^{++} + 4\,\sigma_T^{-+}\right),
\qquad
\mathcal{B}_+ = -\frac{2\,T\hat{n}_0}{5}\,\sigma_T^{-+},
  \label{eq:B_plus}\\[6pt]
\mathcal{A}_- &= \frac{2\,T\hat{n}_0}{5}
  \left(3\,\sigma_T^{--} + 4\,\sigma_T^{+-}\right),
\qquad
\mathcal{B}_- = -\frac{2\,T\hat{n}_0}{5}\,\sigma_T^{+-}.
  \label{eq:B_minus}
\end{align}
The opposite signs in the magnetic terms reflect the opposite electric charges of the two species.

At this point, it is no longer convenient to keep working with $\pi_+^{\mu\nu}$ and $\pi_-^{\mu\nu}$ separately. The physically more transparent variables are their sum and difference,
\begin{equation}
\pi^{\mu\nu} \equiv \pi_+^{\mu\nu}+\pi_-^{\mu\nu},
\qquad
\delta\pi^{\mu\nu} \equiv \pi_+^{\mu\nu}-\pi_-^{\mu\nu}.
\label{eq:total_relative_defs_again}
\end{equation}
The first is the total shear-stress tensor that enters the total energy-momentum tensor, while the second is the relative shear-stress tensor that measures the difference in anisotropic stress carried by the two species.

To obtain the equations for these variables, we simply add and subtract the two species equations. Adding Eqs.~\eqref{eq:species_plus_closed_schematic} and \eqref{eq:species_minus_closed_schematic} yields the evolution equation for the total shear tensor. The magnetic terms do not cancel because they come with opposite sign in the species equations and therefore combine into a coupling to the relative shear tensor:
\begin{equation}
b^{\lambda\langle\mu}\pi_{+\,\lambda}^{\nu\rangle}
-
b^{\lambda\langle\mu}\pi_{-\,\lambda}^{\nu\rangle}
=
b^{\lambda\langle\mu}\delta\pi_{\lambda}^{\nu\rangle}.
\label{eq:magnetic_addition_identity}
\end{equation}
Similarly, subtracting the two species equations gives the evolution equation for the relative shear tensor, and now the magnetic terms combine into a coupling to the total shear tensor:
\begin{equation}
b^{\lambda\langle\mu}\pi_{+\,\lambda}^{\nu\rangle}
+
b^{\lambda\langle\mu}\pi_{-\,\lambda}^{\nu\rangle}
=
b^{\lambda\langle\mu}\pi_{\lambda}^{\nu\rangle}.
\label{eq:magnetic_subtraction_identity}
\end{equation}
After carrying out this algebra, the closed equations become~\cite{Kushwah:2024zgd}
\begin{align}
\Delta^{\mu\nu}_{\alpha\beta}\dot{\pi}^{\alpha\beta}
+
\Sigma\,\pi^{\mu\nu}
+
\frac{2|q|B}{5T}\,
b^{\lambda\langle\mu}\delta\pi^{\nu\rangle}_{\ \lambda}
&=
\frac{8}{15}\epsilon\,\sigma^{\mu\nu}
-
\frac{4}{3}\pi^{\mu\nu}\theta
-
\frac{10}{7}\sigma^{\lambda\langle\mu}\pi^{\nu\rangle}_{\ \lambda}
-
2\omega^{\lambda\langle\nu}\pi^{\mu\rangle}_{\ \lambda},
\label{eq:total_shear_after_14m}
\\
\Delta^{\mu\nu}_{\alpha\beta}\delta\dot{\pi}^{\alpha\beta}
+
\Sigma'\,\delta\pi^{\mu\nu}
+
\frac{2|q|B}{5T}\,
b^{\lambda\langle\mu}\pi^{\nu\rangle}_{\ \lambda}
&=
-
\frac{4}{3}\delta\pi^{\mu\nu}\theta
-
\frac{10}{7}\sigma^{\lambda\langle\mu}\delta\pi^{\nu\rangle}_{\ \lambda}
-
2\omega^{\lambda\langle\nu}\delta\pi^{\mu\rangle}_{\ \lambda}.
\label{eq:relative_shear_after_14m}
\end{align}
These are the coupled total and relative shear equations obtained after the 14-moment closure. The two effective microscopic rates are
\begin{equation}
\Sigma
=
\frac{3\hat n_0}{5}
\left(
\sigma_T^{+-}+\sigma_T
\right),
\qquad
\Sigma'
=
\frac{\hat n_0}{5}
\left(
5\sigma_T^{+-}+3\sigma_T
\right),
\label{eq:Sigma_Sigmaprime_defs}
\end{equation}
where $\hat n_0$ is the equilibrium density of each species at vanishing electric-charge chemical potential, $n_0^+=n_0^- \equiv \hat n_0$, $\sigma_T^{+-}$ is the total cross section for inter-species scattering, and $\sigma_T$ denotes the common same-species cross section, defined by $\sigma_T^{++}=\sigma_T^{--}\equiv \sigma_T$. In the limit of vanishing magnetic field, $\Sigma$ reduces to the inverse relaxation time of the usual shear sector,
\begin{equation}
\Sigma=\frac{1}{\tau_\pi}=\frac{\epsilon+P_0}{5\eta}.
\label{eq:Sigma_taupi_eta}
\end{equation}
Thus, $\Sigma$ controls the relaxation of the total shear tensor, while $\Sigma'$ controls the relaxation of the relative shear tensor. the only difference is that the coefficient multiplying $\sigma_T^{+-}$ is larger in $\Sigma'$. Hence, relative shear is damped by a rate that gives greater weight to inter-species collisions.

Therefore, the 14-moment approximation does more than simply close the hierarchy. After closure, it reveals that the natural hydrodynamic variables of the two-component plasma are not the two species tensors separately, but the pair $(\pi^{\mu\nu},\delta\pi^{\mu\nu})$. Only after reaching this stage does it become natural to decompose these tensors further with respect to the magnetic-field direction and study the longitudinal, semi-transverse, and transverse sectors independently.

 \subsection{How does the two-component theory differ from the single-component case?}
\label{subsec:one_vs_two_component_eom}

It is useful to pause here and compare the structure of the equations of motion with the one-component theory discussed in the previous chapter (Sec.~\ref{sec:brief_note_em_rmhd}). In the one-component case, the shear sector is described by a single shear-stress tensor $\pi^{\mu\nu}$. After the 14-moment approximation, its equation of motion for massless particles has the generic transient form
\begin{equation}
\Delta^{\mu\nu}_{\alpha\beta}\dot{\pi}^{\alpha\beta}
+
\frac{1}{\tau_\pi}\pi^{\mu\nu} 
+
\Omega_B\, b^{\lambda\langle\mu}\pi_{\lambda}^{\nu\rangle}
=
2\eta\, \sigma^{\mu\nu}
-
\frac{4}{3}\pi^{\mu\nu}\theta
-
\frac{10}{7}\sigma^{\lambda\langle\mu}\pi^{\nu\rangle}_{\ \lambda}
-
2\omega^{\lambda\langle\nu}\pi^{\mu\rangle}_{\ \lambda}
,
\label{eq:ch5_one_component_schematic}
\end{equation}
with $\Omega_B \equiv 2|q|B/(5T)$. The important point is that the magnetic field acts on the same tensor $\pi^{\mu\nu}$. Thus, even though the magnetic field changes the relaxation structure of the shear sector, the equation still contains only one dissipative tensor and one microscopic relaxation scale.

In the present two-component theory, this structure changes already at the level of the unprojected equations of motion. As shown in the previous sections, using Eqs.~\eqref{eq:species_plus_closed_schematic} and~\eqref{eq:species_minus_closed_schematic}, we can write,
\begin{align}
\Delta^{\mu\nu}_{\alpha\beta}\dot{\pi}^{\alpha\beta}
+
\Sigma\,\pi^{\mu\nu}
+
\Omega_B\,b^{\lambda\langle\mu}\delta\pi^{\nu\rangle}_{\ \lambda}
&=
\frac{8}{15}\epsilon\,\sigma^{\mu\nu}
-
\frac{4}{3}\pi^{\mu\nu}\theta
-
\frac{10}{7}\sigma^{\lambda\langle\mu}\pi^{\nu\rangle}_{\ \lambda}
-
2\omega^{\lambda\langle\nu}\pi^{\mu\rangle}_{\ \lambda},
\label{eq:ch5_total_shear_compare_expanded}
\end{align}
whereas the relative shear equation has the structure
\begin{align}
\Delta^{\mu\nu}_{\alpha\beta}\delta\dot{\pi}^{\alpha\beta}
+
\Sigma'\,\delta\pi^{\mu\nu}
+
\Omega_B\,b^{\lambda\langle\mu}\pi^{\nu\rangle}_{\ \lambda}
&=
-
\frac{4}{3}\delta\pi^{\mu\nu}\theta
-
\frac{10}{7}\sigma^{\lambda\langle\mu}\delta\pi^{\nu\rangle}_{\ \lambda}
-
2\omega^{\lambda\langle\nu}\delta\pi^{\mu\rangle}_{\ \lambda}.
\label{eq:ch5_relative_shear_compare_expanded}
\end{align}
The contrast with Eq.~\eqref{eq:ch5_one_component_schematic} is now explicit. In the one-component theory, the magnetic field leads to a new contribution to the equation of motion for the shear-stress tensor, without introducing any additional field. On the other hand, in the two-component theory, the magnetic field couples the total shear tensor $\pi^{\mu\nu}$ to another degree of freedom that we here denoted as the relative shear tensor $\delta\pi^{\mu\nu}$. Therefore, even if the total shear tensor is the quantity that enters the total energy-momentum tensor, its evolution cannot be determined independently unless the relative shear tensor is also controlled.  This coupling, driven by the magnetic field, is the defining feature of the two-component theory and the foundation on which the rest of the analysis rests.

One may question whether such coupled dynamics can be reduced to a traditional relaxation-type equation for the total shear-stress tensor in a given regime. We shall address this question carefully in the remainder of this Chapter. We shall actually demonstrate that this will not always be possible: the coupling between the fields induced by the magnetic field can lead to oscillatory dynamics that cannot be capture by simple relaxation-type equations.

\section{Magnetic-field-adapted projections}

The presence of a magnetic field breaks the spatial isotropy that is normally assumed in relativistic hydrodynamics. In the absence of a magnetic field, the only preferred timelike direction is the fluid four-velocity $u^\mu$, and all spatial directions orthogonal to $u^\mu$ are equivalent. Once a magnetic field is present, this is no longer true: in addition to the fluid velocity, the system now contains a distinguished spacelike direction, namely the direction of the magnetic field itself. The main purpose of the new projections introduced in this section is to make this anisotropy explicit at the tensorial level.

This is not merely a matter of notation. If one continues to work only with the usual projector $\Delta^{\mu\nu}=g^{\mu\nu}-u^\mu u^\nu$, then all spatial directions orthogonal to $u^\mu$ are still treated on equal footing. That is too coarse for magnetized fluids, because the dynamics along the magnetic-field direction and the dynamics in the plane perpendicular to it are generally different. Therefore, before deriving or interpreting the equations of motion, we first build a set of projectors and basis tensors adapted to the magnetic field.

\subsection{Magnetic direction and induced spatial splitting}

We begin by re-introducing the normalized spacelike four-vector along the magnetic field,
\begin{equation}
b^\mu \equiv \frac{B^\mu}{B},
\qquad
B \equiv \sqrt{-B^\mu B_\mu},
\qquad
b^\mu b_\mu = -1,
\qquad
u^\mu b_\mu = 0.
\label{eq:chap5_b_def}
\end{equation}
Thus, $b^\mu$ identifies the direction of the magnetic field in a covariant way. Since $B^\mu$ is orthogonal to the fluid velocity, $b^\mu$ is also orthogonal to $u^\mu$. In the local rest frame of the fluid, $u^\mu=(1,0,0,0)$ and $b^\mu$ reduces to an ordinary unit spatial vector pointing along the magnetic field.

At this point, space is naturally split into two pieces:
1) the one-dimensional subspace along $b^\mu$, which we shall call the longitudinal direction, and
2) the two-dimensional subspace orthogonal to both $u^\mu$ and $b^\mu$, which we shall call the transverse plane.

The usual projector $\Delta^{\mu\nu}$ removes the time direction, but it does not separate the magnetic direction from the plane orthogonal to it. To do that, one introduces a second projector,
\begin{equation}
\Xi^{\mu\nu}
\equiv
g^{\mu\nu}-u^\mu u^\nu+b^\mu b^\nu
=
\Delta^{\mu\nu}+b^\mu b^\nu.
\label{eq:chap5_Xi_def}
\end{equation}
This tensor projects any four-vector onto the two-dimensional subspace orthogonal to both $u^\mu$ and $b^\mu$.

It is useful to verify explicitly that $\Xi^{\mu\nu}$ has the desired properties:
\begin{equation}
\Xi^{\mu\nu}u_\nu = 0,
\qquad
\Xi^{\mu\nu}b_\nu = 0,
\qquad
\Xi^\mu_{\ \alpha}\Xi^{\alpha\nu}=\Xi^{\mu\nu},
\qquad
\Xi^\mu_{\ \mu}=2.
\label{eq:chap5_Xi_props}
\end{equation}
The first two relations show that $\Xi^{\mu\nu}$ annihilates both the time direction and the magnetic direction, while the third shows that it is indeed a projector. The trace $\Xi^\mu_{\ \mu}=2$ confirms that it projects onto a two-dimensional subspace.

Therefore, from Eq.~\eqref{eq:chap5_Xi_def}, the magnetic field induces the following decomposition of the spatial sector: $\Delta^{\mu\nu} = \Xi^{\mu\nu} - b^\mu b^\nu.$
This identity is central. It says that the entire space orthogonal to $u^\mu$ can be decomposed into a longitudinal part, carried by $b^\mu b^\nu$, and a transverse part, carried by $\Xi^{\mu\nu}$.

Since the main dissipative quantity of interest in this chapter is the shear-stress tensor, which is symmetric, traceless, and orthogonal to $u^\mu$, we also need a projector that acts on rank-two tensors in the subspace orthogonal to both $u^\mu$ and $b^\mu$. This is defined as
\begin{equation}
\Xi^{\mu\nu}_{\alpha\beta}
\equiv
\frac{1}{2}
\left(
\Xi^\mu_{\ \alpha}\Xi^\nu_{\ \beta}
+
\Xi^\mu_{\ \beta}\Xi^\nu_{\ \alpha}
-
\Xi^{\mu\nu}\Xi_{\alpha\beta}
\right).
\label{eq:chap5_Xi2_def}
\end{equation}
This is the analogue of the usual double, symmetric, traceless projector $\Delta^{\mu\nu}_{\alpha\beta}$, except that it acts entirely in the two-dimensional plane orthogonal to both $u^\mu$ and $b^\mu$.

Its role is the following. Suppose $A^{\mu\nu}$ is any symmetric tensor orthogonal to $u^\mu$. Then
\begin{equation}
A_{\indep}^{\mu\nu}
\equiv
\Xi^{\mu\nu}_{\alpha\beta}A^{\alpha\beta}
\label{eq:chap5_Aperpperp_def}
\end{equation}
is the part of $A^{\mu\nu}$ that is not only orthogonal to $u^\mu$, but also orthogonal to the magnetic direction in both indices and traceless within the transverse plane.

In particular,
\begin{equation}
u_\mu A_{\indep}^{\mu\nu}=0,
\qquad
b_\mu A_{\indep}^{\mu\nu}=0,
\qquad
\Xi_{\mu\nu}A_{\indep}^{\mu\nu}=0.
\label{eq:chap5_Aperpperp_props}
\end{equation}
Thus, $A_{\indep}^{\mu\nu}$ lives entirely in the two-dimensional transverse plane and contains the genuinely transverse traceless part of the tensor.

\subsection{Decomposition of the shear-stress tensor}

We now apply these projectors to a generic traceless second-rank tensor orthogonal to $u^\mu$, such as the shear-stress tensor $\pi^{\mu\nu}$. Because $\pi^{\mu\nu}$ has five independent components, and because the magnetic field splits the spatial subspace into one longitudinal direction plus a transverse plane, it is natural to decompose $\pi^{\mu\nu}$ into pieces adapted to that splitting:
\begin{equation}
\pi^{\mu\nu}
=
\pi_\parallel
\left(
b^\mu b^\nu + \frac{1}{2}\Xi^{\mu\nu}
\right)
+
2\pi_\perp^{(\mu} b^{\nu)}
+
\pi_{\indep}^{\mu\nu}.
\label{eq:chap5_pi_decomp}
\end{equation}
The three objects appearing here are defined by
\begin{equation}
\pi_\parallel \equiv b_\alpha b_\beta \pi^{\alpha\beta},
\label{eq:chap5_pi_parallel_def}
\end{equation}
\begin{equation}
\pi_\perp^\mu
\equiv
-\Xi^\mu_{\ \alpha}b_\beta \pi^{\alpha\beta},
\label{eq:chap5_pi_perp_def}
\end{equation}
\begin{equation}
\pi_{\indep}^{\mu\nu}
\equiv
\Xi^{\mu\nu}_{\alpha\beta}\pi^{\alpha\beta}.
\label{eq:chap5_pi_perpperp_def}
\end{equation}
The scalar $\pi_\parallel$ measures the component of the shear-stress tensor along the magnetic direction. Since it is obtained by contracting both indices with $b^\mu$, it is the longitudinal part of the anisotropic stress. The vector $\pi_\perp^\mu$ contains one index effectively along the magnetic direction and one index in the plane orthogonal to it. For this reason it is often called the semi-transverse part. The tensor $\pi_{\indep}^{\mu\nu}$ is entirely orthogonal to both $u^\mu$ and $b^\mu$ and is traceless in the transverse plane. It is the purely transverse part of the shear tensor.

The advantage of writing the decomposition in the form \eqref{eq:chap5_pi_decomp} is that it is complete and irreducible with respect to the magnetic direction. Nothing is lost and nothing is double counted. Indeed, the five independent components of $\pi^{\mu\nu}$ are reorganized as
\begin{equation}
5 = 1 + 2 + 2,
\label{eq:chap5_5split}
\end{equation}
namely one longitudinal scalar, two independent components in $\pi_\perp^\mu$, and two independent components in $\pi_{\indep}^{\mu\nu}$.

Exactly the same decomposition can be applied to the relative shear tensor:
\begin{equation}
\delta\pi^{\mu\nu}
=
\delta\pi_\parallel
\left(
b^\mu b^\nu + \frac{1}{2}\Xi^{\mu\nu}
\right)
+
2\delta\pi_\perp^{(\mu} b^{\nu)}
+
\delta\pi_{\indep}^{\mu\nu}.
\label{eq:chap5_deltapi_decomp}
\end{equation}
This is important because later both $\pi^{\mu\nu}$ and $\delta\pi^{\mu\nu}$ will be projected onto the same magnetic-field-adapted basis.

\subsection{Antisymmetric tensor associated with the magnetic field}

Besides the projectors above, it is convenient to recall the antisymmetric tensor
\begin{equation}
b^{\mu\nu}
\equiv
-\epsilon^{\mu\nu\alpha\beta}u_\alpha b_\beta.
\label{eq:chap5_bmunu_def}
\end{equation}
This object acts within the two-dimensional plane orthogonal to $u^\mu$ and $b^\mu$. Geometrically, it generates rotations in that plane. This is why it appears naturally in the magnetic coupling terms of the shear equations.

To see this more explicitly, we consider an orthonormal basis $(u^\mu,b^\mu,x^\mu,y^\mu)$ such that in the local rest frame
\begin{equation}
u^\mu=(1,0,0,0),
\qquad
x^\mu=(0,1,0,0),
\qquad
y^\mu=(0,0,1,0),
\qquad
b^\mu=(0,0,0,1).
\label{eq:chap5_rest_basis}
\end{equation}
Then $x^\mu$ and $y^\mu$ span the plane perpendicular to the magnetic field. In this basis,
\begin{equation}
b^{\mu\nu}=x^\mu y^\nu - y^\mu x^\nu.
\label{eq:chap5_bmunu_xy}
\end{equation}
This makes the interpretation transparent: $b^{\mu\nu}$ is the antisymmetric tensor that rotates $x^\mu$ into $y^\mu$ and vice versa.

\subsection{Complex basis in the transverse plane}

Although $x^\mu$ and $y^\mu$ are perfectly valid basis vectors for the transverse plane, it is more convenient to work with the complex combinations \cite{Kushwah:2024zgd}
\begin{equation}
\ell_\pm^\mu
\equiv
\frac{1}{\sqrt{2}}\left(x^\mu \pm i y^\mu\right).
\label{eq:chap5_lpm_def}
\end{equation}
These vectors satisfy
\begin{equation}
\ell_\pm^\mu \ell_{\pm\,\mu}=0,
\label{eq:chap5_lpm_null}
\end{equation}
\begin{equation}
\ell_\mp^\mu \ell_{\pm\,\mu}=-1,
\label{eq:chap5_lpm_inner}
\end{equation}
as well as
\begin{equation}
u_\mu \ell_\pm^\mu = 0,
\qquad
b_\mu \ell_\pm^\mu = 0.
\label{eq:chap5_lpm_orth}
\end{equation}
Thus, $\ell^\mu_\pm$ are basis vectors for the same transverse plane, but they are adapted to the action of $b^{\mu\nu}$.

Indeed, using Eq.~\eqref{eq:chap5_bmunu_xy}, one finds
\begin{equation}
b^{\mu\nu}\ell_{\pm\,\nu}
=
\pm i\,\ell_\pm^\mu.
\label{eq:chap5_lpm_eigen}
\end{equation}
This is the main reason for introducing this complex basis. The tensor $b^{\mu\nu}$, which represents magnetic rotation in the transverse plane, becomes diagonal on $\ell^\mu_\pm$. As a result, the magnetic coupling terms in the equations of motion simplify drastically once the semi-transverse and transverse parts of the shear tensor are projected onto this basis.

\subsection{New projections for vectors and rank-two tensors}

Any four-vector $A_\perp^\mu$ orthogonal to both $u^\mu$ and $b^\mu$ can now be expanded in the $\ell_\pm^\mu$ basis:
\begin{equation}
A_\perp^\mu
=
A_\perp^+ \ell_+^\mu + A_\perp^- \ell_-^\mu,
\label{eq:chap5_Aperp_expand}
\end{equation}
with scalar coefficients
\begin{equation}
A_\perp^\pm
\equiv
-\ell_\mp^\mu A_{\perp\,\mu}.
\label{eq:chap5_Aperp_pm_def}
\end{equation}
This is simply the decomposition of a transverse two-vector into components of definite magnetic rotation.

Likewise, any symmetric traceless rank-two tensor $A_{\indep}^{\mu\nu}$ orthogonal to both $u^\mu$ and $b^\mu$ can be decomposed as
\begin{equation}
A_{\indep}^{\mu\nu}
=
A_{\indep}^+ \ell_+^\mu \ell_+^\nu
+
A_{\indep}^- \ell_-^\mu \ell_-^\nu,
\label{eq:chap5_Aperpperp_expand}
\end{equation}
where
\begin{equation}
A_{\indep}^\pm
\equiv
\ell_\mp^\mu \ell_\mp^\nu A_{\indep\,\mu\nu}.
\label{eq:chap5_Aperpperp_pm_def}
\end{equation}
Therefore, once the decomposition is performed, a vector orthogonal to $u^\mu$ and $b^\mu$ is replaced by two scalar amplitudes $A_\perp^\pm$, and a transverse traceless tensor is likewise replaced by two scalar amplitudes $A_{\indep}^\pm$.

Applying these definitions to the shear-stress tensor, we obtain
\begin{equation}
\pi_\perp^\mu
=
\pi_\perp^+ \ell_+^\mu + \pi_\perp^- \ell_-^\mu,
\qquad
\pi_\perp^\pm
=
-\ell_\mp^\mu \pi_{\perp\,\mu},
\label{eq:chap5_pi_perp_pm}
\end{equation}
and
\begin{equation}
\pi_{\indep}^{\mu\nu}
=
\pi_{\indep}^+ \ell_+^\mu \ell_+^\nu
+
\pi_{\indep}^- \ell_-^\mu \ell_-^\nu,
\qquad
\pi_{\indep}^\pm
=
\ell_\mp^\mu \ell_\mp^\nu \pi_{\indep\,\mu\nu}.
\label{eq:chap5_pi_perpperp_pm}
\end{equation}
An identical construction holds for the relative shear tensor:
\begin{equation}
\delta\pi_\perp^\mu
=
\delta\pi_\perp^+ \ell_+^\mu + \delta\pi_\perp^- \ell_-^\mu,
\qquad
\delta\pi_{\indep}^{\mu\nu}
=
\delta\pi_{\indep}^+ \ell_+^\mu \ell_+^\nu
+
\delta\pi_{\indep}^- \ell_-^\mu \ell_-^\nu.
\label{eq:chap5_deltapi_pm_modes}
\end{equation}

\subsection{Why these projections are useful}

At first glance, this construction may look more complicated than the original tensor notation. In fact, it does the opposite -- it converts tensor equations into scalar equations adapted to the symmetry breaking caused by the magnetic field.

Before the magnetic field is introduced, one often thinks of $\pi^{\mu\nu}$ as one object with five equivalent components. After the magnetic field is introduced, those five components are no longer equivalent. The new projections reveal exactly how the tensor splits:
\begin{equation}
\pi^{\mu\nu}
\quad \longrightarrow \quad
\left\{
\pi_\parallel,\;
\pi_\perp^+,\;
\pi_\perp^-,\;
\pi_{\indep}^+,\;
\pi_{\indep}^-
\right\}.
\label{eq:chap5_pi_five_scalars}
\end{equation}
Thus, the original shear tensor is rewritten in terms of five scalar degrees of freedom. This is extremely useful because the equations of motion can now be projected onto these scalar amplitudes one by one.

For this reason, the introduction of $\Xi^{\mu\nu}$, $\Xi^{\mu\nu}_{\alpha\beta}$, $b^{\mu\nu}$, and $\ell_\pm^\mu$ is not a decorative reformulation. It is a better language in which the shear dynamics of a magnetized two-component plasma can be better understood. In the following sections, these projections will allow us to derive separate evolution equations for the longitudinal, semi-transverse, and transverse components of both the total and relative shear tensors.

\section{Equations of motion for individual components}

After introducing the magnetic-field-adapted projections, the next natural step is to apply them to the coupled equations of motion for the total and relative shear tensors. The point is simple: the equations for $\pi^{\mu\nu}$ and $\delta\pi^{\mu\nu}$ are still tensor equations, but the magnetic field has already told us that not all tensor components are dynamically equivalent. Therefore, instead of keeping the full tensor notation all the way through, we now project the equations onto the longitudinal, semi-transverse, and fully transverse sectors defined earlier.

The logic is always the same. One starts from the coupled equations
\begin{align}
\Delta^{\mu\nu}_{\alpha\beta}\dot{\pi}^{\alpha\beta}
+
\Sigma\, \pi^{\mu\nu}
+
\frac{2|q|B}{5T}\,
b^{\lambda\langle\mu}\delta\pi^{\nu\rangle}_{\ \lambda}
&=
\frac{8}{15}\epsilon \sigma^{\mu\nu}
-
\frac{4}{3}\pi^{\mu\nu}\theta
-
\frac{10}{7}\sigma^{\lambda\langle\mu}\pi^{\nu\rangle}_{\ \lambda}
-
2\omega^{\lambda\langle\nu}\pi^{\mu\rangle}_{\ \lambda},
\label{eq:comp_start_total}
\\
\Delta^{\mu\nu}_{\alpha\beta}\delta\dot{\pi}^{\alpha\beta}
+
\Sigma' \delta\pi^{\mu\nu}
+
\frac{2|q|B}{5T}\,
b^{\lambda\langle\mu}\pi^{\nu\rangle}_{\ \lambda}
&=
-
\frac{4}{3}\delta\pi^{\mu\nu}\theta
-
\frac{10}{7}\sigma^{\lambda\langle\mu}\delta\pi^{\nu\rangle}_{\ \lambda}
-
2\omega^{\lambda\langle\nu}\delta\pi^{\mu\rangle}_{\ \lambda},
\label{eq:comp_start_relative}
\end{align}
and then contracts them with the appropriate projector. The only difference from one sector to another is which contraction is used.

\subsection{Scalar projection}

We begin with the longitudinal scalar component. This is obtained by contracting the tensor equations with $b_\mu b_\nu$. Since
\begin{equation}
\pi_\parallel \equiv b_\mu b_\nu \pi^{\mu\nu},
\qquad
\delta\pi_\parallel \equiv b_\mu b_\nu \delta\pi^{\mu\nu},
\end{equation}
this contraction isolates the component of the shear tensor aligned with the magnetic-field direction.

The first point to note is that the time derivative does not act only on $\pi^{\mu\nu}$, but also on the projector $b_\mu b_\nu$. Therefore,
\begin{equation}
b_\mu b_\nu \dot{\pi}^{\mu\nu}
=
\dot{\pi}_\parallel
+
\pi_\perp^\mu \dot{b}_\mu,
\label{eq:scalar_projection_time}
\end{equation}
and similarly
\begin{equation}
b_\mu b_\nu \delta\dot{\pi}^{\mu\nu}
=
\delta\dot{\pi}_\parallel
+
\delta\pi_\perp^\mu \dot{b}_\mu.
\label{eq:scalar_projection_time_relative}
\end{equation}
This explains why the scalar equations contain terms proportional to $\dot b^\mu$. Even though the magnetic term proportional to $B$ itself drops out in this projection, the scalar sector still feels the time dependence of the magnetic-field direction.

The magnetic coupling term vanishes in this projection,
\begin{equation}
b_\mu b_\nu\, b^{\lambda\langle\mu}A^{\nu\rangle}_{\ \lambda}=0,
\label{eq:scalar_Bterm_zero}
\end{equation}
because $b^{\mu\nu}$ is antisymmetric and orthogonal to $b^\mu$. Here A can be understood as $\pi^{\mu\nu}$ and $\delta\pi^{\mu\nu}$. This is why the scalar sector is not directly coupled by the explicit $B$-dependent term.

The remaining terms are then decomposed using the magnetic splitting of $\sigma^{\mu\nu}$, $\omega^{\mu\nu}$, and the nonlinear products of projected tensors. Carrying out this contraction yields
\begin{align}
\dot{\pi}_\parallel + \pi_\perp^\mu \dot b_\mu + \Sigma \pi_\parallel
&=
\frac{8}{15}\epsilon \sigma_\parallel
-
\frac{4}{3}\pi_\parallel \theta
-
\frac{10}{7}
\left(
-\frac{1}{2}\pi_\parallel \sigma_\parallel
+
\frac{1}{3}\sigma_\perp^\mu \pi_{\perp\mu}
+
\frac{1}{3}\pi_{\indep}^{\alpha\beta}\sigma_{\indep\,\alpha\beta}
\right)
\nonumber\\
&\quad
-
\frac{2}{3}
\left(
\omega_\perp^\mu \pi_{\perp\mu}
+
\omega_{\indep}^{\alpha\beta}\pi_{\indep\,\alpha\beta}
\right),
\label{eq:scalar_total_final}
\\
\delta\dot{\pi}_\parallel + \delta\pi_\perp^\mu \dot b_\mu + \Sigma' \delta\pi_\parallel
&=
-
\frac{4}{3}\delta\pi_\parallel \theta
-
\frac{10}{7}
\left(
-\frac{1}{2}\delta\pi_\parallel \sigma_\parallel
+
\frac{1}{3}\sigma_\perp^\mu \delta\pi_{\perp\mu}
+
\frac{1}{3}\delta\pi_{\indep}^{\alpha\beta}\sigma_{\indep\,\alpha\beta}
\right)
\nonumber\\
&\quad
-
\frac{2}{3}
\left(
\omega_\perp^\mu \delta\pi_{\perp\mu}
+
\omega_{\indep}^{\alpha\beta}\delta\pi_{\indep\,\alpha\beta}
\right).
\label{eq:scalar_relative_final}
\end{align}

These equations are already instructive. The explicit magnetic term proportional to $B$ vanishes, so the scalar sector is not directly mixed by the magnetic field. The only magnetic dependence appears through $\dot b^\mu$, that is, through changes in the direction of the magnetic field itself. Also, since we are working at vanishing chemical potential, the equation for $\delta\pi_\parallel$ has no Navier-Stokes source term proportional to $\sigma_\parallel$. This means that $\delta\pi_\parallel$ starts effectively at higher order in gradients.

\subsection{Vector projection}

We now move to the semi-transverse component. This is obtained by contracting the tensor equations with $b_\mu \Xi^\lambda_{\ \nu}$. The corresponding projected variables are
\begin{equation}
\pi_\perp^\lambda \equiv -\Xi^\lambda_{\ \alpha}b_\beta \pi^{\alpha\beta},
\qquad
\delta\pi_\perp^\lambda \equiv -\Xi^\lambda_{\ \alpha}b_\beta \delta\pi^{\alpha\beta}.
\end{equation}
This projection isolates the part of the shear tensor with one leg along the magnetic direction and one leg in the transverse plane.

Again, the derivative of the projector produces extra terms. One finds
\begin{equation}
b_\mu \Xi^\lambda_{\ \nu}\dot{\pi}^{\mu\nu}
=
\Xi^\lambda_{\ \nu}\dot{\pi}_\perp^\nu
+
\left(
\frac{3}{2}\pi_\parallel \Xi^{\lambda\nu}
+
\pi_{\indep}^{\lambda\nu}
\right)\dot b_\nu,
\label{eq:vector_time_projection}
\end{equation}
and similarly
\begin{equation}
b_\mu \Xi^\lambda_{\ \nu}\delta\dot{\pi}^{\mu\nu}
=
\Xi^\lambda_{\ \nu}\delta\dot{\pi}_\perp^\nu
+
\left(
\frac{3}{2}\delta\pi_\parallel \Xi^{\lambda\nu}
+
\delta\pi_{\indep}^{\lambda\nu}
\right)\dot b_\nu.
\label{eq:vector_time_projection_relative}
\end{equation}

Unlike the scalar projection, the magnetic term now survives:
\begin{equation}
b_\mu \Xi^\lambda_{\ \nu}\,
b^{\alpha\langle\mu}A^{\nu\rangle}_{\ \alpha}
\neq 0.
\end{equation}
This is precisely why the vector sector is the first one in which the magnetic field explicitly couples total and relative shear.

Carrying out the full projection gives
\begin{align}
&\Xi^\lambda_{\ \nu}\dot{\pi}_\perp^\nu
+
\left(
\frac{3}{2}\pi_\parallel \Xi^{\lambda\nu}
+
\pi_{\indep}^{\lambda\nu}
\right)\dot b_\nu
+
\Sigma \pi_\perp^\lambda
+
\frac{B|q|}{5T}\,b^{\lambda\nu}\delta\pi_{\perp\nu}
\nonumber \\&  =
\frac{8}{15}\epsilon \sigma_\perp^\lambda
-
\frac{4}{3}\pi_\perp^\lambda \theta
+
\frac{5}{14}
\left(
\pi_\perp^\lambda \sigma_\parallel
+
\sigma_\perp^\lambda \pi_\parallel
\right)
-
\frac{5}{7}
\left(
\sigma_\perp^{\ \nu}\pi_{\indep\,\lambda\nu}
+
\pi_\perp^{\ \nu}\sigma_{\indep\,\lambda\nu}
\right) 
\nonumber \\ 
& \qquad +
\frac{1}{2}\pi_\parallel \omega_\perp^\lambda -
\pi_{\indep\,\lambda\nu}\omega_\perp^\nu
 -
\pi_\perp^{\ \nu}\omega_{\indep\,\lambda\nu},
\label{eq:vector_total_real}
\\
& \Xi^\lambda_{\ \nu}\delta\dot{\pi}_\perp^\nu
+
\left(
\frac{3}{2}\delta\pi_\parallel \Xi^{\lambda\nu}
+
\delta\pi_{\indep}^{\lambda\nu}
\right)\dot b_\nu
+
\Sigma' \delta\pi_\perp^\lambda
+
\frac{B|q|}{5T}\,b^{\lambda\nu}\pi_{\perp\nu}
\nonumber\\ & =
-
\frac{4}{3}\delta\pi_\perp^\lambda \theta
+
\frac{5}{14}
\left(
\delta\pi_\perp^\lambda \sigma_\parallel
+
\sigma_\perp^\lambda \delta\pi_\parallel
\right)
-
\frac{5}{7}
\left(
\sigma_\perp^{\ \nu}\delta\pi_{\indep\,\lambda\nu}
+
\delta\pi_\perp^{\ \nu}\sigma_{\indep\,\lambda\nu}
\right)
+
\frac{1}{2}\delta\pi_\parallel \omega_\perp^\lambda
\nonumber \\ 
& \qquad -
\delta\pi_{\indep\,\lambda\nu}\omega_\perp^\nu
 -
\delta\pi_\perp^{\ \nu}\omega_{\indep\,\lambda\nu}.
\label{eq:vector_relative_real}
\end{align}

These are already the equations of motion for the vector sector, but they are still written as transverse-vector equations. To turn them into linearly independent scalar equations, we project once more with the complex basis vectors $\ell^\lambda_\mp$. Using
\begin{equation}
\pi_\perp^\mu=\pi_\perp^+\ell_+^\mu+\pi_\perp^-\ell_-^\mu,
\qquad
\delta\pi_\perp^\mu=\delta\pi_\perp^+\ell_+^\mu+\delta\pi_\perp^-\ell_-^\mu,
\end{equation}
together with
\begin{equation}
b^{\mu\nu}\ell_{\pm\,\nu}=\pm i \ell_\pm^\mu,
\end{equation}
the vector equations become
\begin{align}
& \dot{\pi}_\perp^{\mp}
+
\pi_\perp^{\mp}\,\ell_\nu^{\mp}\dot{\ell}_\pm^\nu
-
\left(
\frac{3}{2}\pi_\parallel \ell_\pm^\nu
-
\pi_{\indep}^{\mp}\ell_\mp^\nu
\right)\dot b_\nu
+
\Sigma \pi_\perp^{\mp}
\mp i\frac{B|q|}{5T}\,\delta\pi_\perp^{\mp}
\nonumber \\ &=
\frac{8}{15}\epsilon \sigma_\perp^{\mp}
-
\frac{4}{3}\pi_\perp^{\mp}\theta
+
\frac{5}{14}
\left(
\pi_\perp^{\mp}\sigma_\parallel
+
\pi_\parallel \sigma_\perp^{\mp}
\right)
+
\frac{5}{7}
\left(
\pi_\perp^{\pm}\sigma_{\indep}^{\mp}
+
\pi_{\indep}^{\mp}\sigma_\perp^{\pm}
\right) +
\frac{1}{2}\pi_\parallel \omega_\perp^{\mp}
+
\omega_\perp^{\pm}\pi_{\indep}^{\mp}
+
\omega_{\indep}^{\mp}\pi_\perp^{\pm},
\label{eq:vector_total_pm}
\end{align}
\begin{align} 
& \delta\dot{\pi}_\perp^{\mp}
+
\delta\pi_\perp^{\mp}\,\ell_\nu^{\mp}\dot{\ell}_\pm^\nu
-
\left(
\frac{3}{2}\delta\pi_\parallel \ell_\pm^\nu
-
\delta\pi_{\indep}^{\mp}\ell_\mp^\nu
\right)\dot b_\nu
+
\Sigma' \delta\pi_\perp^{\mp}
\mp i\frac{B|q|}{5T}\,\pi_\perp^{\mp}
\nonumber \\ & =
-
\frac{4}{3}\delta\pi_\perp^{\mp}\theta
+
\frac{5}{14}
\left(
\delta\pi_\perp^{\mp}\sigma_\parallel
+
\delta\pi_\parallel \sigma_\perp^{\mp}
\right)
+
\frac{5}{7}
\left(
\delta\pi_\perp^{\pm}\sigma_{\indep}^{\mp}
+
\delta\pi_{\indep}^{\mp}\sigma_\perp^{\pm}
\right) +
\frac{1}{2}\delta\pi_\parallel \omega_\perp^{\mp}
+
\omega_\perp^{\pm}\delta\pi_{\indep}^{\mp}
+
\omega_{\indep}^{\mp}\delta\pi_\perp^{\pm}.
\label{eq:vector_relative_pm}
\end{align}

This is the cleanest form of the semi-transverse sector. The vector equations are now split into two scalar equations labeled by the $\pm$ magnetic helicity basis. In this form the explicit magnetic coupling becomes diagonal.

\subsection{Tensor projection}

We now turn to the purely transverse traceless component. This is obtained by acting on the tensor equations with the rank-two projector $\Xi^{\lambda\rho}_{\alpha\beta}$. By definition,
\begin{equation}
\pi_{\indep}^{\lambda\rho}
=
\Xi^{\lambda\rho}_{\alpha\beta}\pi^{\alpha\beta},
\qquad
\delta\pi_{\indep}^{\lambda\rho}
=
\Xi^{\lambda\rho}_{\alpha\beta}\delta\pi^{\alpha\beta}.
\end{equation}
This projection isolates the part of the shear tensor that lives completely inside the plane orthogonal to both $u^\mu$ and $b^\mu$, and is traceless in that plane.

Again, the time derivative of the projector produces extra terms. One finds
\begin{equation}
\Xi^{\lambda\rho}_{\alpha\beta}\dot{\pi}^{\alpha\beta}
=
\dot{\pi}_{\indep}^{\lambda\rho}
+
2\Xi^{\lambda\rho}_{\alpha\beta}\pi_\perp^\alpha \dot b^\beta,
\label{eq:tensor_time_projection}
\end{equation}
and
\begin{equation}
\Xi^{\lambda\rho}_{\alpha\beta}\delta\dot{\pi}^{\alpha\beta}
=
\delta\dot{\pi}_{\indep}^{\lambda\rho}
+
2\Xi^{\lambda\rho}_{\alpha\beta}\delta\pi_\perp^\alpha \dot b^\beta.
\label{eq:tensor_time_projection_relative}
\end{equation}
Projecting Eqs.~\eqref{eq:comp_start_total} and \eqref{eq:comp_start_relative} with $\Xi^{\lambda\rho}_{\alpha\beta}$ then gives
\begin{align}
& \dot{\pi}_{\indep}^{\lambda\rho}
+
2\Xi^{\lambda\rho}_{\alpha\beta}\pi_\perp^\alpha \dot b^\beta
+
\Sigma \pi_{\indep}^{\lambda\rho}
+
\frac{B|q|}{5T}
\left(
b^\lambda_{\ \alpha}\delta\pi_{\indep}^{\rho\alpha}
+
b^\rho_{\ \alpha}\delta\pi_{\indep}^{\lambda\alpha}
\right)
\nonumber \\ & =
\frac{8}{15}\epsilon \sigma_{\indep}^{\lambda\rho}
-
\frac{4}{3}\pi_{\indep}^{\lambda\rho}\theta
-
\frac{5}{7}
\left(
\pi_\parallel \sigma_{\indep}^{\lambda\rho}
+
\sigma_\parallel \pi_{\indep}^{\lambda\rho}
\right)
-
\frac{5}{7}
\Big[
\pi_{\indep}^{\lambda\alpha}\sigma_{\indep\,\alpha}^{\ \ \rho}
+
\pi_{\indep}^{\rho\alpha}\sigma_{\indep\,\alpha}^{\ \ \lambda}
-
\pi_\perp^\lambda \sigma_\perp^\rho
-
\pi_\perp^\rho \sigma_\perp^\lambda
\nonumber\\
&\quad
+
\Xi^{\lambda\rho}
\left(
\pi_\perp^\alpha \sigma_{\perp\alpha}
-
\pi_{\indep}^{\alpha\beta}\sigma_{\indep\,\alpha\beta}
\right)
\Big]
-
\pi_\parallel \omega_{\indep}^{\lambda\rho}
+
\pi_\perp^\lambda \omega_\perp^\rho
+
\pi_\perp^\rho \omega_\perp^\lambda
-
\pi_{\indep}^{\lambda\alpha}\omega_{\indep\,\alpha}^{\ \ \rho}
-
\pi_{\indep}^{\rho\alpha}\omega_{\indep\,\alpha}^{\ \ \lambda}
\nonumber\\
&\quad  -
\Xi^{\lambda\rho}
\left(
\pi_\perp^\alpha \omega_{\perp\alpha} 
-
\pi_{\indep}^{\alpha\beta}\omega_{\indep\,\alpha\beta}
\right),
\label{eq:tensor_total_real}
\end{align}
\begin{align}
& \delta\dot{\pi}_{\indep}^{\lambda\rho}
+
2\Xi^{\lambda\rho}_{\alpha\beta}\delta\pi_\perp^\alpha \dot b^\beta
+
\Sigma' \delta\pi_{\indep}^{\lambda\rho}
+
\frac{B|q|}{5T}
\left(
b^\lambda_{\ \alpha}\pi_{\indep}^{\rho\alpha}
+
b^\rho_{\ \alpha}\pi_{\indep}^{\lambda\alpha}
\right)
\nonumber \\ &=
-
\frac{4}{3}\delta\pi_{\indep}^{\lambda\rho}\theta
-
\frac{5}{7}
\left(
\delta\pi_\parallel \sigma_{\indep}^{\lambda\rho}
+
\sigma_\parallel \delta\pi_{\indep}^{\lambda\rho}
\right)
\frac{5}{7}
\Big[
\delta\pi_{\indep}^{\lambda\alpha}\sigma_{\indep\,\alpha}^{\ \ \rho}
+
\delta\pi_{\indep}^{\rho\alpha}\sigma_{\indep\,\alpha}^{\ \ \lambda}
-
\delta\pi_\perp^\lambda \sigma_\perp^\rho
-
\delta\pi_\perp^\rho \sigma_\perp^\lambda
\nonumber\\
&\quad
+
\Xi^{\lambda\rho}
\left(
\delta\pi_\perp^\alpha \sigma_{\perp\alpha}
-
\delta\pi_{\indep}^{\alpha\beta}\sigma_{\indep\,\alpha\beta}
\right)
\Big]
-
\delta\pi_\parallel \omega_{\indep}^{\lambda\rho}
+
\delta\pi_\perp^\lambda \omega_\perp^\rho
+
\delta\pi_\perp^\rho \omega_\perp^\lambda
-
\delta\pi_{\indep}^{\lambda\alpha}\omega_{\indep\,\alpha}^{\ \ \rho}
-
\delta\pi_{\indep}^{\rho\alpha}\omega_{\indep\,\alpha}^{\ \ \lambda}
\nonumber\\
&\quad -
\Xi^{\lambda\rho}
\left(
\delta\pi_\perp^\alpha \omega_{\perp\alpha}
-
\delta\pi_{\indep}^{\alpha\beta}\omega_{\indep\,\alpha\beta}
\right).
\label{eq:tensor_relative_real}
\end{align}

These are the tensor-sector equations in projected-tensor form. As with the vector sector, the magnetic term still mixes components inside the transverse plane. So one more projection is useful.

To diagonalize that mixing, we contract with $\ell_\mp^\lambda \ell_\mp^\rho$. Using
\begin{equation}
\pi_{\indep}^{\mu\nu}
=
\pi_{\indep}^+ \ell_+^\mu \ell_+^\nu
+
\pi_{\indep}^- \ell_-^\mu \ell_-^\nu,
\qquad
\delta\pi_{\indep}^{\mu\nu}
=
\delta\pi_{\indep}^+ \ell_+^\mu \ell_+^\nu
+
\delta\pi_{\indep}^- \ell_-^\mu \ell_-^\nu,
\end{equation}
we obtain the linearly independent equations
\begin{align}
& \dot{\pi}_{\indep}^{\mp}
+
2\pi_{\indep}^{\mp}\ell_\beta^{\mp}\dot{\ell}_\pm^\beta
-
2\pi_\perp^{\mp}\ell_\beta^\pm \dot b^\beta
+
\Sigma \pi_{\indep}^{\mp}
\pm i\frac{2B|q|}{5T}\,\delta\pi_{\indep}^{\mp}
\nonumber \\ &=
\frac{8}{15}\epsilon \sigma_{\indep}^{\mp}
-
\frac{4}{3}\pi_{\indep}^{\mp}\theta
-
\frac{5}{7}
\left(
\pi_\parallel \sigma_{\indep}^{\mp}
+
\sigma_\parallel \pi_{\indep}^{\mp}
\right)
+
\frac{10}{7}\pi_\perp^{\mp}\sigma_\perp^{\mp}
-
\pi_\parallel \omega_{\indep}^{\mp}
+
2\pi_\perp^{\mp}\omega_\perp^{\mp},
\label{eq:tensor_total_pm}
\\
& \delta\dot{\pi}_{\indep}^{\mp}
+
2\delta\pi_{\indep}^{\mp}\ell_\beta^{\mp}\dot{\ell}_\pm^\beta
-
2\delta\pi_\perp^{\mp}\ell_\beta^\pm \dot b^\beta
+
\Sigma' \delta\pi_{\indep}^{\mp}
\pm i\frac{2B|q|}{5T}\,\pi_{\indep}^{\mp}
\nonumber \\ &=
-
\frac{4}{3}\delta\pi_{\indep}^{\mp}\theta
-
\frac{5}{7}
\left(
\delta\pi_\parallel \sigma_{\indep}^{\mp}
+
\sigma_\parallel \delta\pi_{\indep}^{\mp}
\right)
+
\frac{10}{7}\delta\pi_\perp^{\mp}\sigma_\perp^{\mp}
-
\delta\pi_\parallel \omega_{\indep}^{\mp}
+
2\delta\pi_\perp^{\mp}\omega_\perp^{\mp}.
\label{eq:tensor_relative_pm}
\end{align}


The main achievement of this decomposition is now visible. The original tensor equations for $\pi^{\mu\nu}$ and $\delta\pi^{\mu\nu}$ have been converted into equations for five scalar amplitudes:
\begin{equation}
\pi_\parallel,\qquad
\pi_\perp^\pm,\qquad
\pi_{\indep}^\pm,
\end{equation}
and likewise for the relative sector,
\begin{equation}
\delta\pi_\parallel,\qquad
\delta\pi_\perp^\pm,\qquad
\delta\pi_{\indep}^\pm.
\end{equation}
This is the cleanest way to see how the magnetic field reorganizes the shear dynamics. The scalar projection is not directly mixed by the explicit $B$-term, while the vector and tensor sectors are magnetically mixed and become simplest only after the extra $\ell_\pm^\mu$ projections are performed.

So the big picture is that the magnetic field does not merely change a transport coefficient inside one single shear equation. It splits the shear tensor into dynamically inequivalent sectors. The derivation above shows how that splitting happens at the level of projections, and the final equations show that each sector obeys a different equation of motion.

\section{Linear regime}

Before proceeding to the truncation scheme and to an Israel--Stewart-like reformulation, it is useful to study the linear regime of the coupled equations for the total and relative shear-stress tensors. This step is important because, first, it isolates the genuinely new physics introduced by the magnetic field without the clutter of nonlinear gradient couplings. Secondly, it already shows that the transient dynamics of the system is qualitatively different from the usual purely relaxational behavior expected in standard dissipative hydrodynamics~\cite{Israel:1979wp}.

 The idea is to linearize the equations of motion around a static homogeneous equilibrium state characterized by constant temperature $T$ and constant magnetic field $B$. In this limit, the longitudinal component is not particularly interesting. It does not couple to the other of the stress tensor and therefore behaves in the same qualitative way as in ordinary transient hydrodynamics without magnetic fields. For this reason, we focus only on the semi-transverse and transverse sectors, where the magnetic field produces genuinely new dynamics.

\subsection{Semi-transverse sector}

We start from the projected equations for the semi-transverse modes $\pi_\perp^\mp$ and $\delta\pi_\perp^\mp$. In the linear regime, the equations then reduce to
\begin{align}
\dot{\pi}_\perp^{\mp}
+
\Sigma\,\pi_\perp^{\mp}
\mp i\frac{|q|B}{5T}\,\delta\pi_\perp^{\mp}
&=
\frac{8}{15}\epsilon\,\sigma_\perp^{\mp},
\label{eq:lin_semitrans_pi}
\\
\delta\dot{\pi}_\perp^{\mp}
+
\Sigma'\,\delta\pi_\perp^{\mp}
\mp i\frac{|q|B}{5T}\,\pi_\perp^{\mp}
&=
0.
\label{eq:lin_semitrans_dpi}
\end{align}
These two equations already contain the essential physics. In the absence of a magnetic field, the total and relative semi-transverse components relax independently, each with their own microscopic relaxation rate, $\Sigma$ and $\Sigma'$. Once the magnetic field is turned on, however, the two sectors are no longer independent. The magnetic field mixes $\pi_\perp^\mp$ and $\delta\pi_\perp^\mp$, and because the coupling is purely imaginary in the $\ell_\pm^\mu$ basis, it acts mathematically as a rotation in the two-dimensional space spanned by the total and relative shear modes.

It is convenient to introduce the frequency scale
\begin{equation}
\Omega \equiv \frac{|q|B}{5T}.
\label{eq:Omega_def_linear}
\end{equation}
This quantity has dimensions of inverse time and measures the strength of the magnetic coupling relative to the thermal scale.

To obtain an equation involving only the total shear mode, we differentiate Eq.~\eqref{eq:lin_semitrans_pi} with respect to time:
\begin{equation}
\ddot{\pi}_\perp^{\mp}
+
\Sigma\,\dot{\pi}_\perp^{\mp}
\mp i\Omega\,\delta\dot{\pi}_\perp^{\mp}
=
\frac{8}{15}\epsilon\,\dot{\sigma}_\perp^{\mp}.
\label{eq:lin_semitrans_diff}
\end{equation}
We then eliminate $\delta\dot{\pi}_\perp^\mp$ using Eq.~\eqref{eq:lin_semitrans_dpi},
\begin{equation}
\delta\dot{\pi}_\perp^{\mp}
=
-\Sigma'\delta\pi_\perp^{\mp}
\pm i\Omega\,\pi_\perp^{\mp}.
\label{eq:lin_semitrans_dpi_dot}
\end{equation}
From Eq.~\eqref{eq:lin_semitrans_pi}, we also isolate $\delta\pi_\perp^\mp$,
\begin{equation}
\mp i\Omega\,\delta\pi_\perp^{\mp}
=
\frac{8}{15}\epsilon\,\sigma_\perp^{\mp}
-
\dot{\pi}_\perp^{\mp}
-
\Sigma\,\pi_\perp^{\mp}.
\label{eq:lin_semitrans_delta_isolation}
\end{equation}
Substituting Eqs.~\eqref{eq:lin_semitrans_dpi_dot} and \eqref{eq:lin_semitrans_delta_isolation} into Eq.~\eqref{eq:lin_semitrans_diff}, after straightforward algebra, one finds
\begin{equation}
\ddot{\pi}_\perp^{\mp}
+
(\Sigma+\Sigma')\dot{\pi}_\perp^{\mp}
+
(\Sigma\Sigma'+\Omega^2)\pi_\perp^{\mp}
=
\frac{8}{15}\epsilon\,\Sigma'\sigma_\perp^{\mp}
+
\frac{8}{15}\epsilon\,\dot{\sigma}_\perp^{\mp}.
\label{eq:lin_semitrans_oscillator}
\end{equation}

This is the equation of a forced damped harmonic oscillator. The damping coefficient is $\Sigma+\Sigma'$, while the effective natural frequency is controlled by $\Omega$. This is already a major departure from the usual Israel--Stewart picture~\cite{Israel:1979wp}, where transient dynamics is purely relaxational. Here, even in the linear regime, the magnetic field converts the coupled shear system into an oscillator.

\subsection{Transverse sector}

The same reasoning applies to the purely transverse modes $\pi_{\indep}^\mp$ and $\delta\pi_{\indep}^\mp$. Their linearized equations are
\begin{align}
\dot{\pi}_{\indep}^{\mp}
+
\Sigma\,\pi_{\indep}^{\mp}
\pm i\frac{2|q|B}{5T}\,\delta\pi_{\indep}^{\mp}
&=
\frac{8}{15}\epsilon\,\sigma_{\indep}^{\mp},
\label{eq:lin_trans_pi}
\\
\delta\dot{\pi}_{\indep}^{\mp}
+
\Sigma'\,\delta\pi_{\indep}^{\mp}
\pm i\frac{2|q|B}{5T}\,\pi_{\indep}^{\mp}
&=
0.
\label{eq:lin_trans_dpi}
\end{align}
The only structural difference from the semi-transverse sector is that the magnetic coupling now appears with twice the strength, that is, with $2\Omega$ instead of $\Omega$.

Repeating the same differentiation-and-elimination procedure as above, one obtains
\begin{equation}
\ddot{\pi}_{\indep}^{\mp}
+
(\Sigma+\Sigma')\dot{\pi}_{\indep}^{\mp}
+
(\Sigma\Sigma'+4\Omega^2)\pi_{\indep}^{\mp}
=
\frac{8}{15}\epsilon\,\Sigma'\sigma_{\indep}^{\mp}
+
\frac{8}{15}\epsilon\,\dot{\sigma}_{\indep}^{\mp}.
\label{eq:lin_trans_oscillator}
\end{equation}

Therefore, the transverse modes are also forced damped oscillators, but with a stronger magnetic contribution to the restoring term. This means that the transverse sector enters the oscillatory regime more easily than the semi-transverse sector.

\subsection{Homogeneous limit and dispersion relations}

The structure of the linear regime becomes even clearer in the homogeneous limit. In this limit, the source terms are set to zero,
\begin{equation}
\sigma_\perp^\mp = 0,
\qquad
\sigma_{\indep}^\mp = 0,
\qquad
\dot{\sigma}_\perp^\mp = 0,
\qquad
\dot{\sigma}_{\indep}^\mp = 0,
\end{equation}
so the equations describe the intrinsic nonhydrodynamic relaxation of the modes toward equilibrium.

For the semi-transverse sector, inserting the ansatz
\begin{equation}
\pi_\perp^\mp \sim e^{-i\omega t}
\end{equation}
into Eq.~\eqref{eq:lin_semitrans_oscillator} yields
\begin{equation}
-\omega^2
+
i(\Sigma+\Sigma')\omega
+
(\Sigma\Sigma'+\Omega^2)
=
0.
\label{eq:lin_semitrans_dispersion}
\end{equation}
Solving this quadratic equation gives
\begin{equation}
\omega
=
\frac{i}{2}
\left[
\Sigma+\Sigma'
\pm
\sqrt{(\Sigma-\Sigma')^2-4\Omega^2}
\right].
\label{eq:lin_semitrans_roots}
\end{equation}

In the limit $B\to 0$, so that $\Omega\to 0$, this reduces to the two purely damped modes
\begin{equation}
\omega = i\Sigma,
\qquad
\omega=i\Sigma'.
\end{equation}
Thus, when the magnetic field vanishes, the system simply relaxes exponentially with rates $\Sigma$ and $\Sigma'$, exactly as expected.

However, once the magnetic field is finite, the square root in Eq.~\eqref{eq:lin_semitrans_roots} can become imaginary. This happens when
\begin{equation}
4\Omega^2 > (\Sigma-\Sigma')^2.
\label{eq:lin_semitrans_osc_condition}
\end{equation}
In that case, the mode frequencies acquire a nonzero real part, and the system no longer relaxes monotonically to equilibrium. Instead, it relaxes with oscillations.

Using the explicit forms
\begin{equation}
\Sigma
=
\frac{3\hat n_0}{5}
\left(
\sigma_T^{+-}+\sigma_T
\right),
\qquad
\Sigma'
=
\frac{\hat n_0}{5}
\left(
5\sigma_T^{+-}+3\sigma_T
\right),
\end{equation}
the oscillation condition can be rewritten as
\begin{equation}
\Omega > \frac{\Sigma'-\Sigma}{2}
\qquad \Longrightarrow \qquad
\frac{|q|B}{T} > \hat n_0 \sigma_T^{+-}.
\label{eq:lin_semitrans_osc_condition_micro}
\end{equation}
So the onset of oscillatory dynamics is controlled by the competition between magnetic rotation and inter-species relaxation.

For the transverse sector, the same calculation gives the dispersion relation
\begin{equation}
\omega
=
\frac{i}{2}
\left[
\Sigma+\Sigma'
\pm
\sqrt{(\Sigma-\Sigma')^2-16\Omega^2}
\right],
\label{eq:lin_trans_roots}
\end{equation}
which means that the oscillatory regime begins when
\begin{equation}
16\Omega^2 > (\Sigma-\Sigma')^2,
\end{equation}
or equivalently,
\begin{equation}
\Omega > \frac{\Sigma'-\Sigma}{4}
\qquad \Longrightarrow \qquad
\frac{2|q|B}{T} > \hat n_0 \sigma_T^{+-}.
\label{eq:lin_trans_osc_condition_micro}
\end{equation}
Thus, the transverse sector reaches the oscillatory regime at a smaller magnetic field than the semi-transverse sector.

\begin{figure}[H]
    \centering
    \includegraphics[width=0.7\linewidth]{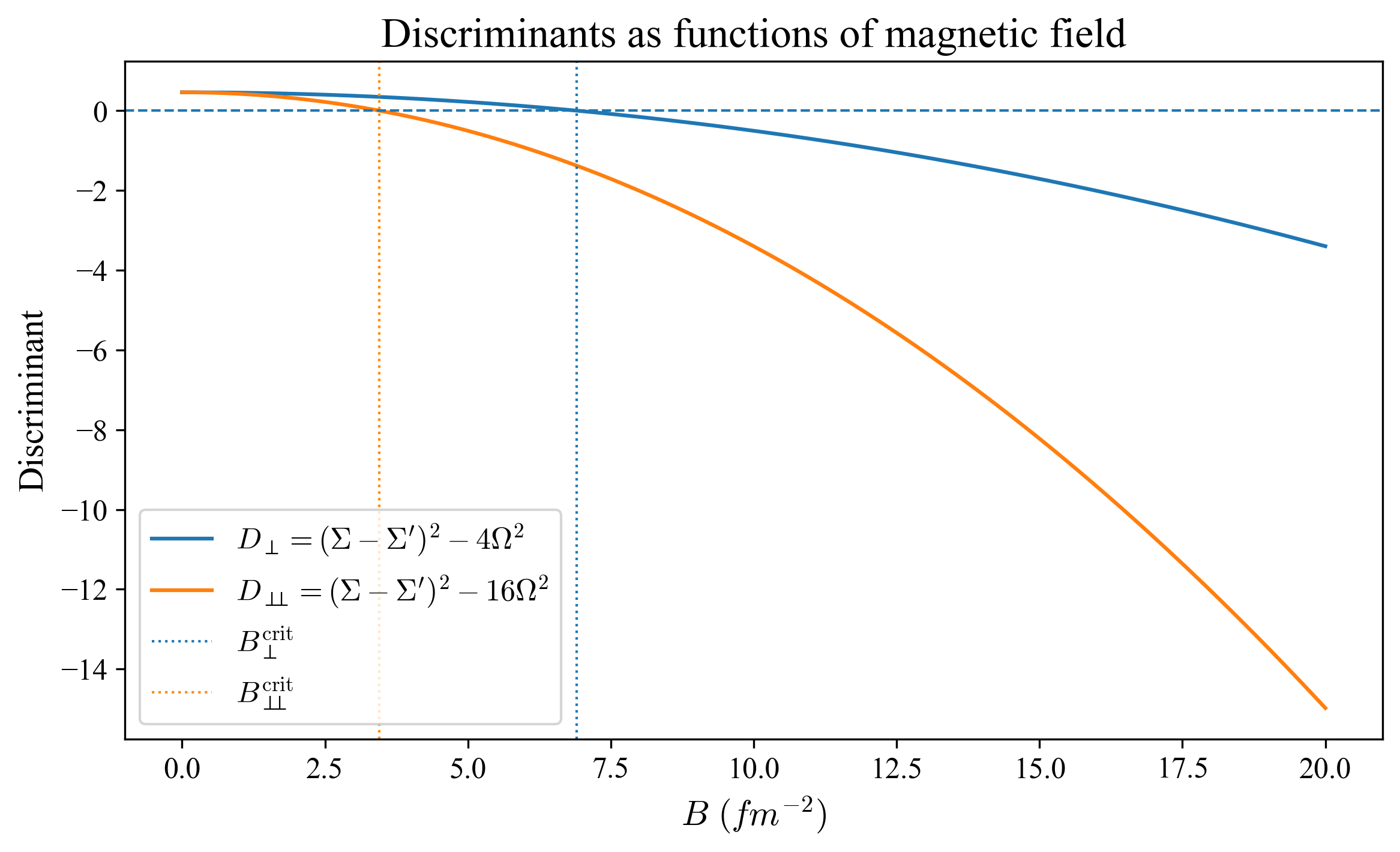}
    \caption[Magnetic-field discriminants]{Magnetic-field dependence of the discriminants $D_{\perp}=(\Sigma-\Sigma')^2-4\Omega^2$ and $D_{\indep}=(\Sigma-\Sigma')^2-16\Omega^2$ in the homogeneous limit. The horizontal dashed line indicates the boundary between overdamped and oscillatory behavior. The vertical dotted lines mark the corresponding critical magnetic fields at which each discriminant vanishes. Since $D_{\indep}$ crosses zero at a smaller value of $B$, the purely transverse modes become oscillatory before the semi-transverse ones.}
    \label{fig:discriminants_magnetic_field}
\end{figure}

\begin{figure}[h!]
    \centering
    \begin{subfigure}{0.48\textwidth}
        \centering
        \includegraphics[width=\linewidth]{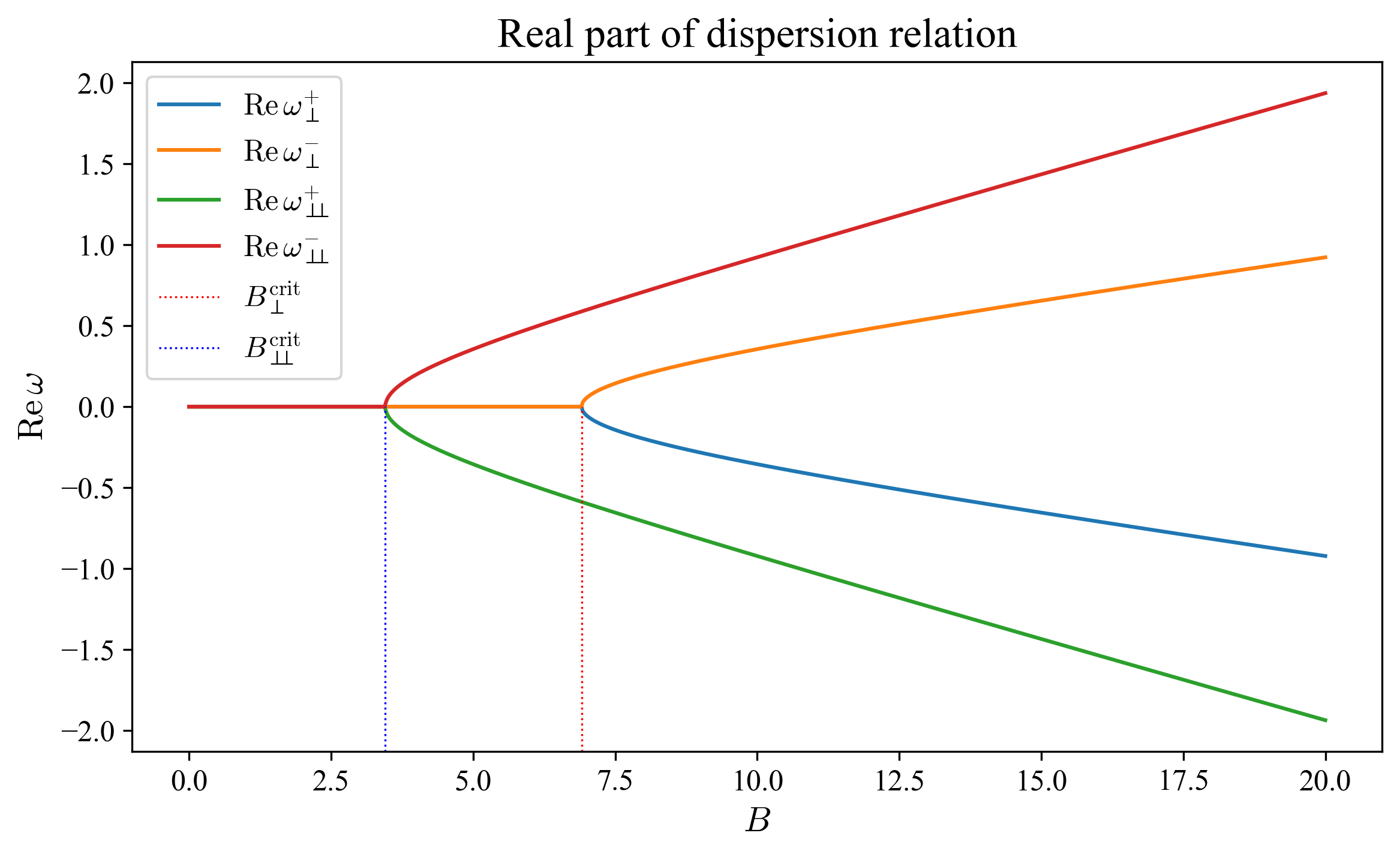}
        \caption[Real part]{Real part}
        \label{fig:dispersion_real}
    \end{subfigure}
    \hfill
    \begin{subfigure}{0.48\textwidth}
        \centering
        \includegraphics[width=\linewidth]{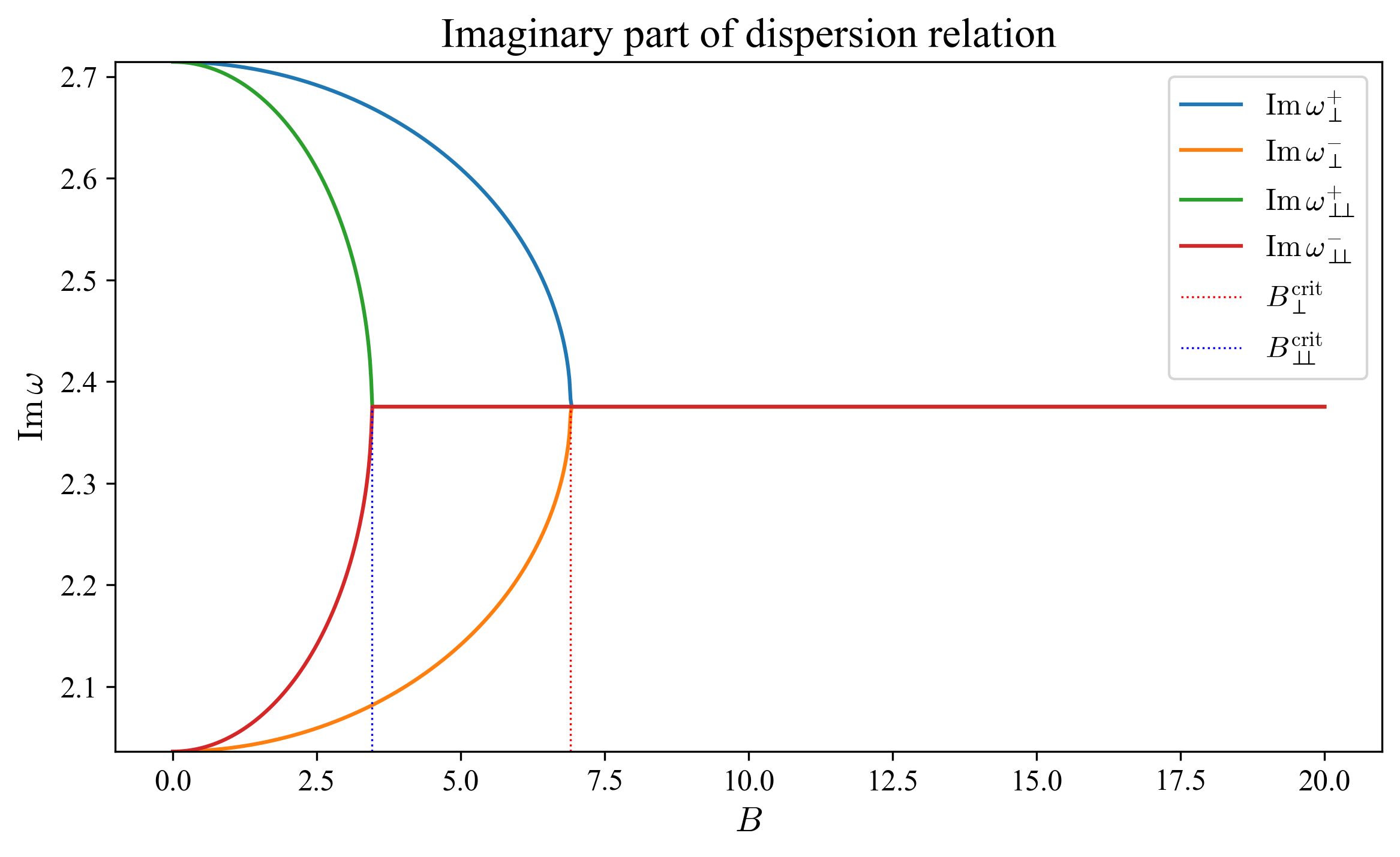}
        \caption[Imaginary part]{Imaginary part}
        \label{fig:dispersion_imag}
    \end{subfigure}

    \caption[Complex frequencies in magnetic field]{Magnetic-field dependence of the complex frequencies in the homogeneous limit. The left panel shows the real parts and the right panel shows the imaginary parts of the semi-transverse and purely transverse modes. Together they show the transition from purely damped relaxation to damped oscillatory behavior as the magnetic field increases.}
    \label{fig:dispersion_relation_magnetic_field}
\end{figure}

Figures~\ref{fig:discriminants_magnetic_field} and \ref{fig:dispersion_relation_magnetic_field} summarize the linear analysis in a compact way. Figure~\ref{fig:discriminants_magnetic_field} shows that both discriminants decrease monotonically with increasing magnetic field, but the transverse discriminant $D_{\indep}=(\Sigma-\Sigma')^2-16\Omega^2$ drops faster than the semi-transverse one $D_{\perp}=(\Sigma-\Sigma')^2-4\Omega^2$. This directly reflects the stronger magnetic coupling in the purely transverse sector, where the restoring term is proportional to $4\Omega^2$ instead of $\Omega^2$. Consequently, $D_{\indep}$ crosses zero first, so the purely transverse modes enter the oscillatory regime at a smaller value of $B$ than the semi-transverse modes. Figure~\ref{fig:dispersion_relation_magnetic_field} shows the same transition from the complementary point of view of the complex frequencies. In the real-part plot, the frequencies remain purely imaginary below the critical field, which corresponds to purely damped relaxation. Once the corresponding discriminant changes sign, the two branches split symmetrically and a nonzero real part develops, signaling the onset of oscillatory behavior. In the imaginary-part plot, one sees at the same time that the damping rates approach each other as the critical field is reached and then merge into a common value once the modes become underdamped. Taken together, the two figures make the same point in two equivalent ways: the discriminant plot identifies where the transition occurs, while the dispersion plot shows how that transition appears directly in the mode spectrum. Most importantly, both figures make it clear that the transverse sector reacts first to the magnetic field and therefore carries the earliest signature of the oscillatory regime.

\subsection{Physical meaning of the linear regime}
The physical lesson of the linear regime is very apparent here. In ordinary transient hydrodynamics~\cite{Israel:1979wp}, the dissipative modes approach equilibrium exponentially. Here, because the magnetic field couples the total and relative shear sectors, the approach to equilibrium can instead resemble that of a damped oscillator.

This is not an artifact of nonlinearities, nor does it require an expanding background. It appears already in the simplest possible setting: a static equilibrium state with a constant magnetic field. That is why this analysis is so important. It shows that the oscillatory behavior is intrinsic to the coupled two-component magnetized system itself.

There are two time scales competing here. One is the usual microscopic relaxation scale, set by $\Sigma^{-1}$ and $\Sigma'^{-1}$. The other is the magnetic time scale, set by $\Omega^{-1}$. When the magnetic time scale becomes comparable to or shorter than the mismatch between the two relaxation scales, the system crosses from purely damped behavior to damped oscillatory behavior.

The transverse sector is more sensitive to this competition because the magnetic field enters there with $2\Omega$, not $\Omega$. This is why the transverse modes oscillate more easily and why, later in the chapter, they are also the first ones to signal the breakdown of a standard Israel--Stewart-like truncation.

\section{Truncation scheme}

We now derive a second-order fluid-dynamical theory from the coupled equations obtained in the previous sections. The main idea is to estimate the magnitude of each term using the leading contribution of an asymptotic gradient expansion. In this way, the relative shear-stress tensor is not kept as an independent dynamical variable. Instead, it is re-expressed order by order in terms of the total shear-stress tensor and its derivatives. This is precisely what allows one to obtain an Israel--Stewart-like description for each projected component of the total shear tensor.

The starting point is the set of coupled equations for the projected components of $\pi^{\mu\nu}$ and $\delta\pi^{\mu\nu}$. The longitudinal component is already decoupled from the relative longitudinal shear and therefore does not require any further truncation. The nontrivial step concerns the semi-transverse and fully transverse sectors, where the magnetic field couples the total and relative shear tensors.

The truncation is based on an asymptotic gradient expansion. To first order in gradients, the relative shear components are algebraically related to the total shear components. One finds
\begin{equation}
\delta\pi_{\perp}^{\mp}
=
\pm i\phi\,\pi_{\perp}^{\mp}
+
\mathcal{O}(2),
\label{eq:delta_pi_perp_first_order}
\end{equation}
\begin{equation}
\delta\pi_{\indep}^{\mp}
=
\pm 2i\phi\,\pi_{\indep}^{\mp}
+
\mathcal{O}(2),
\label{eq:delta_pi_perpperp_first_order}
\end{equation}
where
\begin{equation}
\phi \equiv \frac{\Omega}{\Sigma'}.
\label{eq:phi_definition}
\end{equation}
Here, $\mathcal{O}(2)$ denotes terms that are second order or higher in gradients or dissipative quantities. These relations are already enough to see the basic structure: the relative shear is not independent at leading order, but is induced by the total shear through the magnetic field.

The next step is to iterate these first-order relations back into the coupled equations of motion and keep all terms up to second order. This produces improved expressions for the relative shear components. For the semi-transverse sector, one obtains
\begin{align}
\Sigma'\,\delta\pi_{\perp}^{\mp}
&=
\mp i\phi\,\pi_{\perp}^{\mp}
\left(
\frac{4}{3}\theta
-
\frac{5}{14}\sigma_{\parallel}
\right)
\mp \frac{5i}{7}\phi
\left(
\pi_{\perp}^{\pm}\sigma_{\perp}^{\mp}
+
2\pi_{\indep}^{\mp}\sigma_{\perp}^{\pm}
\right)
\mp i\phi\,\pi_{\perp}^{\mp}\ell_{\nu}^{\mp}\dot{\ell}_{\pm}^{\nu}
\nonumber\\
&\quad
\pm 2i\phi\,\pi_{\indep}^{\mp}\ell_{\nu}^{\mp}\dot{b}^{\nu}
\pm i\phi\Sigma'\,\pi_{\perp}^{\mp}
\mp i\phi\,\dot{\pi}_{\perp}^{\mp}
\mp i\phi
\left(
2\omega_{\perp}^{\pm}\pi_{\indep}^{\mp}
+
\omega_{\indep}^{\mp}\pi_{\perp}^{\pm}
\right)
+
\mathcal{O}(3),
\label{eq:delta_pi_perp_second_order}
\end{align}
while for the fully transverse sector,
\begin{align}
\Sigma'\,\delta\pi_{\indep}^{\mp}
&=
\pm \frac{8i}{3}\phi\,\pi_{\indep}^{\mp}\theta
\pm \frac{10i}{7}\phi\,\pi_{\indep}^{\mp}\sigma_{\parallel}
\pm \frac{10i}{7}\phi\,\pi_{\perp}^{\mp}\sigma_{\perp}^{\mp}
\mp 2i\Omega\,\pi_{\indep}^{\mp}
\pm 2i\phi\,\dot{\pi}_{\indep}^{\mp}
\pm 2i\pi_{\indep}^{\mp}\dot{\phi}
\nonumber\\
&\quad
\pm 4i\phi\,\pi_{\indep}^{\mp}\ell_{\beta}^{\mp}\dot{\ell}_{\pm}^{\beta}
\pm 2i\phi\,\pi_{\perp}^{\mp}\ell_{\beta}^{\pm}\dot{b}^{\beta}
\pm 2i\phi\,\pi_{\perp}^{\mp}\omega_{\perp}^{\mp}
+
\mathcal{O}(3).
\label{eq:delta_pi_perpperp_second_order}
\end{align}
As before, $\mathcal{O}(3)$ denotes terms that are third order or higher in gradients or dissipative quantities.

These two equations are not yet the final hydrodynamic equations. Their purpose is different. They provide the order-by-order elimination of $\delta\pi_{\perp}^{\mp}$ and $\delta\pi_{\indep}^{\mp}$ in favor of the total shear components. Once these expressions are substituted back into the equations of motion for the total shear-stress tensor, and all terms of $\mathcal{O}(3)$ are discarded, one obtains a closed second-order theory for each component of the total shear tensor.

The resulting second-order equations for the semi-transverse components are
\begin{align}
& \left(1-\phi^2\right)\dot{\pi}_{\perp}^{\mp}
+
\left(\Sigma+\phi^2\Sigma'\right)\pi_{\perp}^{\mp}
\nonumber \\ & =
\frac{8}{15}\epsilon\,\sigma_{\perp}^{\mp}
-
\left[
\left(1-\phi^2\right)
\left(
\ell_{\nu}^{\mp}\dot{\ell}_{\pm}^{\nu}
+
\frac{4}{3}\theta
-
\frac{5}{14}\sigma_{\parallel}
\right)
+
\phi\dot{\phi}
\right]\pi_{\perp}^{\mp}
\nonumber\\
&\quad
+
\left(1+\phi^2\right)
\left(
\omega_{\indep}^{\mp}
+
\frac{5}{7}\sigma_{\indep}^{\mp}
\right)\pi_{\perp}^{\pm}
+
\left(1+2\phi^2\right)
\left(
-\ell_{\nu}^{\mp}\dot{b}^{\nu}
+
\omega_{\perp}^{\pm}
+
\frac{5}{7}\sigma_{\perp}^{\pm}
\right)\pi_{\indep}^{\mp}
\nonumber\\
&\quad
+
\left(
\frac{3}{2}\ell_{\nu}^{\pm}\dot{b}^{\nu}
+
\omega_{\perp}^{\mp}
+
\frac{5}{14}\sigma_{\perp}^{\mp}
\right)\pi_{\parallel}.
\label{eq:pi_perp_truncated}
\end{align}
Similarly, the second-order equations for the fully transverse components are
\begin{align}
\left(1-4\phi^2\right)\dot{\pi}_{\indep}^{\mp}
+
\left(\Sigma+4\phi^2\Sigma'\right)\pi_{\indep}^{\mp}
&=
\frac{8}{15}\epsilon\,\sigma_{\indep}^{\mp}
-
\left[
\left(1-4\phi^2\right)
\left(
2\ell_{\beta}^{\mp}\dot{\ell}_{\pm}^{\beta}
+
\frac{4}{3}\theta
+
\frac{5}{7}\sigma_{\parallel}
\right)
-
4\phi\dot{\phi}
\right]\pi_{\indep}^{\mp}
\nonumber\\
&\quad
-
\left(
\frac{5}{7}\sigma_{\indep}^{\mp}
+
\omega_{\indep}^{\mp}
\right)\pi_{\parallel}
+
\left(1+2\phi^2\right)
\left(
2\ell_{\beta}^{\pm}\dot{b}^{\beta}
+
\frac{10}{7}\sigma_{\perp}^{\mp}
+
2\omega_{\perp}^{\mp}
\right)\pi_{\perp}^{\mp}.
\label{eq:pi_perpperp_truncated}
\end{align}
We note that the longitudinal component did not need any such truncation, because it was already independent of the relative longitudinal shear. The nontrivial simplification only concerns the semi-transverse and fully transverse sectors.

These equations also show explicitly that the shear-stress tensor no longer obeys a single relaxation equation. Instead, its different components obey different dynamical equations, with different coefficients and different effective relaxation times. Furthermore, the magnetic field enters the equations in two distinct ways. It appears explicitly through $\phi$, which modifies the coefficients multiplying $\dot{\pi}$ and $\pi$, and it also appears geometrically through $\dot b^\mu$, $\dot\ell_\pm^\mu$, and through the projected vorticity and shear couplings. So the field does not merely renormalize one transport coefficient. It reorganizes the entire transient structure of the theory.

These equations also allow one to identify component-dependent shear viscosities as follows
\begin{equation}
\eta_{\parallel}
=
\frac{4\epsilon}{15\Sigma}
\equiv \eta,
\qquad
\eta_{\perp}
=
\frac{4\epsilon}{15\left(\Sigma+\phi^2\Sigma'\right)},
\qquad
\eta_{\indep}
=
\frac{4\epsilon}{15\left(\Sigma+4\phi^2\Sigma'\right)}.
\label{eq:component_viscosities}
\end{equation}
Thus, the longitudinal shear viscosity remains equal to the usual shear viscosity, while the semi-transverse and transverse viscosities acquire nontrivial magnetic-field dependence.

In the same way, one can read off effective relaxation times for each component,
\begin{equation}
\tau_{\parallel}
=
\frac{1}{\Sigma}
\equiv \tau_{\pi},
\qquad
\tau_{\perp}
=
\frac{1-\phi^2}{\Sigma+\phi^2\Sigma'},
\qquad
\tau_{\indep}
=
\frac{1-4\phi^2}{\Sigma+4\phi^2\Sigma'}.
\label{eq:component_relaxation_times}
\end{equation}

These are among the most important results of the truncation scheme. In the limit $B\to0$, one has $\phi\to0$, and all three relaxation times reduce to the usual Israel--Stewart relaxation time~\cite{Israel:1979wp},
\begin{equation}
\tau_{\parallel}=\tau_{\perp}=\tau_{\indep}=\tau_{\pi}.
\end{equation}
However, once the magnetic field becomes appreciable, the three sectors separate dynamically.

\begin{figure}[t]
    \centering
    \includegraphics[width=\textwidth]{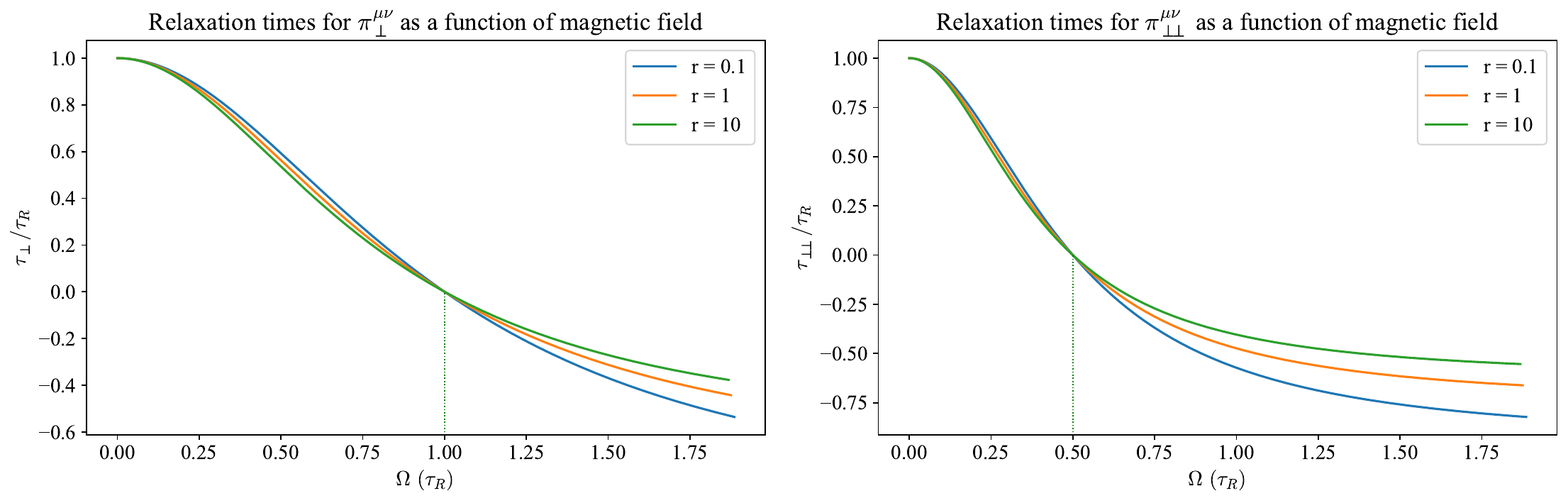}
    \caption[Projected relaxation times]{ Semi-transverse and transverse relaxation time $\tau_{\perp}/\tau_R$ and $\tau_{\indep}/\tau_R$, respectively as a function of $\Omega(\tau_R)$ for different values of $r=\sigma_T^{+-}/\sigma_T$. As the magnetic field increases, both relaxation times decrease and eventually become negative. The transverse sector reaches this unphysical regime earlier, at $\Omega(\tau_R) = 0.5$, while the semi-transverse sector does so only at $\Omega(\tau_R)=1$. This shows that the truncation scheme fails first in the transverse sector and only later in the semi-transverse sector.}
    \label{fig:relaxation_times_truncation}
\end{figure}

A crucial feature now becomes visible. If
\begin{equation}
4\phi^2>1,
\end{equation}
then $\tau_{\indep}$ becomes negative. If the field is increased further so that
\begin{equation}
\phi^2>1,
\end{equation}
then $\tau_{\perp}$ also becomes negative. Negative relaxation times are clearly unphysical. They signal that the truncated second-order theory breaks down for sufficiently strong magnetic fields.

This behavior is shown explicitly in Fig.~\ref{fig:relaxation_times_truncation}. The right panel, which displays $\tau_{\indep}/\tau_R$, shows that the purely transverse relaxation time crosses zero already at a smaller value of the magnetic field. By contrast, the left panel shows that $\tau_{\perp}/\tau_R$ remains positive up to a larger value of $\Omega(\tau_R)$ and only later changes sign. Therefore, the figure makes visually clear what is already encoded in Eqs.~\eqref{eq:component_relaxation_times}: the first sector to become pathological is the fully transverse sector, while the semi-transverse sector breaks down only at stronger magnetic field. The different curves, corresponding to different values of $r=\sigma_T^{+-}/\sigma_T$, modify the detailed slope of the decay but do not change this ordering. In all cases, $\tau_{\indep}$ reaches the unphysical regime before $\tau_{\perp}$.

This breakdown is not accidental. Since $\phi\propto\Omega\tau_\pi$, a value $\phi\gtrsim1$ means that the magnetic oscillation frequency is no longer small compared to the inverse relaxation time. In that regime, the oscillatory dynamics identified in the linear analysis cannot be neglected, and a relaxation-type truncation is no longer justified. The same conclusion is reflected in the plot: as the magnetic field increases, both effective relaxation times are driven downward, eventually crossing zero, which indicates that the attempt to describe the corresponding sectors by ordinary Israel--Stewart-type relaxation equations has ceased to be meaningful. Therefore, the truncation scheme derived here is reliable only when $\phi$ remains sufficiently small.

This gives the physical meaning of the entire construction. The truncation scheme is the step that turns the fundamental coupled equations into an Israel--Stewart-like theory for each projected component. It works when the magnetic field is weak enough that oscillatory effects remain subleading. Once that condition is lost, one must return to the underlying coupled equations for the total and relative shear tensors instead of using the truncated second-order theory. In this sense, Fig.~\ref{fig:relaxation_times_truncation} marks directly the domain of validity of the truncation itself.

\section{Bjorken flow}

The truncation scheme derived in the previous section provides a component-wise Israel--Stewart-like description for the shear sector, but at that stage the discussion is still rather formal. In order to see what these equations actually imply for the time evolution of the system, it is useful to study them in a highly symmetric flow configuration. The natural choice is Bjorken flow~\cite{Bjorken:1982qr}. This choice is motivated both physically and technically. Physically, Bjorken flow is the standard idealized description of the boost-invariant longitudinal expansion of matter produced after a relativistic heavy-ion collision. Technically, it is symmetric enough that the full tensor equations reduce to a small number of scalar evolution equations, while still retaining nontrivial dissipative dynamics. It therefore provides the cleanest setting in which one can isolate how the magnetic field breaks the degeneracy of the shear sector and how the resulting longitudinal, semi-transverse, and purely transverse components evolve differently in time.

\subsection{Setup}
\subsubsection{What is Bjorken flow}

The standard Bjorken picture describes a fluid that expands longitudinally along the beam axis and is invariant under boosts in the longitudinal direction. In addition, one usually assumes translational invariance and rotational symmetry in the transverse plane. Thus, the fluid variables do not depend on the transverse coordinates and depend on the longitudinal coordinates only through the proper time.

It is convenient to work in Milne coordinates~\cite{Bjorken:1982qr},
\begin{equation}
x^\mu = (\tau,x,y,\xi),
\end{equation}
defined in terms of Minkowski coordinates by
\begin{equation}
\tau \equiv \sqrt{t^2-z^2},
\qquad
\xi \equiv \frac{1}{2}\ln\left(\frac{t+z}{t-z}\right).
\end{equation}
In these coordinates, surfaces of constant $\tau$ describe the proper-time evolution of the longitudinally expanding system, while $\xi$ is the space-time rapidity.

The metric in Milne coordinates is
\begin{equation}
g_{\mu\nu}
=
\mathrm{diag}(1,-1,-1,-\tau^2),
\label{eq:bjorken_metric}
\end{equation}
and the only nonvanishing Christoffel symbols are
\begin{equation}
\Gamma^\tau_{\xi\xi}=\tau,
\qquad
\Gamma^\xi_{\tau\xi}=\Gamma^\xi_{\xi\tau}=\frac{1}{\tau}.
\label{eq:bjorken_christoffel}
\end{equation}
Therefore, once one works in Milne coordinates, ordinary derivatives must be replaced by covariant derivatives.

In the Bjorken frame, the 4-velocity becomes static,
\begin{equation}
u^\mu = (1,0,0,0),
\label{eq:bjorken_velocity}
\end{equation}
which means that the fluid moves only through the geometry of the coordinates, not through an explicit spatial velocity field. This is one of the main reasons why Bjorken flow is so useful: the expansion is nontrivial, but the velocity field itself is extremely simple.

The expansion scalar becomes
\begin{equation}
\theta = D_\mu u^\mu = \frac{1}{\tau}.
\label{eq:bjorken_theta}
\end{equation}
Moreover, the shear tensor generated by this expansion is
\begin{equation}
\sigma^{\mu\nu}
=
\mathrm{diag}
\left(
0,
\frac{1}{3\tau},
\frac{1}{3\tau},
-\frac{2}{3\tau^3}
\right),
\label{eq:bjorken_shear_tensor}
\end{equation}
or, equivalently, in mixed spatial form, one may think of it as carrying the familiar pattern of longitudinal dilution and transverse compression associated with boost-invariant expansion.

In ordinary viscous hydrodynamics without a magnetic field, Bjorken flow is already enough to produce nontrivial shear dynamics. The novelty in the present chapter is that, once a magnetic field is included, the projected shear sectors no longer remain degenerate. Thus, Bjorken flow becomes a clean probe of how the magnetic field breaks that degeneracy dynamically.

\subsubsection{Choice of magnetic-field direction and adapted basis}

To connect Bjorken flow to the present problem, we now choose the magnetic field to lie in the transverse plane, rather than along the beam axis. This is the natural choice for the problem under consideration, since the magnetic field produced in relativistic heavy-ion collisions is predominantly transverse.

In the Milne basis, we choose the magnetic field to be in the $x$--direction,
\begin{equation}
b^\mu = (0,1,0,0).
\label{eq:bjorken_b_choice}
\end{equation}
In the previous sections we introduced basis vectors orthogonal to $b^\mu$, denoted by $x^\mu$ and $y^\mu$. These basis vectors may be taken as
\begin{equation}
x^\mu = (0,0,1,0),
\qquad
y^\mu = \left(0,0,0,\frac{1}{\tau}\right).
\label{eq:bjorken_basis_vectors}
\end{equation}
Here, we must be careful with this notation: $x^\mu$ actually describes the $y$--direction and $y^\mu$ describes the beam direction (hence the factor $1/\tau$).
With this choice, the $x$-direction is singled out by the magnetic field, while the remaining transverse plane of the magnetic decomposition is spanned by the $y$-direction and the longitudinal rapidity direction.

\subsubsection{Projected shear tensor in Bjorken flow}

The shear tensor of the background flow has the following projected components with respect to the magnetic direction:
\begin{equation}
\sigma_\parallel \equiv b_\mu b_\nu \sigma^{\mu\nu} = \frac{1}{3\tau},
\label{eq:bjorken_sigma_parallel}
\end{equation}
\begin{equation}
\sigma_\perp^\pm \equiv \ell_\mp^\mu b^\nu \sigma_{\mu\nu} = 0,
\label{eq:bjorken_sigma_perp}
\end{equation}
\begin{equation}
\sigma_{\indep}^\pm
\equiv
\ell_\pm^\mu \ell_\pm^\nu \sigma_{\mu\nu}
=
\frac{1}{2\tau}.
\label{eq:bjorken_sigma_perpperp}
\end{equation}
Thus, Bjorken flow generates a longitudinal projected shear component and a purely transverse projected shear component, but no semi-transverse one. This is why the semi-transverse sector can be consistently set to zero in this flow, provided its initial value vanishes.

The shear-stress tensor itself can then be decomposed as
\begin{equation}
\pi^{\mu\nu}
=
\pi_\parallel\, b^\mu b^\nu
+
\left(
\pi_{\indep}^{++}
+
\pi_{\indep}^{--}
-
\frac{\pi_\parallel}{2}
\right)x^\mu x^\nu
-
\left(
\pi_{\indep}^{++}
+
\pi_{\indep}^{--}
+
\frac{\pi_\parallel}{2}
\right)y^\mu y^\nu .
\label{eq:bjorken_pi_decomp}
\end{equation}
It is convenient to define the combination
\begin{equation}
\pi_{\indep}
\equiv
\pi_{\indep}^{++}
+
\pi_{\indep}^{--},
\label{eq:bjorken_pi_perpperp_def}
\end{equation}
so that the full dynamics is described by the pair $(\pi_\parallel,\pi_{\indep})$.

This is one of the main simplifications induced by Bjorken symmetry. The full tensorial dynamics collapses to two independent scalar amplitudes. However, unlike in ordinary Bjorken flow, those amplitudes are no longer equal or trivially related. The magnetic field has split them into genuinely distinct dynamical sectors.

\subsection{Equation of state and reduced equations in Bjorken flow}
To close the system, one needs an equation of state. For the massless two-component gas considered here, the energy density is taken to be
\begin{equation}
\epsilon
=
\frac{3\times 2\times 2\times 3}{\pi^2}\,T^4.
\label{eq:bjorken_eos}
\end{equation}
This relation reflects the conformal equation of state of a massless classical gas, $\epsilon = 3g\,T^4/\pi^2$, where the leading factor of $3$ is the standard kinematic coefficient for a massless ideal gas and $g = 2 \times 2 \times 3 = 12$ is the internal degeneracy, counting the two charged species (particles and antiparticles), the two spin states, and the three quark colours. Therefore, the temperature can always be expressed in terms of the energy density, and vice versa.


Using all the ingredients above, the fluid-dynamical equations reduce to a closed system for $\epsilon(\tau)$, $\pi_\parallel(\tau)$, and $\pi_{\indep}(\tau)$. The energy-conservation equation becomes
\begin{equation}
\frac{d\epsilon}{d\tau}
=
\frac{\pi_\parallel}{2\tau}
+
\frac{\pi_{\indep}}{2\tau}
-
\frac{4\epsilon}{3\tau}.
\label{eq:bjorken_energy_eq}
\end{equation}
This equation has a very transparent interpretation. The last term is the usual ideal Bjorken cooling of a conformal fluid, while the first two terms represent the work done by the anisotropic shear sectors. Thus, in the magnetized case, the cooling of the system is not governed by one single shear correction, but by two distinct projected components.

The evolution equation for the longitudinal shear component reads
\begin{equation}
\frac{d}{d\tau}\left(\frac{\pi_\parallel}{\epsilon}\right)
+
\frac{1}{\tau_\pi}\frac{\pi_\parallel}{\epsilon}
=
\frac{8}{45\tau}
+
\frac{5}{21\tau}\frac{\pi_\parallel}{\epsilon}
-
\frac{5}{21\tau}\frac{\pi_{\indep}}{\epsilon}
-
\left(\frac{\pi_\parallel}{\epsilon}\right)^2
\left(
\frac{\pi_\parallel+\pi_{\indep}}{2\tau}
\right).
\label{eq:bjorken_pi_parallel_eq}
\end{equation}
Similarly, the evolution equation for the purely transverse sector is
\begin{align}
\frac{d}{d\tau}\left(\frac{\pi_{\indep}}{\epsilon}\right)
+
\frac{1}{\tau_{\indep}}\frac{\pi_{\indep}}{\epsilon}
&=
\frac{1}{1-4\phi^2}\frac{8}{15\tau}
-
\frac{5}{21\tau}\frac{\pi_{\indep}}{\epsilon}
-
\frac{1}{1-4\phi^2}\frac{5}{7\tau}\frac{\pi_\parallel}{\epsilon}
+
\frac{4\phi\dot\phi}{1-4\phi^2}\frac{\pi_{\indep}}{\epsilon}
\nonumber\\
&\quad
-
\left(\frac{\pi_{\indep}}{\epsilon}\right)^2
\left(
\frac{\pi_\parallel+\pi_{\indep}}{2\tau}
\right).
\label{eq:bjorken_pi_perpperp_eq}
\end{align}
In addition, one needs
\begin{equation}
\dot\phi
=
-\phi\left(\frac{1}{\tau}+2\frac{\dot T}{T}\right),
\label{eq:bjorken_phi_dot}
\end{equation}
together with
\begin{equation}
\frac{\dot T}{T}
=
\frac{1}{4\tau}
\left(
\frac{\pi_\parallel+\pi_{\indep}}{2\epsilon}
-
\frac{4}{3}
\right).
\label{eq:bjorken_T_dot}
\end{equation}

Now, first, Eq.~\eqref{eq:bjorken_energy_eq} shows that the energy density is sensitive to the sum of the two projected shear sectors. Therefore, even though the magnetic field splits the shear tensor dynamically, the bulk cooling of the system still depends on their combined effect.

Second, Eqs.~\eqref{eq:bjorken_pi_parallel_eq} and \eqref{eq:bjorken_pi_perpperp_eq} are not the same equation with different labels. Their Navier--Stokes sources are different, their nonlinear self-couplings are different, and most importantly the transverse equation carries explicit $\phi$-dependence through the factors $(1-4\phi^2)^{-1}$. Thus, the effect of the magnetic field is that it changes the very structure of the evolution equation.

Third, the denominator $1-4\phi^2$ is precisely the same structure that already appeared in the truncation-scheme analysis. Therefore, Bjorken flow provides a dynamical realization of the limitation identified earlier: as $\phi$ approaches $1/2$, the purely transverse sector becomes increasingly sensitive to the magnetic field, and the truncated description approaches its domain of breakdown.

\subsubsection{Connection with ordinary Bjorken flow}

The relation to the usual Bjorken-flow equations is seen simply by taking the vanishing-field limit,
\begin{equation}
B\to 0
\qquad \Longrightarrow \qquad
\phi\to 0.
\end{equation}
In this limit, the magnetic splitting of the shear sector disappears, so that the component-dependent transport coefficients collapse to their isotropic values,
\begin{equation}
\eta_{\parallel}=\eta_{\perp}=\eta_{\indep}\equiv\eta,
\qquad
\tau_{\parallel}=\tau_{\perp}=\tau_{\indep}\equiv\tau_\pi.
\label{eq:bjorken_isotropic_limit}
\end{equation}
Therefore, the special role of $\pi_{\indep}$ is lost, all explicit $\phi$-dependent terms vanish, and the equations reduce to the standard transient Bjorken system.

Thus, the present construction may be viewed as a magnetically deformed Bjorken flow. The spacetime symmetries are still those of Bjorken flow, but the internal dissipative structure of the fluid is no longer isotropic. This is exactly the effect we wanted to isolate -- not just a change in the background expansion, but rather a change in how the shear-stress tensor responds to that expansion.

\subsection{Time evolution of the magnetic field}

In order to solve the Bjorken-flow equations, one still needs an equation of motion for
the magnetic field. At this point, the magnetic field is no longer just a background
direction that defines the projections. Its magnitude also evolves in time and therefore
feeds back into the transport coefficients through the parameter $\phi$.

The evolution of the magnetic field is obtained from Maxwell's equations. In the present
Bjorken-flow setup, this leads to
\begin{equation}
\dot B + B\theta = 0.
\label{eq:bjorken_Bdot_eq}
\end{equation}
Since for Bjorken flow
\begin{equation}
\theta=\frac{1}{\tau},
\end{equation}
one immediately finds
\begin{equation}
\dot B + \frac{B}{\tau}=0,
\end{equation}
whose solution is~\cite{Roy:2015kma}
\begin{equation}
B(\tau)=B_0\frac{\tau_0}{\tau}.
\label{eq:bjorken_B_solution}
\end{equation}
Thus, within this simple setup, the magnetic field decays inversely with proper time.

This result is physically natural. The system expands longitudinally, and the magnetic
flux is diluted by the expansion. As a consequence, the magnetic field is strongest at
early times and becomes progressively less important as the system cools and dilutes.
This also means that any actual magnetic impact on the shear sector is expected to be most visible at early times, precisely where the system is farthest from the late-time Navier--Stokes regime.

\subsection{Truncated theory: First numerical results}

To proceed numerically, one chooses initial conditions at a starting proper time $\tau_0$.
Following the setup used throughout this analysis, one may take the system to be initially
in local equilibrium,
\begin{equation}
\pi_\parallel(\tau_0)=0,
\qquad
\pi_{\indep}(\tau_0)=0,
\label{eq:bjorken_initial_shear}
\end{equation}
with given initial energy density $\epsilon_0\equiv \epsilon(\tau_0)$ and magnetic field $B_0\equiv B(\tau_0)$. In the numerical examples, one usually fixes $\tau_0=0.1~\mathrm{fm}$ and $\epsilon_0(\tau_0)=1000~\mathrm{fm}^{-4}$. For reference, a value of $B_0 = 30~\mathrm{fm}^{-2}$ corresponds to magnetic field strength of $\sim 10^{19}\, G$, which is the magnitude estimated for LHC, while $B_0 \sim 50~\mathrm{fm}^{-2}$ corresponds to field strengths $\sim 10^{20}\, G$~\cite{Skokov:2009qp}.

It is also convenient to choose the microscopic rates such that
\begin{equation}
\Sigma'=\frac{4}{3}\Sigma,
\label{eq:sigmaprime_choice_bjorken}
\end{equation}
which corresponds to the choice $r\equiv \sigma_T^{+-}/\sigma_T=1$. Once $\Sigma$ is fixed
through the desired value of $\eta/s$, the system is completely determined. The resulting system of coupled ODEs is solved numerically in Python using a Runge--Kutta integration scheme.

The first thing one may ask is whether the truncated equations already show any visible magnetic effect during the Bjorken expansion. The answer is yes, but only within a rather restricted window. When one solves Eqs.~\eqref{eq:bjorken_energy_eq}--\eqref{eq:bjorken_pi_perpperp_eq}
for moderate values of the initial magnetic field, the longitudinal and transverse components
of the shear-stress tensor are modified at early times, especially around $\tau\sim\tau_R$,
where $\tau_R$ denotes the microscopic relaxation time scale. This is the regime where the
magnetic field is still strong enough to matter, but the truncation has not yet broken down.

At this stage, it is useful to include the first pair of plots, namely the time evolution of
the longitudinal and transverse components of the shear-stress tensor obtained from the
truncated equations for several values of the initial magnetic field $B_0$. These plots make
two points immediately visible.
\begin{figure}[H]
    \centering
    \includegraphics[width=1\linewidth]{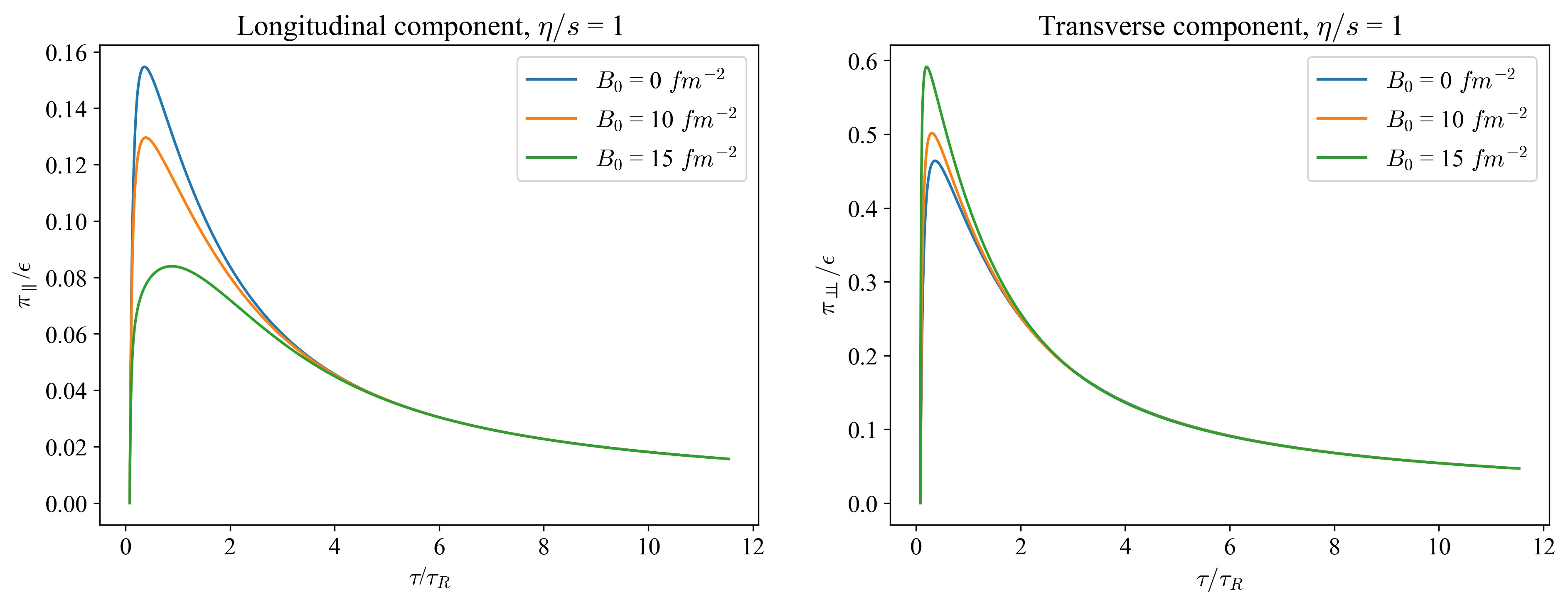}
    \caption[Truncated Bjorken evolution]{Time evolution of the longitudinal and transverse components of the shear-stress tensor in Bjorken flow, obtained from the truncated Israel--Stewart-like equations, for several values of the initial magnetic field $B_0$. All simulations are performed with $\eta/s=1$. The magnetic field produces only a modest deformation of the relaxation pattern, especially in the transverse sector, but no clear oscillatory behavior appears. This illustrates the limited range of validity of the truncated description in the magnetized regime.}
    \label{fig:bjorken_truncated_eta1}
\end{figure}

As shown in Fig.~\ref{fig:bjorken_truncated_eta1}, the truncated Israel--Stewart-like theory predicts only a modest magnetic modification of the Bjorken-flow evolution. The longitudinal component is only weakly affected as the initial magnetic field is increased, while the transverse component shows a somewhat stronger response, but still no pronounced oscillatory behavior. This is already an important indication that the truncation is too restrictive in the regime where the magnetic field should matter most. In particular, the transverse sector, which is the one most strongly coupled to the magnetic field, is forced into a single relaxation-type equation. As a result, the mechanism responsible for the oscillatory exchange between total and relative shear is effectively removed.

This is precisely why the truncated theory must be interpreted with caution. Formally, it resembles an Israel--Stewart theory for each projected component, but physically it is not a proper description once the magnetic coupling becomes important. The point is not merely that the quantitative results may become inaccurate. Rather, the very structure of the transient dynamics is altered by the truncation, since the relative shear has been eliminated in favor of a relaxation-type closure. Therefore, Fig.~\ref{fig:bjorken_truncated_eta1} should be read mainly as a demonstration of the limitations of the truncated approach: for weak enough fields it reproduces a small deformation of the usual viscous Bjorken evolution, but it fails to capture the qualitatively new oscillatory behavior that emerges in the full coupled equations. This is why one must go back to the fundamental coupled system if the goal is to describe genuinely magnetic effects in the shear sector.

\subsubsection{Why one must go beyond the truncation scheme}

The limitation of the truncated theory is not a minor technical issue. It is in fact the central physical message of the present analysis. From the previous section, we already know that the effective relaxation times of the projected shear sectors can become negative. In Bjorken flow, this means that the reduced Israel--Stewart-like equations cease to be trustworthy precisely in the regime where the magnetic field should have the strongest effect. In other words, the truncated equations remain reliable mainly when the magnetic field is weak enough that it does not qualitatively alter the dynamics.

This point is important. The truncation scheme was introduced in order to obtain a closed second-order theory for the projected components of the total shear-stress tensor. However, this closure was achieved by eliminating the relative shear sector. As long as the magnetic coupling is weak, this may still provide a reasonable approximation. But once the magnetic field becomes sufficiently strong, the relative shear is no longer a small correction. It becomes an essential part of the transient dynamics, and forcing the system into a single relaxation-type equation is no longer justified.

Therefore, if one wants to study genuinely magnetic effects, one must return to the more fundamental coupled equations, where the relative shear has not been integrated out and no second-order relaxation-type truncation has been imposed. This is the next logical step. The goal is to recover, in an expanding background, the oscillatory dynamics already suggested by the linear analysis.

\subsection{Full coupled system}
\subsubsection{Fundamental coupled equations in Bjorken flow}

When no truncation scheme is imposed, the relevant Bjorken-flow equations take the form
\begin{equation}
\frac{d\epsilon}{d\tau}
=
\frac{\pi_\parallel}{2\tau}
+
\frac{\pi_{\indep}}{2\tau}
-
\frac{4\epsilon}{3\tau},
\label{eq:bjorken_exact_energy}
\end{equation}
\begin{equation}
\frac{d}{d\tau}\left(\frac{\pi_\parallel}{\epsilon}\right)
+
\Sigma\,\frac{\pi_\parallel}{\epsilon}
=
\frac{8}{45\tau}
+
\frac{5}{21\tau}\frac{\pi_\parallel}{\epsilon}
-
\frac{5}{21\tau}\frac{\pi_{\indep}}{\epsilon}
-
\left(\frac{\pi_\parallel}{\epsilon}\right)^2
\left(
\frac{\pi_\parallel+\pi_{\indep}}{2\tau}
\right),
\label{eq:bjorken_exact_parallel}
\end{equation}
\begin{align}
\frac{d}{d\tau}\left(\frac{\pi_{\indep}}{\epsilon}\right)
+
\Sigma\,\frac{\pi_{\indep}}{\epsilon}
-
\frac{2|q|B}{5T}\frac{\delta\hat\pi_{\indep}}{\epsilon}
&=
\frac{8}{15\tau}
-
\frac{5}{7\tau}\frac{\pi_\parallel}{\epsilon}
-
\frac{5}{21\tau}\frac{\pi_{\indep}}{\epsilon}
-
\left(\frac{\pi_{\indep}}{\epsilon}\right)^2
\left(
\frac{\pi_\parallel+\pi_{\indep}}{2\tau}
\right),
\label{eq:bjorken_exact_transverse}
\\
\frac{d}{d\tau}\left(\frac{\delta\hat\pi_{\indep}}{\epsilon}\right)
+
\Sigma'\,\frac{\delta\hat\pi_{\indep}}{\epsilon}
+
\frac{2|q|B}{5T}\frac{\pi_{\indep}}{\epsilon}
&=
-
\frac{5}{21\tau}\frac{\delta\hat\pi_{\indep}}{\epsilon}
-
\left(\frac{\delta\hat\pi_{\indep}}{\epsilon}\right)^2
\left(
\frac{\pi_\parallel+\pi_{\indep}}{2\tau}
\right),
\label{eq:bjorken_exact_relative}
\end{align}
where
\begin{equation}
\pi_{\indep}\equiv \pi_{\indep}^-+\pi_{\indep}^+,
\qquad
\delta\pi_{\indep}\equiv \delta\pi_{\indep}^- - \delta\pi_{\indep}^+,
\qquad
\delta\pi_{\indep}=i\,\delta\hat\pi_{\indep}.
\label{eq:bjorken_exact_defs}
\end{equation}

These equations are considerably more informative than the truncated system. The essential new ingredient is that the relative $\indep$ shear now appears explicitly as an independent dynamical variable. The magnetic field couples it directly to the total $\indep$ shear through the terms proportional to $2|q|B/(5T)$. This is exactly the structure needed to generate oscillatory behavior.

The point is simple but important. In the truncated description, the $\indep$ sector is forced into a single relaxation-type equation. In the full theory, by contrast, the $\indep$ sector belongs to a coupled two-variable system. It is this coupled structure that allows the dynamics to rotate between total and relative shear, which is the physical origin of the oscillations anticipated earlier in the linear analysis.

\begin{figure}
    \centering
    \includegraphics[width=1\linewidth]{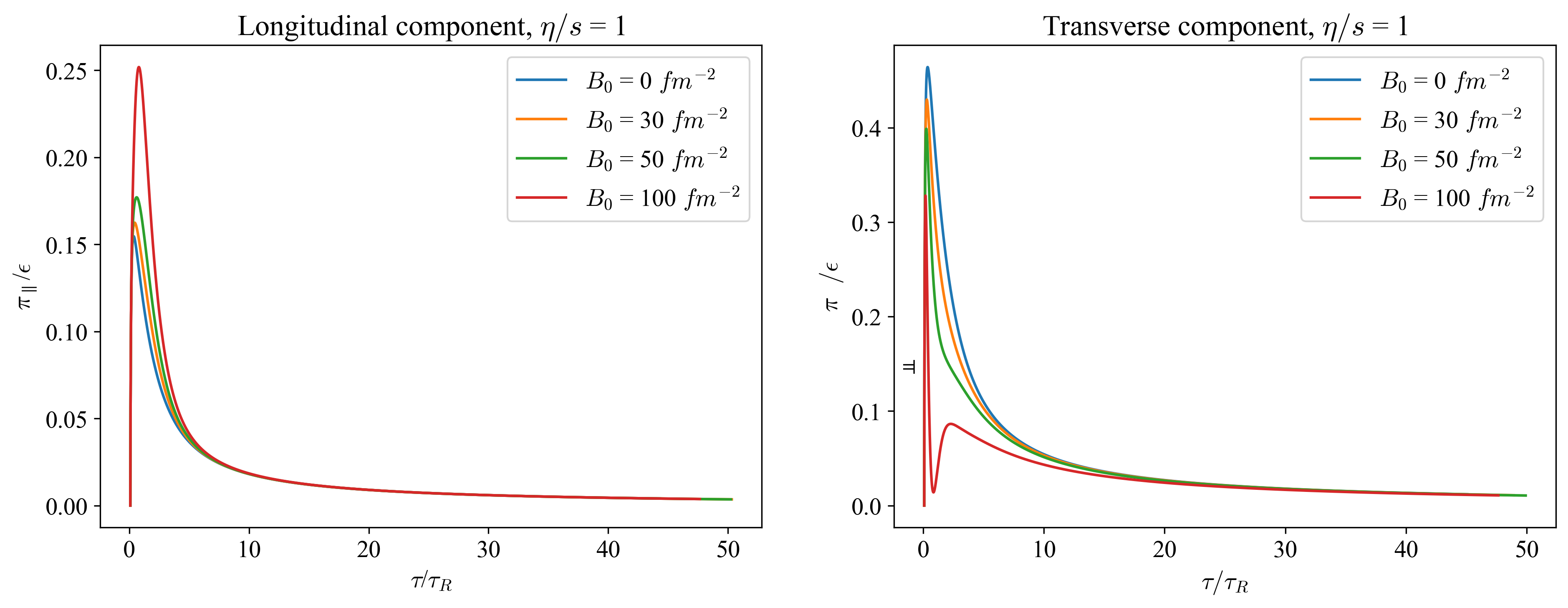}
    \caption[Full Bjorken evolution for $\eta/s=1$]{Time evolution of the longitudinal and $\indep$ components of the shear-stress tensor in Bjorken flow, obtained from the full coupled equations for the total and relative shear sectors, for several values of the initial magnetic field $B_0$. All simulations are performed with $\eta/s=1$. The magnetic field modifies both sectors, but the effect remains comparatively modest because the microscopic relaxation time is still short.}
    \label{fig:bjorken_full_eta1}
\end{figure}

\begin{figure}
    \centering
    \includegraphics[width=1\linewidth]{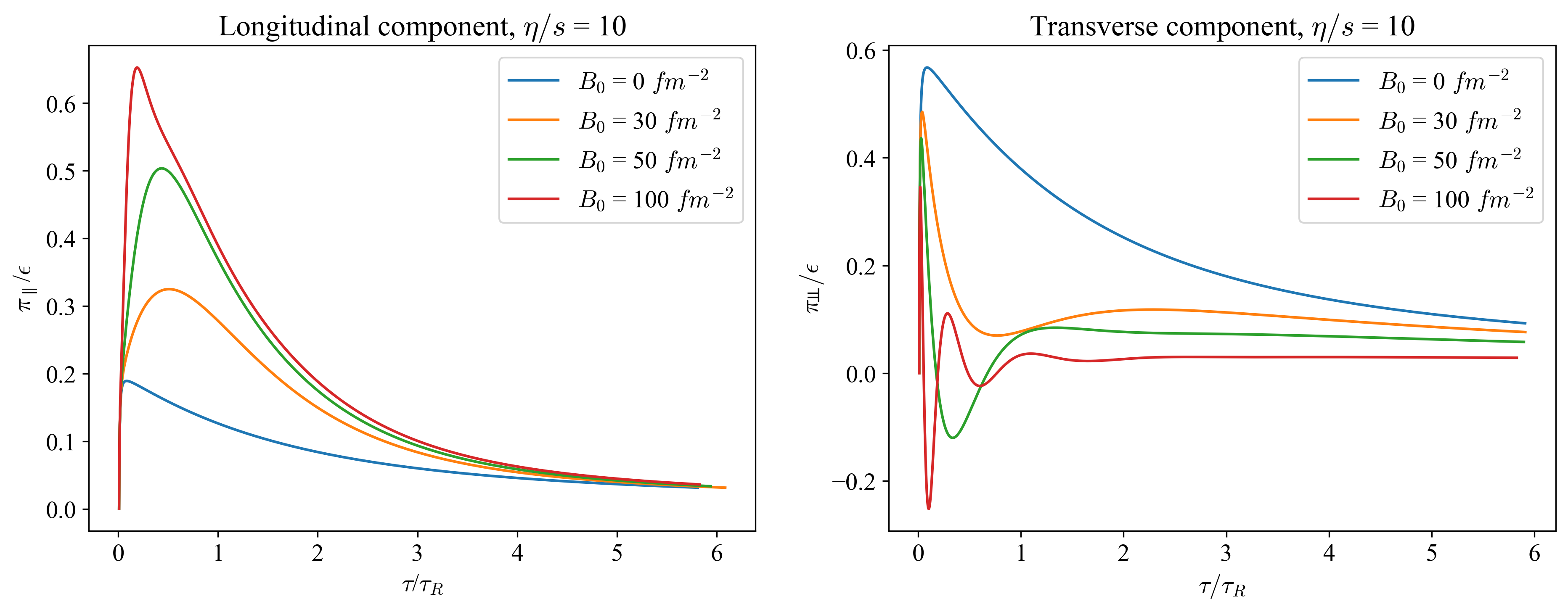}
    \caption[Full Bjorken evolution for $\eta/s=10$]{Time evolution of the longitudinal and $\indep$ components of the shear-stress tensor in Bjorken flow, obtained from the full coupled equations for the total and relative shear sectors, for several values of the initial magnetic field $B_0$. All simulations are performed with $\eta/s=10$. In this case the $\indep$ sector develops clearly visible damped oscillations, while the longitudinal component remains comparatively less affected.}
    \label{fig:bjorken_full_eta10}
\end{figure}

\subsubsection{Oscillatory dynamics in an expanding plasma}

The qualitative difference between the truncated and untruncated theories becomes manifest once Eqs.~\eqref{eq:bjorken_exact_energy}--\eqref{eq:bjorken_exact_relative} are solved for sufficiently large magnetic fields and for different values of $\eta/s$.

The main result is that the $\indep$ component of the shear-stress tensor no longer relaxes monotonically to zero. Instead, for sufficiently large magnetic field, it exhibits damped oscillations during its approach to the late-time regime. The longitudinal component, on the other hand, does not display this behavior nearly as strongly and remains much closer to the usual relaxational pattern. This is already visible in Fig.~\ref{fig:bjorken_full_eta1}, where the magnetic field changes the evolution of both components, but the effect is still moderate. The same pattern becomes much clearer in Fig.~\ref{fig:bjorken_full_eta10}, where the $\indep$ component shows pronounced oscillatory behavior.

The interpretation is straightforward. The oscillations appear primarily in the $\indep$ sector because this is the sector that couples directly to the magnetic field through the relative-shear variable $\delta\hat\pi_{\indep}$. They do not appear in the truncated theory because the very mechanism responsible for them has been removed by construction. Once the relative shear is eliminated, the possibility of this coupled oscillatory exchange is lost.

A second important point concerns the role of viscosity. Comparing Figs.~\ref{fig:bjorken_full_eta1} and \ref{fig:bjorken_full_eta10}, one sees that the oscillations become much more visible as $\eta/s$ increases. This happens because a larger value of $\eta/s$ corresponds to a larger microscopic relaxation time. The system then remains away from its asymptotic regime for a longer interval, and the magnetic field has more time to imprint its oscillatory dynamics before it decays away. By contrast, for smaller $\eta/s$, the system relaxes more quickly, so the $\indep$ sector has less time to complete oscillations before the magnetic field becomes weak.

Therefore, in the full Bjorken problem, stronger oscillations are expected when both of the following are true:
\begin{equation}
B_0 \ \text{is sufficiently large},
\qquad
\eta/s \ \text{is sufficiently large}.
\end{equation}
This is exactly the trend displayed by the numerical solutions.

\subsubsection{Role of the parameter $\phi$}

The oscillatory dynamics becomes easier to interpret once one tracks the time dependence of the dimensionless parameter $\phi$. As discussed earlier, $\phi$ measures the strength of the magnetic coupling relative to the microscopic relaxation scale. In the $\indep$ sector, the truncation scheme already suggested that the dynamics becomes problematic when $\phi$ exceeds the critical value $1/2$.

\begin{figure}
    \centering
    \includegraphics[width=0.7\linewidth]{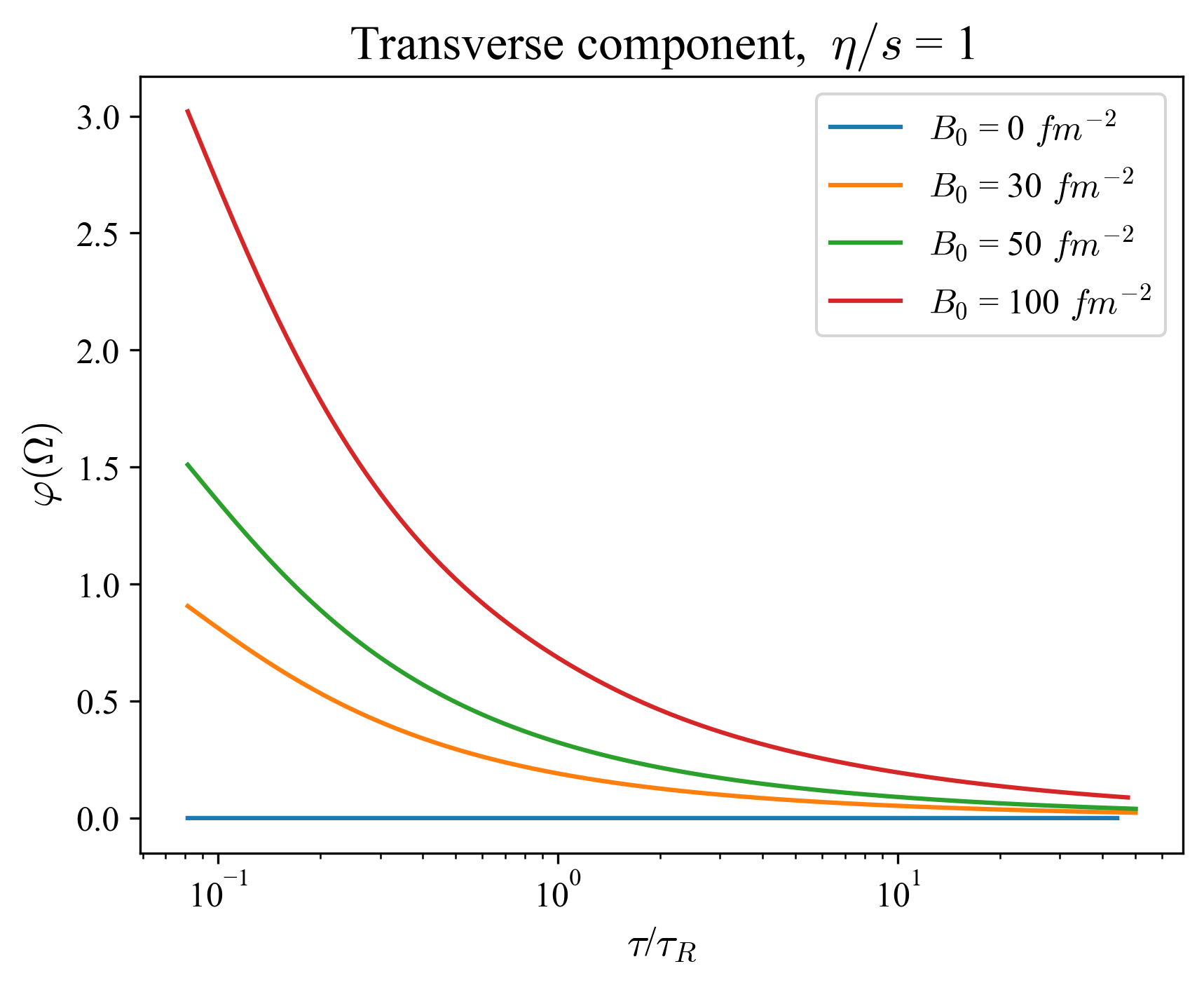}
    \caption[Time evolution of $\phi$]{Time evolution of the dimensionless parameter $\phi$ as a function of $\tau/\tau_R$ for the same parameter sets used in the Bjorken-flow simulations. The horizontal line at $\phi=1/2$ indicates the scale associated with the onset of oscillatory behavior in the $\indep$ sector.}
    \label{fig:phi_bjorken}
\end{figure}

Figure~\ref{fig:phi_bjorken} provides a direct way to correlate the oscillations seen in Figs.~\ref{fig:bjorken_full_eta1} and \ref{fig:bjorken_full_eta10} with the strength of the magnetic coupling. When $\phi$ exceeds $1/2$ only briefly and at very early times, one expects at most a weak hint of oscillatory behavior. By contrast, when $\phi$ remains above $1/2$ for a substantial part of the evolution, the $\indep$ component has enough time to develop clearly visible oscillations.

Thus, the $\phi$-plot is not merely supplementary. It gives the quantitative reason why one set of parameters displays pronounced oscillations while another does not. It is the direct link between the formal criterion extracted from the linear analysis and the actual time-domain behavior seen in the Bjorken solutions.

This is also the right place to emphasize that the criterion
\begin{equation}
\phi>\frac{1}{2}
\end{equation}
should not be interpreted as a perfectly sharp nonlinear boundary. In the full Bjorken problem, the decay of the magnetic field and the time dependence of the background energy density smooth out the transition. Still, the basic logic remains the same: once $\phi$ becomes sufficiently large for sufficiently long, the $\indep$ sector develops oscillatory transient dynamics.

\subsubsection{Approach to the Navier--Stokes limit}

Even in the full coupled theory, the magnetic field decays with time according to Eq.~\eqref{eq:bjorken_B_solution}. Therefore, at sufficiently late times, one expects the system to gradually forget about the magnetic field and to approach the conventional Navier--Stokes limit.
\begin{figure}
    \centering
    \includegraphics[width=1\linewidth]{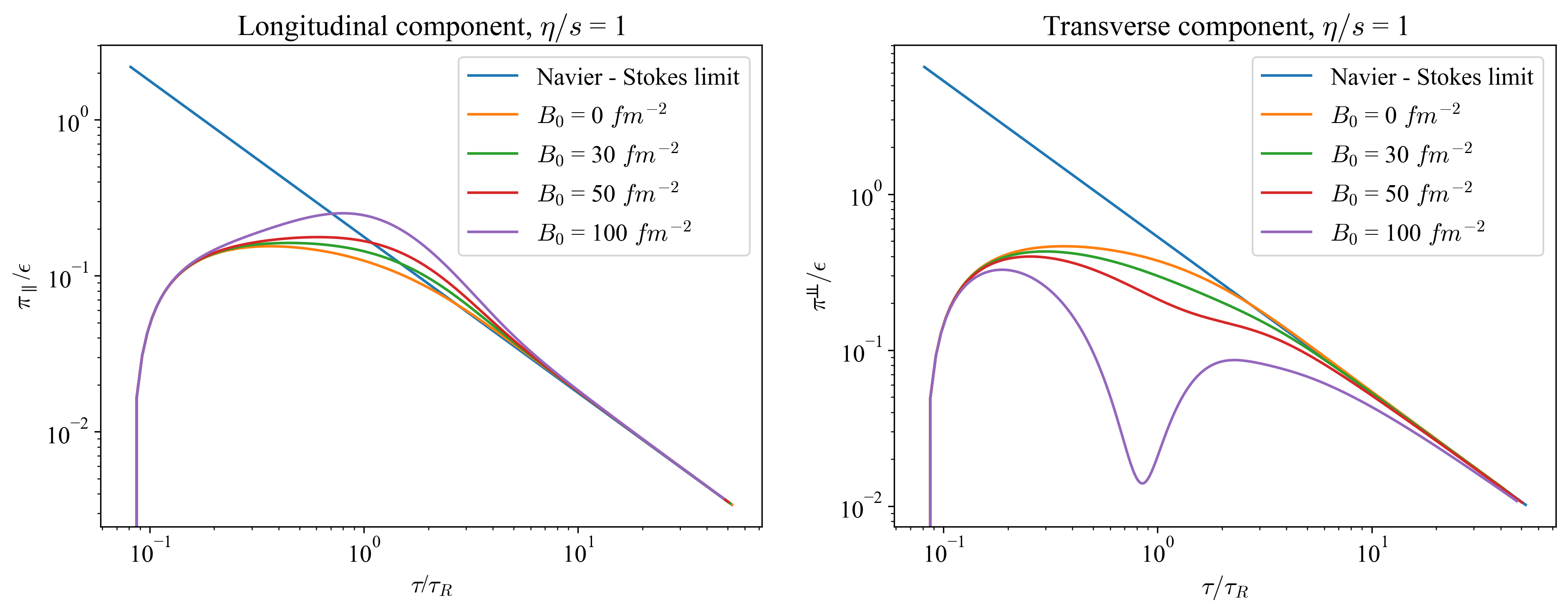}
    \caption[Approach to Navier--Stokes behavior]{Longitudinal and $\indep$ components of the shear-stress tensor as functions of $\tau/\tau_R$, compared with their respective Navier--Stokes limits, for several values of the initial magnetic field $B_0$. At late times both sectors approach their Navier--Stokes behavior, although the $\indep$ component may do so through damped oscillations.}
    \label{fig:bjorken_navier_stokes}
\end{figure}
This expectation is confirmed in Fig.~\ref{fig:bjorken_navier_stokes}, where the numerical solutions for $\pi_\parallel$ and $\pi_{\indep}$ are compared with their corresponding Navier--Stokes values. The longitudinal component approaches its Navier--Stokes limit in the usual manner, while the $\indep$ component approaches its own late-time limit through damped oscillations whenever the magnetic field was initially strong enough.

This comparison is important because it clarifies the status of the oscillations. They do not represent a new late-time equilibrium state. Rather, they are a transient phenomenon that appears during the system's approach to the usual late-time viscous regime. In other words, the oscillations are a feature of transient magnetized dynamics, not of the final asymptotic state.

This is precisely why the Bjorken analysis is so useful. It separates the early-time magnetic physics from the late-time viscous attractor in a very clean way. The system is strongly affected by the magnetic field only while the field is still large. Once the field has decayed sufficiently, the familiar hydrodynamic intuition is recovered.

\subsubsection{Interpretation of the numerical results}

Taken together, the Bjorken-flow results lead to a clear picture. For weak enough magnetic fields, the component-wise Israel--Stewart-like truncation is formally consistent, but in practice the magnetic field produces only a modest deformation of the dynamics. This is why the truncated simulations showed only limited deviations from ordinary relaxation. For stronger magnetic fields, however, the truncated equations break down because the projected relaxation times become negative. This breakdown is not a numerical artifact. It is a genuine signal that the underlying dynamics has left the regime where a relaxation-type closure is adequate.

When one then returns to the full coupled equations, the missing dynamics reappears in the form of oscillations of the $\indep$ shear sector. These oscillations are strongest when the magnetic coupling remains important for a sufficiently long time, which typically requires both larger $B_0$ and larger $\eta/s$. This is exactly what is seen when comparing Fig.~\ref{fig:bjorken_full_eta1} with Fig.~\ref{fig:bjorken_full_eta10}, and it is quantitatively supported by Fig.~\ref{fig:phi_bjorken}.

Finally, because the magnetic field decays as $1/\tau$, all of these genuinely magnetic effects are transient. At late times the solutions approach the conventional Navier--Stokes behavior, even if the intermediate-time evolution was qualitatively different, as illustrated in Fig.~\ref{fig:bjorken_navier_stokes}. The main lesson is therefore not that the magnetic field changes the final state, but that it qualitatively changes the transient path by which the system approaches that state.

\section{Summary}

This chapter took the first step towards a realistic kinetic description of a magnetized plasma by treating the two charged constituents as independent microscopic components. The central result is that this seemingly modest generalization fundamentally changes the structure of the dissipative theory: the magnetic field no longer influences just a single shear equation, but instead couples the total shear-stress to a new relative degree of freedom that is invisible to the energy-momentum tensor $(\delta \pi^{\mu\nu})$. This coupling turns the shear sector into an oscillatory system: a qualitative departure from the purely relaxation picture of standard Israel--Stewart theory \cite{Israel:1976tn} that persists even in the simplest expanding geometry.

The Bjorken-flow analysis showed that a naive truncation of the coupled system, which attempts to recover a standard second-order theory, breaks down precisely where the magnetic field is most important. Only by keeping the relative shear as an independent dynamical variable does the oscillatory transient behavior emerge, and only then does one see how the magnetic field imprints itself on the evolution before eventually decaying away.

What is still missing, however, is the electric field. Throughout this chapter the non-resistive limit was imposed, meaning that the charge transport and electromagnetic induction were suppressed from the outset. A real plasma resists this suppression: finite conductivity allows the electric field to develop, carry current, and dissipate 
energy. The next chapter lifts this restriction, promoting the electric field to an independent dynamical variable and thereby exposing the full resistive structure of the two-component theory.
\chapter{Resistive relativistic magnetohydrodynamics for a two-component plasma}
\label{chap:resistive_rmhd_two_component}

In the previous chapter, we developed the non-resistive relativistic magnetohydrodynamic description of a two-component plasma from kinetic theory and examined in detail how the magnetic field modifies the dissipative sector. In particular, the covariant decomposition of the electromagnetic field, the structure of the macroscopic conservation laws, the Landau frame choice, and the use of the 14-moment approximation were established there. The purpose of the present chapter is to extend that framework to the resistive case, where the electric current becomes an independent dissipative degree of freedom that must be evolved together with the shear-stress tensor. The resistive case has been studied previously for single-component systems~\cite{Denicol:2019rmhd,Panda:2021res,Dash:2022xkz}, and here we extend those developments to the two-component plasma discussed in the previous chapter~\cite{Kushwah:2025jsb}.

The physical motivation for this extension is straightforward: the non-resistive limit assumes infinite electrical conductivity, which is never realized in practice. While this may often provide a reasonable leading-order approximation, corrections associated with finite conductivity will inevitably appear and must be understood theoretically. Physically, finite conductivity relaxes the ideal magnetohydrodynamic freezing condition and allows charge diffusion and electric-field dynamics to develop. As a consequence, the plasma can sustain a non-trivial charge-diffusion current that must be incorporated as an independent dissipative degree of freedom.

We remark that although the derivation is performed in the presence of both electric and magnetic fields, the explicit applications studied later in the chapter will focus only on the electric field, i.e., the analysis is specialized to homogeneous and Bjorken expanding systems including only the effects of the electric field. This makes it possible to isolate the genuinely new features of the resistive sector, namely the relaxation of the current, its non-linear backreaction, and the generation of a sizable anisotropy even in situations where no non-trivial flow profile is imposed. 

The structure of the chapter is therefore as follows. We first present the macroscopic conservation laws and electromagnetic equations relevant for the resistive two-component plasma. We then introduce the corresponding kinetic theory description and apply the 14-moment approximation to close the hierarchy of moments. From this, we obtain the coupled equations of motion for the dissipative quantities. Finally, we specialize these equations to simple physical settings in order to isolate the role of the electric field in the resistive dynamics. 

\section{Revisiting Maxwell's equations and conservation laws}

The macroscopic setup for the resistive theory is built on the same electromagnetic and fluid decompositions established in the previous Chapter. We collect the key equations here for reference without repeating their derivation. The energy-momentum tensor is decomposed in terms of the fluid velocity as,
\begin{equation}
T^{\mu\nu}
=
\epsilon u^\mu u^\nu
-
(P+\Pi)\Delta^{\mu\nu}
+
2h^{(\mu}u^{\nu)}
+
\pi^{\mu\nu},
\label{eq:ch6_T_fluid}
\end{equation}
and the Faraday tensor as
\begin{equation}
F^{\mu\nu}
=
E^\mu u^\nu
-
E^\nu u^\mu
+
\epsilon^{\mu\nu\alpha\beta}u_\alpha B_\beta,
\label{eq:ch6_F}
\end{equation}
where $E^\mu=u_\nu F^{\mu\nu}$, $2B^\mu=\epsilon^{\mu\nu\alpha\beta}F_{\nu\alpha}u_\beta$, $b^\mu\equiv B^\mu/B$, and $b^{\mu\nu}\equiv -\epsilon^{\mu\nu\alpha\beta}u_\alpha b_\beta$ are all as defined in Chapter~\ref{ch:two_component_rmhd}. Although later applications focus on the electric field alone, the formal setup retains both $E^\mu$ and $B^\mu$.

Recall that the electromagnetic field evolves according to Maxwell's equations,
\begin{equation}
\partial_\mu F^{\mu\nu}=J^\nu,
\qquad
\partial_\mu \tilde{F}^{\mu\nu}=0,
\label{eq:ch6_Maxwell}
\end{equation}
with $\tilde{F}_{\mu\nu}$ being the Hodge dual and $J^\mu$ the electric net-charge four-current. The electric-charge four-current is decomposed as
\begin{equation}
J^\mu=n_q u^\mu+V_q^\mu,
\label{eq:ch6_J}
\end{equation}
with $n_q=J^\mu u_\mu$ the charge density in the local rest frame and
$V_q^\mu=\Delta^{\mu\nu}J_\nu$ the charge-diffusion four-current.
In addition to the electric current, the total particle four-current is 
\begin{equation}
N^\mu=n u^\mu+V^\mu,
\label{eq:ch6_N}
\end{equation}
with $n=u_\mu N^\mu$ being the total particle density and $V^\mu=\Delta^{\mu\nu}N_\nu$
the particle-diffusion four-current.

If only elastic collisions are present, then both the charge current and the total 
particle current are conserved,
\begin{equation}
\partial_\mu J^\mu=0,
\qquad
\partial_\mu N^\mu=0.
\label{eq:ch6_charge_particle_cons}
\end{equation}
On the other hand, the fluid part of the energy-momentum tensor is not conserved by 
itself, since the fluid exchanges energy and momentum with the electromagnetic field. 
Therefore,
\begin{equation}
\partial_\mu T^{\mu\nu}=F^{\nu\lambda}J_\lambda.
\label{eq:ch6_Tf_cons}
\end{equation}
It is convenient to project this equation along and orthogonal to the fluid four-velocity. Contracting Eq.~\eqref{eq:ch6_Tf_cons} with $u_\nu$, and using the 
decomposition~\eqref{eq:ch6_T_fluid} together with the identities $u_\nu \dot{u}^\nu = 0$, $u_\nu \nabla^\nu f = 0$, and $u_\nu \pi^{\mu\nu}=0$, the 
left-hand side reduces to 
\begin{equation}
    \dot{\epsilon}+(\epsilon+P+\Pi)\theta - \pi^{\mu\nu}\sigma_{\mu\nu}.
\end{equation}

For the right-hand side, one uses $u_\nu F^{\nu\lambda} = -E^\lambda$ together with 
$J_\lambda = n_q u_\lambda + V_{q\lambda}$ to obtain $u_\nu F^{\nu\lambda}J_\lambda 
= -E^\lambda J_\lambda$. The resulting energy equation is
\begin{equation}
u_\nu\partial_\mu T^{\mu\nu}
=
\dot{\epsilon}
+
(\epsilon+P)\theta
-
\pi^{\mu\nu}\sigma_{\mu\nu}
+
E^\mu J_\mu
=
0,
\label{eq:ch6_energy_proj}
\end{equation}
where the bulk viscous pressure drops out since $\Pi=0$ for the massless system 
considered here.

For the orthogonal projection, one contracts Eq.~\eqref{eq:ch6_Tf_cons} with 
$\Delta^\alpha_{\ \nu}$ and uses the decomposition~\eqref{eq:ch6_T_fluid} to 
evaluate the left-hand side. The standard identities $\Delta^\alpha_{\ \nu}u^\nu=0$ 
and $u_\nu \pi^{\mu\nu}=0$ reduce the fluid contributions to pressure-gradient 
and acceleration terms. For the right-hand side, one substitutes the decompositions 
\eqref{eq:ch6_F} and \eqref{eq:ch6_J} and uses 
$\Delta^\alpha_{\ \nu}(E^\nu u^\lambda - E^\lambda u^\nu)J_\lambda = E^\alpha n_q$
together with
$\Delta^\alpha_{\ \nu}\epsilon^{\nu\lambda\rho\sigma}u_\rho B_\sigma J_\lambda 
= -\epsilon^{\mu\alpha\lambda\rho}u_\lambda B_\rho V_{q,\mu}$,
which follows from the antisymmetry of $\epsilon^{\mu\nu\alpha\beta}$ and the 
orthogonality of $V_q^\mu$ to $u^\mu$. This yields
\begin{equation}
\Delta^\alpha_{\ \nu}\partial_\mu T^{\mu\nu}
=
(\epsilon+P+\Pi)\dot{u}^\alpha
+
\nabla^\alpha(P+\Pi)
+
\Delta^\alpha_{\ \mu}\partial_\nu\pi^{\mu\nu}
-
E^\alpha n_q
+
\epsilon^{\mu\alpha\lambda\rho}u_\lambda B_\rho V_{q,\mu}
=
0.
\label{eq:ch6_momentum_proj}
\end{equation}
These are the projected fluid equations of motion, written using the notation 
established in previous chapters.

At this point, the macroscopic structure of the resistive problem is fixed. The equations above are not yet closed, since one still needs an equation of state and dynamical equations for the dissipative quantities. Here, we again restrict ourselves to an ultra-relativistic gas, for which
\begin{equation}
P=\frac{\epsilon}{3}.
\label{eq:ch6_eos}
\end{equation}
The remaining task is then to derive the evolution equations for the dissipative currents from kinetic theory.

\section{Kinetic theory and 14-moment approximation}
We now turn to the microscopic description underlying the resistive two-component theory. As in the previous chapter, we consider a relativistic dilute gas composed of massless charged particles carrying charges $+q$ and $-q$, which we label by the indices $+$ and $-$, respectively. The state of the system is characterized by the single-particle distribution functions $f_k^+$ and $f_k^-$. Their evolution is governed by the Boltzmann-Vlasov equation for each species~\cite{CercignaniKremer2002,DeGroot},
\begin{equation}
k^\mu \partial_\mu f_k^i
+ |q| F^{\mu\nu} k_\nu \frac{\partial f_k^i}{\partial k^\mu} = \sum_j \,C[f^i,f^j].
\label{eq:ch6_BV}
\end{equation}
Here $k^\mu$ denotes the particle four-momentum and $C[f_k^i,f_k^j]$ is the collision kernel.

We continue with the assumption of binary elastic collisions, in which case the collision kernel retains the same form. For the sake of completeness, we repeat it below,
\begin{align}
C[f^i,f^j]
&\equiv
\gamma_{ij} \int dK' \, dP \, dP' \,
W_{KK'\leftrightarrow PP'}^{ii}
\left(
f_p^i f_{p'}^j - f_k^i f_{k'}^j
\right),
\label{eq:ch6_collision}
\end{align}
where $\gamma^{ij}=1-\delta^{ij}/2$. Furthermore, $ W_{kk'\to pp'}^{ij}$ is the transition rate and is written in terms of the total cross sections as
\begin{equation}
W_{kk'\to pp'}^{ij}
=
s \, \sigma_T^{ij} \, (2\pi)^5
\delta^{(4)}(k^\mu+k'^\mu-p^\mu-p'^\mu),
\label{eq:ch6_transition_rate}
\end{equation}
with
\begin{equation}
\sigma_T^{ij}
=
\int d\theta \, d\phi \, \sin\theta \, s(\theta,\phi)\,\sigma^{ij}.
\label{eq:ch6_sigmaT}
\end{equation}
Again, for simplicity, we continue with the assumption of the constant cross sections, such that
\begin{equation}
\sigma_T^{++}=\sigma_T^{--}\equiv \sigma_T,
\qquad
\sigma_T^{+-}=\sigma_T^{-+},
\label{eq:ch6_cross_sections}
\end{equation}
so that like-charged particles scatter with the same total cross section and the opposite-charge channel is symmetric under particle exchange.

The species-resolved energy-momentum tensors and particle currents are decomposed in the standard way established in Chapter~\ref{ch:two_component_rmhd},
\begin{align}
T_\pm^{\mu\nu}
&=
\epsilon_\pm u^\mu u^\nu
-
\Delta^{\mu\nu}(P_\pm+\Pi_\pm)
+
h_\pm^\mu u^\nu
+
h_\pm^\nu u^\mu
+
\pi_\pm^{\mu\nu},
\nonumber\\
N_\pm^\mu
&=
n_\pm u^\mu + V_\pm^\mu,
\label{eq:ch6_species_decomp}
\end{align}
where $\epsilon_\pm$, $P_\pm$, $\Pi_\pm$, $h_\pm^\mu$, $\pi_\pm^{\mu\nu}$, $n_\pm$, and $V_\pm^\mu$ are extracted by the same projections as before.
As in Chapter~\ref{ch:two_component_rmhd}, the hydrodynamic variables are formed from total and relative combinations of the species currents. The total energy-momentum tensor and total particle current are the sums,
\begin{equation}
T^{\mu\nu} \equiv T_+^{\mu\nu}+T_-^{\mu\nu}, 
\qquad 
N^\mu \equiv N_+^\mu+N_-^\mu = nu^\mu + V^\mu,
\label{eq:ch6_total_fluid_tensor}
\end{equation}
while the electric-charge current is the weighted difference,
\begin{equation}
J^\mu = |q|(N_+^\mu - N_-^\mu) = n_q u^\mu + V_q^\mu,
\label{eq:ch6_charge_current_kinetic}
\end{equation}
with $n_q = |q|(n_+-n_-)$ and $V_q^\mu = |q|(V_+^\mu - V_-^\mu)$. 

In the resistive theory $V_q^\mu$ is driven directly by the electric field. It is this quantity, rather than the total diffusion current $V^\mu$, that forms the central dynamical variable in the resistive sector.

\subsection{Matching conditions}

The matching conditions follow the same Landau-frame prescription derived in detail in Chapter~\ref{ch:two_component_rmhd}. The local equilibrium distributions are
\begin{equation}
f_{0k}^\pm = \exp\left(\alpha^\pm - \beta_0 E_k\right),
\qquad
\alpha^\pm = \pm \frac{|q|\mu}{T},
\qquad
\beta_0 = \frac{1}{T},
\qquad
E_k = u_\mu k^\mu.
\label{eq:ch6_f0}
\end{equation}
The Landau frame is defined with respect to the total energy-momentum tensor, so the hydrodynamic fields $T$, $\mu$, and $u^\mu$ are fixed by requiring that non-equilibrium corrections leave the total energy density and net-charge density at their local equilibrium values,
\begin{equation}
\delta\epsilon_+ + \delta\epsilon_- = 0,
\qquad
\delta n_+ - \delta n_- = 0.
\label{eq:ch6_landau_match}
\end{equation}
The Landau-frame condition also requires the total energy-diffusion current to vanish,
\begin{equation}
h^\mu \equiv h_+^\mu + h_-^\mu = 0.
\label{eq:ch6_total_h_zero}
\end{equation}
Unlike the non-resistive case of the previous chapter, here the chemical potential $\mu$ is not set to zero, so the equilibrium charge density $n_q = |q|(n_{0,+}-n_{0,-})$ need not vanish. Importantly, while $h^\mu = 0$, the difference
\begin{equation}
\delta h^\mu \equiv h_+^\mu - h_-^\mu
\label{eq:ch6_delta_h}
\end{equation}
can be non-zero, and as will be seen below it contributes to the resistive dynamics through the relative energy transport between the two charged sectors.

\subsection{14-moment approximation}

We close the kinetic hierarchy using the 14-moment approximation following 
the same procedure as in Chapter~\ref{ch:two_component_rmhd}. The departure 
from equilibrium is written as $\delta f_k^\pm = f_{0k}^\pm\,\delta y_k^\pm$, 
where $\delta y_k^\pm$ is expanded in irreducible momentum tensors and 
truncated at second order,
\begin{equation}
\delta y_k^\pm
=
\varepsilon^\pm
+
\varepsilon_\alpha^\pm k^\alpha
+
\varepsilon_{\alpha\beta}^\pm\, k^\alpha k^\beta.
\label{eq:ch6_generic_14moment}
\end{equation}
The key difference from the non-resistive case is that the vector sector 
must now be retained, since $V_\pm^\mu$ and $h_\pm^\mu$ are no longer zero. 
For a massless gas the scalar contribution is at least order 2 and does not contribute to our fluid-dynamical description. We thus set $\varepsilon^\pm = 0$. Thus, the non-equilibrium correction is determined entirely by the vector and traceless rank-two tensor sectors. 

The coefficients remaining coefficients are fixed by demanding that $\delta f_k^\pm$ reproduces the definitions of the dissipative currents and the matching conditions imposed, as explained in the previous Chapter -- see Eqs.~\eqref{eq:IS_match_n}-~\eqref{eq:IS_Landau2}. Solving these equations uniquely determines the 
14-moment ansatz,
\begin{equation}
\delta f_k^\pm
=
f_{0k}^\pm \, \hat n_0 T
\left[
-5\, V_\pm^\mu k_\mu
+
h_\pm^\mu k_\mu
+
\frac{u^{(\nu}V_\pm^{\mu)}k_\mu k_\nu}{T}
-
\frac{u^{(\nu}h_\pm^{\mu)}k_\mu k_\nu}{4T^2}
+
\frac{\pi_\pm^{\mu\nu}k_\mu k_\nu}{8T^2}
\right].
\label{eq:ch6_deltaf_14}
\end{equation}

With this result, all higher irreducible moments are no longer independent. 
The relevant ones reduce to
\begin{equation}
\int dK \, E_k^r \, k^{\langle\mu\rangle} f_k^\pm
=
\gamma_{r,\pm}^V \, V_\pm^\mu,
\qquad
\int dK \, E_k^r \, k^{\langle\mu}k^{\nu\rangle} f_k^\pm
=
\gamma_{r,\pm}^\pi \, \pi_\pm^{\mu\nu},
\label{eq:ch6_moment_closures}
\end{equation}
where the proportionality coefficients are
\begin{equation}
\gamma_{r,\pm}^V
=
-
\frac{\mathcal{I}_{41}^\pm \mathcal{I}_{r+2,1}^\pm}{D_{31}^\pm}
+
\frac{\mathcal{I}_{31}^\pm \mathcal{I}_{r+3,1}^\pm}{D_{31}^\pm},
\qquad
\gamma_{r,\pm}^\pi
=
\frac{\mathcal{I}_{r+4,2}^\pm}{\mathcal{I}_{4,2}^\pm},
\label{eq:ch6_gamma_coeffs}
\end{equation}
with $D_{ij}^\pm = \mathcal{I}_{i+1,j}^\pm \mathcal{I}_{i-1,j}^\pm - (\mathcal{I}_{ij}^\pm)^2$.  Here,  $\mathcal{I}^\pm_{ij}$ are simply the thermodynamic integrals stated in Eq.~\eqref{eq:Iij_general}. Through 
these relations, the kinetic description is reduced to a finite set of 
macroscopic variables, and one may proceed to derive their equations of 
motion by taking moments of the Boltzmann--Vlasov equation.

\section{Equations of motion}

We now derive the equations of motion for the macroscopic variables by 
taking moments of the Boltzmann--Vlasov equations. As already seen in the 
previous Chapter, this procedure converts the microscopic kinetic equation 
into evolution equations for the hydrodynamic fields and dissipative 
quantities. The system naturally splits into equations for scalar 
quantities, the particle and charge four-currents, and the tensor sector.

\subsection{Scalar quantities}

We begin with the Lorentz scalars of the theory. These are the net-charge density $n_q$, the total particle density $n$, the total energy density $\epsilon$, and the relative energy density
$\Delta\epsilon \equiv \epsilon_+ - \epsilon_-$. The corresponding equations of motion are obtained by taking the lowest moments of the Boltzmann--Vlasov equations for each species.

\subsubsection{Number density}

The particle four-current of each species is
\begin{equation}
N_\pm^\mu = n_\pm u^\mu + V_\pm^\mu.
\label{eq:ch6_species_current_again}
\end{equation}
Since only elastic collisions are considered, the particle number of each species is separately conserved, so that
\begin{equation}
\partial_\mu N_\pm^\mu = 0.
\label{eq:ch6_species_number_cons}
\end{equation}
Substituting Eq.~\eqref{eq:ch6_species_current_again} into Eq.~\eqref{eq:ch6_species_number_cons} gives
\begin{equation}
\partial_\mu(n_\pm u^\mu + V_\pm^\mu)=0.
\end{equation}
Using
\begin{equation}
\partial_\mu(n_\pm u^\mu)=\dot n_\pm + \theta n_\pm,
\qquad
\partial_\mu V_\pm^\mu = \nabla_\mu V_\pm^\mu - \dot u_\mu V_\pm^\mu,
\label{eq:ch6_vector_div_identity}
\end{equation}
one immediately finds
\begin{equation}
\dot n_\pm = - \theta n_\pm -\nabla_\mu V_\pm^\mu  + \dot u_\mu V_\pm^\mu.
\label{eq:ch6_npm_eom}
\end{equation}
This is the exact equation of motion for the number density of each component.

It is then convenient to form the total and relative combinations. Multiplying the difference of the two species equations by $|q|$ gives the net-charge density equation,
\begin{equation}
\dot n_q \equiv |q|(\dot n_+ - \dot n_-)
=
-\theta n_q
-\nabla_\mu V_q^\mu
+\dot u_\mu V_q^\mu,
\label{eq:ch6_nq_eom}
\end{equation}
while summing them yields the total particle density equation,
\begin{equation}
\dot n \equiv \dot n_+ + \dot n_-
=
-\theta n
-\nabla_\mu V^\mu
+\dot u_\mu V^\mu.
\label{eq:ch6_n_eom}
\end{equation}
Thus, the scalar densities $n_q$ and $n$ are driven by the divergence of the corresponding diffusion currents and by the local expansion of the fluid. 

\subsubsection{Energy density}

We next derive the equations of motion for the energy densities of the two species. Their kinetic definition is
\begin{equation}
\epsilon_\pm = \int dK\, E_k^2 f_k^\pm,
\end{equation}
with $E_k=u_\mu k^\mu$. Taking the comoving derivative gives
\begin{equation}
\dot\epsilon_\pm = \frac{d}{d\tau}\int dK\, E_k^2 f_k^\pm.
\label{eq:ch6_epspm_start}
\end{equation}
Evaluating this derivative with the Boltzmann--Vlasov equation and decomposing the resulting moments in terms of the hydrodynamic variables leads to
\begin{equation}
\dot\epsilon_\pm
=
\mathcal C_\pm[f]
-
\nabla_\mu h_\pm^\mu
-
\frac{4}{3}\theta \epsilon_\pm
+
\sigma_{\mu\nu}\pi_\pm^{\mu\nu}
-
q_\pm E_\nu V_\pm^\nu
+
2\dot u_\mu h_\pm^\mu.
\label{eq:ch6_epspm_eom}
\end{equation}
Here $q_+=|q|$ and $q_-=-|q|$, while
\begin{equation}
\mathcal C_\pm[f]
=
\int dK\, dK' \, dP \, dP'\,
E_k\,W^\pm_{kk'\leftrightarrow pp'}
\left(f_p^\pm f_{p'}^\mp - f_k^\pm f_{k'}^\mp\right)
\label{eq:ch6_Cpm}
\end{equation}
denotes the corresponding moment of the collision term.

Summing Eq.~\eqref{eq:ch6_epspm_eom} over the two species gives the equation for the total energy density,
\begin{equation}
\dot\epsilon \equiv \dot\epsilon_+ + \dot\epsilon_-
=
-\frac{4}{3}\theta \epsilon
+
\sigma_{\mu\nu}\pi^{\mu\nu}
-
E_\nu V_q^\nu.
\label{eq:ch6_eps_eom}
\end{equation}
To obtain this result, one uses the definitions
\begin{equation}
\epsilon=\epsilon_+ + \epsilon_-,
\qquad
\pi^{\mu\nu}=\pi_+^{\mu\nu}+\pi_-^{\mu\nu},
\qquad
V_q^\mu = |q|(V_+^\mu - V_-^\mu),
\label{eq:ch6_total_defs_again}
\end{equation}
together with the Landau-frame condition $h^\mu \equiv h_+^\mu + h_-^\mu = 0$, and the fact that the sum of the collision moments vanishes,
\begin{equation}
\mathcal C_+[f] + \mathcal C_-[f] = 0,
\label{eq:ch6_collision_energy_cons}
\end{equation}
as a consequence of energy conservation in microscopic collisions. The term
\begin{equation}
-E_\nu V_q^\nu
\end{equation}
represents the work done by the electric field on the plasma and is the direct scalar manifestation of resistive dissipation. This is the total energy-density equation used later in the homogeneous and Bjorken analyses. 

\subsubsection{Relative energy density}

Finally, we consider the difference between the energy-density equations of the two species. Recall
\begin{equation}
\Delta\epsilon \equiv \epsilon_+ - \epsilon_-,
\qquad
\delta h^\mu \equiv h_+^\mu - h_-^\mu,
\qquad
\delta\pi^{\mu\nu} \equiv \pi_+^{\mu\nu} - \pi_-^{\mu\nu},
\label{eq:ch6_delta_defs_rel_energy}
\end{equation}
and thus subtracting the $(-)$ equation from the $(+)$ equation, one obtains an evolution equation for $\Delta\epsilon$. In this combination, unlike in the total energy-density equation, the collision term does not cancel. This is because the difference between the two species is not protected by a conservation law and hence collisions can redistribute energy between the positively and negatively charged sectors even though the total energy is conserved.

More precisely, defining
\begin{equation}
\delta \mathcal C \equiv \mathcal C_+[f] - \mathcal C_-[f],
\label{eq:ch6_deltaCepsilon_def}
\end{equation}
the difference of the two species equations gives
\begin{equation}
\Delta\dot\epsilon
=
\delta \mathcal C
-
\nabla_\mu \delta h^\mu
-
\frac{4}{3}\theta\,\Delta\epsilon
+
\sigma_{\mu\nu}\delta\pi^{\mu\nu}
-
|q|E_\nu V^\nu
+
2\dot u_\mu \delta h^\mu.
\label{eq:ch6_deltaeps_eom_intermediate}
\end{equation}
The collision term $\delta \mathcal C$ is obtained by taking the difference between the corresponding energy moments of the collision kernels for the two species. Using the 14-moment form of $\delta f_k^\pm$, linearizing the collision integral around local equilibrium, and evaluating the resulting momentum integrals for constant cross sections, one finds the exact expression,
\begin{equation}
\mathcal C_+[f] - \mathcal C_-[f]
=
\frac{\sigma_T^{+-}}{2}
\left(
\frac{n_q}{|q|}\,\epsilon
-
n\,\Delta\epsilon
\right).
\label{eq:ch6_deltaCepsilon_final}
\end{equation}
Substituting this result into Eq.~\eqref{eq:ch6_deltaeps_eom_intermediate} yields
\begin{equation}
\Delta\dot\epsilon
=
\frac{\sigma_T^{+-}}{2}
\left(
\frac{n_q}{|q|}\,\epsilon
-
n\,\Delta\epsilon
\right)
-
\nabla_\mu \delta h^\mu
-
\frac{4}{3}\theta\,\Delta\epsilon
+
\sigma_{\mu\nu}\delta\pi^{\mu\nu}
-
|q|E_\nu V^\nu
+
2\dot u_\mu \delta h^\mu.
\label{eq:ch6_deltaeps_eom}
\end{equation}

This equation captures the imbalance in energy transport and dissipation between the positively and negatively charged sectors. The term proportional to $\sigma_T^{+-}$ explicitly reflects the inter-species coupling through collisions, while the terms involving $\delta h^\mu$ and $\delta\pi^{\mu\nu}$ encode the effects of relative heat and shear transport. The electric field enters here through the total particle-diffusion current $V^\mu$, rather than through the charge-diffusion current $V_q^\mu$.

In the hydrodynamic limit, $\sigma_T^{+-}\to\infty$, one may construct a perturbative solution in powers of the inverse cross section, normalized by $T^2$. At leading order, only the collision term contributes, so Eq.~\eqref{eq:ch6_deltaeps_eom} gives
\begin{equation}
\Delta\epsilon
\approx
\frac{n_q}{n|q|}\,\epsilon
+
\mathcal O\!\left(\frac{T^2}{\sigma_T^{+-}}\right).
\label{eq:ch6_deltaeps_hydro0}
\end{equation}
Thus, at zeroth order in the hydrodynamic expansion, the relative energy density is fixed algebraically by the total energy density and the net-charge density.

The next correction is obtained iteratively by substituting Eq.~\eqref{eq:ch6_deltaeps_hydro0} back into the full equation of motion and retaining terms of order $T^2/\sigma_T^{+-}$. This yields
\begin{equation}
\Delta\epsilon
\approx
\frac{n_q}{n|q|}\,\epsilon
+
\frac{1}{\sigma_T^{+-} n}\frac{n_q}{n|q|}E_\nu V_q^\nu
-
\frac{|q|}{\sigma_T^{+-} n}E_\nu V^\nu
+
\mathcal O\!\left[\left(\frac{T^2}{\sigma_T^{+-}}\right)^2\right].
\label{eq:ch6_deltaeps_hydro1}
\end{equation}
In the following, $\Delta\epsilon$ will be approximated by this compact expression. This makes explicit that, in the hydrodynamic regime, the relative energy density is not an independent slow degree of freedom, but is instead slaved to the conserved densities and to the electric-field-induced diffusive currents.

\subsection{Energy-diffusion current}

We now consider the equation of motion for the energy-diffusion four-current of each species. It is obtained from the projected first moment of the Boltzmann--Vlasov equation. Starting from the kinetic definition
\begin{equation}
h_\pm^\mu \equiv u_\alpha T_\pm^{\langle\mu\rangle\alpha}
=
u_\alpha \Delta^\mu_{\ \beta}\int dK\, k^\alpha k^\beta f_k^\pm,
\label{eq:ch6_hpm_def_again}
\end{equation}
and following the same projected first-moment procedure as in 
Chapter~\ref{ch:two_component_rmhd}, one takes the comoving derivative,  substitutes the Boltzmann--Vlasov 
equation~\eqref{eq:ch6_BV}, and applies the 14-moment closure. This gives
\begin{align}
    \dot{h}_\pm^{\langle \mu \rangle} &= \mathcal{C}^\mu_{\delta h, \pm}[f]
    - h^\nu_\pm \omega_\nu{}^\mu 
    + \dot{u}_\nu \pi^{\mu \nu}_\pm 
    - \Delta^\mu{}_\alpha \nabla_\beta \pi^{\alpha \beta}_\pm
    -\frac{4}{3}  h^\mu_\pm \theta 
    - h_\pm^\nu \sigma^\mu{}_\nu
    \nonumber \\ & \quad
    - \frac{4}{3}\dot{u}^\mu \epsilon_\pm 
    +\frac{1}{3} \nabla^\mu\epsilon_\pm + q_\pm E^\mu n_\pm - q_\pm B b^{\mu\nu}V_{\nu}^\pm .
\label{eq:ch6_hpm_eom}
\end{align}
The collision term, $\mathcal{C}^\mu_{\delta h, \, \pm}[f] \equiv \int dK k^{\langle\mu\rangle} \left(C[f^\pm, f^\pm] + C[f^\pm, f^\mp]\right)$ is obtained by inserting the 14-moment approximation into the linearized collision kernel, as shown in Appendix~\ref{app:collision_integrals_resistive}. The result contains contributions proportional to the total and inter-species cross sections, which determine the relaxation of the diffusion current.

What enters the resistive theory most directly is the difference 
$\delta h^\mu \equiv h_+^\mu - h_-^\mu$, which remains non-zero in general and  measures the relative energy transport between the two charged sectors. Subtracting the two species equations gives its equation of motion,

\begin{align}
 \delta\dot h^{\langle\mu\rangle}
+
\frac{\sigma_T^{+-}}{2}\,n\,\delta h^\mu
\nonumber  & =
\frac{\sigma_T^{+-}}{6}
\left(
\Delta\epsilon\,V^\mu
-
\epsilon\,\frac{V_q^\mu}{|q|}
\right)
+
|q|E^\mu n
-
B\,b^{\mu\nu}V_{q,\nu}
-
\frac{4}{3}\theta\,\delta h^\mu
-
\delta h_\nu \sigma^{\mu\nu}
\nonumber 
\\ & \quad  -
\frac{4}{3}\dot u^\mu \Delta\epsilon
+
\frac{1}{3}\nabla^\mu \Delta\epsilon
-
\delta h_\nu \omega^{\mu\nu}
+
\dot u_\nu \delta\pi^{\mu\nu}
-
\Delta^\mu_{\ \alpha}\nabla_\beta \delta\pi^{\alpha\beta}.
\label{eq:ch6_deltah_eom}
\end{align}

This equation shows that inter-species collisions damp $\delta h^\mu$ through the term proportional to $\sigma_T^{+-}n$, while the gradients, the shear and vorticity tensors, the electromagnetic field, and the relative energy density act as sources. In particular, the electric field and magnetic field couple directly to the relative heat flow, and the relative shear-stress tensor $\delta\pi^{\mu\nu}$ feeds back into its evolution through the projected derivative and acceleration terms.

It is convenient to rewrite Eq.~\eqref{eq:ch6_deltah_eom} in the form
\begin{equation}
\delta\dot h^{\langle\mu\rangle}
=
\frac{\sigma_T^{+-}}{2}
\left(
-n\,\delta h^\mu
+
\frac{\Delta\epsilon}{3}V^\mu
-
\frac{\epsilon}{3}\frac{V_q^\mu}{|q|}
\right)
+
S^\mu,
\label{eq:ch6_deltah_rewritten}
\end{equation}
where all terms independent of the cross section have been grouped into
\begin{equation}
S^\mu
\equiv
-\delta h_\nu \omega^{\mu\nu}
+\dot u_\nu \delta\pi^{\mu\nu}
-\Delta^\mu_{\ \alpha}\nabla_\beta \delta\pi^{\alpha\beta}
-\frac{4}{3}\theta\,\delta h^\mu
-\delta h_\nu \sigma^{\mu\nu}
-\frac{4}{3}\dot u^\mu \Delta\epsilon
+\frac{1}{3}\nabla^\mu \Delta\epsilon
+|q|E^\mu n
-
B\,b^{\mu\nu}V_{q,\nu}.
\label{eq:ch6_Smu}
\end{equation}
This form makes the hydrodynamic limit transparent. When
\begin{equation}
\sigma_T^{+-}\to\infty,
\end{equation}
collisions dominate and drive the system rapidly toward the leading-order algebraic solution obtained by neglecting $\delta\dot h^{\langle\mu\rangle}$ and $S^\mu$. One then finds
\begin{equation}
\delta h^\mu
\approx
\frac{\Delta\epsilon}{3n}V^\mu
-
\frac{\epsilon}{3n}\frac{V_q^\mu}{|q|}
\equiv
\delta h_0^\mu.
\label{eq:ch6_deltah0}
\end{equation}
Thus, in the hydrodynamic regime the relative heat-diffusion current is not an independent slow variable, but is fixed algebraically by the relative energy density together with the total and charge diffusion currents.

The next correction is obtained iteratively by substituting the leading-order solution $\delta h_0^\mu$ back into Eq.~\eqref{eq:ch6_deltah_rewritten}. Concretely, this means that every occurrence of $\delta h^\mu$ on the right-hand side of Eq.~\eqref{eq:ch6_deltah_eom} is replaced by $\delta h_0^\mu$, and the resulting expression is then solved to first order in $1/\sigma_T^{+-}$. In this way, $\delta h^\mu$ is expanded systematically around its hydrodynamic limit as
\begin{equation}
\delta h^\mu
\approx
\delta h_0^\mu
+
\frac{1}{\sigma_T^{+-}n}
\left(
-\delta\dot h_0^{\langle\mu\rangle}
+
S^\mu\big|_{\delta h\to \delta h_0}
\right),
\label{eq:ch6_deltah_iter}
\end{equation}
where $S^\mu|_{\delta h\to \delta h_0}$ denotes the expression in Eq.~\eqref{eq:ch6_Smu} evaluated with $\delta h^\mu$ replaced by $\delta h_0^\mu$. Carrying out this substitution explicitly yields
\begin{align}
\delta h^\mu \approx\;&
-\frac{\epsilon}{3|q|\,n}V_q^\mu
+
\frac{\epsilon\,n_q}{3|q|\,n^2}V^\mu
+
\frac{2\epsilon\,\theta}{3|q|\,\sigma_T^{+-}n^2}V_q^\mu
+
\frac{2\epsilon}{3|q|\,\sigma_T^{+-}n^2}\dot V_q^{\langle\mu\rangle}
-
\frac{2\epsilon\,n_q}{3|q|\,\sigma_T^{+-}n^3}\dot V^{\langle\mu\rangle}
+
\frac{2|q|}{\sigma_T^{+-}}E^\mu
\nonumber\\
& -
\frac{2B}{\sigma_T^{+-}n}b^{\mu\nu}V_{q,\nu}
+
\left(\omega^\mu_{\ \nu}+\sigma^\mu_{\ \nu}\right)
\left(
\frac{2\epsilon}{3|q|\,\sigma_T^{+-}n^2}V_q^\nu
-
\frac{2\epsilon\,n_q}{3|q|\,\sigma_T^{+-}n^3}V^\nu
\right)
-
\frac{2}{3|q|\,\sigma_T^{+-}n^2}E_\nu V^\nu\,V^\mu
\nonumber\\
& +
\frac{4n_q}{3|q|\,\sigma_T^{+-}n^3}E_\nu V_q^\nu\,V^\mu
-
\frac{2}{3|q|\,\sigma_T^{+-}n^2}E_\nu V_q^\nu\,V_q^\mu
-
\frac{8\epsilon\,n_q}{3|q|\,\sigma_T^{+-}n^2}\dot u^\mu
+
\frac{2}{\sigma_T^{+-}n}\dot u_\nu \delta\pi^{\mu\nu}
\nonumber \\ & -
\frac{2}{\sigma_T^{+-}n}\Delta^\mu_{\ \alpha}\nabla_\beta \delta\pi^{\alpha\beta}
+
\frac{2}{3|q|\,\sigma_T^{+-}n}\nabla^\mu\!\left(\frac{n_q}{n}\epsilon\right).
\label{eq:ch6_deltah_final}
\end{align}

Equation~\eqref{eq:ch6_deltah_final} is the explicit first-order hydrodynamic approximation for the relative heat-diffusion current. Its meaning is simple: the first two terms reproduce the leading-order algebraic result $\delta h_0^\mu$, while all remaining terms are corrections suppressed by $1/\sigma_T^{+-}$. These subleading contributions encode the fact that the relaxation of $\delta h^\mu$ is not exactly instantaneous. They contain time derivatives of the diffusion currents, couplings to the electric and magnetic fields, vorticity and shear corrections, acceleration terms, and gradients involving the relative shear-stress tensor.

It is therefore useful to state the approximation explicitly. In the hydrodynamic regime, one assumes that the inter-species collision rate is large, or equivalently that
\begin{equation}
\frac{T^2}{\sigma_T^{+-}} \ll 1.
\label{eq:ch6_hydro_assumption}
\end{equation}
Under this assumption, $\delta h^\mu$ is treated as a fast mode and expanded in powers of $1/\sigma_T^{+-}$. To leading order,
\begin{equation}
\delta h^\mu \simeq \delta h_0^\mu
=
-\frac{\epsilon}{3|q|\,n}V_q^\mu
+
\frac{\epsilon\,n_q}{3|q|\,n^2}V^\mu.
\label{eq:ch6_deltah_LO}
\end{equation}
while to first subleading order one uses the full expression in Eq.~\eqref{eq:ch6_deltah_final}. In practice, this means that $\delta h^\mu$ is no longer evolved as an independent slow degree of freedom, but is instead replaced by its approximate constitutive form in terms of $V^\mu$, $V_q^\mu$, $\delta\pi^{\mu\nu}$, and the electromagnetic field. This approximation will be used to simplify the equations of motion for the particle and net-charge four-currents,
that will be derived next.

\subsection{Net-charge four-current and particle four-current}

We now turn to the evolution equations for the diffusion currents, obtained 
from the projected first moment of the Boltzmann--Vlasov equation. For each 
species, one starts from
\begin{equation}
V_\pm^\mu \equiv \Delta^\mu_{\ \nu}\int dK\,k^\nu f_k^\pm,
\label{eq:ch6_Vpm_def_again}
\end{equation}
and takes the comoving derivative using Eq.~\eqref{eq:ch6_BV}. After 
projecting the first moment, decomposing into irreducible tensors, and 
applying the 14-moment closure, one obtains
\begin{align}
\dot V_\pm^{\langle\mu\rangle}
&=
\mathcal C_{V,\pm}^\mu[f]
+\frac{2}{3}qE^\mu\int dK\,f_k^\pm
+\frac{1}{20T^2}qE_\nu\pi_\pm^{\mu\nu}
+\frac{2}{3T}qB\,b^{\mu\nu}V_{\pm,\nu}
-\dot u^\mu n_\pm
+\frac{1}{3}\nabla^\mu n_\pm
\nonumber\\
&\quad
-
V_\pm^\mu\theta
-
\frac{3}{5}V_\pm^\nu\sigma^\mu_{\ \nu}
-
V_\pm^\nu\omega^\mu_{\ \nu}
-
\Delta^\mu_{\ \alpha}\nabla_\beta\frac{\pi_\pm^{\alpha\beta}}{5T},
\label{eq:ch6_Vpm_eom}
\end{align}
where $\mathcal{C}_{V,\pm}^\mu[f] \equiv \int dK\,E_k^{-1}k^{\langle\mu\rangle}
\left(C[f^\pm, f^\pm] + C[f^\pm, f^\mp]\right)$ is the projected vector moment of the collision kernel. 
Its evaluation is carried out in Appendix~\ref{app:collision_integrals_resistive}. 
The collision term is first computed separately for each species by linearizing 
the gain-minus-loss structure around local equilibrium and inserting the 
14-moment ansatz. For the intra-species channel, momentum conservation forces 
the scalar and vector parts of the ansatz to drop out, leaving only the rank-two 
tensor sector. For the inter-species channel this cancellation does not occur 
and the vector sector contributes as well. The species results are then combined 
into the total and relative diffusion currents $V^\mu$ and $V_q^\mu$.

Combining the two species according to
\begin{equation}
V^\mu \equiv V_+^\mu + V_-^\mu,
\qquad
V_q^\mu \equiv |q|\left(V_+^\mu - V_-^\mu\right),
\label{eq:ch6_V_Vq_defs_again}
\end{equation}
one obtains the evolution equation for the net-charge diffusion current,
\begin{align}
& \dot V_q^{\langle\mu\rangle}
+
\frac{2}{9}n\left(3\sigma_T^{+-}+\sigma_T\right)V_q^\mu
\nonumber\\ \quad & =
-\,V_q^\mu\theta
-\frac{3}{5}V_q^\nu\sigma^\mu_{\ \nu}
+\frac{2}{9}n_q\left(3\sigma_T^{+-}-\sigma_T\right)V^\mu
-V_q^\nu\omega^\mu_{\ \nu}
-|q|\Delta^\mu_{\ \alpha}\nabla_\beta\frac{\delta\pi^{\alpha\beta}}{5T}-\dot u^\mu n_q
\nonumber\\
&\quad
+\frac{1}{3}\nabla^\mu n_q
+\frac{1}{3}|q|^2\frac{n}{T}E^\mu
+\frac{1}{20T^2}|q|^2E_\nu\pi^{\mu\nu}
+\frac{2}{3T}|q|^2B\,b^{\mu\nu}V_\nu 
+\frac{1}{18}|q|\,n\,\sigma_T\beta_0\,\delta h^\mu.
\label{eq:ch6_Vq_eom_raw}
\end{align}
Proceeding in the same way for the total particle-diffusion current gives
\begin{align}
\dot V^{\langle\mu\rangle}
+
\frac{2}{9}n\left(\sigma_T+\frac{\sigma_T^{+-}}{2}\right)V^\mu
&=
-\,V^\mu\theta
-\frac{3}{5}V^\nu\sigma^\mu_{\ \nu}
-\frac{2n_q}{9|q|}\left(\sigma_T-\frac{\sigma_T^{+-}}{2}\right)V_q^\mu
-\dot u^\mu n
+\frac{1}{3}\nabla^\mu n
\nonumber\\
&\quad -V^\nu\omega^\mu_{\ \nu}
-\Delta^\mu_{\ \alpha}\nabla_\beta\frac{\pi^{\alpha\beta}}{5T}
+\frac{1}{3}|q|\frac{n_q}{T}E^\mu
+\frac{1}{20T^2}|q|E^\nu\delta\pi^\mu_{\ \nu}
\nonumber \\ &\quad 
+\frac{2}{3T}|q|B\,b^{\mu\nu}V_{q,\nu} 
-\frac{2n_q}{9|q|}\left(\sigma_T+2\sigma_T^{+-}\right)\beta_0\,\delta h^\mu.
\label{eq:ch6_V_eom_raw}
\end{align}

The structure of Eqs.~\eqref{eq:ch6_Vq_eom_raw} and \eqref{eq:ch6_V_eom_raw} makes the coupling pattern explicit. The two currents do not evolve independently: $V^\mu$ and $V_q^\mu$ appear simultaneously in both equations, the electric field provides a direct driving term, the magnetic field mixes components through $b^{\mu\nu}$, and the tensor sectors $\pi^{\mu\nu}$ and $\delta\pi^{\mu\nu}$ feed back into the current dynamics. In addition, both equations still contain the relative heat current $\delta h^\mu$. 

At this stage, it is convenient to eliminate $\delta h^\mu$ in favor of the diffusion currents $V^\mu$ and $V_q^\mu$. This is done by substituting the first-order hydrodynamic approximation derived previously for $\delta h^\mu$, namely Eq.~\eqref{eq:ch6_deltah_final}, which is valid up to $\mathcal O((\sigma_T^{+-\,})^ {-1})$. After this substitution, one obtains a closed set of equations for the independent diffusion currents. The net-charge current obeys
\begin{align}
& (1-\alpha_{V_q})\dot V_q^{\langle\mu\rangle}
+\Gamma_{V_q}V_q^\mu
+\kappa_V V^\mu
\nonumber\\ & =
\mathcal G_E E^\mu
+\frac{|q|^2}{20T^2}E_\nu\pi^{\nu\mu}
-(1-\alpha_{V_q})\theta V_q^\mu
+\left(\alpha_{V_q}-\frac{3}{5}\right)\sigma^\mu_{\ \nu}V_q^\nu
-\kappa_V\sigma^\mu_{\ \nu}V^\nu
+\gamma_V V^\nu V^\mu
\nonumber\\
&\quad
+\frac{2|q|^2}{3T}B\,b^{\mu\nu}V_\nu
-\frac{|q|\alpha_{V_q}}{T}B\,b^{\mu\nu}V_{q,\nu}
-\Gamma_{V_qV_q}E_\nu V_q^\nu V_q^\mu
-|q|^2\Gamma_{V_qV_q}E_\nu V^\nu V^\mu
+\Gamma_{\rm mag}E_\nu V_q^\nu V^\mu
\nonumber\\
&\quad
-(1-\alpha_{V_q})V_q^\nu\omega^\mu_{\ \nu}
-\kappa_VV^\nu\omega^\mu_{\ \nu}
-(1+4\alpha_{V_q})n_q\dot u^\mu
+\frac{1}{3}\nabla^\mu n_q
+\frac{\alpha_{V_q}}{3T}\nabla^\mu\!\left(\frac{n_q}{n}\epsilon\right)
\nonumber\\
&\quad 
-\frac{|q|\alpha_{V_q}}{T}\Delta^\mu_{\ \alpha}\nabla_\beta\delta\pi^{\alpha\beta} 
-\frac{|q|\alpha_{V_q}}{T}\dot u_\nu\delta\pi^{\mu\nu},
\label{eq:ch6_Vq_eom_closed}
\end{align}
while the total particle-diffusion current satisfies
\begin{align}
& (1-\alpha_V)\dot V^{\langle\mu\rangle}
+\Gamma_V V^\mu
+\kappa_{V_q}V_q^\mu
\nonumber\\
&=
\mathcal D_E E^\mu
+\frac{|q|}{20T^2}E^\nu\delta\pi^\mu_{\ \nu}
+\frac{5}{9}|q|\sigma_T^{+-}n_q V_q^\mu
-\theta V^\mu
-\kappa_{V_q}\theta V_q^\mu
-\left(\frac{3}{5}-\alpha_V\right)\sigma^\mu_{\ \nu}V^\nu
\nonumber\\
&\quad -\kappa_{V_q}\sigma^\mu_{\ \nu}V_q^\nu +\mathcal H_{BV}B\,b^{\mu\nu}V_{q,\nu}
+\Gamma_{VV}E_\nu V^\nu V^\mu
-\Gamma_{\rm mix}E_\nu V_q^\nu V^\mu
+\frac{1}{|q|^2}\Gamma_{VV}E_\nu V_q^\nu V_q^\mu
\nonumber\\
&\quad 
-\kappa_{V_q}V_q^\nu\omega^\mu_{\ \nu}
-\frac{1}{3T}\kappa_{V_q}\nabla^\mu\!\left(\frac{n_q}{n}\epsilon\right)
-(1-4\alpha_V)n\dot u^\mu
-\Delta^\mu_{\ \alpha}\nabla_\beta\frac{\pi^{\alpha\beta}}{5T}
-\frac{|q|}{T}\kappa_{V_q}\dot u_\nu\delta\pi^{\mu\nu}
\nonumber \\ &\quad 
+\frac{1}{3}\nabla^\mu n
-(1-\alpha_V)V^\nu\omega^\mu_{\ \nu} +\frac{|q|}{T}\kappa_{V_q}\Delta^\mu_{\ \alpha}\nabla_\beta\delta\pi^{\alpha\beta}.
\label{eq:ch6_V_eom_closed}
\end{align}

The corresponding transport coefficients are
\begin{align}
\alpha_{V_q} &\equiv \frac{1}{9}\frac{\sigma_T}{\sigma_T^{+-}},
&
\alpha_V &\equiv \frac{4}{9|q|^2}\frac{n_q^2}{n^2}\frac{\sigma_T+2\sigma_T^{+-}}{\sigma_T^{+-}},
\label{eq:ch6_coeffs_a}
\\
\Gamma_{V_q} &\equiv \frac{n}{3}\left(2\sigma_T^{+-}+\frac{5}{6}\sigma_T\right),
&
\Gamma_V &\equiv \frac{2n}{9}\left(\sigma_T+\frac{\sigma_T^{+-}}{2}\right)+\frac{2}{9|q|^2}\frac{n_q^2}{n}\left(\sigma_T+2\sigma_T^{+-}\right),
\label{eq:ch6_coeffs_b}
\\
\mathcal G_E &\equiv \frac{1}{3}|q|^2\frac{n}{T}+\frac{|q|^2}{9}\frac{\sigma_T}{\sigma_T^{+-}}\frac{n}{T},
&
\mathcal D_E &\equiv \frac{1}{3}|q|\frac{n_q}{T}-\frac{4}{9}\frac{n_q}{T}\frac{\sigma_T+2\sigma_T^{+-}}{\sigma_T^{+-}},
\label{eq:ch6_coeffs_c}
\\
\kappa_V &\equiv \frac{1}{9}\frac{\sigma_T}{\sigma_T^{+-}}\frac{n_q}{n},
&
\kappa_{V_q} &\equiv \frac{4}{9|q|^2}\frac{n_q}{n}\frac{\sigma_T+2\sigma_T^{+-}}{\sigma_T^{+-}},
\label{eq:ch6_coeffs_d}
\\
\gamma_V &\equiv \frac{n_q}{3}\left(2\sigma_T^{+-}-\frac{\sigma_T}{2}\right),
&
\gamma_{V_q} &\equiv \frac{1}{9|q|^2}n_q\left(5\sigma_T^{+-}-\frac{3}{2}\sigma_T\right),
\label{eq:ch6_coeffs_e}
\\
\Gamma_{V_qV_q} &\equiv \frac{1}{27T}\frac{\sigma_T}{\sigma_T^{+-}}\frac{1}{n},
&
\Gamma_{VV} &\equiv \frac{4}{27T}\frac{n_q}{n^2}\frac{\sigma_T+2\sigma_T^{+-}}{\sigma_T^{+-}},
\label{eq:ch6_coeffs_f}
\\
\Gamma_{\rm mag} &\equiv \frac{2}{27T}\frac{\sigma_T}{\sigma_T^{+-}}\frac{n_q}{n^2},
&
\Gamma_{\rm mix} &\equiv \frac{8}{27T\,n^3|q|^2}n_q^2\frac{\sigma_T+2\sigma_T^{+-}}{\sigma_T^{+-}},
\label{eq:ch6_coeffs_g}
\\
\mathcal H_{BV} &\equiv \frac{1}{T}\left(\frac{2}{3}|q|+\frac{4}{9|q|}\frac{n_q}{n}\frac{\sigma_T+2\sigma_T^{+-}}{\sigma_T^{+-}}\right).
\label{eq:ch6_coeffs_h}
\end{align}

The structure of these equations reflects the full complexity of the resistive 
two-component theory. The collision term enters both directly, through the 
relaxation rates multiplying $V_q^\mu$ and $V^\mu$, and indirectly through 
the coefficients generated when $\delta h^\mu$ is eliminated. The two currents 
remain coupled even after that elimination. The electromagnetic sector is richer 
than a simple Ohmic term: the electric field appears linearly, but also multiplied 
by $\pi^{\mu\nu}$, $\delta\pi^{\mu\nu}$, and quadratic combinations of the 
currents, while the magnetic field mixes components through $b^{\mu\nu}$, and 
gradient terms involving the shear tensors make explicit that charge transport and 
viscous anisotropy are dynamically intertwined.

\subsection{Shear-stress tensor}

We now derive the equations of motion for the shear-stress tensor. By now we know that for each species, the comoving derivative of the shear-stress tensor is simply
\begin{equation}
\dot{\pi}^{\mu\nu}_\pm
=
\Delta^{\mu\nu}_{\alpha\beta}
\frac{d}{d\tau}
\int dK\, k^{\langle\alpha}k^{\beta\rangle} f_k^\pm .
\label{eq:ch6_pipm_def}
\end{equation}
This quantity measures the anisotropic part of the momentum flux in the local rest frame and therefore encodes the viscous response of each component of the plasma.

As in the previous Chapter, it is convenient to work with the degrees of freedom given by the total shear stress tensor, $\pi^{\mu\nu}$, and relative shear stress tensor $\delta\pi^{\mu\nu}$. The equations of motion follow from taking the corresponding moment of the kinetic equation, decomposing the resulting moments into irreducible tensors, and using the 14-moment closure to express higher moments in terms of the hydrodynamic variables and dissipative currents. The collision term is treated in the same way as before: one expands $f_k^\pm=f_{0k}^\pm+\delta f_k^\pm$, linearizes the kernel in $\delta f_k^\pm$, and evaluates the projected rank-two moment. In this sector the collision integral produces relaxation terms for both $\pi^{\mu\nu}$ and $\delta\pi^{\mu\nu}$, together with mixing terms between them. The magnetic field also induces a coupling between the total and relative shear sectors through the tensor $b^{\mu\nu}$.

After carrying out these steps\footnote{detailed derivation in Appendix~\ref{app:collision_integrals_shear}}, one obtains evolution of shear stress tensor,
\begin{align}
& \dot{\pi}^{\langle\mu\nu\rangle}
+
\Sigma_\pi \pi^{\mu\nu}
-
\hat{\Sigma}_\pi \delta\pi^{\mu\nu}
+
2\,\Omega\, b^{\lambda\langle\mu}\delta\pi^{\nu\rangle}{}_\lambda
\label{eq:ch6_pi_eom} \\
\quad & =
\frac{8}{5}V_q^{\langle\mu}E^{\nu\rangle}
+
\frac{8}{15}\epsilon\,\sigma^{\mu\nu}
-
\frac{4}{3}\pi^{\mu\nu}\theta
 -
\frac{10}{7}\sigma^{\lambda\langle\mu}\pi^{\nu\rangle}{}_\lambda
-
2\omega^{\lambda\langle\nu}\pi^{\mu\rangle}{}_\lambda
-
2\dot{u}^{\langle\mu}V^{\nu\rangle}
+
\frac{2}{5}\nabla^{\langle\mu}V^{\nu\rangle},\nonumber
\end{align}
and similarly, the evolution of relative shear stress tensor
\begin{align}
& \delta\dot{\pi}^{\langle\mu\nu\rangle}
+
\Sigma_{\delta\pi}\delta\pi^{\mu\nu}
-
\hat{\Sigma}_{\delta\pi}\pi^{\mu\nu}
+
2\,\Omega\, b^{\lambda\langle\mu}\pi^{\nu\rangle}{}_\lambda
\label{eq:ch6_deltapi_eom} \\ 
\quad & =
\frac{8}{5}V^{\langle\mu}E^{\nu\rangle}
+
\frac{8}{15}\delta\epsilon\,\sigma^{\mu\nu}
-
\frac{4}{3}\delta\pi^{\mu\nu}\theta -
\frac{10}{7}\sigma^{\lambda\langle\mu}\delta\pi^{\nu\rangle}{}_\lambda
-
2\omega^{\lambda\langle\nu}\delta\pi^{\mu\rangle}{}_\lambda
-
\frac{2}{|q|}\dot{u}^{\langle\mu}V_q^{\nu\rangle}
+
\frac{2}{5|q|}\nabla^{\langle\mu}V_q^{\nu\rangle}.
\nonumber
\end{align}
where, $\Omega \equiv |q|B/(5T)$ is the magnetic-field-induced frequency, defined in previous chapter. The coefficients appearing in Eqs.~\eqref{eq:ch6_pi_eom} and \eqref{eq:ch6_deltapi_eom} are
\begin{align}
\Sigma_\pi & = \frac{3n}{10}\left(\sigma_T^{+-}+\sigma_T\right),
\qquad
\hat{\Sigma}_\pi = \frac{3n_q}{10|q|}\left(\sigma_T^{+-}-\sigma_T\right),
\nonumber \\ 
\Sigma_{\delta\pi} & = \frac{n}{10}\left(5\sigma_T^{+-}+3\sigma_T\right),
\quad
\hat{\Sigma}_{\delta\pi} = \frac{n_q}{10|q|}\left(5\sigma_T^{+-}-3\sigma_T\right).
\label{eq:ch6_sigma_pi_coeffs}
\end{align}

Several features of these equations are worth emphasizing. First, the total and relative shear-stress tensors are coupled even when the magnetic field is set to zero. This happens because we did not impose that the chemical potential is zero. The coefficients $\hat{\Sigma}_\pi$ and $\hat{\Sigma}_{\delta\pi}$ encode the mixing generated by the difference between the two charged components, while the term proportional to $\Omega$ describes the magnetic-field-induced rotation between $\pi^{\mu\nu}$ and $\delta\pi^{\mu\nu}$. Second, the electric field introduces new source terms proportional to $V_q^{\langle\mu}E^{\nu\rangle}$ and $V^{\langle\mu}E^{\nu\rangle}$. Therefore, even in the absence of magnetic effects, an electric field can generate anisotropic stress through its coupling to the diffusion currents. Third, the remaining terms have the same general interpretation as in ordinary transient hydrodynamics: $\sigma^{\mu\nu}$ drives viscous stresses, $\theta$ dilutes them through expansion, $\omega^{\mu\nu}$ rotates them, and gradients and acceleration of the diffusion currents feed into their evolution.

In the limit of vanishing electromagnetic fields and vanishing net-charge density, the mixing terms disappear and Eq.~\eqref{eq:ch6_pi_eom} reduces to the familiar Israel--Stewart-type shear equation~\cite{Israel:1979wp}. In particular,
\begin{equation}
\Sigma_\pi = \frac{1}{\tau_\pi} = \frac{\epsilon+P_0}{5\eta},
\label{eq:ch6_taupi_relation}
\end{equation}
which identifies $\Sigma_\pi$ with the inverse shear relaxation time and relates the microscopic relaxation rate to the macroscopic shear viscosity $\eta$.

For the present resistive theory, however, the new point is precisely that the shear sector does not evolve independently. The total shear-stress tensor couples directly to the net-charge diffusion current, while the relative shear-stress tensor couples to the total particle-diffusion current. Consequently, anisotropic stress may be generated purely by the electric field, even when magnetic effects are absent. Moreover, because the evolution equation for $V_q^\mu$ also depends on $\pi^{\mu\nu}$, the shear-stress tensor feeds back into the relaxation of the current itself. In this sense, the current sector and the viscous sector form a genuinely coupled dynamical system.

\section{Homogeneous case}

We now consider the homogeneous limit of the theory. In this case, all spatial gradients vanish and the electromagnetic fields are taken to be uniform. The dynamics is therefore purely temporal and is driven only by the coupling between the plasma and the electromagnetic field. Since the main purpose here is to isolate the effects of the electric field, we restrict attention to the locally neutral case,
\begin{equation}
n_q = 0.
\label{eq:ch6_hom_nq_zero}
\end{equation}
This eliminates any static background charge density and allows one to focus directly on the relaxation of the dissipative currents generated by the field. We also choose the electric field to point along the $x$-direction,
\begin{equation}
E^\mu \parallel x^\mu.
\label{eq:ch6_hom_E_direction}
\end{equation}

In the homogeneous limit, the equations derived in the previous section simplify considerably. Since all spatial gradients vanish,
\begin{equation}
\nabla^\mu \to 0,
\qquad
\theta = 0,
\qquad
\sigma^{\mu\nu}=0,
\qquad
\omega^{\mu\nu}=0,
\qquad
\dot u^\mu=0.
\label{eq:ch6_hom_kinematics}
\end{equation}
Thus, all terms involving expansion, shear, vorticity, acceleration, and spatial derivatives drop out. What remains is the purely temporal interplay between the electric field and the dissipative currents.

\subsection{Dynamical equations for $B=0$}

We now further specialize to the case of vanishing magnetic field,
\begin{equation}
B^\mu = 0.
\label{eq:ch6_hom_B_zero}
\end{equation}
This is the simplest setting in which one can directly examine how the electric field alone generates the dissipative currents. In this limit, Maxwell's equation reduces to
\begin{equation}
\dot E^\mu = -V_q^\mu.
\label{eq:ch6_hom_Maxwell}
\end{equation}
Thus, the electric field decays in time through the charge-diffusion current that it itself induces.

The scalar equations also simplify. Since the system is homogeneous and locally neutral, the particle number density remains constant while the energy density evolves only through the work done by the electric field on the charge current,
\begin{equation}
\dot\epsilon + E^\mu V_{q,\mu} = 0,
\qquad
\dot n = 0.
\label{eq:ch6_hom_scalar}
\end{equation}
Hence, in this limit there is no dilution by expansion and no transport by spatial gradients; the only scalar dynamics comes from the exchange of energy between the field and the plasma.

The vector and tensor sectors simplify even more strongly. The total particle-diffusion current vanishes identically,
\begin{equation}
V^\mu = 0,
\label{eq:ch6_hom_V_zero}
\end{equation}
and likewise the relative shear-stress tensor vanishes,
\begin{equation}
\delta\pi^{\mu\nu}=0.
\label{eq:ch6_hom_deltapi_zero}
\end{equation}
This leaves only the net-charge diffusion current $V_q^\mu$ and the total shear-stress tensor $\pi^{\mu\nu}$ as non-trivial dissipative quantities.

In this simplified scenario, the net-charge diffusion current then satisfies the following equation of motion,
\begin{equation}
\tau_{V_q}\dot V_q^\mu + V_q^\mu
=
\sigma_E E^\mu
+
\Omega_{E\pi} E_\nu \pi^{\mu\nu}
-
\Gamma_{\mathrm{NL}} E_\nu V_q^\nu V_q^\mu,
\label{eq:ch6_hom_Vq}
\end{equation}
while the shear-stress tensor obeys
\begin{equation}
\tau_\pi \dot\pi^{\mu\nu} + \pi^{\mu\nu}
=
\frac{8}{5}\tau_\pi V_q^{\langle\mu}E^{\nu\rangle}.
\label{eq:ch6_hom_pi}
\end{equation}
The corresponding transport coefficients are
\begin{equation}
\tau_{V_q}=\frac{1-\alpha_{V_q}}{\Gamma_{V_q}},
\qquad
\sigma_E=\frac{\mathcal G_E}{\Gamma_{V_q}},
\qquad
\Omega_{E\pi}=\frac{|q|^2}{20T^2\Gamma_{V_q}},
\qquad
\Gamma_{\mathrm{NL}}=\frac{\Gamma_{V_qV_q}}{\Gamma_{V_q}},
\qquad
\tau_\pi=\frac{1}{\Sigma_\pi}.
\label{eq:ch6_hom_transport}
\end{equation}
These are precisely the combinations of the transport coefficients introduced earlier in the full theory.

Equation~\eqref{eq:ch6_hom_Vq} shows that the electric field drives the charge-diffusion current through the conductivity term $\sigma_E E^\mu$, but the current does not simply follow a linear Ohmic law. There are two additional structures. First, the term proportional to $\Omega_{E\pi}$ couples the current to the shear-stress tensor. Second, the term proportional to $\Gamma_{\mathrm{NL}}$ is nonlinear in $V_q^\mu$ and represents a backreaction of the current on its own evolution. Thus, even in the homogeneous case, the resistive response of the plasma is already more complicated than a simple relaxation towards $V_q^\mu=\sigma_E E^\mu$.

Equation~\eqref{eq:ch6_hom_pi} shows, in turn, that the shear-stress tensor is generated directly by the coupling between the electric field and the charge-diffusion current. This is one of the main qualitative points of the resistive theory: even without any flow gradients, the electric field can generate anisotropic stress dynamically. The homogeneous limit therefore isolates in the clearest possible way the coupled evolution of $V_q^\mu$ and $\pi^{\mu\nu}$.

In the present setup, the symmetry of the problem simplifies the tensor structure further. Since the electric field points along the $x$-direction, the net-charge diffusion current has only an $x$-component,
\begin{equation}
V_q^\mu = (0,V_q^x,0,0),
\label{eq:ch6_hom_Vq_x}
\end{equation}
and the shear-stress tensor becomes diagonal. Its independent information can then be encoded entirely in the $xx$-component,
\begin{equation}
\pi^{\mu\nu}
\;\longrightarrow\;
\pi^{xx},
\label{eq:ch6_hom_pixx}
\end{equation}
with the remaining diagonal components fixed by tracelessness and orthogonality. In this way, the homogeneous problem reduces to a closed set of ordinary differential equations for $E^x(t)$, $V_q^x(t)$, $\pi^{xx}(t)$, and $\epsilon(t)$.

\subsection{Numerical results}

We now present numerical solutions of the coupled evolution equations for the net-charge diffusion four-current $V_q^\mu$ and the shear-stress tensor $\pi^{\mu\nu}$ in the homogeneous case~\cite{Kushwah2025RRmhd}. The system is taken to be initially in equilibrium at time $t_0=0$, with initial energy density
\begin{equation}
\epsilon_0 = 1000~\mathrm{fm}^{-4}.
\label{eq:ch6_num_eps0}
\end{equation}
We then solve the full set of coupled equations derived in the previous subsection for different values of the cross sections, the initial electric field, and the shear viscosity to entropy density ratio.

To assess the importance of the nonlinear couplings, it is useful to compare the full solutions with those obtained from the simpler relaxation-type Ohmic description,
\begin{equation}
\tau_{V_q}\dot V_q^{\langle\mu\rangle} + V_q^\mu = \sigma_E E^\mu.
\label{eq:ch6_linear_ohm}
\end{equation}
This equation is recovered by neglecting the coupling between the net-charge diffusion current and the shear-stress tensor, as well as the nonlinear term in the current equation. In what follows, we refer to Eq.~\eqref{eq:ch6_linear_ohm} as the linear approximation. In this approximation, the current relaxes exponentially toward the instantaneous Ohmic value $V_q^\mu=\sigma_E E^\mu$.

\begin{figure}[t]
    \centering
    \includegraphics[width=0.8\textwidth]{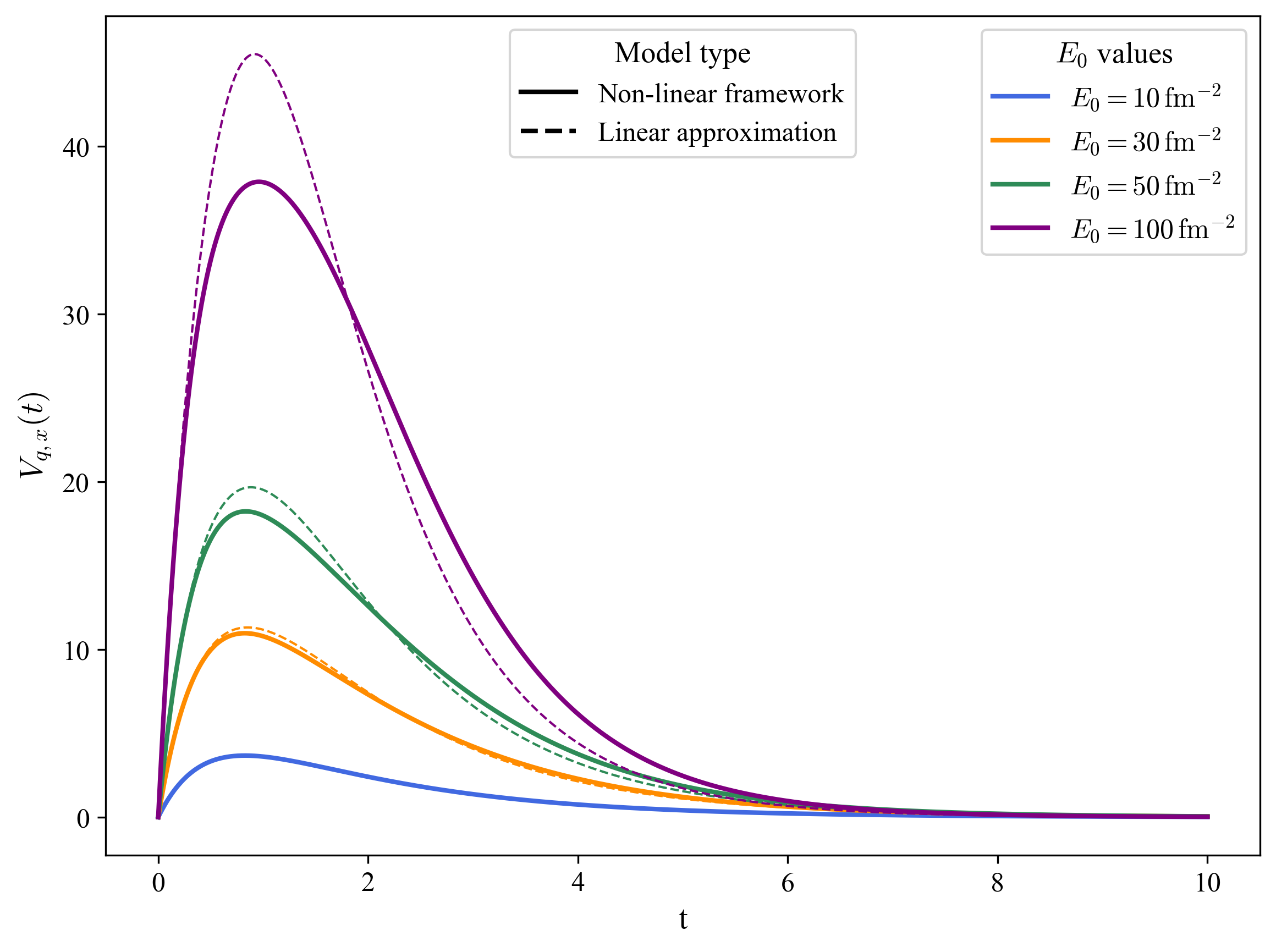}
    \caption[Homogeneous diffusion current]{Time evolution of the unnormalized diffusion current $V_{q,x}(t)$ for different initial electric field strengths $E_0$, with $\eta/s=1$, $|q|=2/3$, and $\sigma_T = 0.1\,\sigma_T^{+-}$. Larger $E_0$ values lead to more pronounced transient peaks before relaxation. All trajectories approach the same asymptotic regime at late times.}
    \label{fig:ch6_hom_vqx_unnorm}
\end{figure}

We begin by examining how the net-charge diffusion current depends on the initial electric field strength. In Fig.~\ref{fig:ch6_hom_vqx_unnorm}, we show the time evolution of the unnormalized current $V_{q,x}(t)$ for several values of $E_0$, keeping $\eta/s=1$, $|q|=2/3$, and $\sigma_T = 0.1\,\sigma_T^{+-}$ fixed. The current initially grows due to the applied electric field and reaches a transient maximum before decreasing at later times. As the initial field strength is increased, the peak becomes more pronounced. Comparing the full result with the linear approximation, one sees that the latter systematically overestimates both the transient amplitude and the decay rate when the initial electric field is large. This shows that the nonlinear backreaction contained in the full evolution equation becomes important precisely in the regime of strong electric fields.

\begin{figure}[t]
    \centering
    \includegraphics[width=0.8\textwidth]{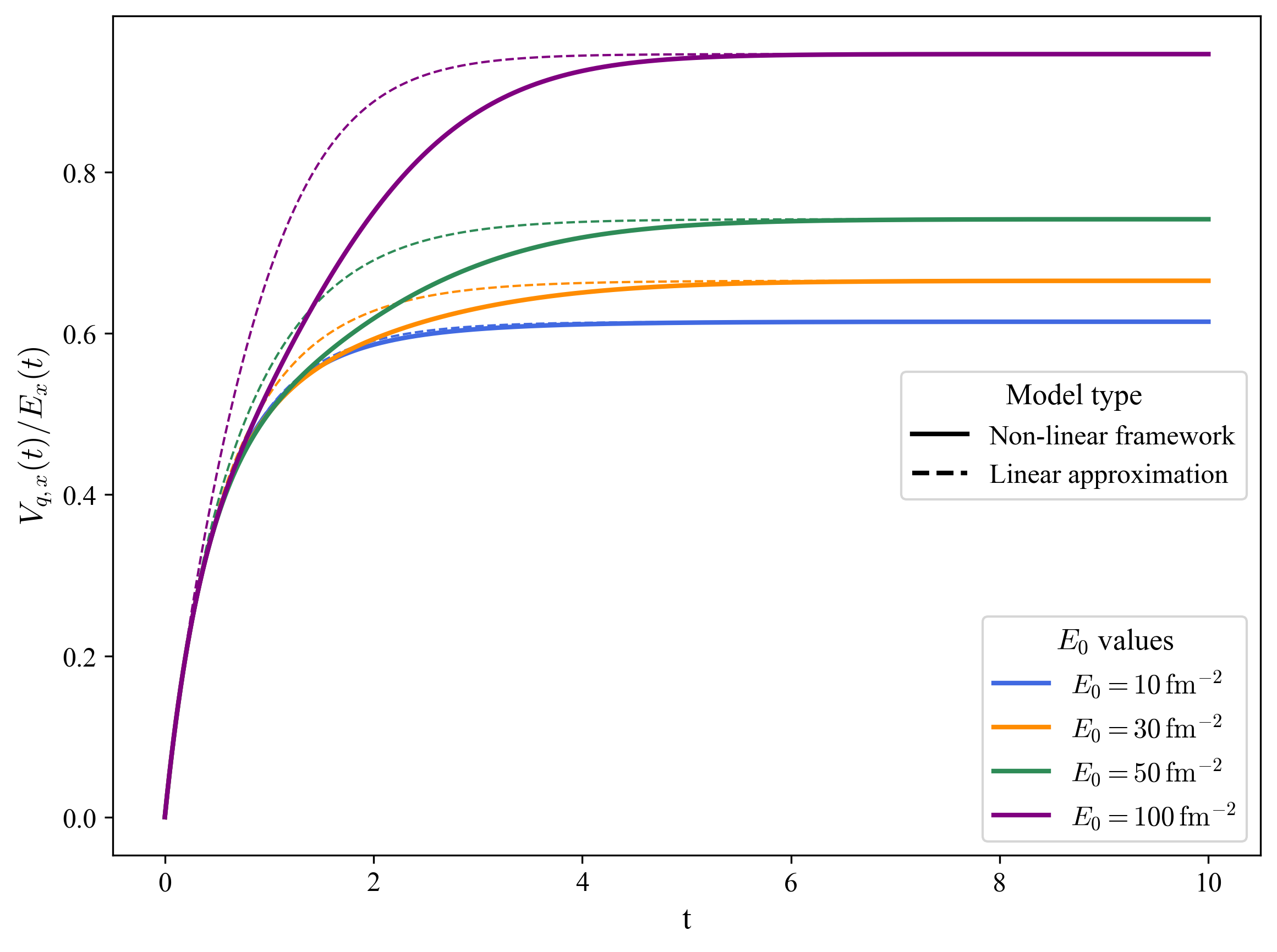}
    \caption[Ohmic relaxation in the homogeneous case]{Time evolution of the normalized current $V_{q,x}/E_x$ for several initial electric field strengths $E_0$, with $\eta/s=1$, $|q|=2/3$, and $\sigma_T = 0.1\,\sigma_T^{+-}$. At late times, all curves approach the same asymptotic value, consistent with the Ohmic limit $V_{q,x}\simeq \sigma_E E_x$. Early-time deviations become more pronounced as the initial field is increased.}
    \label{fig:ch6_hom_vqx_over_ex}
\end{figure}

A more direct way to see the relaxation toward the Ohmic regime is to examine the normalized ratio $V_{q,x}/E_x$. This is shown in Fig.~\ref{fig:ch6_hom_vqx_over_ex} for the same choice of parameters. At late times, all curves converge to the same asymptotic value, consistent with the Ohmic relation $V_{q,x}\simeq \sigma_E E_x$. The differences appear at early times, where stronger electric fields lead to larger deviations from the asymptotic conductivity. Thus, the nonlinear corrections primarily affect the transient regime, while the late-time behavior remains effectively Ohmic.

\begin{figure}[t]
    \centering
    \includegraphics[width=0.8\textwidth]{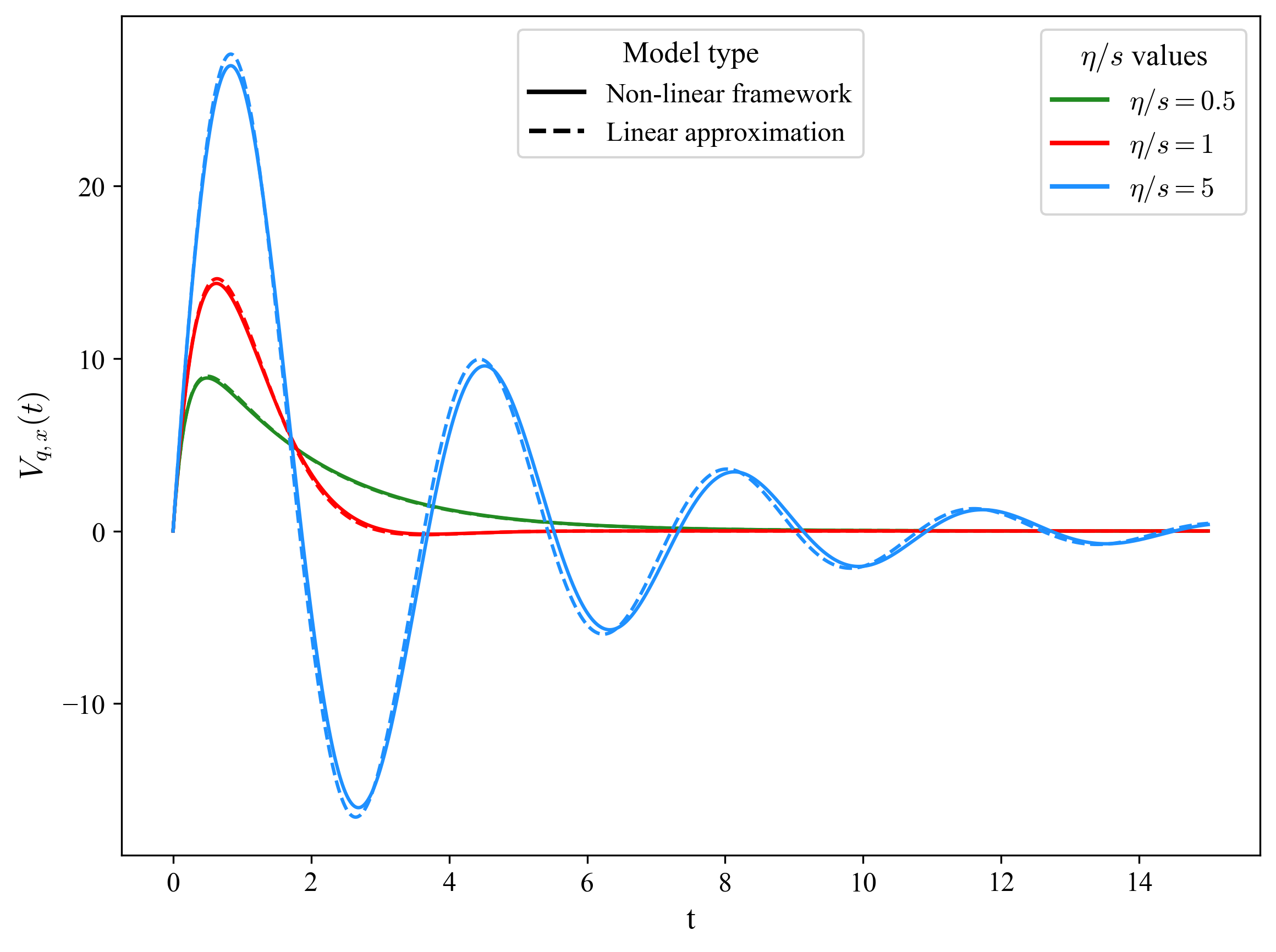}
    \caption[Viscosity dependence of diffusion current]{Time evolution of the diffusion current $V_{q,x}(t)$ for different values of $\eta/s$, with fixed $|q|=1$, $\sigma_T = 0.1\,\sigma_T^{+-}$, and $E_0 = 30~\mathrm{fm}^{-2}$. Increasing $\eta/s$ weakens collisional damping and leads to oscillatory relaxation, whereas smaller $\eta/s$ gives a smooth exponentially damped profile.}
    \label{fig:ch6_hom_vqx_eta_scan}
\end{figure}

Next, we investigate the role of the shear viscosity to entropy density ratio. In Fig.~\ref{fig:ch6_hom_vqx_eta_scan}, we plot the time evolution of $V_{q,x}(t)$ for fixed initial electric field $E_0=30~\mathrm{fm}^{-2}$, fixed $|q|=1$, and fixed ratio $\sigma_T/\sigma_T^{+-}=0.1$, while varying $\eta/s$. For small $\eta/s$, the current exhibits a smooth exponentially damped behavior, characteristic of diffusive relaxation. As $\eta/s$ is increased, the damping weakens and the current develops oscillatory behavior. This signals a transition to an underdamped regime.

This trend can already be understood within the linear approximation. Differentiating Eq.~\eqref{eq:ch6_linear_ohm} with respect to time and using $\dot E^\mu = -V_q^\mu$, one obtains
\begin{equation}
\tau_{V_q}\ddot V_{q,x} + \dot V_{q,x} + \sigma_E V_{q,x} = 0.
\label{eq:ch6_vq_oscillator}
\end{equation}
Assuming a harmonic form $V_{q,x}\propto e^{\omega t}$ yields the dispersion relation
\begin{equation}
\omega_\pm = \frac{-1 \pm \sqrt{1-4\tau_{V_q}\sigma_E}}{2\tau_{V_q}}.
\label{eq:ch6_vq_dispersion}
\end{equation}
The motion becomes oscillatory when
\begin{equation}
4\tau_{V_q}\sigma_E > 1,
\label{eq:ch6_underdamped_condition}
\end{equation}
which corresponds to the underdamped regime of charge transport~\cite{Gavassino_2026}. Since increasing $\eta/s$ effectively reduces collisional damping, it pushes the system toward this regime. The full nonlinear solutions display the same qualitative behavior, but with smaller transient amplitudes and delayed peaks, reflecting the additional feedback encoded in the coupled theory.

\begin{figure}[t]
    \centering
    \includegraphics[width=0.8\textwidth]{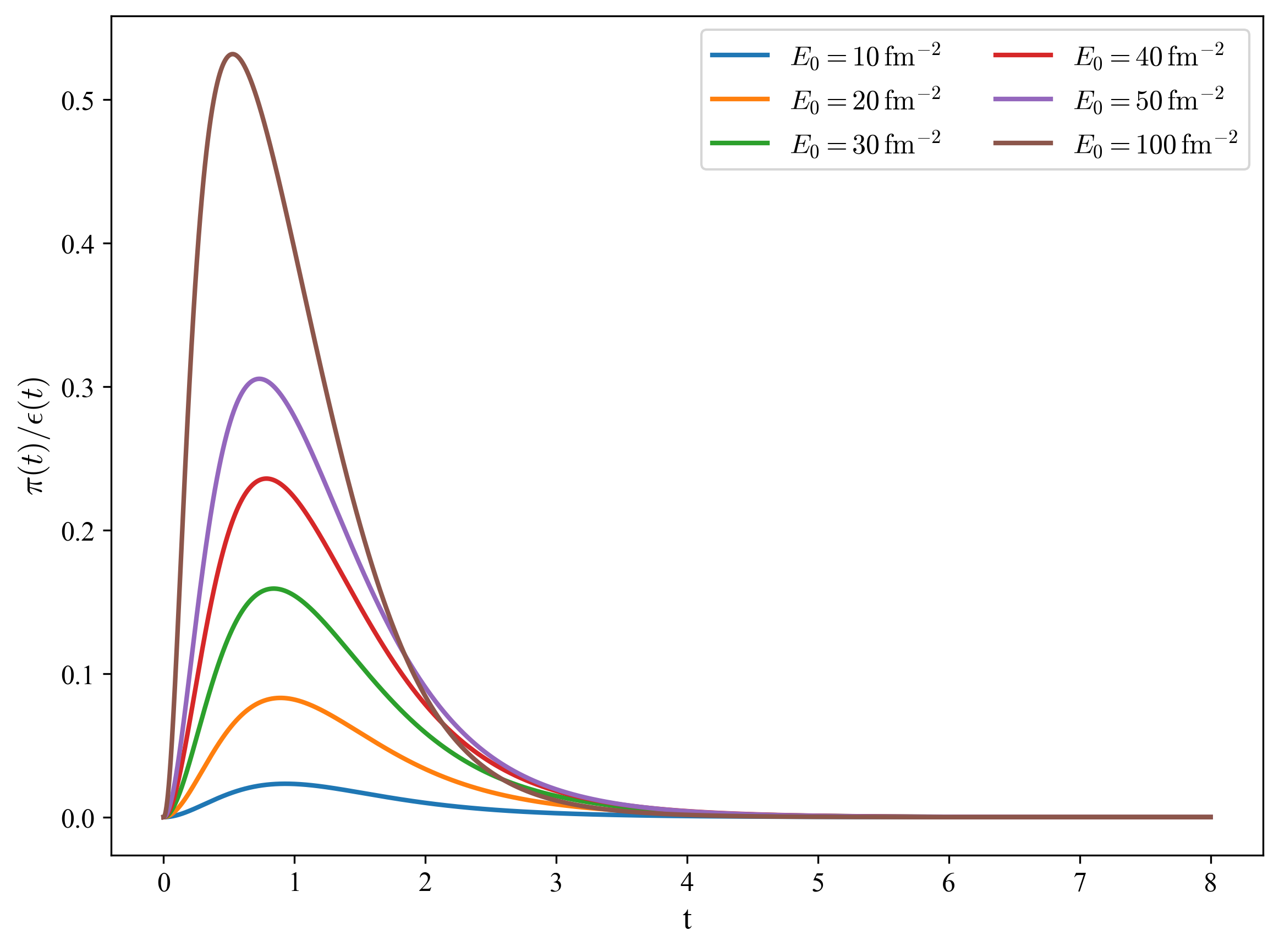}
    \caption[Electric-field-induced shear stress]{Time evolution of the normalized shear-stress component $\pi^{xx}/\epsilon$ for different initial electric field strengths $E_0$, with $\eta/s=1$ and $\sigma_T = 0.1\,\sigma_T^{+-}$. Larger electric fields produce larger transient shear-stress peaks, showing that the electric field alone can generate significant momentum anisotropy. At late times, all curves relax back to zero.}
    \label{fig:ch6_hom_pixx_over_eps}
\end{figure}

We now turn to the shear-stress tensor generated by the electric field. In Fig.~\ref{fig:ch6_hom_pixx_over_eps}, we plot the normalized component $\pi^{xx}(t)/\epsilon(t)$ for several initial electric field strengths, again at fixed $\eta/s=1$ and $\sigma_T/\sigma_T^{+-}=0.1$. The magnitude of the shear-stress tensor increases as the initial electric field is increased. This is one of the main qualitative results of the homogeneous analysis: even in the absence of any flow profile, an electric field alone can produce a sizable momentum anisotropy through its coupling to the net-charge diffusion current. At late times, all curves decay to zero, consistent with the restoration of isotropy as the electric field subsides.

Overall, the homogeneous analysis shows that the usual relaxation-type Ohmic description remains a reasonable approximation when the electric field is not too large. In particular, the late-time conductivity is still governed by the same asymptotic Ohmic limit. However, the full coupled theory becomes necessary whenever one wishes to describe the transient regime in the presence of strong fields, since nonlinear backreaction delays the current peak and reduces its magnitude. At the same time, the electric field generates a non-trivial shear-stress tensor even without fluid expansion, making it clear that resistive effects alone can induce a substantial momentum anisotropy.

\section{Bjorken flow}

We now turn to the case of Bjorken flow~\cite{Bjorken:1982qr}, keeping the electric field but neglecting the magnetic field. In this way, one can examine how the coupling between the net-charge diffusion current and the shear-stress tensor survives in an expanding system. In contrast to the homogeneous case, the longitudinal expansion now becomes an additional source of momentum anisotropy, and it competes with the electric-field-induced effects. As will be seen below, although the same coupling remains present, its quantitative 
impact is considerably weaker because the electric field decays much faster in a rapidly expanding background.

We work in Milne coordinates as introduced in Chapter~\ref{ch:two_component_rmhd}, 
with expansion scalar $\theta=1/\tau$ and shear tensor 
$\sigma^{\mu\nu}=\mathrm{diag}(0,1/3\tau,1/3\tau,-2/3\tau^3)$. The conservation of energy-momentum in the presence of the electric field then gives
\begin{equation}
\dot\epsilon
=
-\frac{4}{3}\theta\,\epsilon
+
\pi^{\mu\nu}\sigma_{\mu\nu}
+
E_\mu V_q^\mu,
\label{eq:ch6_bj_energy_general}
\end{equation}
which, in Milne coordinates, becomes
\begin{equation}
\dot\epsilon
=
-\frac{4}{3\tau}\epsilon
+
\frac{1}{3\tau}
\left(
\pi^{xx}+\pi^{yy}-2\tau^2\pi^{\eta\eta}
\right)
+
E\,V_q^x.
\label{eq:ch6_bj_energy_milne}
\end{equation}
Thus, in contrast to the homogeneous case, the energy density is not only modified by the electric work term $E\,V_q^x$, but is also diluted by the longitudinal expansion and sourced by the shear-stress tensor itself. 

The projected Maxwell equation acquires an additional dilution term from the expanding background,
\begin{equation}
\dot E^{\langle\mu\rangle} + \theta E^\mu = - V_q^\mu,
\label{eq:ch6_bj_maxwell_general}
\end{equation}
or, equivalently, for the $x$-component,
\begin{equation}
\dot E^x + \frac{E^x}{\tau} = -V_q^x.
\label{eq:ch6_bj_maxwell_x}
\end{equation}
The extra $1/\tau$ term has an immediate physical consequence, that is, the electric field decays more rapidly than in the homogeneous case simply because the system is expanding. This is one of the key reasons why the electric-field-induced effects become smaller in Bjorken flow. 

Under Bjorken symmetry, the coupled equations for the net-charge diffusion current and the shear-stress tensor reduce to
\begin{equation}
\tau_{V_q}\dot V_q^\mu + V_q^\mu
=
\sigma_E E^\mu
+
\Omega_{E\pi}E_\nu \pi^{\mu\nu}
-
\tau_{V_q}\theta V_q^\mu
+
\frac{5r-27}{45\Gamma_{V_q}}\sigma^\mu_{\ \nu}V_q^\nu
-
\Gamma_{\mathrm{NL}}(E\!\cdot\!V_q)V_q^\mu,
\label{eq:ch6_bj_Vq_general}
\end{equation}
and
\begin{equation}
\tau_\pi \dot\pi^{\mu\nu} + \pi^{\mu\nu}
=
\frac{8}{5}\tau_\pi V_q^{\langle\mu}E^{\nu\rangle}
+
\frac{8}{15}\epsilon\,\sigma^{\mu\nu}
-
\frac{4}{3}\theta\,\pi^{\mu\nu}
-
\frac{10}{7}\sigma^{\lambda\langle\mu}\pi^{\nu\rangle}_{\ \ \lambda}.
\label{eq:ch6_bj_pi_general}
\end{equation}
Projecting these equations into Milne coordinates and restricting to an electric field along the $x$-direction gives the scalar set used in the numerical analysis. The current equation becomes
\begin{equation}
(1-\alpha_{V_q})\dot V_q^x + \Gamma_{V_q}V_q^x
=
\mathcal G_E E^x
-
\frac{|q|^2}{20T^2}E^x \pi^{xx}
+
\left(
\frac{4\alpha_{V_q}}{3}
-
\frac{6}{5}
\right)\frac{V_q^x}{\tau}
-
\Gamma_{\mathrm{NL}}E^x(V_q^x)^2,
\label{eq:ch6_bj_Vqx}
\end{equation}
while the relevant shear component satisfies
\begin{equation}
\tau_\pi \dot\pi^{xx} + \pi^{xx}
=
\frac{8}{5}\tau_\pi V_q^x E^x
+
\frac{8}{45}\frac{\epsilon}{\tau}
-
\frac{38}{21}\frac{\pi^{xx}}{\tau}.
\label{eq:ch6_bj_pixx}
\end{equation}
These two equations already show the essential structure of the problem. Equation~\eqref{eq:ch6_bj_Vqx} contains the electric driving term, the coupling to $\pi^{xx}$, the explicit dilution term proportional to $1/\tau$, and the nonlinear backreaction proportional to $E^x(V_q^x)^2$. Equation~\eqref{eq:ch6_bj_pixx}, on the other hand, contains both the electric-current source term and the purely hydrodynamic source term proportional to $\epsilon/\tau$. This second term is absent in the homogeneous case and is the dominant source of anisotropy in the expanding system. 

\begin{figure}[t]
    \centering
    \includegraphics[width=\textwidth]{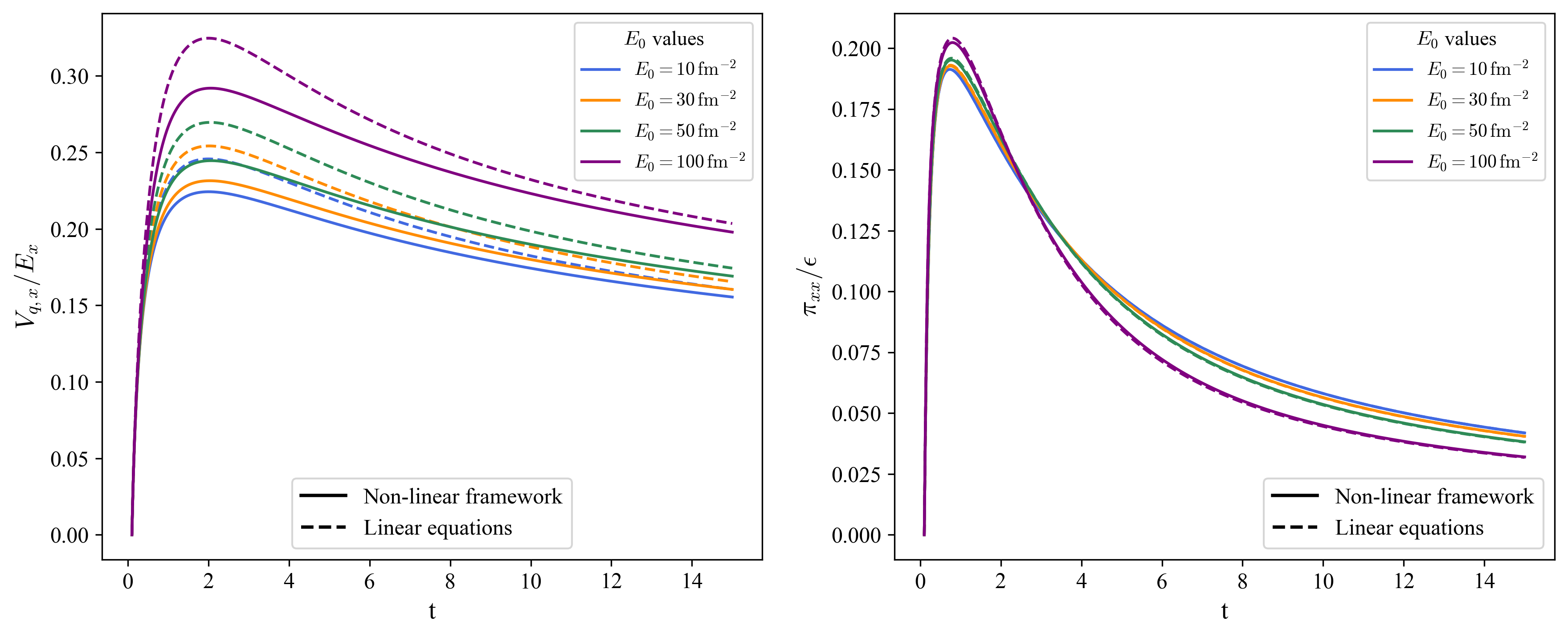}
    \caption[Bjorken evolution with electric field]{Time evolution of the charge-diffusion current $V_q^x$ and the normalized shear-stress component $\pi^{xx}/\epsilon$ in Bjorken flow for different initial electric field strengths $E_0$. The left panel shows $V_q^x$, which exhibits an early-time peak that increases with $E_0$ but remains much less sensitive to the field than in the homogeneous case. The right panel shows $\pi^{xx}/\epsilon$, which displays only a weak dependence on $E_0$, indicating that the dominant source of momentum anisotropy is the hydrodynamic expansion rather than the electric field.}
    \label{fig:ch6_bjorken_current_shear}
\end{figure}

We solve Eqs.~\eqref{eq:ch6_bj_energy_milne}, \eqref{eq:ch6_bj_maxwell_x}, \eqref{eq:ch6_bj_Vqx}, and \eqref{eq:ch6_bj_pixx} for a system initially in equilibrium with
\begin{equation}
\epsilon_0 = 1000~\mathrm{fm}^{-4},
\qquad
\tau_0 = 0.1~\mathrm{fm},
\label{eq:ch6_bj_initial_conditions}
\end{equation}
and vary the initial electric field in the range
\begin{equation}
E_0 = 10\text{--}100~\mathrm{fm}^{-2}.
\label{eq:ch6_bj_E0_range}
\end{equation}
The equations are integrated numerically in Python using a Runge--Kutta scheme. The code is publicly available at~\cite{Kushwah2025RRmhd}. The numerical solutions for $V_q^x$ and $\pi^{xx}/\epsilon$ are shown in Fig.~\ref{fig:ch6_bjorken_current_shear}. The left panel displays the charge-diffusion current $V_q^x$, while the right panel shows the normalized shear-stress component $\pi^{xx}/\epsilon$. The current develops an early-time peak, and this peak increases with the initial electric field $E_0$. However, compared to the homogeneous case, the dependence on $E_0$ is much weaker, because the electric field itself decays rapidly due to the $1/\tau$ dilution in Eq.~\eqref{eq:ch6_bj_maxwell_x}. As a result, the net-charge current does not retain a strong memory of the initial field strength for long. 

The behavior of the shear-stress tensor is qualitatively different. As seen in the right panel of Fig.~\ref{fig:ch6_bjorken_current_shear}, the curves for $\pi^{xx}/\epsilon$ are only weakly sensitive to the initial electric field. This is consistent with Eq.~\eqref{eq:ch6_bj_pixx}: although the electric field enters through the term $V_q^x E^x$, the dominant source of shear is the longitudinal expansion term proportional to $\epsilon/\tau$. Therefore, in Bjorken flow the primary origin of momentum anisotropy is hydrodynamic expansion itself, not the electric field. The electric field still affects the late-time decay, in the sense that larger initial $E_0$ leads to a somewhat faster decrease of $\pi^{xx}/\epsilon$, but the effect remains small even for the largest fields considered. 

Therefore in the Bjorken analysis, the coupling between the net-charge diffusion current and the shear-stress tensor persists under longitudinal expansion, exactly as the equations predict. However, its quantitative impact is significantly reduced relative to the homogeneous case. The reason is that the expansion dilutes the electric field too quickly for it to remain the dominant source of anisotropy. In this regime, the electric field mainly modifies the intermediate- and late-time evolution, while the bulk of the momentum anisotropy is generated by the expanding background itself.

\section{Summary}

This chapter completed the kinetic-theory construction of resistive relativistic 
magnetohydrodynamics for a two-component plasma. The central new ingredient 
relative to the previous chapter is that the electric field is no longer eliminated 
by a non-resistive constraint, but instead evolves dynamically through its coupling 
to the charge-diffusion current $V_q^\mu$. The 14-moment approximation was applied 
to the Boltzmann--Vlasov equation for both charged species, producing coupled 
relaxation equations for $V_q^\mu$, $V^\mu$, $\pi^{\mu\nu}$, and $\delta\pi^{\mu\nu}$, 
with the relative heat current $\delta h^\mu$ eliminated perturbatively in the 
hydrodynamic regime.

The applications to the homogeneous and Bjorken-expanding cases showed that 
the resistive dynamics is qualitatively richer than a simple Ohmic response. In 
the homogeneous case, the electric field generates a non-trivial shear stress 
even without any flow gradients, and nonlinear backreaction modifies the transient 
amplitude of the current at strong fields. In Bjorken flow, the same coupling 
persists but the rapid dilution of the electric field by expansion limits its 
quantitative impact, leaving hydrodynamic expansion as the dominant source of 
momentum anisotropy. Together, these results provide the most complete framework 
developed in this thesis and demonstrate that charge transport, field evolution, 
and viscous anisotropy are genuinely inseparable in a resistive relativistic plasma.
\chapter{Conclusions and Outlook}
\label{ch:conclusions}

This thesis began with a simple but deep observation: the quark-gluon plasma produced in heavy-ion collisions is not the weakly coupled gas that perturbative QCD would suggest at high temperatures. It is a strongly coupled, nearly perfect liquid, and it behaves as one because its constituents interact strongly enough that the system remains close to local thermodynamic equilibrium throughout its rapid expansion. This fact, established through the measurement of large elliptic flow coefficients at RHIC and the LHC, made relativistic viscous hydrodynamics (specifically the causal, second-order Israel--Stewart theory) the indispensable language for describing the QGP's evolution and extracting its transport properties from data.

At the same time, heavy-ion collisions generate magnetic fields of extraordinary strength, reaching $eB \sim m_\pi^2$ at RHIC and $eB \sim 15\,m_\pi^2$ at the LHC. Fields this large are not a small perturbation on the QGP dynamics. They break the rotational symmetry of the plasma, suppress transport across field lines, generate Hall-type responses, and couple directly to the chiral structure of QCD. Any quantitatively faithful description of the QGP therefore requires a theory that simultaneously accounts for viscous dissipation \emph{and} the electromagnetic coupling. Constructing that theory within the rigorous framework of kinetic theory, was the central goal of this thesis.

\section{The tools}

Before any of the new results could be derived, the correct theoretical machinery had to be assembled. Relativistic fluid dynamics rests on two pillars: a clean thermodynamic foundation that specifies what equilibrium looks like, and a kinetic-theory framework that connects the macroscopic transport coefficients to the microscopic collision dynamics.

The thermodynamic foundation, developed in Chapters~\ref{ch:thermo_foundations} and~\ref{chap:dissipative_hydro}, starts from a simple premise: a fluid element is a volume large enough to be in local thermodynamic equilibrium but small enough to resolve macroscopic gradients. From this, the Euler relation, the Gibbs--Duhem relation, and the covariant conservation laws follow. Introducing dissipation into this picture leads immediately to the first conceptual crisis of the field: the most natural relativistic generalization of the Navier--Stokes equations, formulated by Eckart and by Landau and Lifshitz, is both acausal and linearly unstable. The constitutive relations are instantaneous (a gradient anywhere produces a response everywhere), and this violates special relativity at the most basic level. Hiscock and Lindblom showed that the instabilities are not an artefact of any particular frame choice: they are intrinsic to the first-order structure of the theory. No amount of adding more transport coefficients, including anisotropic ones from a magnetic field, can cure this problem at first order.

The cure, as Israel and Stewart understood, is to promote the dissipative currents from algebraic functions of gradients to independent dynamical variables that relax toward their Navier--Stokes values over a finite microscopic timescale. The governing equations become hyperbolic rather than parabolic, perturbations propagate at finite speed, and both causality and stability are restored. This Israel--Stewart philosophy, and the systematic method-of-moments implementation of it developed by Denicol, Niemi, Moln\'{a}r, and Rischke, provided the kinetic-theory scaffold upon which all the subsequent results were built.

Chapter~\ref{chap:kinetic_foundations} laid out this scaffold in detail. The relativistic Boltzmann equation, expanded around local equilibrium, generates an exact but infinite hierarchy of evolution equations for the irreducible moments of the non-equilibrium deviation $\delta f_\mathbf{k}$. The 14-moment truncation closes this hierarchy by retaining only the 14 variables (five equilibrium, three from the diffusion current, five from the shear tensor, and one from the bulk pressure) that correspond to the physical dissipative degrees of freedom of relativistic hydrodynamics. All transport coefficients are expressed as integrals over the equilibrium distribution, connecting macroscopic viscosity and diffusion directly to the microscopic collision cross-section. Extending this entire framework to include the Lorentz force term in the Boltzmann equation is then the key step that makes magnetized dissipative hydrodynamics derivable from first principles.

\section{The magnetic field changes the theory, not just its numbers}

The perhaps most fundamental structural result for RMHD, established at the macroscopic level in Chapter~\ref{chap:dissipative_hydro} and confirmed from kinetic theory in Chapter~\ref{chap:kinetic_foundations}, is the following: a magnetic field does not merely modify the numerical values of the transport coefficients. It changes the mathematical structure of the theory itself.

To see why, recall that in a field-free isotropic plasma, the three-dimensional space orthogonal to the fluid velocity $u^\mu$ has full $SO(3)$ rotational symmetry. There is one shear viscosity, one diffusion coefficient, and one bulk viscosity. But a magnetic field selects a preferred direction $b^\mu = B^\mu / B$ and reduces this symmetry to $SO(2)$ rotations about $b^\mu$. The three-space now has two geometrically distinct subspaces: the direction along the field, and the two-dimensional plane perpendicular to it. Transport along these two subspaces is governed by different physics, and the antisymmetric tensor $b^{\mu\nu} = -\epsilon^{\mu\nu\alpha\beta}u_\alpha b_\beta$ generates a third, Hall-type response in which currents flow perpendicular to both the driving gradient and the field itself.

The consequences are dramatic. The single diffusion coefficient $\kappa$ becomes three: a parallel coefficient $\kappa_\parallel$ for transport along the field, a transverse coefficient $\kappa_\perp < \kappa_\parallel$ for transport across the field (suppressed by cyclotron motion), and a Hall coefficient $\kappa_\times$ for the off-diagonal response. The single shear viscosity $\eta$ is split into five contributions, two of them nondissipative and analogous to a Hall conductivity coefficient. These anisotropic transport tensors are the Navier--Stokes-level manifestation of the symmetry breaking, derived macroscopically in Chapter~\ref{chap:dissipative_hydro}.

But the modification goes deeper than the Navier--Stokes level. When the 14-moment closure is imposed on the moment hierarchy in the presence of a magnetic field, as carried out in Chapter~\ref{chap:kinetic_foundations}, two new terms appear in the second-order relaxation equations that have no analogue in the field-free theory. The diffusion relaxation equation acquires a term $-\delta_{VB}\,qB\,b^\mu_{\ \nu}V_f^\nu$ that rotates the diffusion current in the plane perpendicular to $b^\mu$. The shear relaxation equation acquires a term $-\delta_{\pi B}\,qB\,\Delta^{\mu\nu}_{\alpha\kappa}b^\alpha_{\ \beta}\pi^{\kappa\beta}$ that mixes the components of the shear-stress tensor through the same antisymmetric structure. These are not corrections to the transport coefficients; they are new couplings that alter the \emph{time evolution} of the dissipative currents. For a massless Boltzmann gas, $\delta_{VB} = (15/16)\beta_0\lambda_{\rm mfp}$ and $\delta_{\pi B} = (2/3)\beta_0\lambda_{\rm mfp}$, both controlled by the ratio of mean free path to thermal Larmor radius.

There is an important subtlety here that deserves emphasis. Looking at the relaxation equations, the first-order Navier--Stokes driving terms on the right-hand side, namely $-\zeta\theta$, $\kappa\nabla^\mu\alpha_0$, and $2\eta\sigma^{\mu\nu}$, are the familiar isotropic ones, apparently independent of the magnetic field. So how does the anisotropic first-order transport emerge? The answer is that it emerges dynamically, from the interplay between these isotropic sources and the $qB$-dependent couplings in the relaxation equations. When the relaxation equations are solved in the first-order (Navier--Stokes) limit by dropping time derivatives, the $\delta_{VB}$ and $\delta_{\pi B}$ terms generate the anisotropic constitutive relations. The anisotropy is not put in by hand at the Navier--Stokes level; it is a consequence of the second-order structure. This is a satisfying result: the second-order formulation is not merely a tool for restoring causality, it is also a natural framework in which the anisotropic first-order physics can be understood to originate.

\section{Two species are qualitatively different from one}

The single-component theory, while conceptually complete and technically rich, is too idealized for the QGP or any realistic plasma. Real plasmas contain particles of both signs of charge. In the QGP this means quarks and antiquarks; in nuclear matter it means protons and neutrons or, at higher densities, up and down quarks. The two-component structure introduces physics that simply cannot be captured by a single-species description.

The most important new feature, explored in Chapter~\ref{ch:two_component_rmhd}, is the emergence of two independent conserved currents: the total particle current $N^\mu = N_+^\mu + N_-^\mu$ and the net charge current $J^\mu \propto N_+^\mu - N_-^\mu$. These two currents carry different information. The total current tracks how matter as a whole is transported. The charge current tracks the relative motion of the two species, and it is the charge current, not the total current, that couples to the electromagnetic field. Similarly, both a total shear-stress tensor $\pi^{\mu\nu}$ and a relative shear-stress tensor $\delta\pi^{\mu\nu}$ appear, and the magnetic field couples them through the cyclotron precession of the two species.

The most striking dynamical signature of this coupling was seen in the Bjorken flow analysis. When both species precess around the magnetic field in opposite directions, the transverse component of the shear-stress tensor (the component in the plane perpendicular to $b^\mu$) exhibits damped oscillations. This is not a numerical instability or a truncation artefact. It is actual physics: the two species' shear contributions precess out of phase with each other, and their interference produces an oscillatory envelope in the total transverse shear. These oscillations are absent in the single-component theory and appear precisely because the two species respond differently to the magnetic field.

A technical lesson also emerged from this analysis. When one tries to simplify the coupled shear equations by assigning each projected component its own independent relaxation time, the projected relaxation times can become negative for sufficiently strong magnetic fields. A negative relaxation time corresponds to an unstable mode, which would be pathological. But the instability is not in the full theory: it is in the approximation. The full coupled equations remain well-behaved; it is only the additional step of decoupling the tensor components that breaks down. This provides a concrete and useful diagnostic: when the ratio of cyclotron frequency to inverse relaxation time approaches unity, the component-wise approximation fails and the full coupled system must be used.

\section{Resistance is not futile: the electric field has its own dynamics}

The most general framework of this thesis, developed in Chapter~\ref{chap:resistive_rmhd_two_component}, is the resistive two-component theory. Here, the electric conductivity is finite, the electric field $E^\mu$ is no longer constrained to vanish but evolves dynamically, and the charge diffusion current $V_q^\mu$ becomes a genuine dissipative degree of freedom with its own relaxation equation rather than being slaved to external constraints.

Moving to the resistive setting changes the character of the theory fundamentally. In the non-resistive case, the electromagnetic field modifies the dynamics but does not feed energy directly into or out of the plasma (the magnetic force does no work). In the resistive case, the electric field drives the charge current through an Ohmic term, and this does work on the plasma. The energy balance now contains the term $E_\nu V_q^\nu$, which transfers energy between the electromagnetic field and the plasma. This is real dissipation in the electromagnetic sense: the field does work, the plasma heats up, and entropy is produced.

But the most surprising result from the resistive analysis is not the Ohmic heating. It is the coupling between the charge current and the shear-stress tensor. The relaxation equation for $V_q^\mu$ contains not only the electric driving term $\sigma_E E^\mu$ but also a coupling to $\pi^{\mu\nu}$ through the term $\Omega_{E\pi} E_\nu \pi^{\mu\nu}$. Conversely, the shear relaxation equation contains a source term $(8\tau_\pi/5) V_q^{\langle\mu} E^{\nu\rangle}$ that generates shear stress from the product of the electric field and the charge current. This coupling means that even a perfectly homogeneous plasma with no velocity gradients at all can develop a non-trivial shear stress if it carries a net charge current in an electric field. Momentum anisotropy is generated not by geometry but by electromagnetic coupling.

This was confirmed numerically in the homogeneous case: starting from a perfectly isotropic plasma and applying a uniform electric field in the $x$-direction, the shear-stress component $\pi^{xx}$ grows from zero, reaches a peak proportional to the initial electric field strength, and then decays as the field itself subsides. The larger the initial field, the larger the transient shear peak. This is a qualitative result with no analogue in ordinary viscous hydrodynamics.

Another striking feature of the homogeneous analysis was the transition to underdamped oscillations as $\eta/s$ increases. In ordinary Ohmic relaxation, the charge current simply decays exponentially toward $V_q^\mu = \sigma_E E^\mu$. But once the shear viscosity is large enough that the condition $4\tau_{V_q}\sigma_E > 1$ is satisfied, the coupled current-shear system enters an underdamped regime and oscillates before settling. This is not an instability; it is a transient resonance between the relaxation timescales of the current and the shear tensor. The full nonlinear solutions show the same qualitative behaviour, but with reduced transient amplitudes because the backreaction term $\Gamma_{\rm NL}(E\cdot V_q)V_q^\mu$ partially offsets the linear driving.

In Bjorken flow, the electric field decays rapidly due to the dilution from longitudinal expansion. The additional $1/\tau$ term in the projected Maxwell equation means the field loses much of its strength before it has had time to generate a large current. As a result, the electric-field-induced contribution to the shear stress is much smaller than in the homogeneous case. The dominant source of momentum anisotropy in Bjorken flow remains hydrodynamic expansion itself. The electric field mainly affects the intermediate-time evolution of the current while the shear tensor is driven primarily by the expanding background.

\section{The unified picture}

Stepping back from the individual chapters, a coherent picture emerges. The journey from ideal hydrodynamics to resistive two-component magnetohydrodynamics is a journey of progressive symmetry breaking. Ideal hydrodynamics has full $SO(3)$ symmetry in three-space. Adding a magnetic field breaks this to $SO(2)$. Going to two components breaks the equivalence of the two species and introduces new relative dynamical variables. Going to the resistive setting breaks the constraint $E^\mu = 0$ and promotes the electric field to a full dynamical partner of the plasma.

At each stage, the number of independent transport coefficients grows, the tensor structure becomes richer, and new couplings appear between sectors that were previously decoupled. But the fundamental architecture of the theory, i.e., causal relaxation equations derived from the method of moments and closing an infinite Boltzmann hierarchy at the 14-moment level, remains the same throughout. This \emph{universality} reflects the fact that the Israel--Stewart structure is the minimal consistent framework for relativistic dissipative dynamics, regardless of what forces are acting or how many species are present. The magnetic field modifies the details, but the overall structure remains unchanged.

\section{Outlook}

The framework developed in this thesis opens several directions that follow naturally from what was derived here.

The most immediate theoretical extension is to combine, in a single numerical analysis, the effects that Chapters~\ref{ch:two_component_rmhd} and~\ref{chap:resistive_rmhd_two_component} treated separately. The derivations in Chapter~\ref{chap:resistive_rmhd_two_component} were carried out with both $E^\mu$ and $B^\mu$ present, but the applications were restricted to $B^\mu = 0$ in order to isolate the new resistive physics clearly. The full setting, where a dynamical electric field coexists with a magnetic field, should combine the oscillatory inter-species shear dynamics of Chapter~\ref{ch:two_component_rmhd} with the current--shear feedback loop of Chapter~\ref{chap:resistive_rmhd_two_component}. This is the more physical situation in heavy-ion collisions, where the intense transient magnetic field and the charge currents generated by it are present at the same time during the early QGP evolution.

A second natural step is to implement the two-component resistive equations in a numerical relativistic hydrodynamics code. The equations of motion derived in Chapters~\ref{ch:two_component_rmhd} and~\ref{chap:resistive_rmhd_two_component} are explicit, and in Bjorken geometry they already reduce to a tractable system of coupled ordinary differential equations. A three-dimensional implementation would make it possible to study charge-odd azimuthal correlations associated with the chiral magnetic effect, the directed flow coefficient $v_1$, which is especially sensitive to the early-time magnetic field, and possible modifications of charged-hadron spectra due to anisotropic transport. These are the kinds of observables that will ultimately test whether the theory developed here can describe the QGP dynamics quantitatively.

Throughout this thesis, all particles were taken to be massless. This simplifies the thermodynamic integrals and allows analytic expressions for the transport coefficients, but it also excludes finite-mass effects that become important near the QCD crossover temperature, where pion and kaon masses are comparable to $T$. Extending the 14-moment framework to massive species would change the equilibrium distribution function and the explicit form of the transport coefficients. The qualitative structure of the theory, namely anisotropic dissipation, Hall responses, and inter-species coupling, should remain present, but the quantitative results will change. Such an extension would also be necessary for applications to strongly magnetized astrophysical systems, where particle masses cannot be neglected compared to the local temperature.

The particles considered in this thesis were also spinless. In QCD, quarks carry spin-$\frac{1}{2}$, and the strong magnetic field and vorticity present in heavy-ion collisions can affect polarization observables, such as the global polarization of strange baryons at freeze-out. Including spin in the present kinetic framework would require going beyond the scalar distribution function and using a Wigner-function formulation, where the spin degrees of freedom couple to the electromagnetic field and to vorticity. The resulting dissipative magnetohydrodynamic theory would contain viscous, charge, and polarization degrees of freedom coupled to the magnetic field, making it a richer and more complete description of the strongly magnetized QGP.

Finally, the resistive two-component framework derived here is also relevant to astrophysical plasmas. Neutron star merger remnants contain hot, rapidly rotating, and strongly magnetized matter, where resistive MHD effects can influence the early post-merger dynamics. Magnetar crusts, with surface field strengths of order $10^{15}\,\mathrm{G}$, provide another regime where resistive, anisotropic, and Hall-type transport can play an important role. Applying the present framework to such systems would require adapting the equation of state and particle content to the relevant astrophysical environment, but the basic structure derived in this thesis provides a useful starting point.

\section{A closing thought}

This thesis set out to answer a question that was, at its core, a question about foundations: can we write down a consistent, causal, and stable description of a relativistic plasma that is simultaneously viscous and magnetized, derived from the same kinetic-theory principles that underpin the standard Israel--Stewart theory? The answer is yes, and the construction reveals something that could not have been anticipated from the individual ingredients alone.

The magnetic field is not a spectator that modifies a pre-existing theory. It is an active participant that rewires the theory's internal structure. It creates new geometric objects ($\Xi^{\mu\nu}$, $b^{\mu\nu}$), new transport channels ($\kappa_\times$, $\eta_{03}$, $\eta_{04}$), new second-order couplings ($\delta_{VB}$, $\delta_{\pi B}$), and in the resistive case, a dynamical feedback loop between electromagnetic and mechanical degrees of freedom that has no analogue in field-free physics. The Israel--Stewart structure absorbs all of this naturally, confirming that giving the dissipative currents their own dynamics is not merely a device for restoring causality but is the physically correct way to describe any strongly interacting relativistic fluid, magnetized or otherwise.

The quark-gluon plasma is hot, dense, rapidly expanding, and strongly magnetized. Describing it correctly demands exactly the kind of theory constructed here. The bridge from the formal framework of this thesis to quantitative QGP phenomenology remains to be crossed, but the foundations have been laid.

\appendix
\renewcommand{\chaptername}{Appendix}
\renewcommand{\theequation}{\thechapter\arabic{equation}}
\renewcommand{\chaptermark}[1]{%
  \markboth{Appendix \thechapter.\quad #1}{}%
}

\appchapter{Collision integrals: Shear Stress Tensor}{app:collision_integrals_shear}
\label{app:collision_integrals_shear}

Before evaluating the relevant integrals, we introduce the auxiliary function $\varphi_k$, which characterizes the deviation of the distribution function from its local equilibrium form. In relativistic kinetic theory, the single-particle distribution function $f_k$ is expanded around the equilibrium distribution $f_{0k}$, and the deviation is systematically expressed in terms of hydrodynamic gradients and dissipative corrections:
\begin{equation}
    f_k = f_{0k} \left(1+\varphi_k\right).
\end{equation}
To truncate the infinite hierarchy of moment equations, we adopt the 14-moment approximation originally developed by Israel and Stewart. Within this approximation, the deviation function $\varphi_k$ is expanded up to second order in particle momenta:
\begin{equation}\label{eq: varphi}
    \varphi_k = \varepsilon + \varepsilon_\lambda\,k^\lambda + \varepsilon_{\alpha\beta}\,k^\alpha k^\beta.
\end{equation}

Here, the scalar quantity $\varepsilon$ is proportional to the bulk viscous pressure $\Pi$, and to the off-equilibrium corrections to the energy density and number density, denoted by $\delta\epsilon$ and $\delta n$, respectively. Since all of these contributions are considered higher order in the present analysis, we neglect them in the subsequent calculations. Consequently, the contribution of $\varepsilon$ to $\varphi_k$ is omitted, and only the vector $\varepsilon_\lambda$ and tensor $\varepsilon_{\alpha\beta}$ terms remain relevant. These are linear in the dissipative hydrodynamic variables and are given by:

\begin{align}\label{eq: varepsilons}
    \varepsilon_\mu & =  B_3\,n_\mu + B_4\,h_\mu, \\
    \varepsilon_{\mu\nu} &= 2 D_3\,u_{(\mu}n_{\nu)} + 2 D_4\,u_{(\mu}h_{\nu)} + D_5\,\pi_{\mu\nu}.
\end{align}

We will require the explicit forms of a subset of these coefficients\footnote{The coefficients \(A_i, B_i, D_i\) appearing in the 14-moment expansion are determined by matching the kinetic theory to second-order relativistic hydrodynamics. Their explicit expressions involve thermodynamic integrals over the equilibrium distribution function and are detailed in Refs.~\cite{Israel:1979wp, Denicol_Rischke, Denicol:2012cn}. We follow these standard conventions throughout this work.}:
\begin{align}
    B_3 &= \frac{I_{41}}{2(I_{31}^2 - I_{21} I_{41})}, \qquad B_4 = D_3 = -\frac{I_{31}}{2(I_{31}^2 - I_{21} I_{41})}, \\ 
    D_4 &= \frac{I_{21}}{2(I_{31}^2 - I_{21} I_{41})}, \qquad
    D_5 = \frac{1}{2 J_{42}}.
\end{align}

 \section{Collision integral calculation}
We consider a classical gas of massless particles undergoing elastic two-to-two collisions. The corresponding collision term for the shear stress tensor is
\begin{equation}
\mathcal C_\pi^{\mu\nu}
\equiv
\int dK\,E_k^{-1}
k^{\langle\mu}k^{\nu\rangle}
C[f_k],
\label{eq:app_Cpi_def}
\end{equation}
where $E_k\equiv u_\mu k^\mu$.

\subsection{Linearization of the collision term}

We expand the distribution function around local equilibrium,
\begin{equation}
f_k=f_{0k}\left(1+\varphi_k\right),
\label{eq:app_f_phi}
\end{equation}
where the 14-moment approximation gives
\begin{equation}
\varphi_k
=
\varepsilon
+
\varepsilon_\alpha k^\alpha
+
\varepsilon_{\alpha\beta}k^\alpha k^\beta .
\label{eq:app_phi_general}
\end{equation}
For the shear collision term, the only relevant part is the rank-two contribution,
\begin{equation}
\varphi_k
\supset
\varepsilon_{\alpha\beta}k^\alpha k^\beta,
\qquad
\varepsilon_{\alpha\beta}
=
D_5\,\pi_{\alpha\beta}
=
\frac{1}{2J_{42}}\pi_{\alpha\beta}.
\label{eq:app_phi_shear}
\end{equation}
The scalar part cannot contribute to a traceless rank-two tensor after the final momentum integrations, and the vector part gives an odd-rank tensor which vanishes by isotropy in the local rest frame.

Using detailed balance,
\begin{equation}
f_{0p}f_{0p'}=f_{0k}f_{0k'},
\label{eq:app_detailed_balance}
\end{equation}
the gain-loss structure becomes, to first order in deviations from equilibrium,
\begin{equation}
f_p f_{p'}-f_k f_{k'}
=
f_{0k}f_{0k'}
\left(
\varphi_p+\varphi_{p'}-\varphi_k-\varphi_{k'}
\right)
+
\mathcal O(\varphi^2).
\label{eq:app_gain_loss_linear}
\end{equation}
Substituting Eq.~\eqref{eq:app_phi_shear} into Eq.~\eqref{eq:app_gain_loss_linear}, one obtains
\begin{equation}
\varphi_p+\varphi_{p'}-\varphi_k-\varphi_{k'}
=
\varepsilon_{\alpha\beta}
\left(
p^\alpha p^\beta
+
p'^\alpha p'^\beta
-
k^\alpha k^\beta
-
k'^\alpha k'^\beta
\right).
\label{eq:app_gain_loss_shear}
\end{equation}

Therefore, Eq.~\eqref{eq:app_Cpi_def} becomes
\begin{equation}
\begin{split}
\mathcal C_\pi^{\mu\nu}
&=
\frac{1}{2}
\varepsilon_{\alpha\beta}
\int dK\,dK'\,dP\,dP'\,
E_k^{-1}
k^{\langle\mu}k^{\nu\rangle}
W_{kk'\leftrightarrow pp'}
f_{0k}f_{0k'}
\\
&\quad\times
\left(
p^\alpha p^\beta
+
p'^\alpha p'^\beta
-
k^\alpha k^\beta
-
k'^\alpha k'^\beta
\right).
\end{split}
\label{eq:app_Cpi_linear_start}
\end{equation}
This is the basic object that must be evaluated.

\subsection{Separation into outgoing and incoming pieces}

It is convenient to separate the integral into the part containing the outgoing momenta and the part containing the incoming momenta:
\begin{equation}
\mathcal C_\pi^{\mu\nu}
=
\mathcal C_{\pi,\mathrm{out}}^{\mu\nu}
-
\mathcal C_{\pi,\mathrm{in}}^{\mu\nu},
\label{eq:app_Cpi_out_in}
\end{equation}
with
\begin{equation}
\mathcal C_{\pi,\mathrm{out}}^{\mu\nu}
=
\frac{1}{2}
\varepsilon_{\alpha\beta}
\int dK\,dK'\,
E_k^{-1}
k^{\langle\mu}k^{\nu\rangle}
f_{0k}f_{0k'}
\int dP\,dP'\,
W_{kk'\leftrightarrow pp'}
\left(
p^\alpha p^\beta
+
p'^\alpha p'^\beta
\right),
\label{eq:app_Cpi_out}
\end{equation}
and
\begin{equation}
\mathcal C_{\pi,\mathrm{in}}^{\mu\nu}
=
\frac{1}{2}
\varepsilon_{\alpha\beta}
\int dK\,dK'\,
E_k^{-1}
k^{\langle\mu}k^{\nu\rangle}
f_{0k}f_{0k'}
\left(
k^\alpha k^\beta+k'^\alpha k'^\beta
\right)
\int dP\,dP'\,
W_{kk'\leftrightarrow pp'} .
\label{eq:app_Cpi_in}
\end{equation}

The final-state integrations are done in the center-of-mass frame of the two-particle collision. Since the final-state angular integral is isotropic, the result can only depend on the total incoming momentum
\begin{equation}
P^\mu \equiv k^\mu+k'^\mu
\end{equation}
and the metric tensor. For the constant cross section used here, the identities required are
\begin{equation}
\int dP\,dP'\,
W_{kk'\leftrightarrow pp'}
=
s\,\sigma_T,
\label{eq:app_identity_zero}
\end{equation}
and
\begin{equation}
\int dP\,dP'\,
W_{kk'\leftrightarrow pp'}
\left(
p^\alpha p^\beta+p'^\alpha p'^\beta
\right)
=
\frac{2}{3}s\sigma_T
\left(k^\alpha+k'^\alpha\right)
\left(k^\beta+k'^\beta\right)
-
\frac{1}{12}s^2\sigma_T g^{\alpha\beta}.
\label{eq:app_identity_two}
\end{equation}
The last term in Eq.~\eqref{eq:app_identity_two} does not contribute after contraction with $\varepsilon_{\alpha\beta}$ because the shear tensor is traceless,
\begin{equation}
g^{\alpha\beta}\varepsilon_{\alpha\beta}=0.
\end{equation}

Using Eqs.~\eqref{eq:app_identity_zero} and~\eqref{eq:app_identity_two}, the square bracket in Eq.~\eqref{eq:app_Cpi_linear_start} reduces to
\begin{equation}
\begin{split}
&\int dP\,dP'\,
W_{kk'\leftrightarrow pp'}
\left(
p^\alpha p^\beta+p'^\alpha p'^\beta
-k^\alpha k^\beta-k'^\alpha k'^\beta
\right)
\\
&=
s\sigma_T
\left[
\frac{2}{3}
\left(k^\alpha+k'^\alpha\right)
\left(k^\beta+k'^\beta\right)
-
k^\alpha k^\beta
-
k'^\alpha k'^\beta
\right],
\end{split}
\label{eq:app_out_minus_in_reduced}
\end{equation}
where the trace term has already been dropped.

Expanding the product gives
\begin{equation}
\frac{2}{3}
\left(k^\alpha+k'^\alpha\right)
\left(k^\beta+k'^\beta\right)
-
k^\alpha k^\beta
-
k'^\alpha k'^\beta
=
-\frac{1}{3}
\left(
k^\alpha k^\beta+k'^\alpha k'^\beta
\right)
+
\frac{2}{3}
\left(
k^\alpha k'^\beta+k'^\alpha k^\beta
\right).
\label{eq:app_momentum_algebra}
\end{equation}
For massless particles,
\begin{equation}
s=(k+k')^2=2k\cdot k'.
\end{equation}
Hence
\begin{equation}
\begin{split}
\mathcal C_\pi^{\mu\nu}
&=
\frac{1}{2}\sigma_T\varepsilon_{\alpha\beta}
\int dK\,dK'\,
E_k^{-1}
k^{\langle\mu}k^{\nu\rangle}
f_{0k}f_{0k'}
\,s
\\
&\quad\times
\left[
-\frac{1}{3}
\left(
k^\alpha k^\beta+k'^\alpha k'^\beta
\right)
+
\frac{2}{3}
\left(
k^\alpha k'^\beta+k'^\alpha k^\beta
\right)
\right].
\end{split}
\label{eq:app_Cpi_before_equilibrium_integrals}
\end{equation}

\subsection{Reduction of the remaining equilibrium integrals}

The remaining integrations contain only equilibrium distributions. Since $f_{0k}$ is isotropic in the local rest frame, all tensor integrals can be decomposed in terms of $u^\mu$ and $\Delta^{\mu\nu}$. The useful identities are
\begin{equation}
\int dK\,f_{0k}\,E_k^r\,k^{\langle\mu}k^{\nu\rangle}=0,
\label{eq:app_eq_irreducible_zero}
\end{equation}
and
\begin{equation}
\int dK\,f_{0k}\,E_k^r\,
k^{\langle\mu}k^{\nu\rangle}k^\alpha k^\beta
=
2I_{r+4,2}\,
\Delta^{\mu\nu\alpha\beta}.
\label{eq:app_rank_four_eq_integral}
\end{equation}
Here $I_{nq}$ are the usual thermodynamic integrals. The projection operator satisfies
\begin{equation}
\Delta^{\mu\nu\alpha\beta}\varepsilon_{\alpha\beta}
=
\varepsilon^{\mu\nu}.
\label{eq:app_project_eps}
\end{equation}

Because one factor in Eq.~\eqref{eq:app_Cpi_before_equilibrium_integrals} is already $k^{\langle\mu}k^{\nu\rangle}$, only those terms that leave another rank-two tensor built from $k^\alpha k^\beta$ can survive the final projection. All terms that contain only scalar or vector structures vanish after the traceless projection or by isotropy. Carrying out the contractions gives
\begin{equation}
\mathcal C_\pi^{\mu\nu}
=
-\Gamma_\pi\,\pi^{\mu\nu},
\label{eq:app_Cpi_relax_form}
\end{equation}
where, for a massless classical gas with constant cross section,
\begin{equation}
\Gamma_\pi
=
\frac{3}{5}n_0\sigma_T.
\label{eq:app_Gamma_pi_single}
\end{equation}
Therefore,
\begin{equation}
\mathcal C_\pi^{\mu\nu}
=
-
\frac{3}{5}n_0\sigma_T\,\pi^{\mu\nu}
\label{eq:app_Cpi_single_final}
\end{equation}
This is the single-component shear collision term. It is a relaxation term: collisions drive the shear-stress tensor toward zero in local equilibrium.

\section{Two-component shear collision term}
\label{app:two_component_shear_collision}

We now repeat the calculation for a two-component gas with species $i,j=\pm$. The distribution functions are $f_k^+$ and $f_k^-$, and the collision term for species $i$ is
\begin{equation}
C_i[f]
=
C_{ii}[f_i,f_i]
+
C_{ij}[f_i,f_j],
\qquad
j\neq i.
\label{eq:app_Ci_two_sum}
\end{equation}
The same-species part has the symmetry factor $1/2$,
\begin{equation}
C_{ii}[f_i,f_i]
=
\frac{1}{2}
\int dK'\,dP\,dP'\,
W^{ii}_{kk'\leftrightarrow pp'}
\left(
f_p^i f_{p'}^i-f_k^i f_{k'}^i
\right),
\label{eq:app_Cii}
\end{equation}
whereas the cross-species part has no factor $1/2$,
\begin{equation}
C_{ij}[f_i,f_j]
=
\int dK'\,dP\,dP'\,
W^{ij}_{kk'\leftrightarrow pp'}
\left(
f_p^i f_{p'}^j-f_k^i f_{k'}^j
\right).
\label{eq:app_Cij}
\end{equation}
The corresponding shear collision moment is
\begin{equation}
\mathcal C_i^{\mu\nu}
\equiv
\int dK\,E_k^{-1}
k^{\langle\mu}k^{\nu\rangle}
C_i[f].
\label{eq:app_Ci_shear_def}
\end{equation}
It is useful to write it as
\begin{equation}
\mathcal C_i^{\mu\nu}
=
\mathcal C_{ii}^{\mu\nu}
+
\mathcal C_{ij}^{\mu\nu}.
\label{eq:app_Ci_same_cross}
\end{equation}

\subsection{Same-species contribution}

The same-species term is identical in structure to the single-component calculation. The only change is that the equilibrium density and cross section are those of species $i$:
\begin{equation}
\mathcal C_{ii}^{\mu\nu}
=
-
\frac{3}{5}n_i\sigma_T^{ii}\,\pi_i^{\mu\nu}.
\label{eq:app_Cii_shear_result}
\end{equation}
For the charge-symmetric system considered in the main text,
\begin{equation}
n_+=n_-=\hat n_0,
\qquad
\sigma_T^{++}=\sigma_T^{--}\equiv\sigma_T,
\label{eq:app_symmetric_same_defs}
\end{equation}
and therefore
\begin{equation}
\mathcal C_{++}^{\mu\nu}
=
-
\frac{3}{5}\hat n_0\sigma_T\,\pi_+^{\mu\nu},
\qquad
\mathcal C_{--}^{\mu\nu}
=
-
\frac{3}{5}\hat n_0\sigma_T\,\pi_-^{\mu\nu}.
\label{eq:app_same_species_symmetric}
\end{equation}

\subsection{Cross-species contribution}

The cross-species term is more important because it couples the two shear tensors. For the $+$ species, we have
\begin{equation}
\mathcal C_{+-}^{\mu\nu}
=
\int dK\,dK'\,dP\,dP'\,
E_k^{-1}k^{\langle\mu}k^{\nu\rangle}
W^{+-}_{kk'\leftrightarrow pp'}
\left(
f_p^+f_{p'}^-
-
f_k^+f_{k'}^-
\right).
\label{eq:app_Cplusminus_start}
\end{equation}
Linearizing the distribution functions gives
\begin{equation}
f_p^+f_{p'}^-
-
f_k^+f_{k'}^-
=
f_{0k}^+f_{0k'}^-
\left(
\varphi_p^+
+
\varphi_{p'}^-
-
\varphi_k^+
-
\varphi_{k'}^-
\right).
\label{eq:app_cross_linear_gain_loss}
\end{equation}
The shear part of each deviation is
\begin{equation}
\varphi_k^\pm
\supset
\varepsilon^\pm_{\alpha\beta}k^\alpha k^\beta,
\qquad
\varepsilon^\pm_{\alpha\beta}
=
\frac{1}{2J_{42}^\pm}\pi^\pm_{\alpha\beta}.
\label{eq:app_phi_pm_shear}
\end{equation}
At vanishing charge chemical potential,
\begin{equation}
J_{42}^+=J_{42}^-\equiv \hat J_{42}.
\end{equation}
Then Eq.~\eqref{eq:app_cross_linear_gain_loss} becomes
\begin{equation}
\varphi_p^+
+
\varphi_{p'}^-
-
\varphi_k^+
-
\varphi_{k'}^-
=
\varepsilon^+_{\alpha\beta}
\left(
p^\alpha p^\beta-k^\alpha k^\beta
\right)
+
\varepsilon^-_{\alpha\beta}
\left(
p'^\alpha p'^\beta-k'^\alpha k'^\beta
\right).
\label{eq:app_cross_phi_split}
\end{equation}

Thus the cross collision term separates into two pieces:
\begin{equation}
\mathcal C_{+-}^{\mu\nu}
=
\mathcal C_{+-}^{\mu\nu}[\pi_+]
+
\mathcal C_{+-}^{\mu\nu}[\pi_-],
\label{eq:app_Cplusminus_split}
\end{equation}
where $\mathcal C_{+-}^{\mu\nu}[\pi_+]$ contains $\varepsilon^+_{\alpha\beta}$ and $\mathcal C_{+-}^{\mu\nu}[\pi_-]$ contains $\varepsilon^-_{\alpha\beta}$.

The first piece is
\begin{equation}
\mathcal C_{+-}^{\mu\nu}[\pi_+]
=
\varepsilon^+_{\alpha\beta}
\int dK\,dK'\,dP\,dP'\,
E_k^{-1}k^{\langle\mu}k^{\nu\rangle}
W^{+-}_{kk'\leftrightarrow pp'}
f_{0k}^+f_{0k'}^-
\left(
p^\alpha p^\beta-k^\alpha k^\beta
\right).
\label{eq:app_cross_piece_plus}
\end{equation}
The second piece is
\begin{equation}
\mathcal C_{+-}^{\mu\nu}[\pi_-]
=
\varepsilon^-_{\alpha\beta}
\int dK\,dK'\,dP\,dP'\,
E_k^{-1}k^{\langle\mu}k^{\nu\rangle}
W^{+-}_{kk'\leftrightarrow pp'}
f_{0k}^+f_{0k'}^-
\left(
p'^\alpha p'^\beta-k'^\alpha k'^\beta
\right).
\label{eq:app_cross_piece_minus}
\end{equation}

The evaluation proceeds exactly as before: first perform the $dP\,dP'$ integral using the center-of-mass angular identities, then reduce the remaining $dK\,dK'$ integrals using isotropy in the local rest frame. The difference from the identical-particle case is that the measured particle is always the one with momentum $k^\mu$, whereas its collision partner has momentum $k'^\mu$ and belongs to the other species. Because of this, the terms proportional to $\pi_+^{\mu\nu}$ and $\pi_-^{\mu\nu}$ do not enter with the same coefficient.

The part proportional to $\pi_+^{\mu\nu}$ gives
\begin{equation}
\mathcal C_{+-}^{\mu\nu}[\pi_+]
=
-
\frac{4}{5}\hat n_0\sigma_T^{+-}
\pi_+^{\mu\nu},
\label{eq:app_cross_piece_plus_result}
\end{equation}
while the part proportional to $\pi_-^{\mu\nu}$ gives
\begin{equation}
\mathcal C_{+-}^{\mu\nu}[\pi_-]
=
+
\frac{1}{5}\hat n_0\sigma_T^{+-}
\pi_-^{\mu\nu}.
\label{eq:app_cross_piece_minus_result}
\end{equation}
Therefore,
\begin{equation}
\mathcal C_{+-}^{\mu\nu}
=
-
\frac{\hat n_0\sigma_T^{+-}}{5}
\left(
4\pi_+^{\mu\nu}-\pi_-^{\mu\nu}
\right)
\label{eq:app_Cplusminus_final}
\end{equation}
Similarly, for the negative species,
\begin{equation}
\mathcal C_{-+}^{\mu\nu}
=
-
\frac{\hat n_0\sigma_T^{+-}}{5}
\left(
4\pi_-^{\mu\nu}-\pi_+^{\mu\nu}
\right)
\label{eq:app_Cminusplus_final}
\end{equation}

\subsection{Full species collision terms}

Combining the same-species and cross-species contributions, we obtain
\begin{align}
\mathcal C_+^{\mu\nu}
&=
-
\frac{3}{5}\hat n_0\sigma_T\,\pi_+^{\mu\nu}
-
\frac{\hat n_0\sigma_T^{+-}}{5}
\left(
4\pi_+^{\mu\nu}-\pi_-^{\mu\nu}
\right),
\label{eq:app_Cplus_full}
\\
\mathcal C_-^{\mu\nu}
&=
-
\frac{3}{5}\hat n_0\sigma_T\,\pi_-^{\mu\nu}
-
\frac{\hat n_0\sigma_T^{+-}}{5}
\left(
4\pi_-^{\mu\nu}-\pi_+^{\mu\nu}
\right).
\label{eq:app_Cminus_full}
\end{align}
Equivalently,
\begin{align}
\mathcal C_+^{\mu\nu}
&=
-
\frac{\hat n_0}{5}
\left(
3\sigma_T+4\sigma_T^{+-}
\right)
\pi_+^{\mu\nu}
+
\frac{\hat n_0}{5}
\sigma_T^{+-}
\pi_-^{\mu\nu},
\label{eq:app_Cplus_matrix}
\\
\mathcal C_-^{\mu\nu}
&=
-
\frac{\hat n_0}{5}
\left(
3\sigma_T+4\sigma_T^{+-}
\right)
\pi_-^{\mu\nu}
+
\frac{\hat n_0}{5}
\sigma_T^{+-}
\pi_+^{\mu\nu}.
\label{eq:app_Cminus_matrix}
\end{align}
This shows explicitly that inter-species collisions do not merely add another damping term. They also mix the two species shear tensors.

\section{Total and relative shear basis}
\label{app:total_relative_shear_collision_derivation}

The natural variables used in the main text are the total and relative shear tensors,
\begin{equation}
\pi^{\mu\nu}
\equiv
\pi_+^{\mu\nu}+\pi_-^{\mu\nu},
\qquad
\delta\pi^{\mu\nu}
\equiv
\pi_+^{\mu\nu}-\pi_-^{\mu\nu}.
\label{eq:app_pi_deltapi_defs}
\end{equation}
The corresponding collision terms are
\begin{equation}
\mathcal C_\pi^{\mu\nu}
\equiv
\mathcal C_+^{\mu\nu}
+
\mathcal C_-^{\mu\nu},
\qquad
\mathcal C_{\delta\pi}^{\mu\nu}
\equiv
\mathcal C_+^{\mu\nu}
-
\mathcal C_-^{\mu\nu}.
\label{eq:app_Cpi_Cdeltapi_defs}
\end{equation}

Adding Eqs.~\eqref{eq:app_Cplus_full} and~\eqref{eq:app_Cminus_full}, we find
\begin{equation}
\begin{split}
\mathcal C_\pi^{\mu\nu}
&=
-
\frac{3}{5}\hat n_0\sigma_T
\left(
\pi_+^{\mu\nu}+\pi_-^{\mu\nu}
\right)
-
\frac{\hat n_0\sigma_T^{+-}}{5}
\left[
4\pi_+^{\mu\nu}-\pi_-^{\mu\nu}
+
4\pi_-^{\mu\nu}-\pi_+^{\mu\nu}
\right]
\\
&=
-
\frac{3}{5}\hat n_0\sigma_T\,\pi^{\mu\nu}
-
\frac{3}{5}\hat n_0\sigma_T^{+-}\,\pi^{\mu\nu}.
\end{split}
\end{equation}
Thus
\begin{equation}
\mathcal C_\pi^{\mu\nu}
= -
\Sigma\,\pi^{\mu\nu}\qquad 
\text{with} \qquad
\Sigma
=
\frac{3\hat n_0}{5}
\left(
\sigma_T+\sigma_T^{+-}
\right)
.
\label{eq:app_Sigma_final}
\end{equation}

Subtracting Eqs.~\eqref{eq:app_Cplus_full} and~\eqref{eq:app_Cminus_full}, we obtain
\begin{equation}
\begin{split}
\mathcal C_{\delta\pi}^{\mu\nu}
&=
-
\frac{3}{5}\hat n_0\sigma_T
\left(
\pi_+^{\mu\nu}-\pi_-^{\mu\nu}
\right)
-
\frac{\hat n_0\sigma_T^{+-}}{5}
\left[
4\pi_+^{\mu\nu}-\pi_-^{\mu\nu}
-
4\pi_-^{\mu\nu}+\pi_+^{\mu\nu}
\right]
\\
&=
-
\frac{3}{5}\hat n_0\sigma_T\,\delta\pi^{\mu\nu}
-
\hat n_0\sigma_T^{+-}\,\delta\pi^{\mu\nu}.
\end{split}
\end{equation}
Therefore
\begin{equation}
\mathcal C_{\delta\pi}^{\mu\nu}
= -
\Sigma'\,\delta\pi^{\mu\nu}
\qquad \text{with}\qquad 
\Sigma'
=
\frac{\hat n_0}{5}
\left(
3\sigma_T+5\sigma_T^{+-}
\right).
\label{eq:app_Sigmaprime_final}
\end{equation}

These two rates are not equal. The total shear tensor relaxes with the rate $\Sigma$, while the relative shear tensor relaxes with the rate $\Sigma'$. The difference is entirely due to inter-species scattering. Same-species collisions damp each species shear tensor but do not distinguish the total and relative combinations. Cross-species collisions exchange momentum between the two components and therefore damp the relative shear tensor more strongly.

\appchapter{Collision integrals: Two-component resistive case}{app:collision_integrals_resistive}
\label{app:collision_integrals_resistive}

\section{Collision Term for $\delta h^\mu$}

The collisional contribution to the evolution of the relative energy-diffusion four-current $\delta h^\mu$ arises from the difference of the projected collision integrals of the two species:
\begin{equation}
\begin{split}
    & \int dK \,  k^{\langle \mu \rangle} C[f^+, f^-]  -  \int dK \,  k^{\langle \mu \rangle} C[f^-, f^+] \\
     & = \frac{1}{2} \int dK\, E_k\, k^{\langle\mu\rangle} W^{++}_{kk' \leftrightarrow pp'}\left( f_p^+ f_{p'}^+ - f_k^+ f_{k'}^+ \right) - \frac{1}{2} \int dK\, E_k\, k^{\langle\mu\rangle} W^{--}_{kk' \leftrightarrow pp'}\left( f_p^- f_{p'}^- - f_k^- f_{k'}^- \right) \\
     &\quad + \int dK\, E_k\, k^{\langle\mu\rangle} W^{+-}_{kk' \leftrightarrow pp'}\left( f_p^+ f_{p'}^- - f_k^+ f_{k'}^- \right) - \int dK\, E_k\, k^{\langle\mu\rangle} W^{-+}_{kk' \leftrightarrow pp'}\left( f_p^- f_{p'}^+ - f_k^- f_{k'}^+ \right).
\end{split}
\end{equation}

We now analyze the individual contributions. The first two integrals represent the intra-species scattering terms:

\begin{equation}
\begin{split}
    &\frac{1}{2} \int dK\, E_k\, k^{\langle\mu\rangle} W^{++}_{kk' \leftrightarrow pp'} \left( f_p^+ f_{p'}^+ - f_k^+ f_{k'}^+ \right) 
    - \frac{1}{2} \int dK\, E_k\, k^{\langle\mu\rangle} W^{--}_{kk' \leftrightarrow pp'} \left( f_p^- f_{p'}^- - f_k^- f_{k'}^- \right).
\end{split}
\end{equation}
This contribution vanishes due to the symmetry under \(p \leftrightarrow p'\) and \(k \leftrightarrow k'\), provided that the collision kernels \(W^{ii}\) are symmetric and the integrals are properly antisymmetrized with respect to incoming and outgoing momenta.

We now focus on the inter-species terms:
\begin{equation}
\begin{split}
&\int dK\, E_k\, k^{\langle\mu\rangle} W^{+-}_{kk' \leftrightarrow pp'}\left( f_p^+ f_{p'}^- - f_k^+ f_{k'}^- \right)
- \int dK\, E_k\, k^{\langle\mu\rangle} W^{-+}_{kk' \leftrightarrow pp'}\left( f_p^- f_{p'}^+ - f_k^- f_{k'}^+ \right) \\
&= \int dK\, dK'\, dP\, dP'\, E_k\, \Delta^\mu_\nu\, k^\nu\, W^{+-}_{kk' \leftrightarrow pp'}\left( f_p^+ f_{p'}^- - f_k^+ f_{k'}^- \right) \\
&\quad - \int dK\, dK'\, dP\, dP'\, E_k\, \Delta^\mu_\nu\, k'^\nu\, W^{+-}_{kk' \leftrightarrow pp'}\left( f_p^+ f_{p'}^- - f_k^+ f_{k'}^- \right),
\end{split}
\end{equation}
where we have used \(W^{+-} = W^{-+}\) and relabeled momenta to combine both inter-species terms. To analyze the contribution involving the outgoing distributions \(f_p f_{p'}\), we perform a change of variables \(pp' \leftrightarrow kk'\). This leads to:
\begin{equation}
\begin{split}
    & = \int dK\, dK'\, dP\, dP'\, E_k\, \Delta^\mu_\nu\, p^\nu\, W^{+-}_{kk' \leftrightarrow pp'} f_k^+ f_{k'}^- 
- \int dK\, dK'\, dP\, dP'\, E_k\, \Delta^\mu_\nu\, p'^\nu\, W^{+-}_{kk' \leftrightarrow pp'} f_k^+ f_{k'}^- \\
&  = \int dK\, dK'\, dP\, dP'\, E_k\, \Delta^\mu_\nu\, \left(p^\nu - p'^\nu\right)\, W^{+-}_{kk' \leftrightarrow pp'} f_k^+ f_{k'}^-.
\end{split}
\end{equation}

This integral vanishes identically under the exchange \(p \leftrightarrow p'\). Therefore, the only surviving contribution to the collision term arises from the difference between incoming momentum components, resulting in
\begin{equation}
- \Delta^\mu_\nu \int dK\, dK'\, dP\, dP'\, (k^\nu - k'^\nu)\, W^{+-}_{kk' \leftrightarrow pp'}\, f_k^+ f_{k'}^- 
= -\, \Delta^\mu_\nu \, \sigma_T^{+-} \left( T_+^{\nu\lambda} N^-_\lambda - T_-^{\nu\lambda} N^+_\lambda \right),
\end{equation}
where \(T^{\nu\lambda}_\pm\) and \(N^\mu_\pm\) denote the energy-momentum tensors and particle four-currents of species \(+\) and \(-\), respectively. This result captures the net inter-species momentum exchange, projected orthogonally to the fluid velocity.

\section{Collision Term for $V^\mu_q$}
We begin by evaluating the collision integral corresponding to a single particle species. The relevant expression, after linearizing the collision term in the distribution function perturbations $\varphi_k$, reads:
\begin{equation}
\begin{split}
|q| & \int dK\, dK' \, dP \, dP'\, E_k^{-1} k^{\langle\mu\rangle} (2\pi)^5 s \sigma_T \delta^4(p + p' - k - k') (f_p f_{p'} - f_k f_{k'}) \\
= &\, |q| \int dK\, dK' \, dP \, dP'\, E_k^{-1} k^{\langle\mu\rangle} (2\pi)^5 s \sigma_T \delta^4(\cdots) f_{0k} f_{0k'} (\varphi_p + \varphi_{p'} - \varphi_k - \varphi_{k'}) \\
= &\, |q| \varepsilon_{\alpha\beta} \int dK\, dK' \, dP \, dP'\, E_k^{-1} k^{\langle\mu\rangle} (2\pi)^5 s \sigma_T \delta^4(\cdots) f_{0k} f_{0k'} (p^\alpha p^\beta + p'^\alpha p'^\beta - k^\alpha k^\beta - k'^\alpha k'^\beta)
\end{split}
\end{equation}
where the last step uses a quadratic ansatz in momenta for $\varphi_k = \varepsilon+ \varepsilon_\lambda k^\lambda+ \varepsilon_{\alpha\beta} k^\alpha k^\beta$. But only the second rank tensor contributes because the first two terms vanish under conservation of momentum.

To simplify this expression, we separate the integrations as:
\begin{equation}
\begin{split}
& |q| \varepsilon_{\alpha\beta} \int dK dK' E_k^{-1} k^{\langle\mu\rangle} f_{0k} f_{0k'} \Bigg[ \int dP dP' (2\pi)^5 s \sigma_T \delta^4(\cdots)(p^\alpha p^\beta + p'^\alpha p'^\beta) \\
& \hspace{4cm} - (k^\alpha k^\beta + k'^\alpha k'^\beta) \int dP dP' (2\pi)^5 s \sigma_T \delta^4(\cdots) \Bigg]
\end{split}
\end{equation}

Using the known identity for the  scattering kernel \cite{Denicol:2012cn, Denicol:2012es}:
\begin{align}
\int dP dP' (2\pi)^5 s \sigma_T \delta^4(\cdots) & = s\sigma_T 
\\ \int dP dP' (2\pi)^5 s \sigma_T \delta^4(\cdots) (p^\alpha p^\beta + p'^\alpha p'^\beta)
& = \frac{2}{3} \sigma_T s (k^\alpha + k'^\alpha)(k^\beta + k'^\beta) - \frac{1}{12} s^2 \sigma_T g^{\alpha\beta},
\end{align}
we substitute into the previous expression and perform the algebra:
\begin{equation}
\begin{split}
& = \frac{4}{3} |q| \sigma_T \varepsilon_{\alpha\beta} \int dK dK' E_k^{-1} k^{\langle\mu\rangle} k^\lambda k'_{\lambda} (k^\alpha k^\beta + k'^\alpha k^\beta + k'^\beta k^\alpha + k'^\alpha k'^\beta) f_{0k} f_{0k'} \\
& \quad - 2 |q| \sigma_T \varepsilon_{\alpha\beta} \int dK dK' E_k^{-1} k^{\langle\mu\rangle} k^\lambda k'_{\lambda} (k^\alpha k^\beta + k'^\alpha k'^\beta) f_{0k} f_{0k'}
\end{split}
\end{equation}

At this point, we evaluate each term using the definitions of the particle number density, equilibrium integrals, and moment decompositions in the Landau frame. After evaluating the contractions with $u^\mu$ and $\Delta^{\mu\nu}$, and performing all simplifications, the full collision integral reduces to:
\begin{equation}
-\frac{8}{9} |q| \sigma_T n_0^2 T^2 \varepsilon_{\alpha\beta} (u^\beta \Delta^{\mu\alpha} + u^\alpha \Delta^{\mu\beta})
\end{equation}

Recognizing the linear combinations of the vector moments in the decomposition of $\delta h^\mu$, this yields:
\begin{equation}\label{eq: Sngl_part_ coll_term}
-\frac{8}{9} |q| \sigma_T n_0^2 T^2 (2 D_3 V^\mu + 2 D_4 h^\mu) = -\frac{8}{9} |q| \sigma_T n_0 (V^\mu - \frac{1}{4\,T} h^\mu)
\end{equation}
where the second equality uses the known relations among the moment coefficients.

Finally for $V^\mu_q$ the single particle contribution will look like:
\begin{equation}
    =-\frac{4}{9}\sigma_T \left( n_q\, V^\mu  + n\,V_q-|q|n\frac{\delta h^\mu}{4T} \right) 
\end{equation}

The cross-species contribution to the collision integral simplifies significantly upon interchanging $p \leftrightarrow p'$ and using the symmetry $\sigma_T^{+-} = \sigma_T^{-+}$. Starting from the antisymmetric combination of gain and loss terms:
\begin{multline}
    |q| \int dK\, dK'\, dP\, dP'\, E_k^{-1} k^{\langle\mu\rangle} (2\pi)^5 s \sigma_T^{+-} \delta^4(p + p' - k - k') \left(f_p^+ f_{p'}^- - f_k^+ f_{k'}^- - f_p^+ f_{p'}^- + f_k^- f_{k'}^+\right) \\
    = |q| \int dK\, dK'\, dP\, dP'\, E_k^{-1} k^{\langle\mu\rangle} (2\pi)^5 s \sigma_T^{+-} \delta^4(p + p' - k - k') \left(f_k^- f_{k'}^+ - f_k^+ f_{k'}^- \right)
\end{multline}

This expression separates into two terms involving moments of the distribution functions:
\begin{equation}
     = 2|q|\sigma_T^{+-} \int dK\, E_k^{-1} k^{\langle\mu\rangle} k^\nu f_k^- \int dK'\, k'_\nu f_{k'}^+ 
    - 2|q|\sigma_T^{+-} \int dK\, E_k^{-1} k^{\langle\mu\rangle} k^\nu f_k^+ \int dK'\, k'_\nu f_{k'}^-
\end{equation}
   
Decomposing the momentum products in terms of fluid velocity and spatial projectors yields:
\begin{multline}
    = 2|q| \sigma_T^{+-} \int dK\, E_k^{-1} 
    \left( k^{\langle\mu}k^{\nu\rangle} - \frac{E_k^2}{3} \Delta^{\mu\nu} + E_k\, k^{\langle\mu\rangle} u^\nu \right) f_k^- 
    \int dK'\, \left(k'_{\langle\nu\rangle} + u_\nu E_{k'}\right) f_{k'}^+ \\
    - 2|q| \sigma_T^{+-} \int dK\, E_k^{-1} 
    \left( k^{\langle\mu}k^{\nu\rangle} - \frac{E_k^2}{3} \Delta^{\mu\nu} + E_k\, k^{\langle\mu\rangle} u^\nu \right) f_k^+ 
    \int dK'\, \left(k'_{\langle\nu\rangle} + u_\nu E_{k'}\right) f_{k'}^-
\end{multline}

Identifying the relevant moments of the distribution functions as hydrodynamic fields, we express the result compactly as:
\begin{equation}
    = 2|q| \sigma_T^{+-} 
    \left( \gamma^{\pi}_{-1} \left(\pi^{\mu\nu}_- V^+_\nu - \pi^{\mu\nu}_+ V^-_\nu \right) 
    + \frac{4}{3} \left(n^+ V^\mu_- - n^- V^\mu_+ \right) \right)
\end{equation}

In the linearized limit, expressing in terms of the net-charge and total particle diffusion currents:

\begin{equation}
    = \frac{4}{3}  \sigma_T^{+-} 
    \left( n_q\, V^\mu  
    -  n\,V_q^\mu  \right)
\end{equation}

\section{Collision term for $V^\mu$}

The single–particle contribution to the collision term, obtained from Eq.~\eqref{eq: Sngl_part_ coll_term}, can be expressed for $V^\mu$ as
\begin{equation}
    =-\frac{4}{9}\sigma_T \left(n V^\mu  + \frac{V_q}{|q|^2}n_q-\frac{\delta h^\mu}{4|q|T} n_q\right) 
\end{equation}

For the cross–species contribution, we follow the same steps as for $V_q^\mu$ and write the corresponding integral as
\begin{multline}
    |q| \int dK\, dK'\, dP\, dP'\, E_k^{-1} k^{\langle\mu\rangle} (2\pi)^5 s \sigma_T^{+-} \delta^4(p + p' - k - k') \left(f_p^+ f_{p'}^- - f_k^+ f_{k'}^- + f_p^+ f_{p'}^- - f_k^- f_{k'}^+\right) \\
    = |q| \int dK\, dK'\, dP\, dP'\, E_k^{-1} k^{\langle\mu\rangle} (2\pi)^5 s \sigma_T^{+-} \delta^4(p + p' - k - k') \left(2f_p^+ f_{p'}^- -f_k^- f_{k'}^+ - f_k^+ f_{k'}^- \right)
\end{multline}

This expression separates naturally into three integrals:
\begin{equation}
\begin{split}
    & = 2|q| \sigma_T^{+-}\int dK E_k^{-1} k^{\langle\mu\rangle}k_\nu s W^{+-}_{kk'\leftrightarrow pp'} f_{0k}^+ f_{0k'}^- \left(\varphi_p^+ + \varphi_p'^-\right)\\
    & \quad -2|q|\sigma_T^{+-} \int dK\, E_k^{-1} k^{\langle\mu\rangle} k^\nu f_k^- \int dK'\, k'_\nu f_{k'}^+ 
    - 2|q|\sigma_T^{+-} \int dK\, E_k^{-1} k^{\langle\mu\rangle} k^\nu f_k^+ \int dK'\, k'_\nu f_{k'}^-
\end{split} 
\end{equation}

In these expressions, we have used the linearized expansion of the single–particle distribution functions and the equilibrium identity
\[
f_{0k}^\pm f_{0k'}^\mp = f_{0p}^\pm f_{0p'}^\mp.
\]
Substituting the explicit forms of $\varphi$ from Eqs.~\eqref{eq: varphi} and \eqref{eq: varepsilons}, and evaluating the corresponding moment integrals analogous to those for the net charge diffusion four–current, we obtain the following compact form for the collision integral of the net charge four–current:

\begin{equation}
    -\frac{2}{9}\sigma_T^{+-}nV^\mu+ \frac{2}{9}\sigma_T^{+-} \frac{V^\mu_q}{|q|^2}n_q + \frac{8}{9} \sigma_T^{+-} \delta h^{\mu}\frac{n_q}{|q|T}
\end{equation}

\bibliographystyle{unsrtnat}
\bibliography{ref}

\end{document}